\documentclass[10pt]{article}

\usepackage[utf8]{inputenc}    
\usepackage[T1]{fontenc}       
\usepackage{lipsum}            
\usepackage[a4paper, top=0.75in, bottom=1in, left=0.75in, right=0.75in]{geometry}
\usepackage{xcolor}
\usepackage{graphicx}          
\usepackage{amsmath, amssymb}  
\usepackage{hyperref}          
\usepackage[numbers]{natbib}
\usepackage{authblk}
\usepackage[font=small,labelfont=bf,skip=4pt]{caption}
\usepackage{placeins}
\usepackage{hyperref}
\usepackage{subcaption}
\usepackage{bm}
\usepackage{booktabs}
\usepackage{lscape}
\usepackage{apalike}
\usepackage{amsmath}
\usepackage{amsfonts}
\usepackage{graphicx}
\usepackage{subcaption}
\usepackage{algorithm}
\usepackage{algpseudocode}
\usepackage{tabulary}
\usepackage{booktabs}
\usepackage{amssymb}
\usepackage{placeins}
\usepackage{adjustbox}  
\makeatletter
\renewcommand\@biblabel[1]{ #1.  }  
\makeatother

\title{Modeling roles and trade-offs in multiplex networks}
\author[a,*]{Nikolaos Nakis}
\author[b]{Sune Lehmann}
\author[a]{Nicholas A. Christakis}
\author[b]{Morten Mørup}

\affil[a]{Human Nature Lab, Yale University, 06511 New Haven, CT, USA}
\affil[b]{Department of Applied Mathematics and Computer Science, Technical University of Denmark, 2800 Lyngby, Denmark}
\affil[*]{Corresponding author e-mail: nicolaos.nakis@gmail.com}

\begin{document}

\maketitle

\begin{abstract}
A multiplex social network captures multiple types of social relations among the same set of people, with each layer representing a distinct type of relationship. Understanding the structure of such systems allows us to identify how social exchanges may be driven by a person's own attributes and actions (independence), the status or resources of others (dependence), and mutual influence between entities (interdependence). Characterizing structure in multiplex networks is challenging, as the distinct layers can reflect different yet complementary roles, with interdependence emerging across multiple scales.
Here, we introduce the Multiplex Latent Trade-off Model (\textsc{MLT}), a framework for extracting roles in multiplex social networks that accounts for independence, dependence, and interdependence. \textsc{MLT} defines roles as trade-offs, requiring each node to distribute its source and target roles across layers while simultaneously distributing community memberships within hierarchical, multi-scale structures.
Applying the \textsc{MLT} approach to 176 real-world multiplex networks, composed of social, health, and economic layers, from villages in western Honduras, we see core social exchange principles emerging, while also revealing local, layer-specific, and multi-scale communities. Link prediction analyses reveal that modeling interdependence yields the greatest performance gains in the social layer, with subtler effects in health and economic layers. This suggests that social ties are structurally embedded, whereas health and economic ties are primarily shaped by individual status and behavioral engagement. Our findings offer new insights into the structure of human social systems.
\end{abstract}

Multiplex networks \citep{jalili2017link, atkisson2020understanding, li2015multiple,ferriani2013social, bargigli2015multiplex,didier2015identifying, bennett2015detection, valdeolivas2019random,halu2019multiplex, zhao2014immunization, montes2020identification,Kivela_2014,de_dom} extend classical networks by incorporating multiple types of relationships between entities, each represented as a distinct layer. This multi-layered perspective allows more nuanced insights into complex systems, and especially social networks. Studying the multiplexity of human networks can enhance our understanding of the structure of human interactions, collective action, and diverse social phenomena. 

Social life is structured by complex exchanges across multiple domains, where relational dynamics are often characterized by three fundamental modes of social exchange: independence, dependence, and interdependence. In the context of directed tie formation in multiplex networks, we refer to \textit{independence} as occurring when a person initiates connections primarily based on their own intrinsic attributes, resources, or motivations, regardless of the recipient’s characteristics or position. \textit{Dependence} implies that the target’s attributes, such as status, resources, or attractiveness, significantly influence the likelihood of receiving a tie, making the sender reliant on the recipient. \textit{Interdependence}, in turn, captures the complex, mutual influence between individuals, extending beyond sender and receiver attributes alone.

The definitions of the three modes of social exchange are inspired by ideas from Social Exchange Theory \citep{Emerson1962PowerDependenceR, Blau1964ExchangeAP, Kelley1978InterpersonalRA, cook, homans}. While we maintain the classical definitions of dependence and interdependence from Social Exchange Theory, our definition of \textit{independence} diverges slightly from its traditional meaning, defined as the absence of relational influence \citep{Emerson1962PowerDependenceR, Kelley1978InterpersonalRA}, by instead interpreting it as sender-driven, unilateral tie formation. This aligns with an effort-based view \cite{cropanzano2005social}, which we adopt to better reflect actor-level agency in tie initiation by explicitly modeling how actors’ intrinsic (sender-focused), extrinsic (receiver-focused), and relational (dyadic) attributes interact in multiplex contexts. Although many studies have modeled various aspects of these exchange dynamics \citep{set_future}, the literature has not fully addressed how independence, dependence, and interdependence manifest within multiplex networks mirrored by human behavior.

Roles in network science were first formalized through the concept of structural equivalence, introduced by White, Boorman, and Breiger \cite{white_roles}, who defined a role as a set of actors exhibiting identical patterns of ties to all other actors across multiple relations. Holland, Laskey, and Leinhardt \cite{HOLLAND1983109} extended this idea by introducing the stochastic blockmodel (SBM), a probabilistic generalization in which actors within the same block are stochastically equivalent, that is, they share the same distribution over tie probabilities, thus embedding structural roles within a statistical framework. These notions of roles connect to Social Exchange Theory, which conceptualizes social roles as patterned expectations that structure obligations, reciprocity norms, and behavioral regularities in repeated interactions \cite{Blau1964ExchangeAP, Emerson1962PowerDependenceR}. From this perspective, roles can be seen as the structural expressions of exchange dynamics that emerge from recurring patterns of interaction.

Individuals generally can maintain only a limited number of stable relationships throughout their lives, with varying costs depending on the connection type (e.g., social or economic) \cite{cropanzano2005social, cook2013social}. Early attempts to quantify this limit focused on the correlation between neocortex volume and group size in primates \cite{dunbar1992neocortex,dunbar1993coevolution}, whereas later work has highlighted non‑cognitive constraints, most notably the limited time people can devote to maintaining ties \cite{MIRITELLO201389,time2}. This limitation creates trade-offs that also show up in multiplex networks: being active in one type of relationship often comes at the cost of reduced activity in another, reflecting how people prioritize across exchange domains.
These constraints on social exchange, raise important questions: 
\begin{enumerate}

\item \emph{How do individuals allocate their relational budget across different types of relationships, and what are the layer-specific structures of interdependence?}
\item \emph{To what extent are the observed social exchanges driven solely by the parties' own efforts, attributes, status, or resources (i.e., independence and dependence) as opposed to mutual interdependence?}
\end{enumerate}

Although numerous models have been developed to study multiplex networks, they are unable to explicitly account for  key aspects of human behavior in terms of roles, trade-offs and exchange shaped by dependence, independence and interdependence. Existing approaches, such as ensemble and growth models \cite{battiston2017new}, centrality measures \cite{rahmede2018centralities, sola2013eigenvector}, diffusion models \cite{gomez2013diffusion}, embedding techniques \cite{embedding1, embedding2, embedding3}, and deep learning methods \cite{deep1, deep2, deep3}, primarily focus on structural or functional properties of networks. None of these approaches have been designed to understand the trade-offs inherent in multiplexity, particularly in the context of human behavior. 


Despite the extensive use of Archetypal Analysis for modeling trade-offs in terms of convex combinations of distinct archetypal characteristics in data, its applications \cite{alcacer2025surveyarchetypalanalysis} have primarily been focused on the modeling of biological systems. These studies have demonstrated Pareto optimality in phenotype space \cite{shoval2012evolutionary} and optimized circuit efficiency \cite{szekely2013evolutionary}. Although recent efforts have applied Archetypal Analysis to relational data, such as exploring polarization in social networks \cite{nakis2023characterizing, nakis2024signed}, these approaches remain inadequate for studying the complexities of multiplex networks. Hence, we build on Archetypal Analysis to handle multiplex networks, addressing a key limitation in modeling trade-offs across multiple layers of social interaction, bridging archetypes and the classical notion of roles in network science.

Many social networks exhibit structure at multiple scales, motivating hierarchical representations of network topology \cite{Clauset_2008,hierarchicalrepr,hbdm,herlau}. In the context of multiplex networks, hierarchical stochastic block models have been proposed to capture both intra-layer structure and inter-layer dependencies \cite{pei_hier,Stanley_2016,hmulti}. While these models offer principled approaches for uncovering latent groupings and multiscale regularities, they often overlook role-based trade-offs and patterns of social exchange that shape real-world social systems. We address this limitation by introducing a framework that integrates hierarchical representations with role dynamics and trade-off structures across layers.

An important problem in graph representation learning and social networks is the analysis and interpretation of the structural properties of such complex systems. Link prediction is a powerful quantitative measure for capturing such properties, because its ability to accurately predict missing or future links allows us to infer the presence of meaningful structural patterns such as homophily, transitivity, and clustering. Foundational work on link prediction \citep{LP1, LP2, LP3} has shown that topological features and community-based information can help us uncover latent relational structures within networks. Consequently, link prediction performance can serve as an indirect proxy for the extent to which a graph is characterized by regularities or community structure, making it an informative experimental setup for understanding the architecture of real-world graphs and their underlying properties.

Here, we use data from 176 multiplex social networks of isolated villages in the western part of Honduras and demonstrate that our model robustly captures trade-offs and core principles of social exchange, while facilitating the extraction of local, layer-specific, and multi-scale structures. Using link prediction, we find that interdependence is structurally embedded in the social layer, whereas, in health and economic layers, social exchange depends on behavioral roles and levels of engagement. As such, the social layer exhibits the greatest benefit from modeling interdependence, while this effect is much more subtle in the health and economic domains. This suggests that instrumental (health and economic) ties are primarily driven by node-level status and patterns of dependence and independence.

\begin{figure}[h!]

\centering
\begin{subfigure}[t]{1\textwidth}
    \includegraphics[width=1\textwidth]{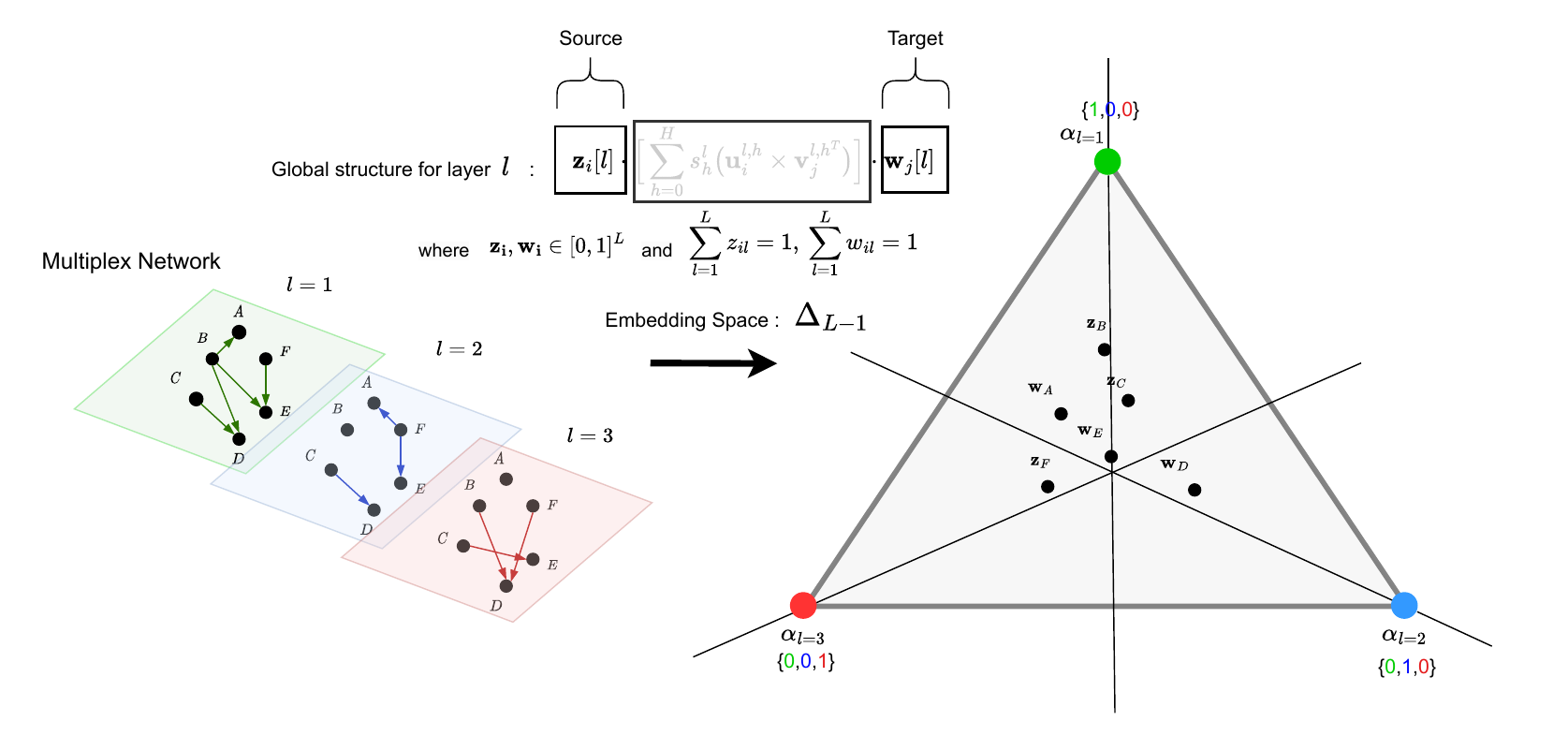}
    \caption{\textbf{Global Trade-Off Structure}: Given a multiplex network as input, our model projects it onto the $(L-1)$-Simplex $\bm{\Delta}_{L-1}$, where $L$ is the number of layers in the network. Each corner of the simplex corresponds to the distinct profile of a specific layer. By construction, a relatively high interdependence activity (high number of ties) in a particular layer pushes a node closer to the corresponding corner, resulting in a role characterization of the node.}
\end{subfigure}
\hfill
\begin{subfigure}[t]{1\textwidth}
    \includegraphics[width=1\textwidth]{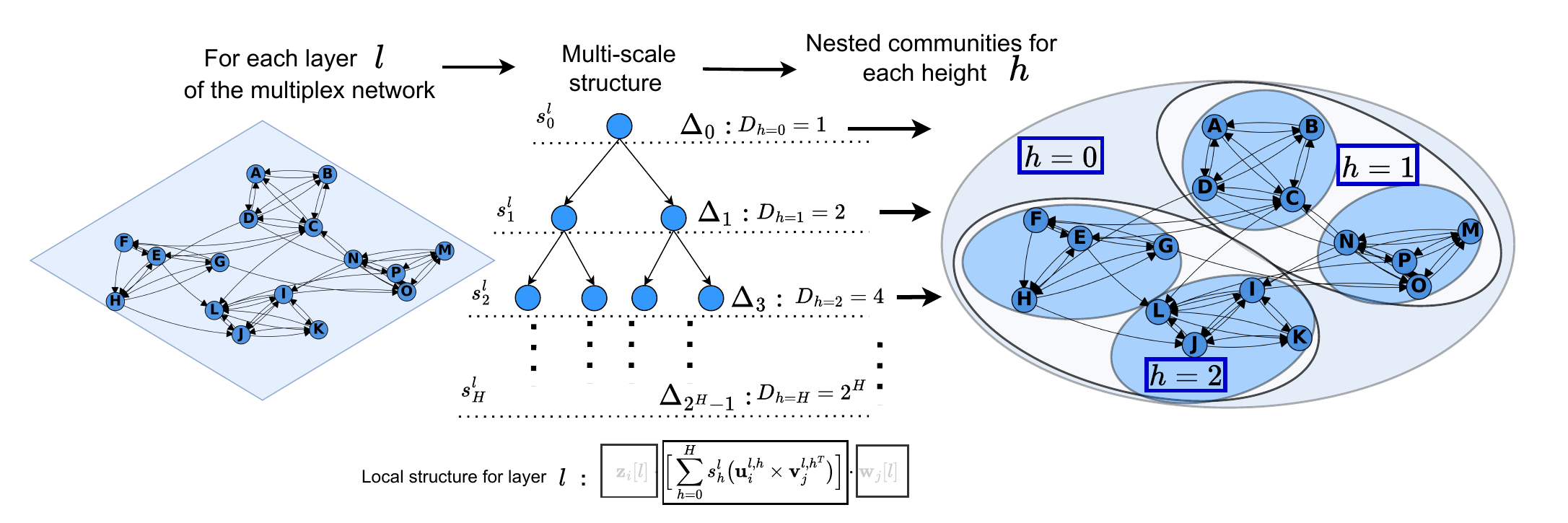}
    \caption{\textbf{Local Multi-scale Community Structure.} For a given layer of the multiplex network, the model learns a multi-scale representation of its structure. For each level $h$ of the hierarchy ($H$ total layers), the layer’s structure is projected onto a $(D_{h}-1)$-simplex $\bm{\Delta}_{D_h-1}$, where $D_h$ denotes the number of latent dimensions at level $h$. Each vertex of the simplex corresponds to a latent community, enabling the characterization of layer-specific and hierarchical community structure. The contribution of each level $h$ to the overall link formation is weighted by the layer-specific strength $s_{h}^l$. This projection facilitates the identification of local communities by capturing relationships between nodes in the context of their interactions within that specific layer at different levels of resolution.}
\end{subfigure}

\caption{\textbf{Model Framework.} This figure illustrates the MLT modeling approach of  multiplex networks accounting for both global trade-offs defining roles across layers and local layer-specific trade-offs defining multi-scale hierarchical community structures. Panel (a) illustrates the global trade-off structure, capturing the layer-specific representation of node behavior across the network via the embeddings $\bm{z}_i$ and $\bm{w}_j$—i.e., the outer components of the interdependence term (not tinted). Panel (b) shows the local multi-scale community structure, identifying latent communities within each individual layer at multiple resolutions as encoded in $\bm{u}_{i}^{l,h}$ and  $\bm{v}_{j}^{l,h}$ embeddings—i.e., the inner components of the interdependence term (not tinted).}
\label{fig:model_overview}
\end{figure}

We now introduce the Multiplex Latent Trade-off Model (\textsc{MLT}) (for a more detailed description see Materials and Methods \ref{sec:materials}). Using a Bernoulli likelihood, the model is expressed via the log-odds $r^l_{ij}$ of observing a link between entity $i$ and $j$ for relational layer $l$ as: 
\begin{equation}
r^l_{ij} = \beta_i^l + \gamma_j^l +  \mathbf{z}_i[l] \cdot \Big[\sum_{h=0}^{H}s^l_h\big(\bm{u}_i^{l,h} \times \bm{v}_j^{l,h^T}\big)\Big] \cdot \mathbf{w}_j[l],
\end{equation}
where $\mathbf{z}_i$ and $\mathbf{w}_i$ are embeddings on the $(L-1)$-simplex (with $L$ layers) that capture the multiplexity and archetypal roles of nodes as sources and targets across the layers; comparatively high interdependent activity in a layer places a node towards the simplex corner representing this layer (see Figure \ref{fig:model_overview} (a)). In such a mixed-membership formulation across layer-specific roles, stochastic equivalence generalizes the classical notion \cite{white_roles, HOLLAND1983109} by capturing similarity in role-based interaction patterns. Nodes are stochastically equivalent to the extent that they share similar role proportions. For pure-role nodes with hard assignments at the simplex corners, this reduces to classical stochastic equivalence, apart from node-specific degree effects which in turn relates to the degree corrected SBM \cite{SBM_newman}.

Similarly, $\bm{u}_i^{l,h}$ and $\bm{v}_i^{l,h}$ are the layer-specific multi-scale embeddings of the $l$'th layer at the $h$'th level of the multi-scale structure/hierarchy which lie on the $(D_{h}-1)$-simplex (with $D_{h}=2^h$ defining the number of communities at the h'th level of the hierarchy), and describe participation in latent groups for each hierarchical level (see Figure \ref{fig:model_overview} (b)) in which each split along the hierarchy redistributes the membership of each node at the previous level of the hierarchy to the two subcommunities formed by the split. In addition, the parameter $s^l_h \geq 0$ describes the strength with which links are generated at the $h$-th level of the hierarchy. We reparameterize this as $s^l_h = s \cdot \pi^l_h$, where $\sum_h \pi^l_h = 1$ and $\pi^l_h \geq 0$. Here, $s \geq 0$ is a global strength parameter that controls the overall influence of the hierarchical structure across the model, while the $\pi^l_h$ values determine how this strength is distributed across hierarchy levels within each layer. This reparameterization shares a single global strength $s$ across all layers, constraining each layer to have the same overall capacity for hierarchical expression. It enables trade-offs in how different layers allocate representational emphasis across hierarchy levels while maintaining a globally consistent capacity. Finally, the parameters $\beta_i^l$ and $\gamma_j^l$ account for node- and layer-specific node status and degree heterogeneity. 

In this sense, our model captures trade-offs across layers: relatively high values of interdependent node activity (i.e., ties) in a specific layer pushes the node toward the corner of the $(L-1)$-simplex corresponding to that layer. Additionally, the model accounts for local, layer-specific structure by imposing hierarchically structured layer-specific simplices, allowing for the discovery of local latent and multi-scale patterns within each layer independently. 

Under the lens of social exchange, and considering a directed tie from node $i\rightarrow j$, we interpret the terms in the \textsc{MLT} model as follows: \textit{Independence} is represented by the sender bias $\beta_i^l$. A higher value of $\beta_i^l$ indicates that node $i$ initiates connections primarily based on its intrinsic attributes, resources, or incentives, independent of node $j$'s characteristics or position. \textit{Dependence} corresponds to the receiver bias $\gamma_j^l$. A higher $\gamma_j^l$ implies that node $j$'s characteristics, status, or resources significantly influence the likelihood of receiving ties from node $i$, making node $i$ dependent on node $j$'s attractiveness or receptivity. \textit{Interdependence} is encapsulated by the dyadic interaction term:
    \begin{equation}
    \eta_{ij}^l = \mathbf{z}_i[l] \cdot \left[\sum_{h=0}^{H}s_h^l\left(\bm{u}_i^{l,h}\times\bm{v}_j^{l,h^T}\right)\right]\cdot\mathbf{w}_j[l].
    \end{equation}
    This term captures the complex and mutual influence between nodes $i$ and $j$, extending beyond sender and receiver attributes alone. Importantly, interdependence as defined here differs from reciprocity, as generally $\eta_{ij}\neq\eta_{ji}$, explicitly capturing directional asymmetries and highlighting mutual but asymmetric contributions of both actors toward tie formation.

Within the framework of social exchange, ties may emerge either from individual-level factors, such as a person's ability to initiate or attract ties, or from mutual, role-based structure. To isolate the influence of individual status alone, we introduce a simplified variant, the \textit{bias model} ($r^l_{ij} = \beta_i^l + \gamma_j^l$), which includes only sender (independence) and receiver (dependence) effects. This model formalizes the idea of unidirectional, status-driven exchange and serves as a conceptual baseline to the full \textsc{MLT} model, which helps us quantify the effects and nature of interdependence through the dyadic relational structure. Overall, the \textit{bias model} represents a null assumption in which network structure can be predicted solely from individual node-level tendencies, such as activity or popularity, without accounting for dyadic or relational dependencies.

\section*{Results}
We apply our model on a sociocentric network study of $22,584$ people in $176$ geographically isolated villages in western Honduras. We start by presenting the results corresponding to the proposed \textsc{MLT} model, focusing on its ability to capture trade-offs and characterizing social exchange in real social multiplex networks. Each network contains three layers, describing the social, health, and economic ties (see Data \ref{sec:real_nets}). Specifically, we highlight how the \textsc{MLT} is able to characterize trade-offs through the global structure layer embeddings, $\mathbf{z}_i$ and $\mathbf{w}_i$; and how observed patterns can reflect underlying social exchange through the design of modeling terms specific to each village network. Additionally, we demonstrate that the model is capable of capturing layer-specific local multi-scale structures of the multiplex network, by analyzing the local embeddings, $\mathbf{u}_i^l$ and $\mathbf{v}_j^l$, for each network. 


\begin{figure}[!htbp]

\centering
  \centering
  \adjustbox{valign=b}{%
    \begin{minipage}[t]{0.62\textwidth}
      \begin{subfigure}[t]{\linewidth}
        \includegraphics[width=1\linewidth]{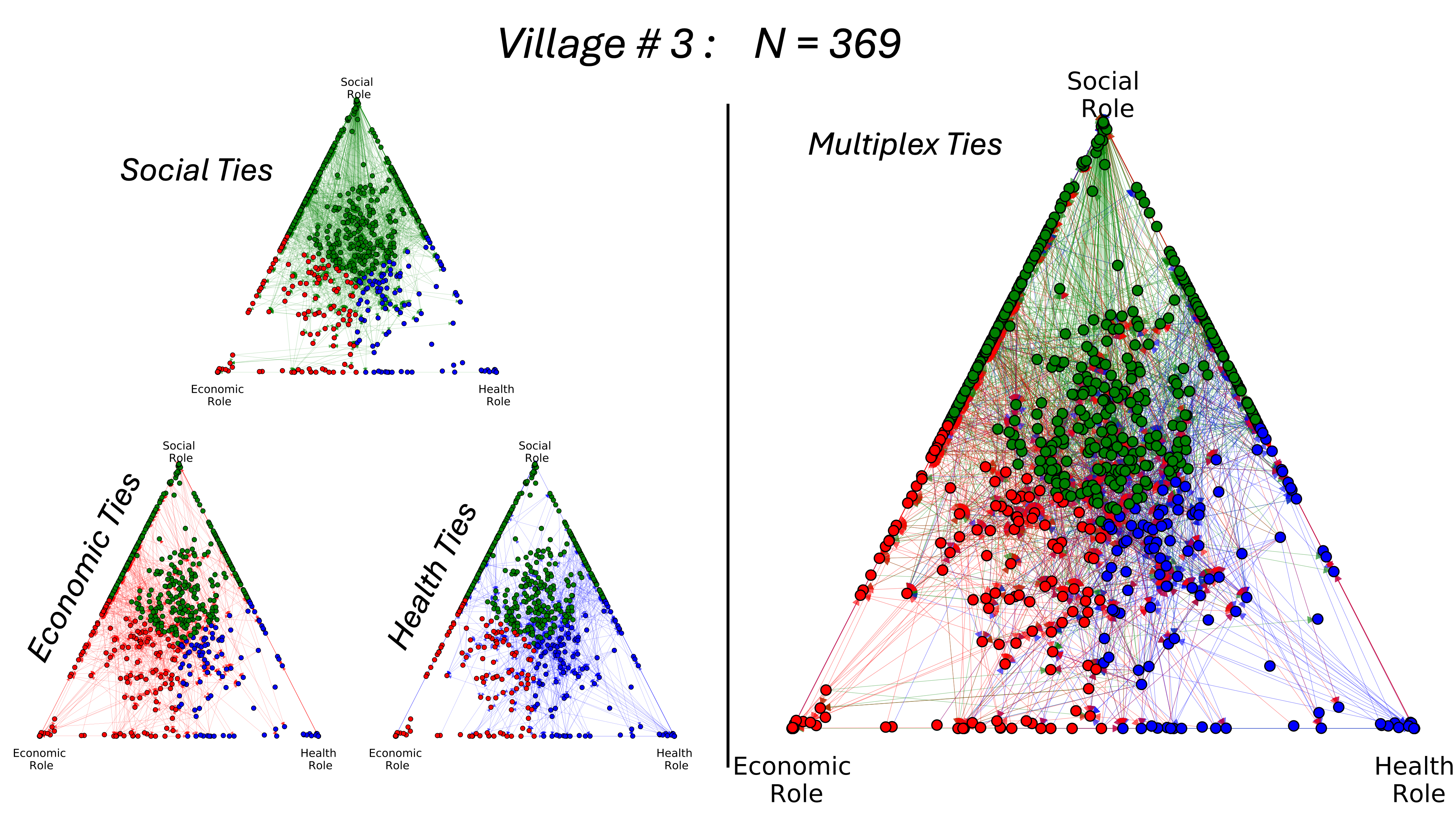}
        \caption{\small Role simplex visualizations for village $\# 3$. Each point represents a node, colored by dominant role. Left: tie-specific roles. Right: multiplex network with all ties overlaid. }
      \end{subfigure}
      \vspace{1em}

      \begin{subfigure}[t]{\linewidth}
        \includegraphics[width=1\linewidth]{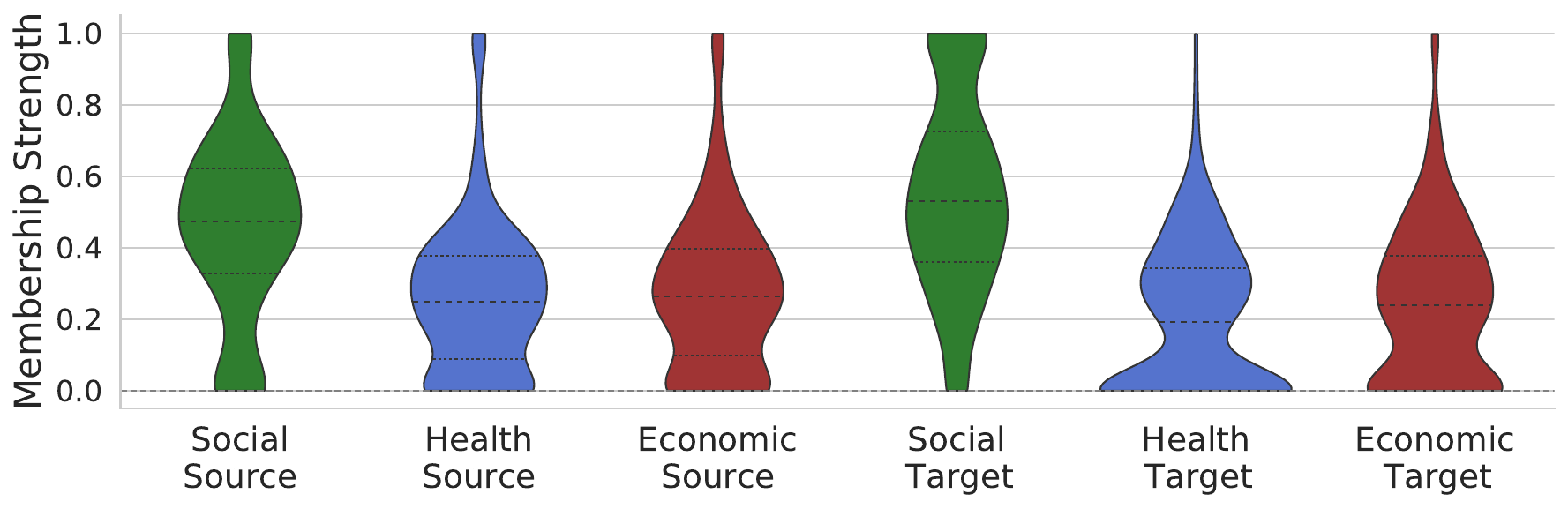}
        \caption{\small Violin plots of node-role membership strengths for source and target positions across domains. }
      \end{subfigure}
    \end{minipage}
  }\hfill
  \adjustbox{valign=b}{%
    \begin{minipage}[t]{0.36\textwidth}

\makebox[1\linewidth]{\centering \textbf{Multi-scale Layer Structure:}}
\\[1ex]
 \makebox[0.32\linewidth]{\centering \textbf{Social}} \hfill
  \makebox[0.32\linewidth]{\centering \textbf{Health}} \hfill
  \makebox[0.32\linewidth]{\centering \textbf{Economic}} \\[-1ex]
    
\centering

  \begin{minipage}[c]{\linewidth}
    \begin{minipage}[c]{0.04\linewidth}
      \centering
      \rotatebox{90}{\small \textbf{$D_{h=1}=2$}}
    \end{minipage}%
    \hfill
     \captionsetup[subfigure]{skip=-5pt,justification=centering}
    \begin{subfigure}[c]{0.28\linewidth}
\includegraphics[width=\linewidth]{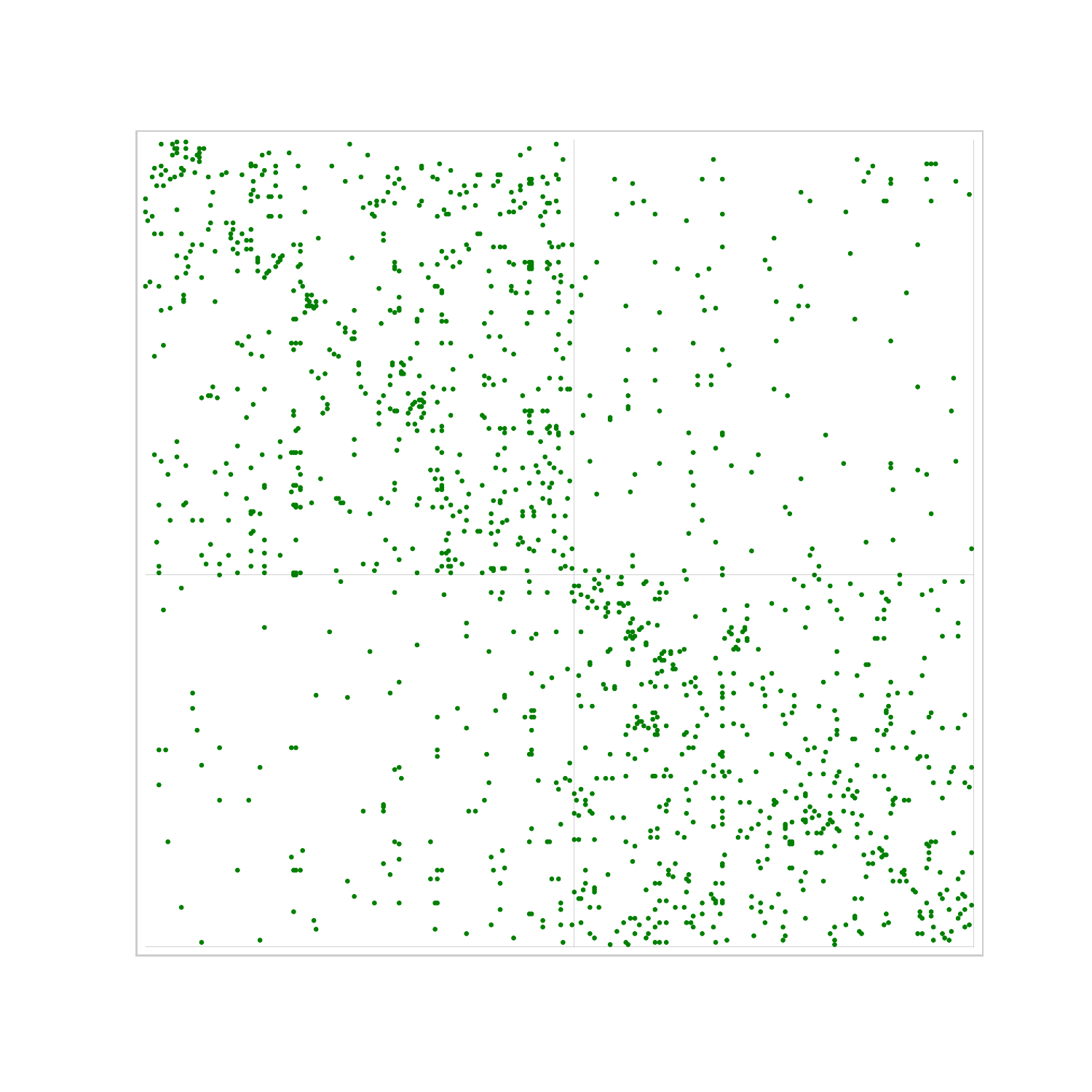} \caption{\tiny\\$s_{h=1}^{l=1}=0.001$}
    \end{subfigure}\hfill
     \captionsetup[subfigure]{skip=-5pt,justification=centering}
    \begin{subfigure}[c]{0.28\linewidth}
\includegraphics[width=\linewidth]{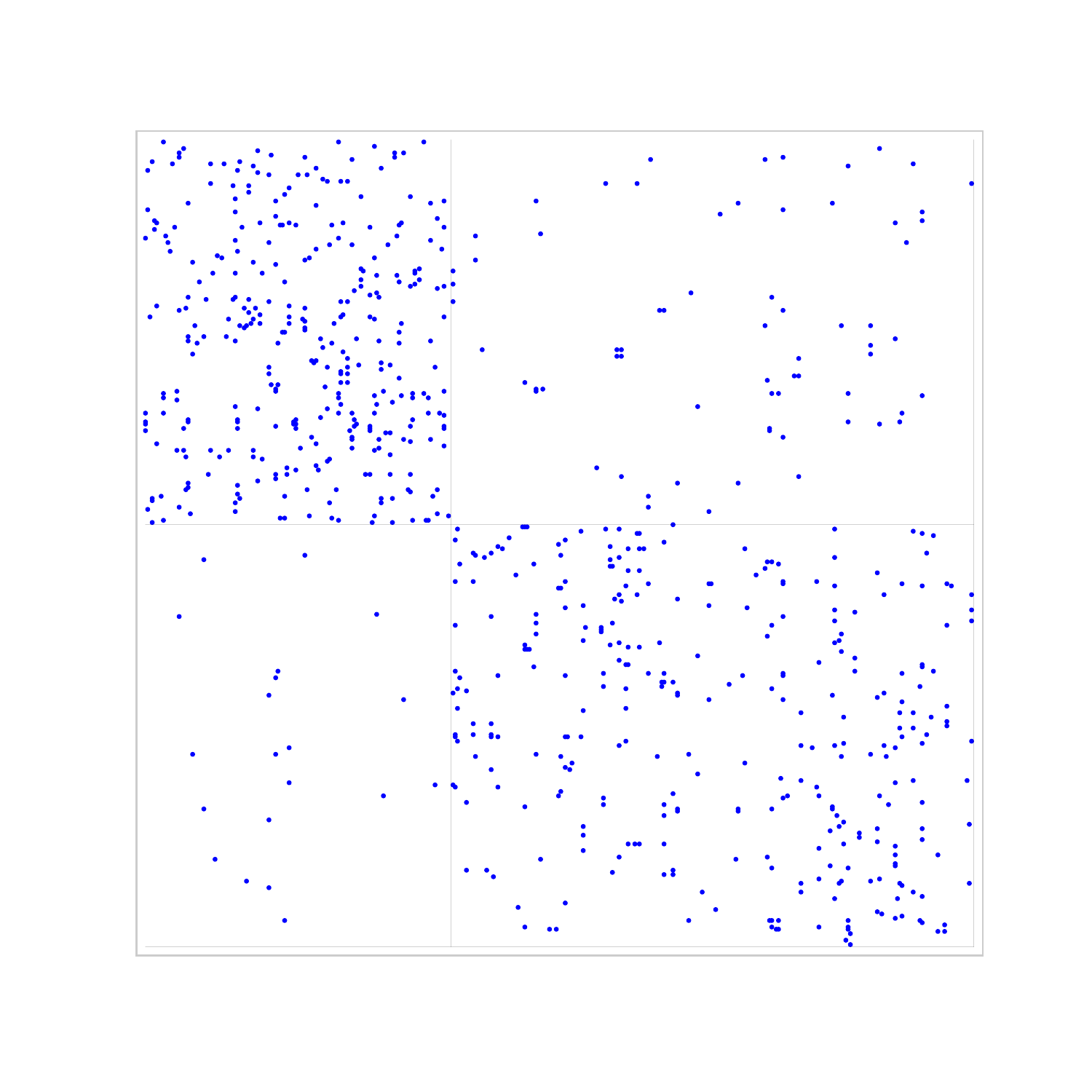} \caption{\tiny\\$s_{h=1}^{l=2}=0.003$}
    \end{subfigure}\hfill
    \captionsetup[subfigure]{skip=-5pt,justification=centering}
    \begin{subfigure}[c]{0.28\linewidth}
\includegraphics[width=\linewidth]{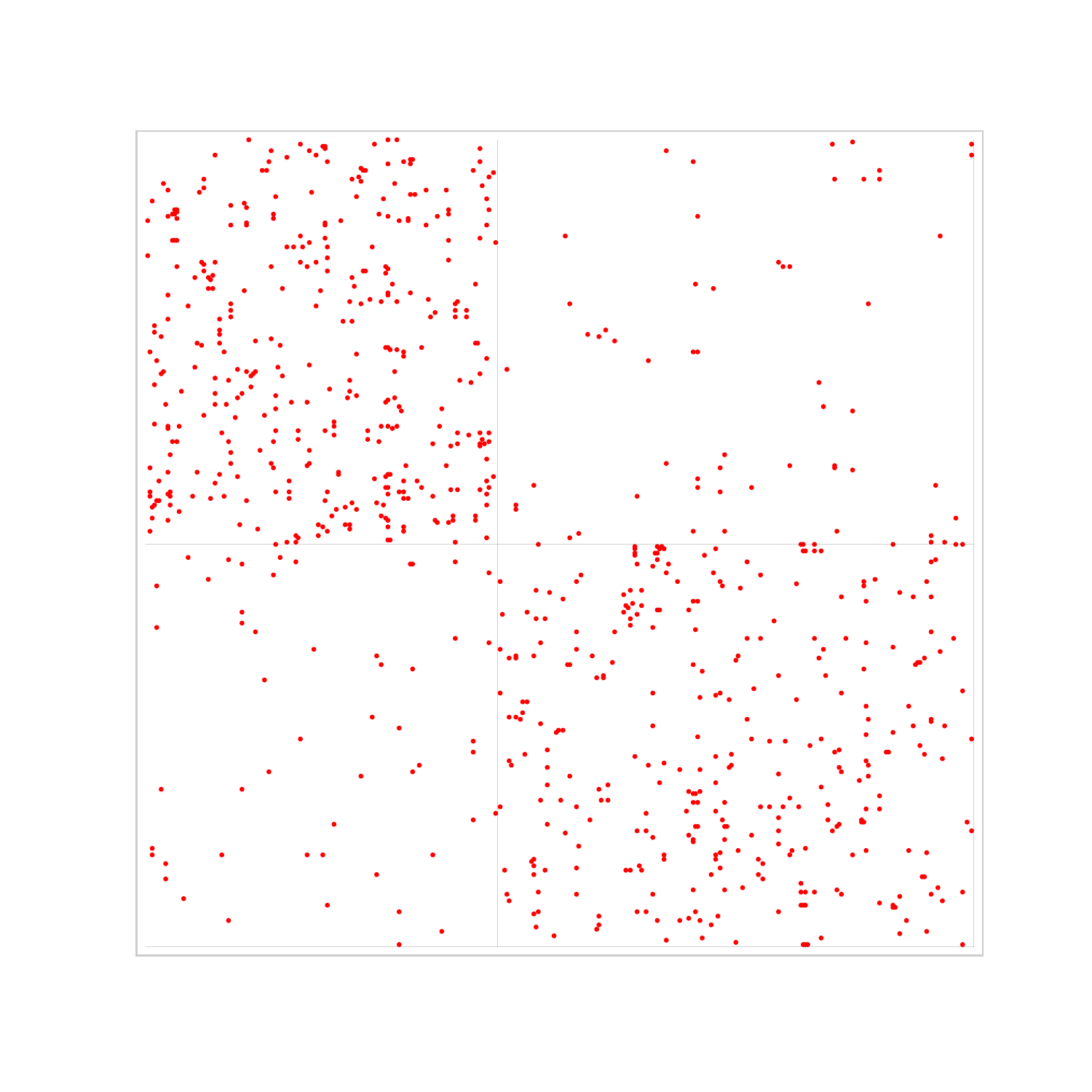}
      \caption{\tiny\\$s_{h=1}^{l=3}=0.004$}
    \end{subfigure}
  \end{minipage}

  \begin{minipage}[c]{\linewidth}
    \begin{minipage}[c]{0.04\linewidth}
      \centering
      \rotatebox{90}{\small \textbf{$D_{h=2}=4$}}
    \end{minipage}%
    \hfill
     \captionsetup[subfigure]{skip=-5pt,justification=centering}
    \begin{subfigure}[c]{0.28\linewidth}
      \includegraphics[width=\linewidth]{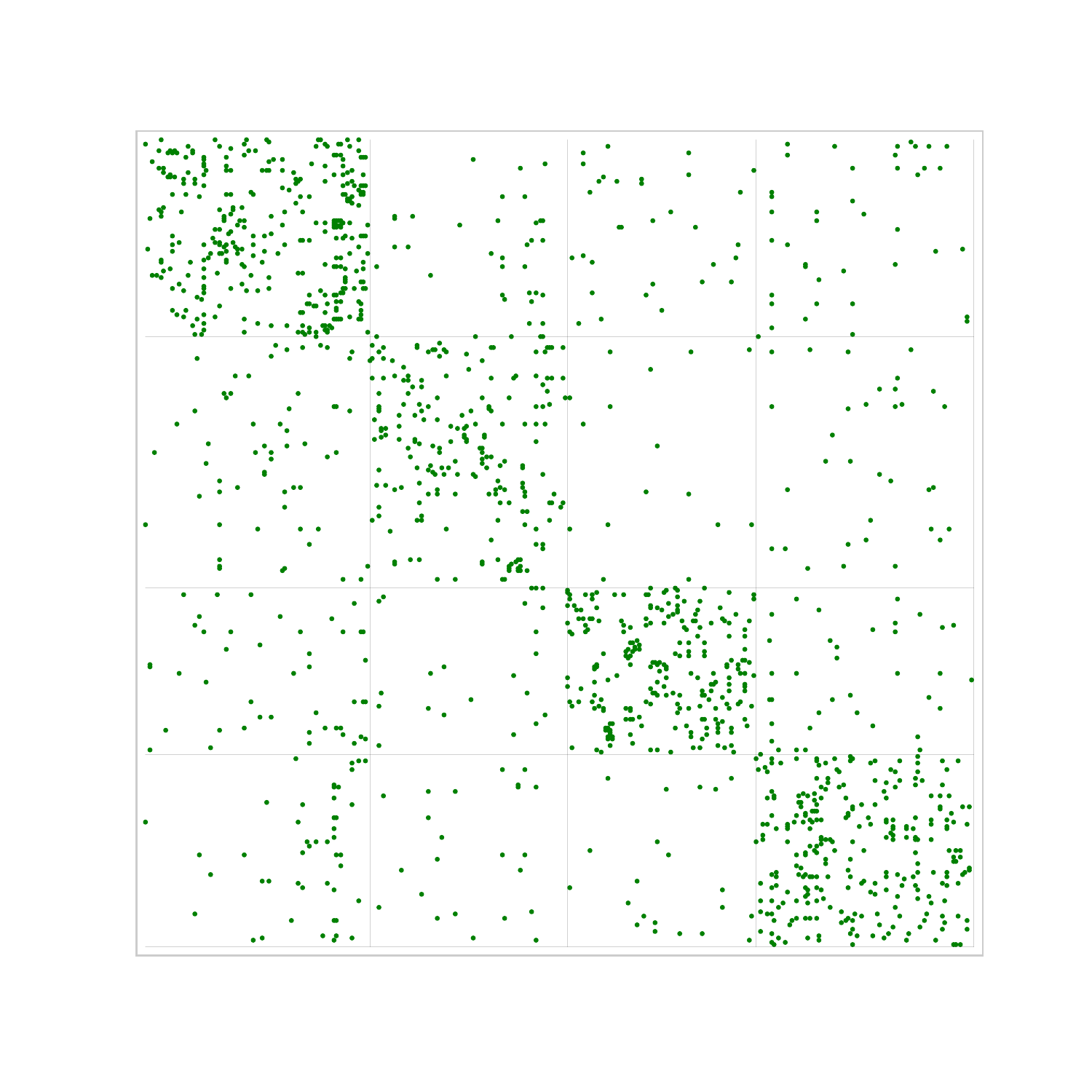}
       \caption{\tiny\\$s_{h=2}^{l=1}=0.013$}
    \end{subfigure}\hfill
     \captionsetup[subfigure]{skip=-5pt,justification=centering}
    \begin{subfigure}[c]{0.28\linewidth}
      \includegraphics[width=\linewidth]{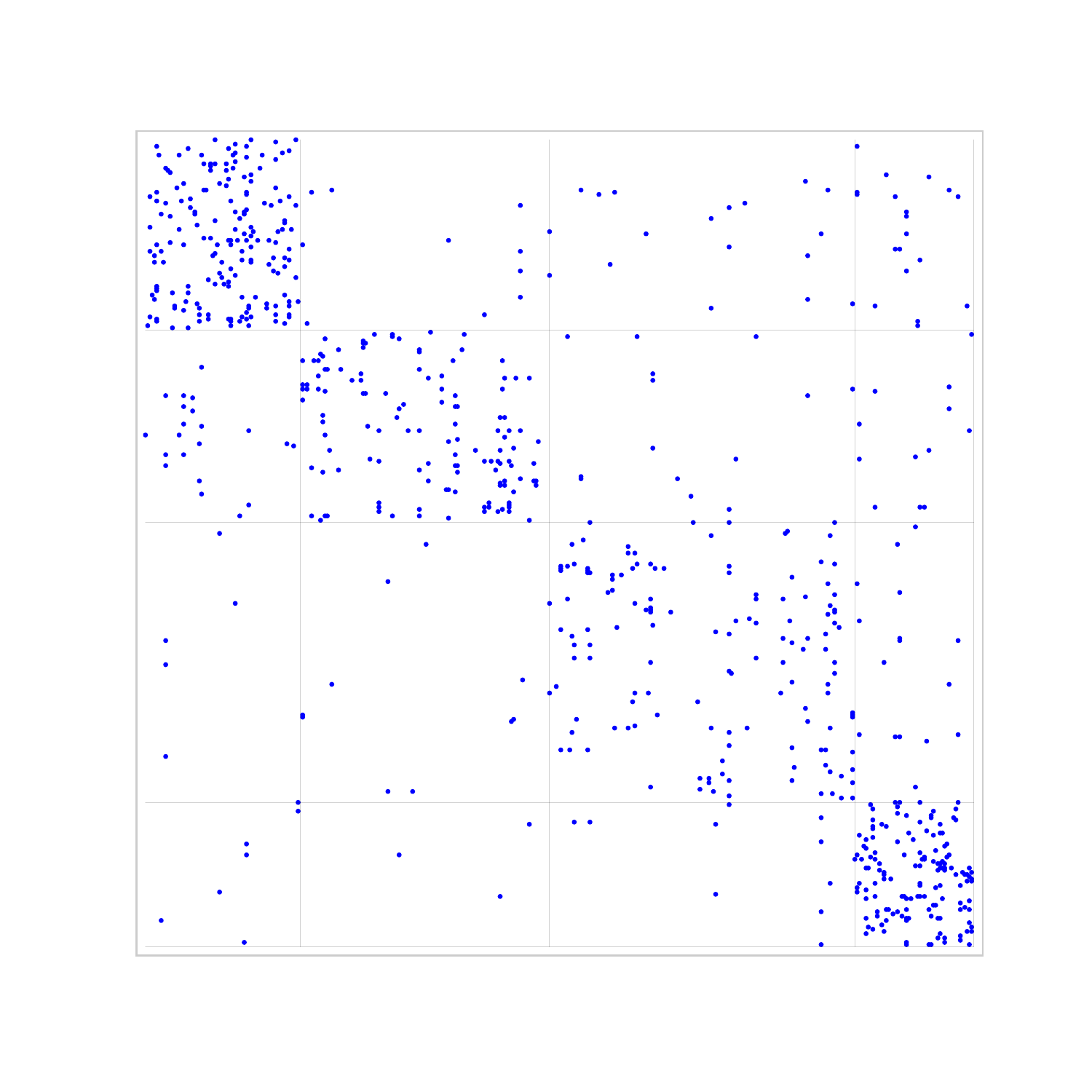}
       \caption{\tiny\\$s_{h=2}^{l=2}=0.028$}
    \end{subfigure}\hfill
     \captionsetup[subfigure]{skip=-5pt,justification=centering}
    \begin{subfigure}[c]{0.28\linewidth}
      \includegraphics[width=\linewidth]{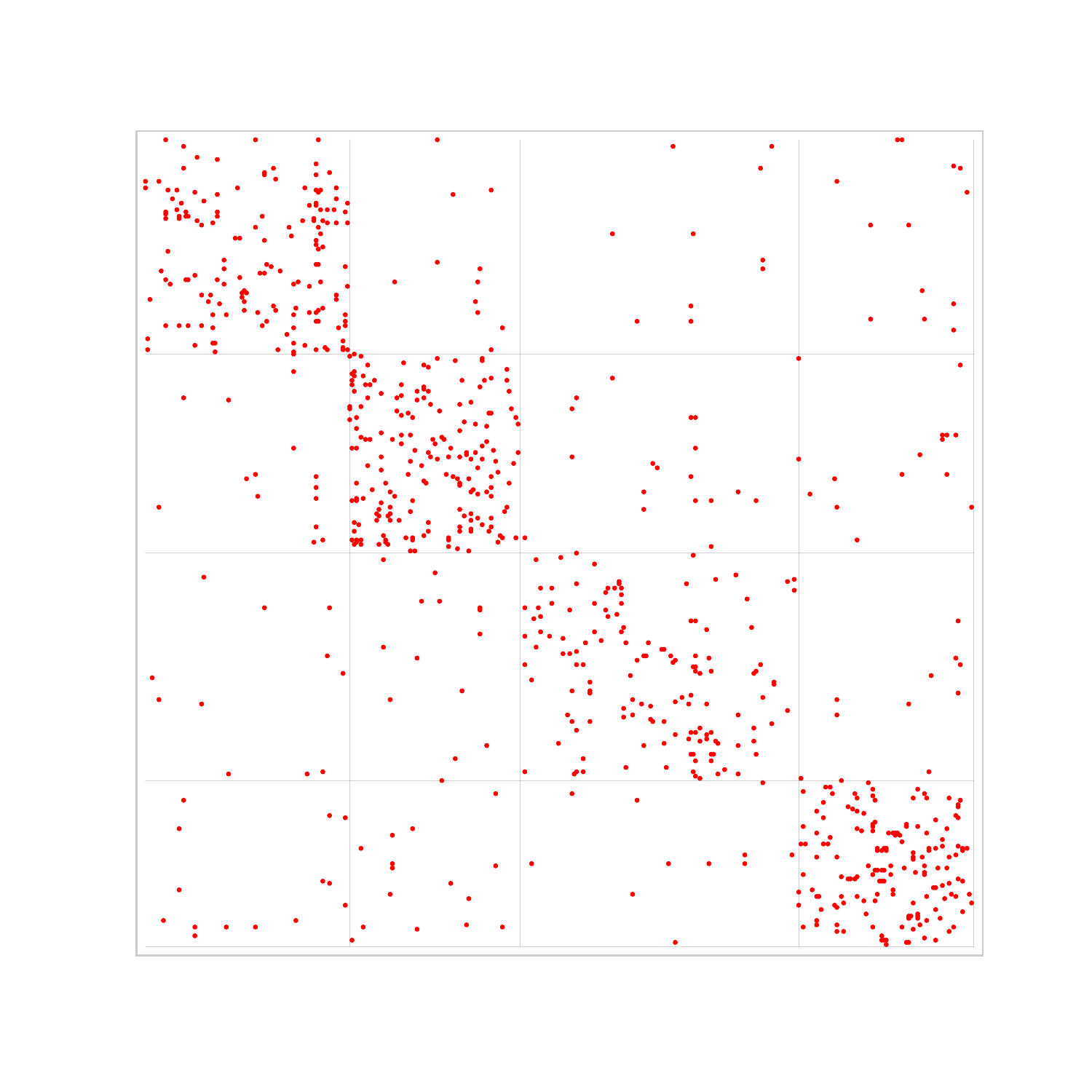}
       \caption{\tiny\\$s_{h=2}^{l=3}=0.073$}
    \end{subfigure}
  \end{minipage}
  
  \begin{minipage}[c]{\linewidth}
    \begin{minipage}[c]{0.04\linewidth}
      \centering
      \rotatebox{90}{\small \textbf{$D_{h=3}=8$}}
    \end{minipage}%
    \hfill
     \captionsetup[subfigure]{skip=-5pt,justification=centering}
    \begin{subfigure}[c]{0.28\linewidth}
      \includegraphics[width=\linewidth]{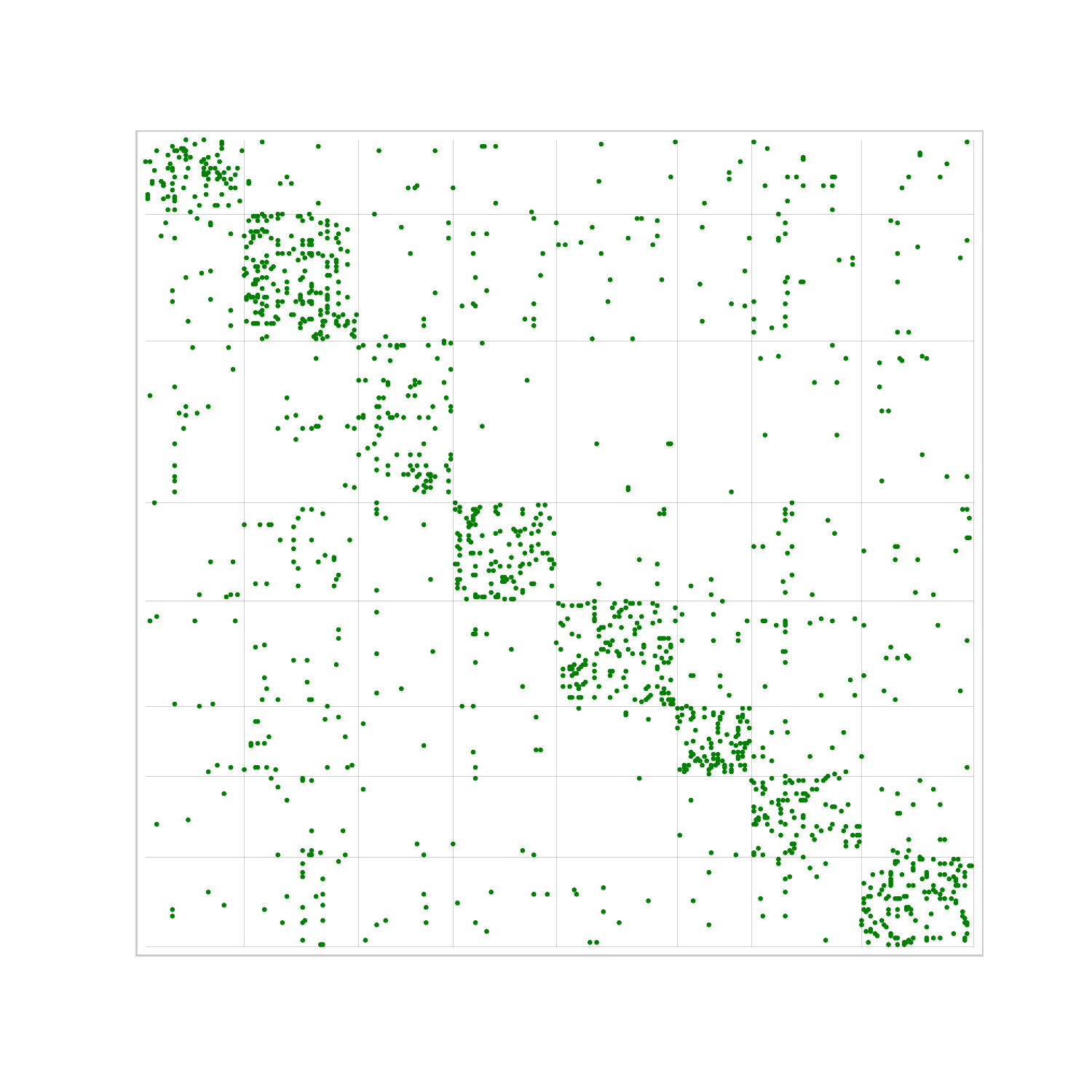}
       \caption{\tiny\\$s_{h=3}^{l=1}=0.497$}
    \end{subfigure}\hfill
     \captionsetup[subfigure]{skip=-5pt,justification=centering}
    \begin{subfigure}[c]{0.28\linewidth}
      \includegraphics[width=\linewidth]{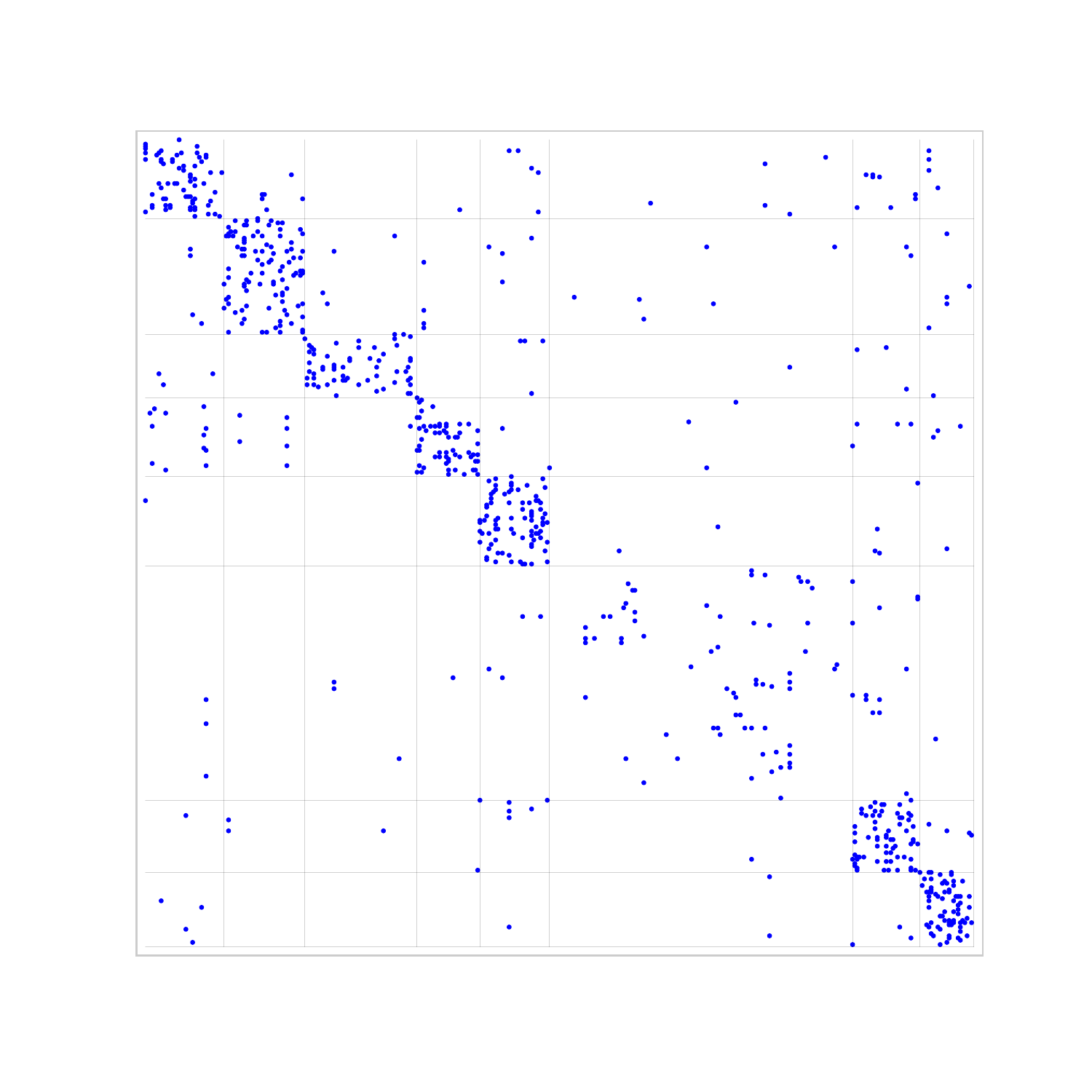}
       \caption{\tiny\\$s_{h=3}^{l=2}=6.212$}
    \end{subfigure}\hfill
     \captionsetup[subfigure]{skip=-5pt,justification=centering}
    \begin{subfigure}[c]{0.28\linewidth}
      \includegraphics[width=\linewidth]{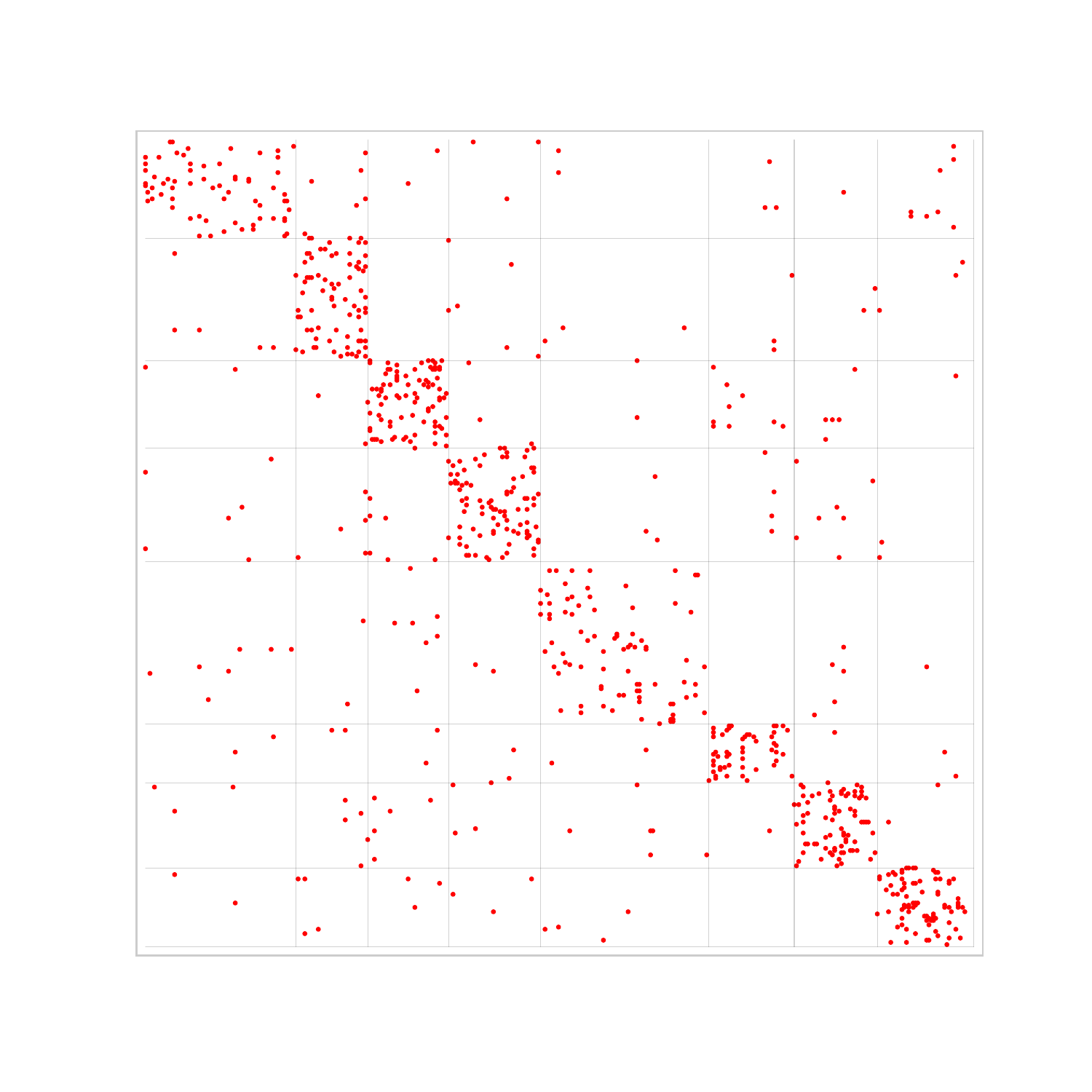}
       \caption{\tiny\\$s_{h=3}^{l=3}=4.893$}
    \end{subfigure}
  \end{minipage}
  
  \begin{minipage}[c]{\linewidth}
    \begin{minipage}[c]{0.04\linewidth}
      \centering
      \rotatebox{90}{\small \textbf{$D_{h=4}=16$}}
    \end{minipage}%
    \hfill \captionsetup[subfigure]{skip=-5pt,justification=centering}
    \begin{subfigure}[c]{0.28\linewidth}
      \includegraphics[width=\linewidth]{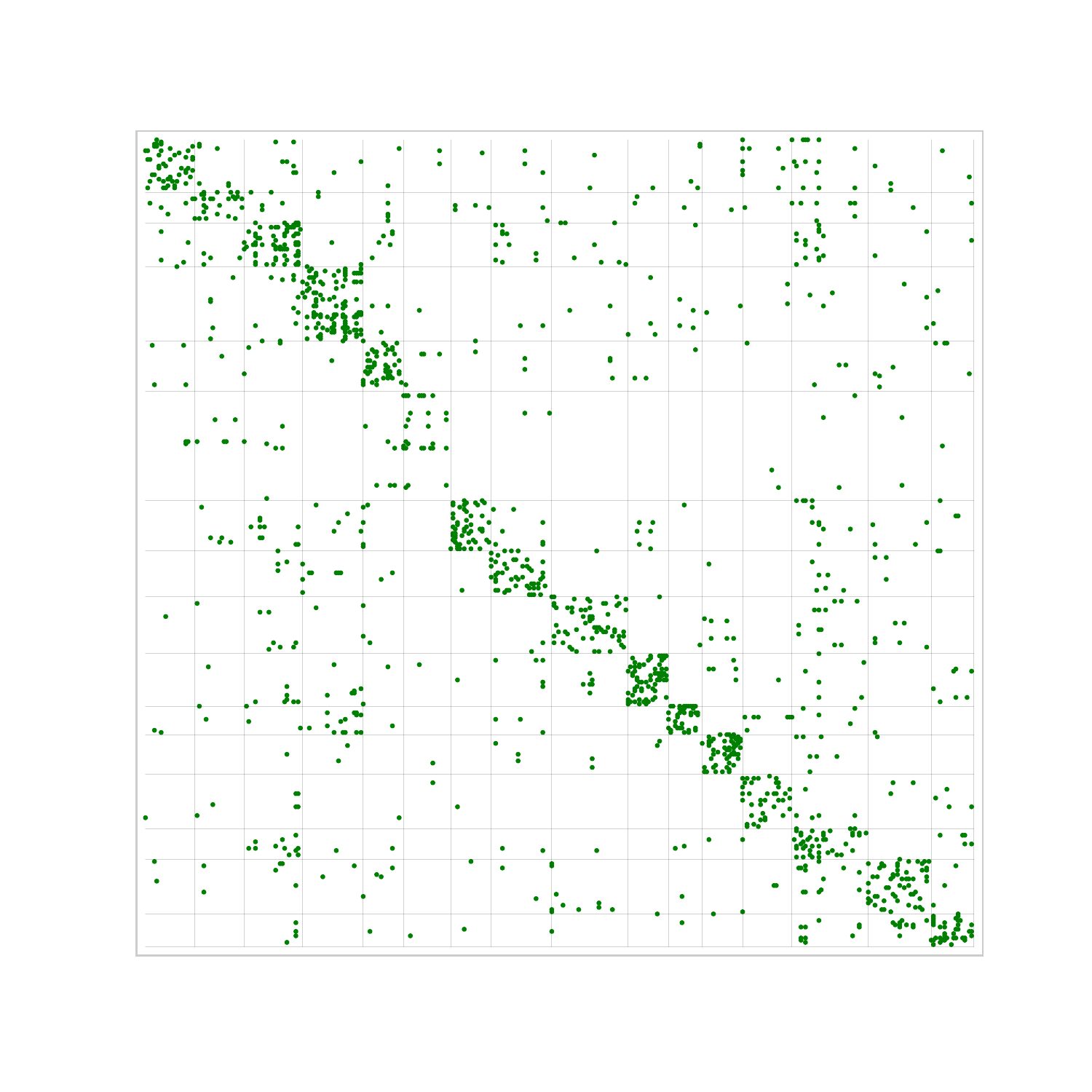} \caption{\tiny\\$s_{h=4}^{l=1}=18.892$}
    \end{subfigure}\hfill
     \captionsetup[subfigure]{skip=-5pt,justification=centering}
    \begin{subfigure}[c]{0.28\linewidth}
      \includegraphics[width=\linewidth]{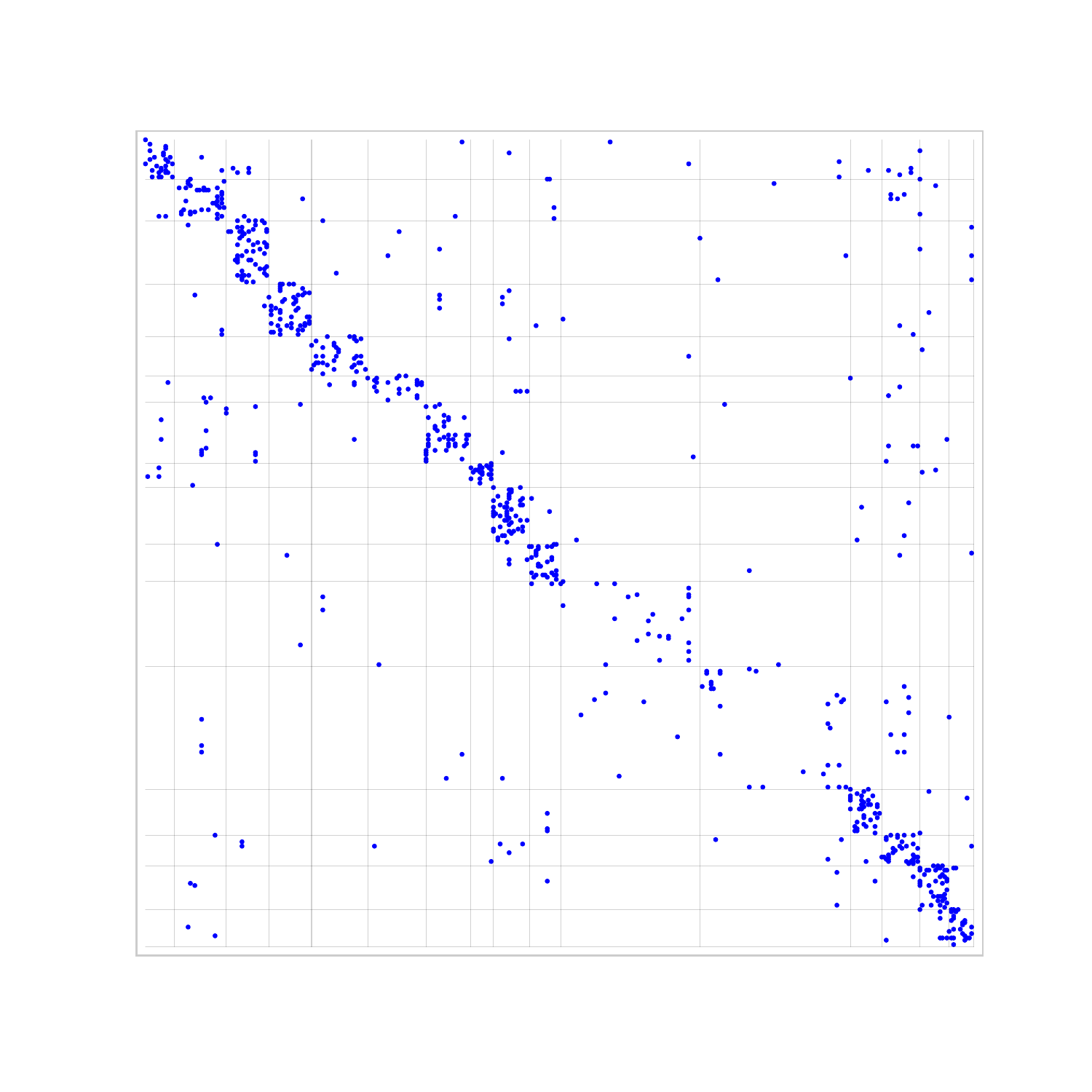} \caption{\tiny\\$s_{h=4}^{l=2}=15.824$}
    \end{subfigure}\hfill
     \captionsetup[subfigure]{skip=-5pt,justification=centering}
    \begin{subfigure}[c]{0.28\linewidth}
      \includegraphics[width=\linewidth]{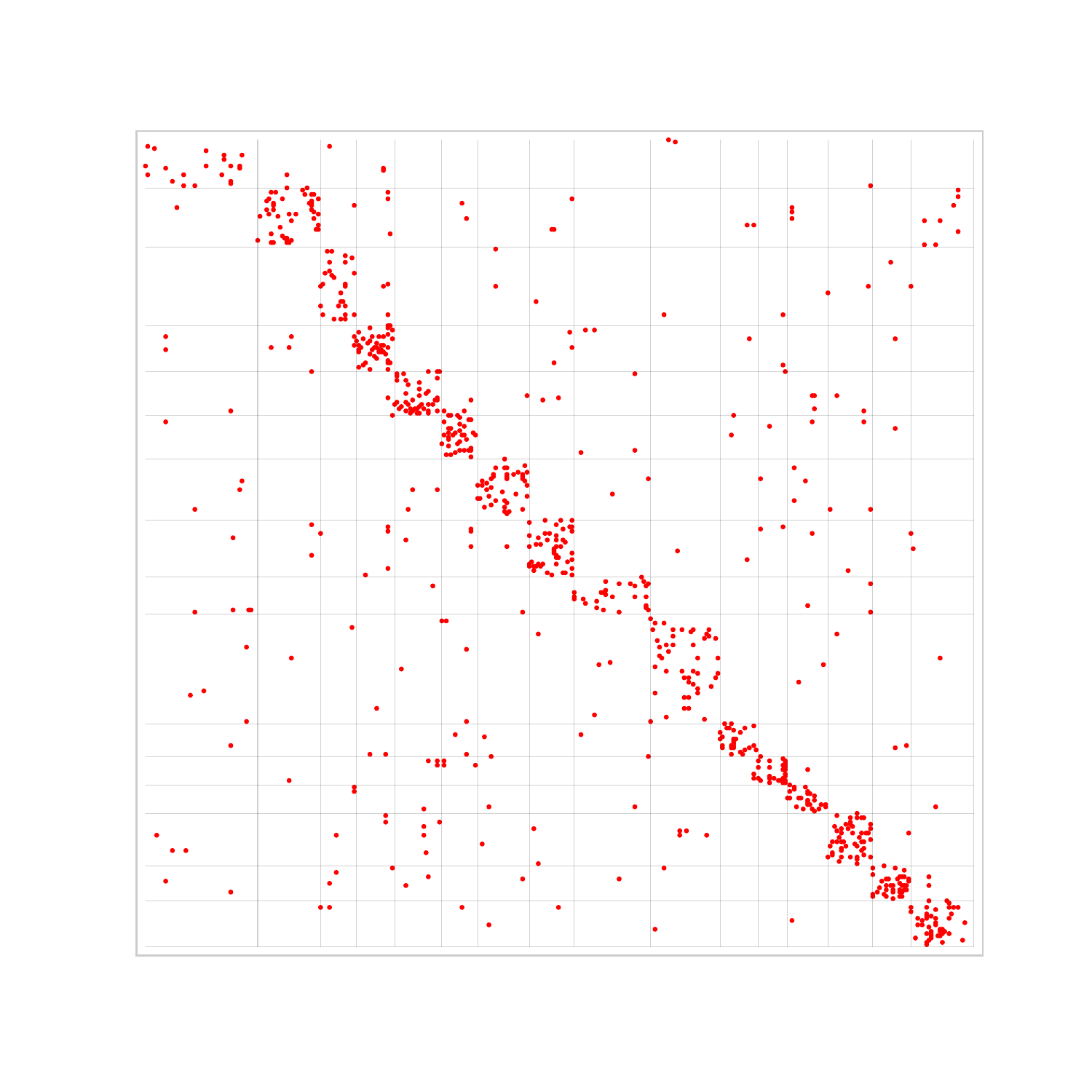} \caption{\tiny\\$s_{h=4}^{l=3}=8.140$}
    \end{subfigure}
  \end{minipage}

  \begin{minipage}[c]{\linewidth}
    \begin{minipage}[c]{0.04\linewidth}
      \centering
      \rotatebox{90}{\small \textbf{$D_{h=5}=32$}}
    \end{minipage}%
    \hfill  \captionsetup[subfigure]{skip=-5pt,justification=centering}
    \begin{subfigure}[c]{0.28\linewidth}
      \includegraphics[width=\linewidth]{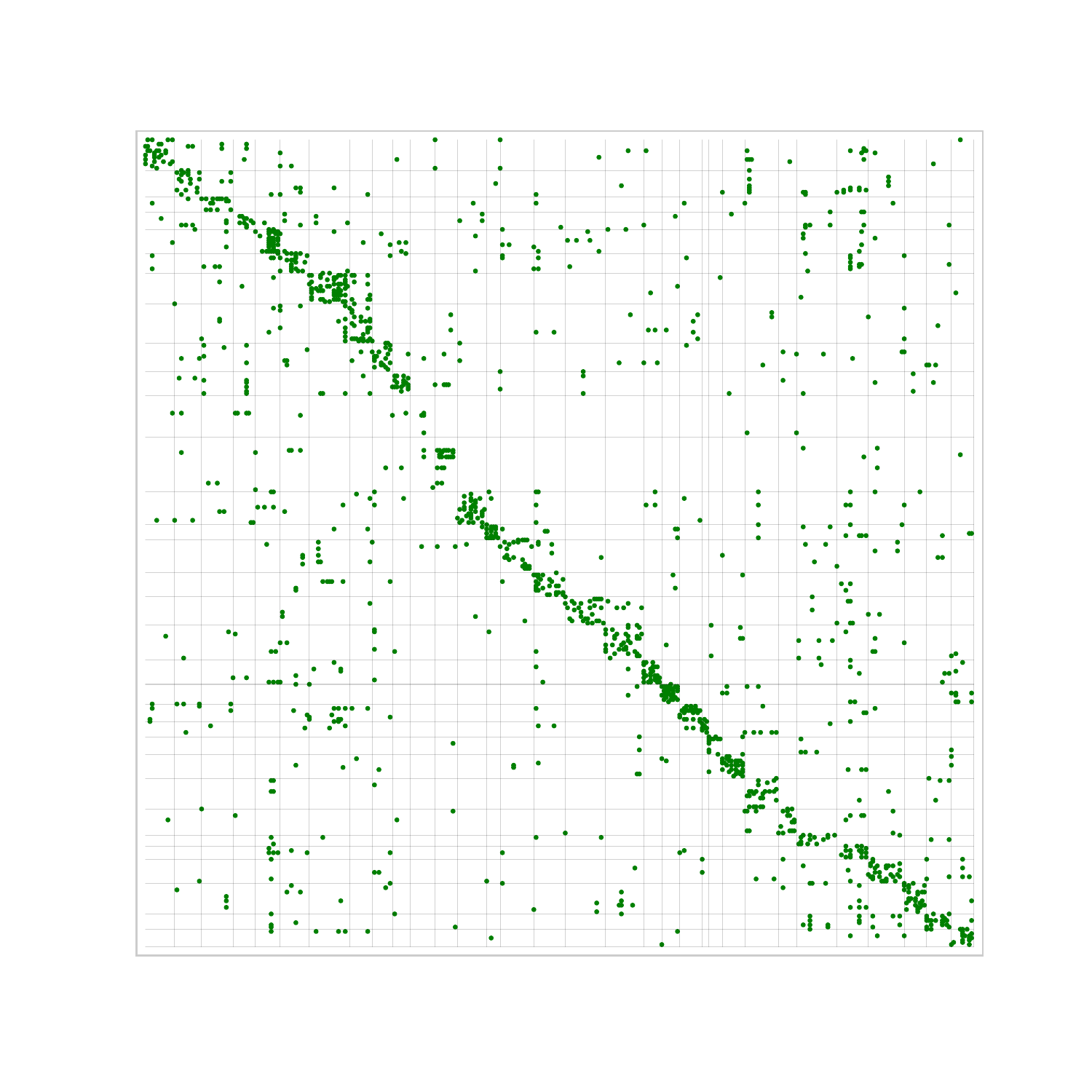}
      \caption{\tiny\\ $s_{h=5}^{l=1}=77.042$}
    \end{subfigure}\hfill  \captionsetup[subfigure]{skip=-5pt,justification=centering}
    \begin{subfigure}[c]{0.28\linewidth}
      \includegraphics[width=\linewidth]{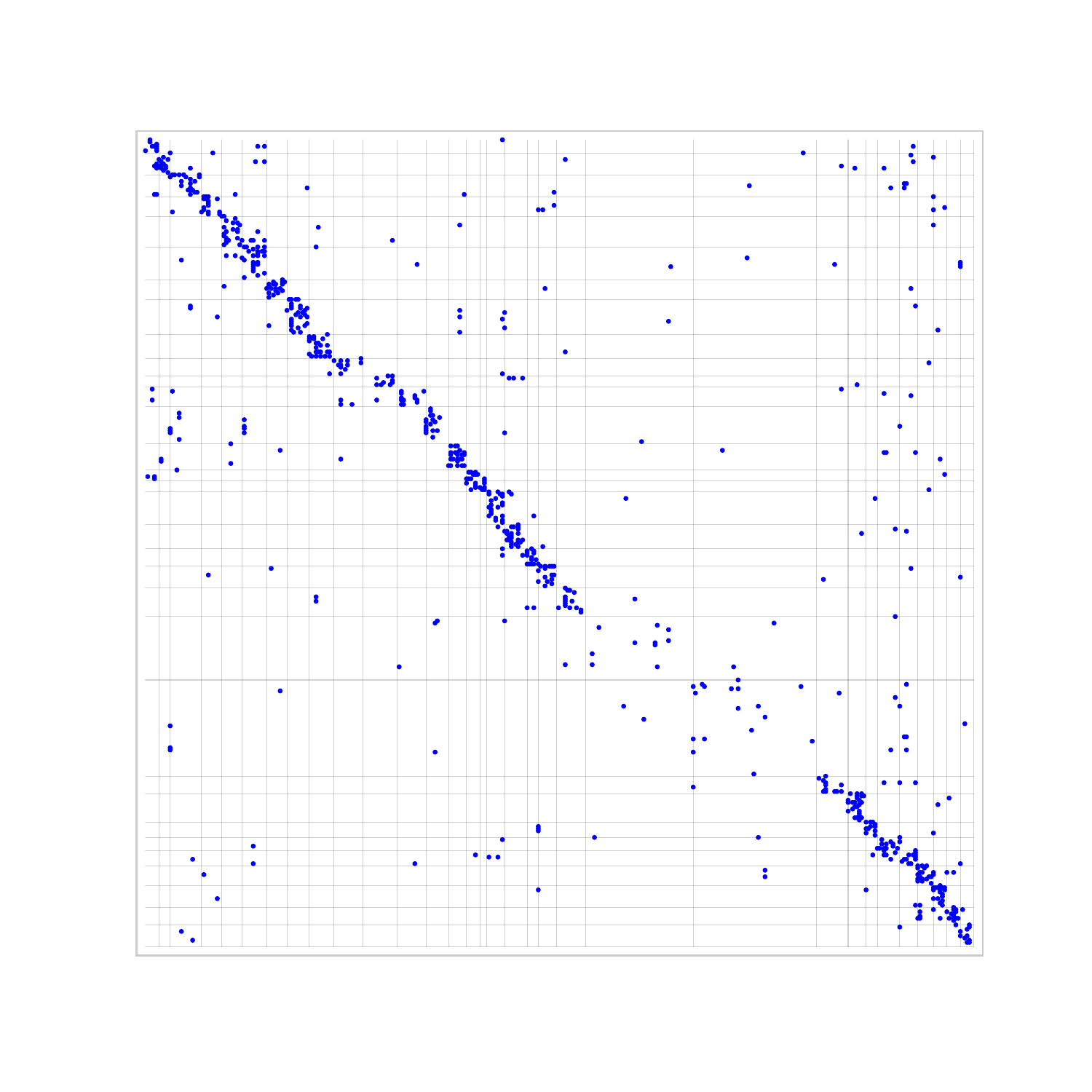}
      \caption{\tiny\\$s_{h=5}^{l=2}=17.048$}
    \end{subfigure}\hfill  \captionsetup[subfigure]{skip=-5pt,justification=centering}
    \begin{subfigure}[c]{0.28\linewidth}
      \includegraphics[width=\linewidth]{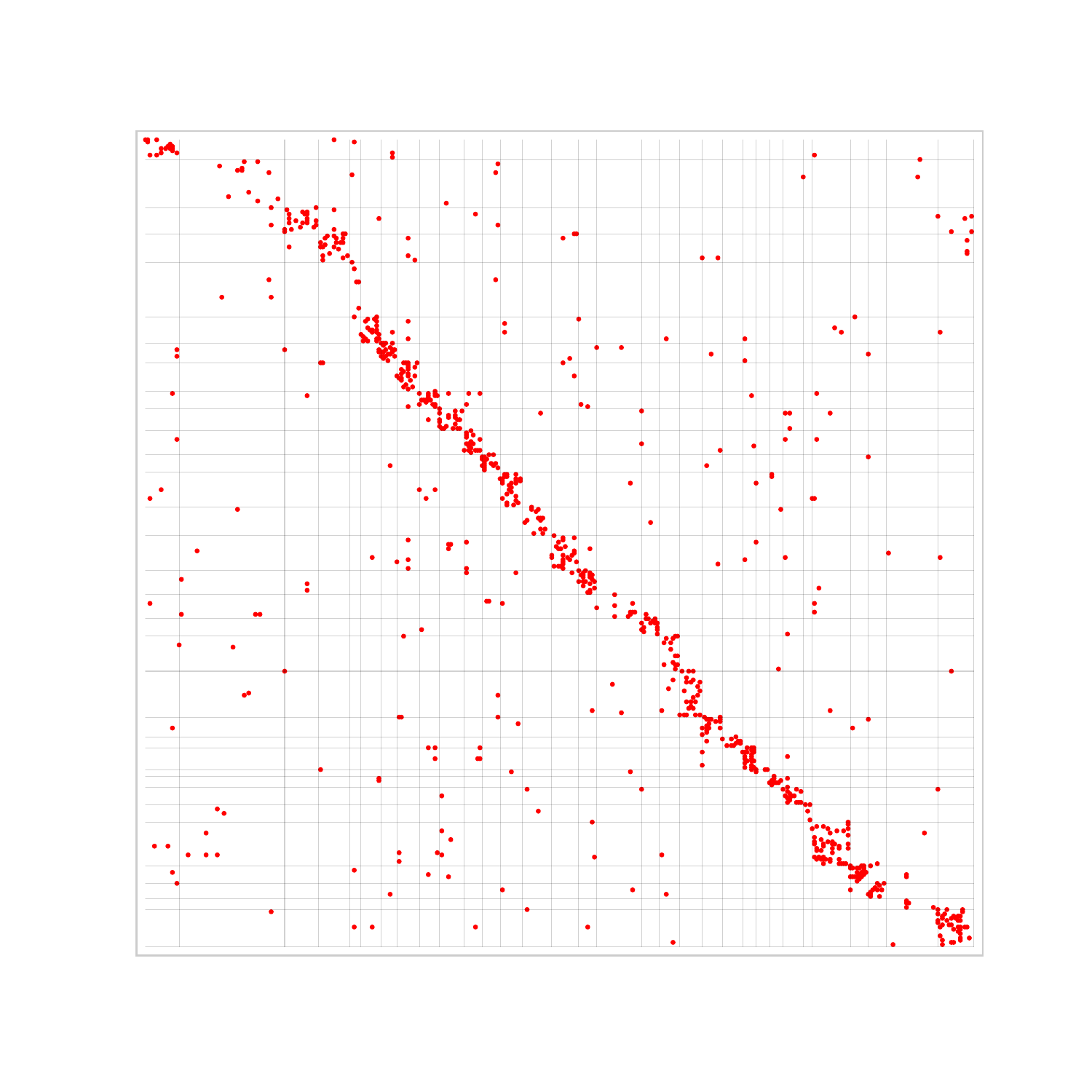}
      \caption{\tiny\\ $s_{h=5}^{l=3}=12.786$}
    \end{subfigure}
  \end{minipage}

      \hfill 
    \end{minipage}
  }

  \centering
  \begin{subfigure}[t]{0.49\textwidth}
    \includegraphics[width=1\linewidth]{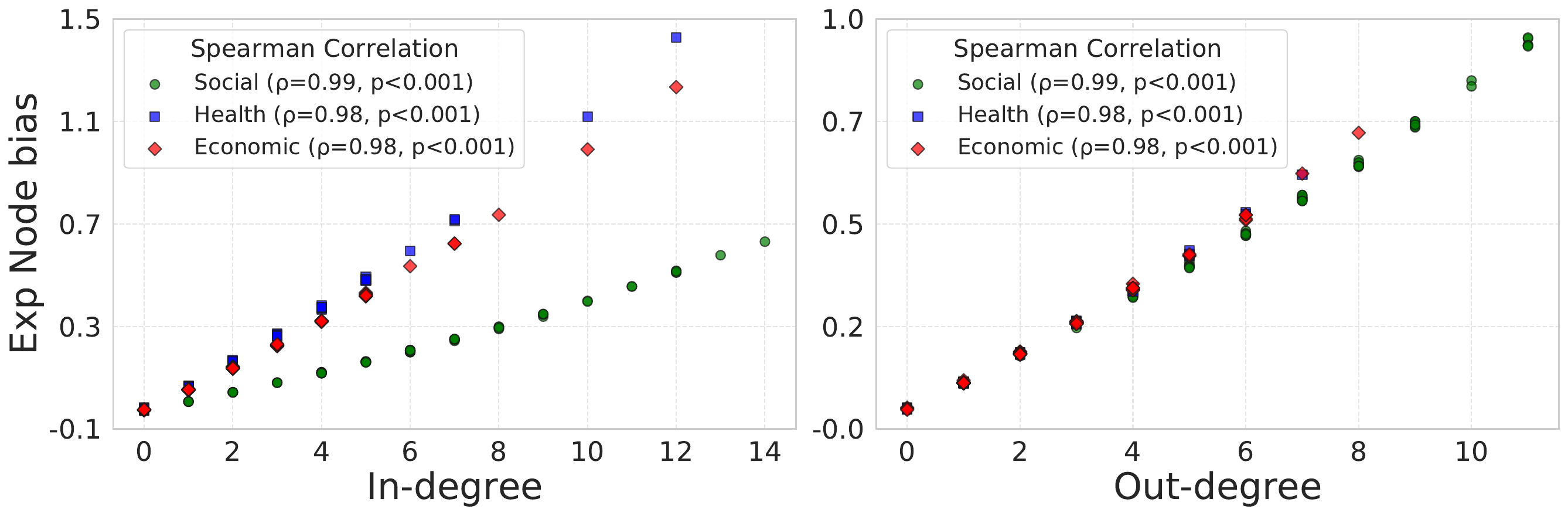}
    \caption{\small Bias model (dependence and independence): inferred node biases vs. in- and out-degree across layers.  }
  \end{subfigure}\hfill
  \begin{subfigure}[t]{0.49\textwidth}
    \includegraphics[width=1\linewidth]{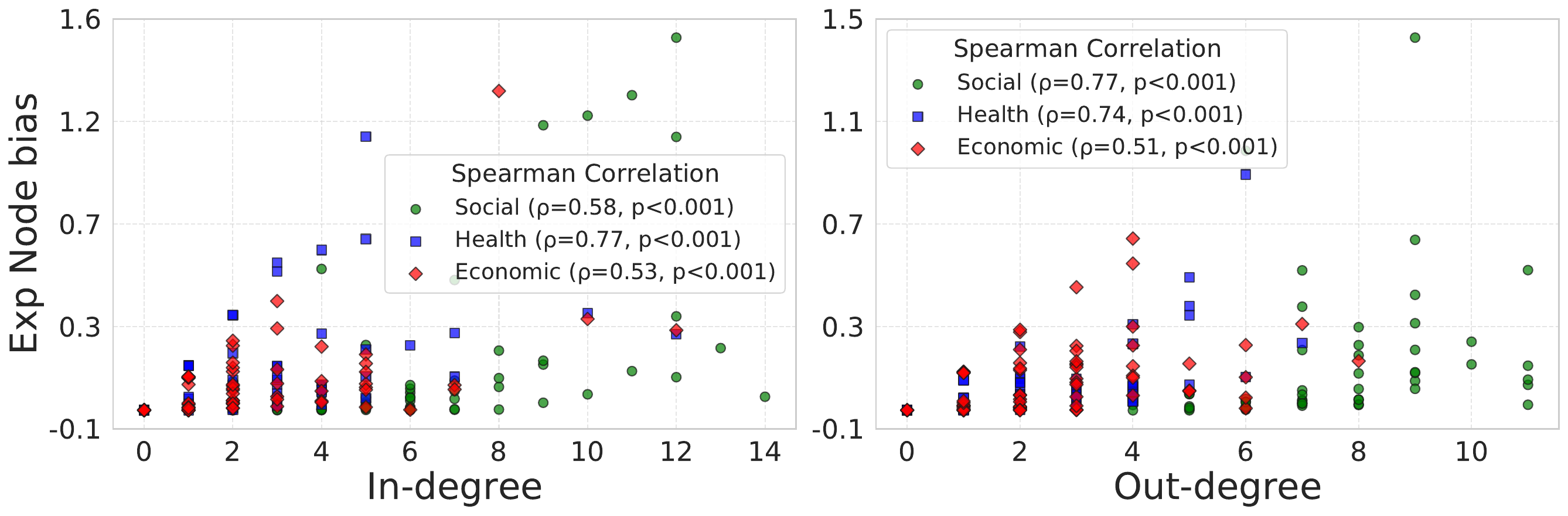}
    \caption{\small Full model (adds interdependence): inferred node biases vs. in- and out-degree across layers. }
  \end{subfigure}
  \begin{subfigure}[t]{0.165\textwidth}
    \includegraphics[width=\linewidth]{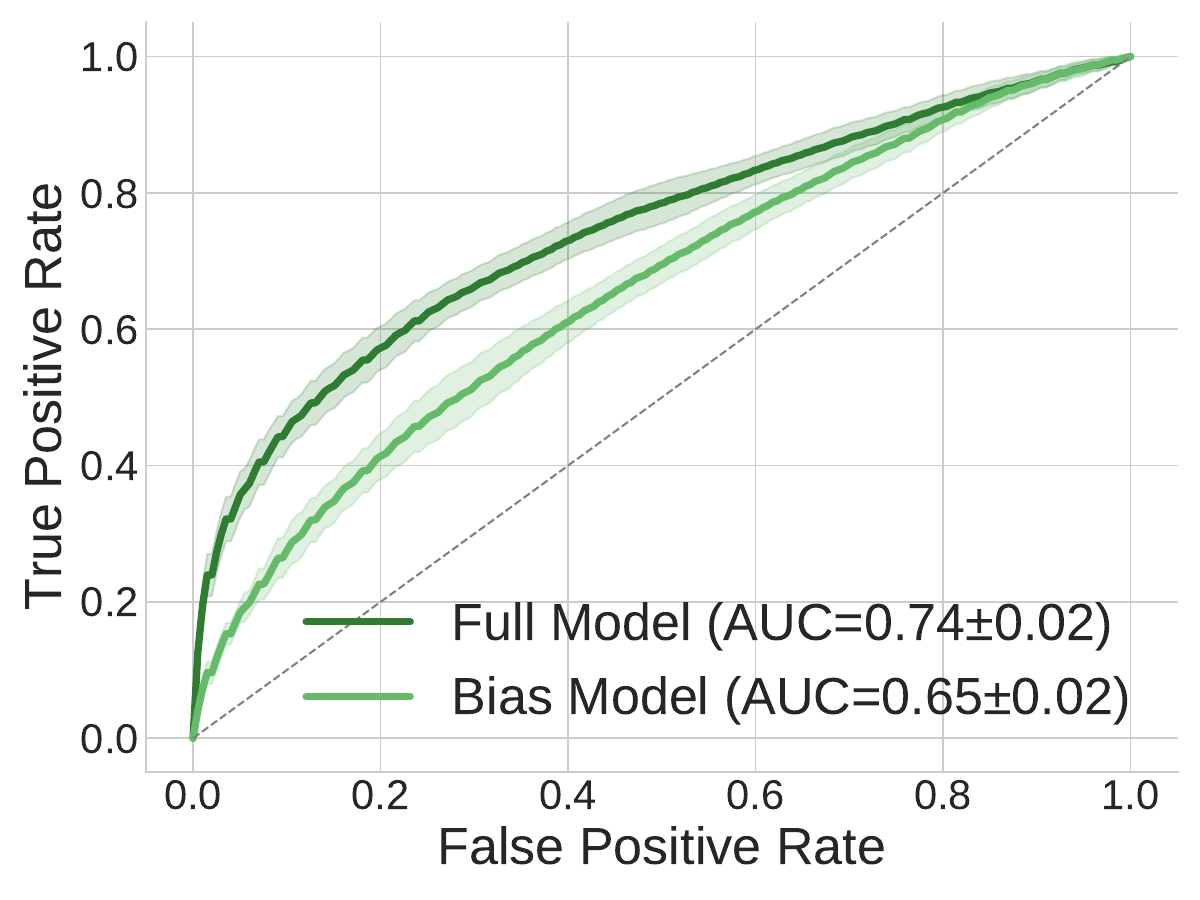}
    \caption{\small ROC Social}
  \end{subfigure}\hfill
  \begin{subfigure}[t]{0.165\textwidth}
    \includegraphics[width=\linewidth]{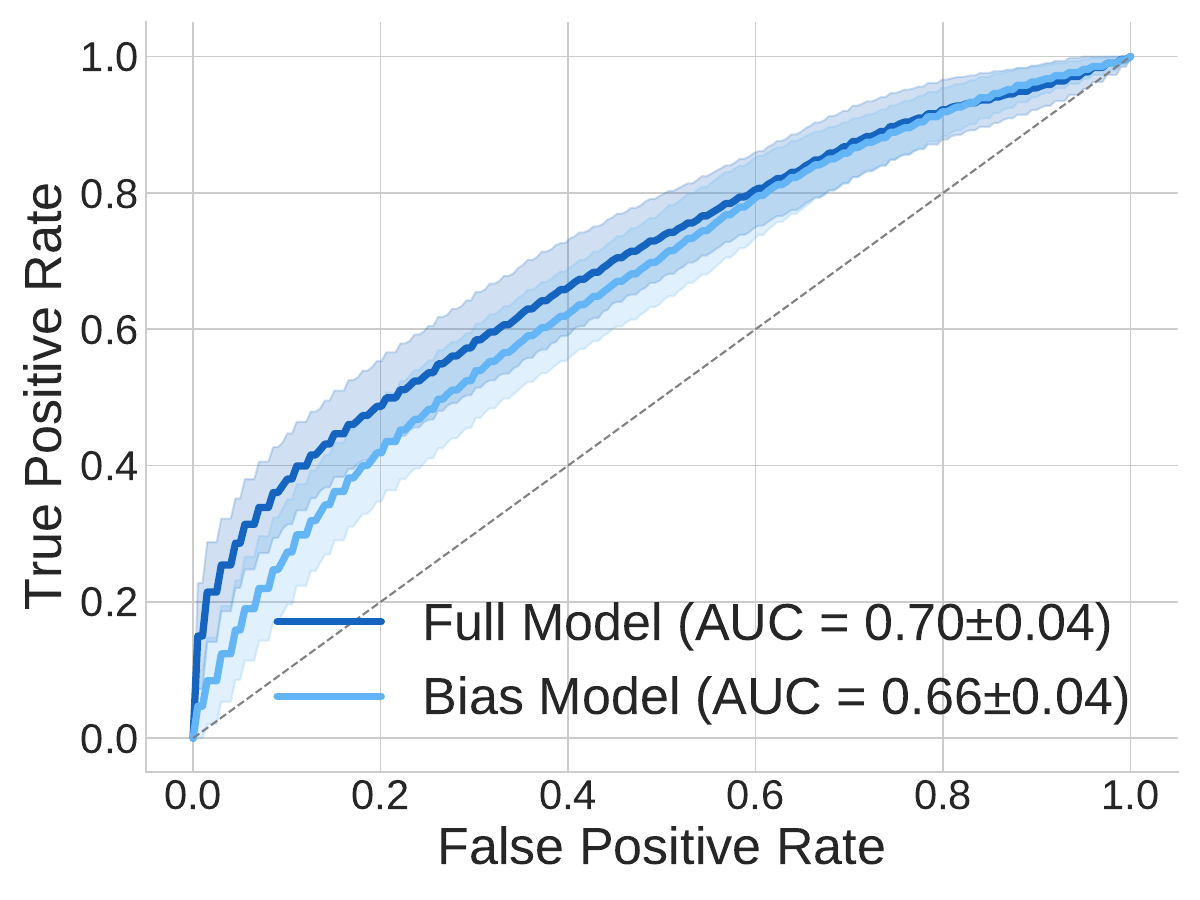}
    \caption{\small ROC Health}
  \end{subfigure}\hfill
  \begin{subfigure}[t]{0.165\textwidth}
    \includegraphics[width=\linewidth]{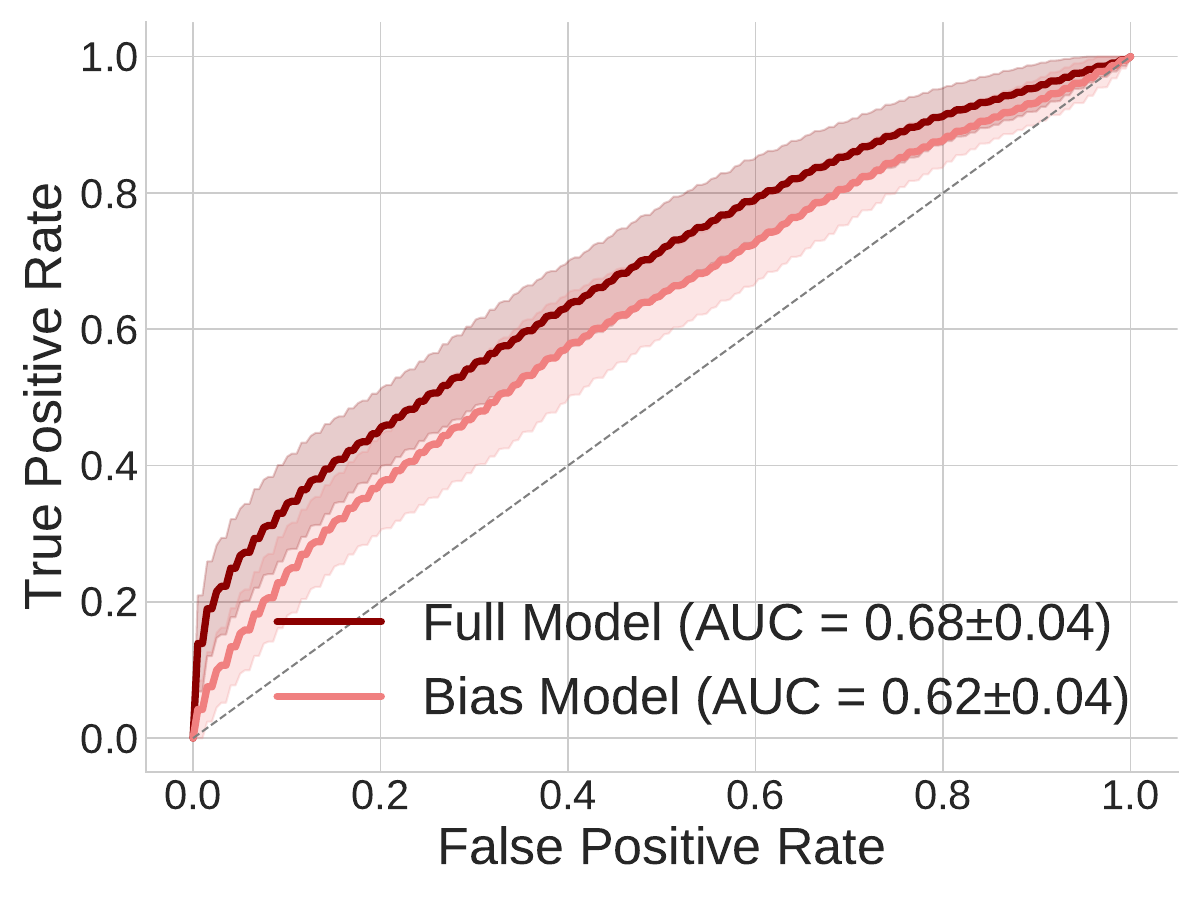}
    \caption{\small ROC Economic}
  \end{subfigure}
  \begin{subfigure}[t]{0.165\textwidth}
    \includegraphics[width=\linewidth]{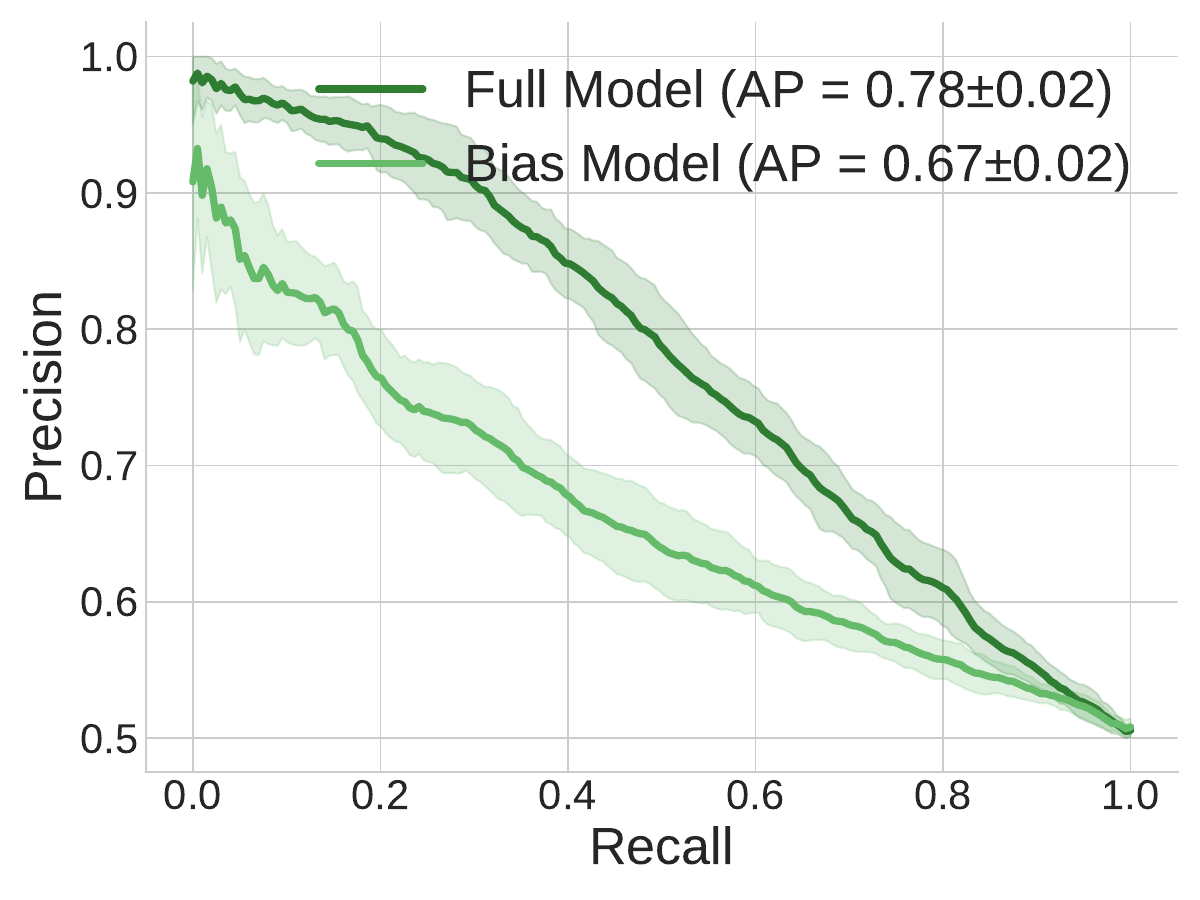}
    \caption{\small PR Social}
  \end{subfigure}\hfill
  \begin{subfigure}[t]{0.165\textwidth}
    \includegraphics[width=\linewidth]{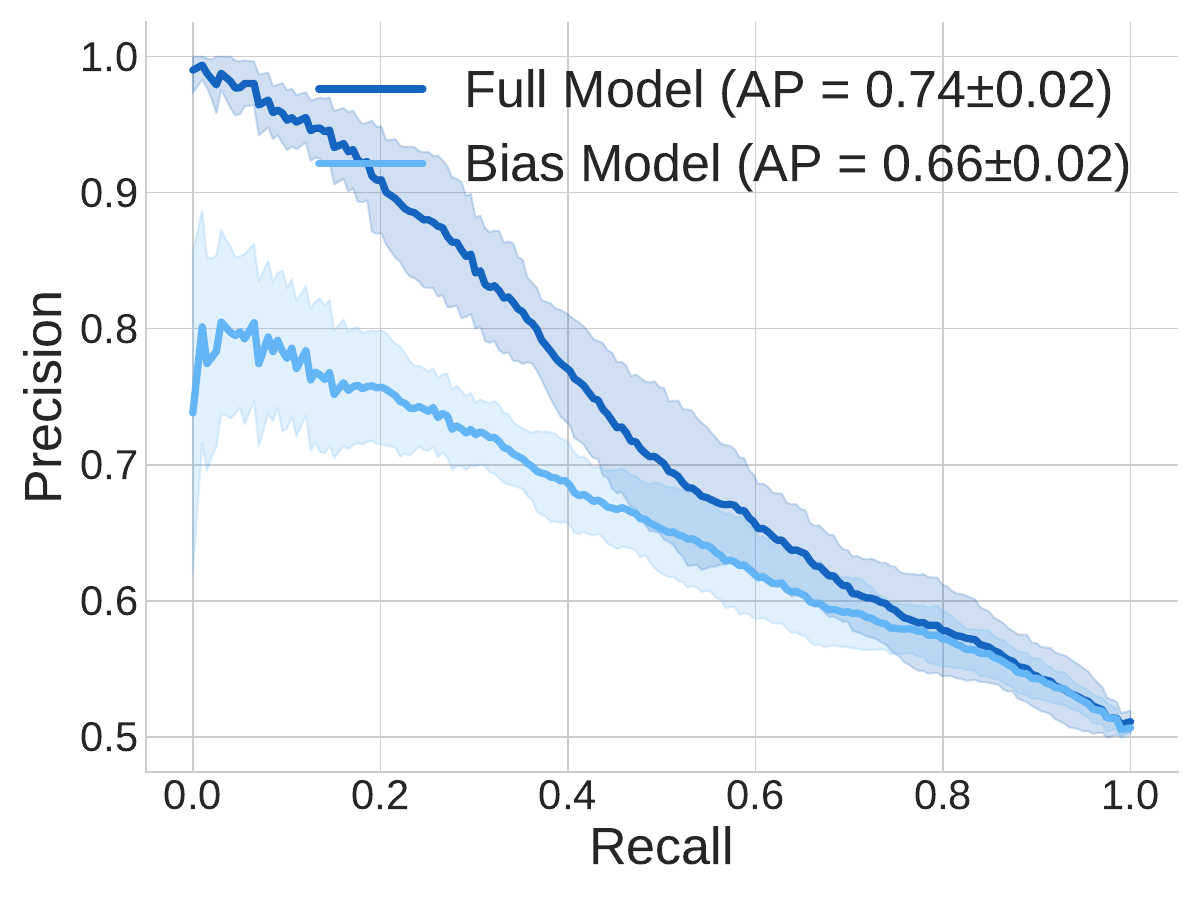}
    \caption{\small PR Health}
  \end{subfigure}\hfill
  \begin{subfigure}[t]{0.165\textwidth}
    \includegraphics[width=\linewidth]{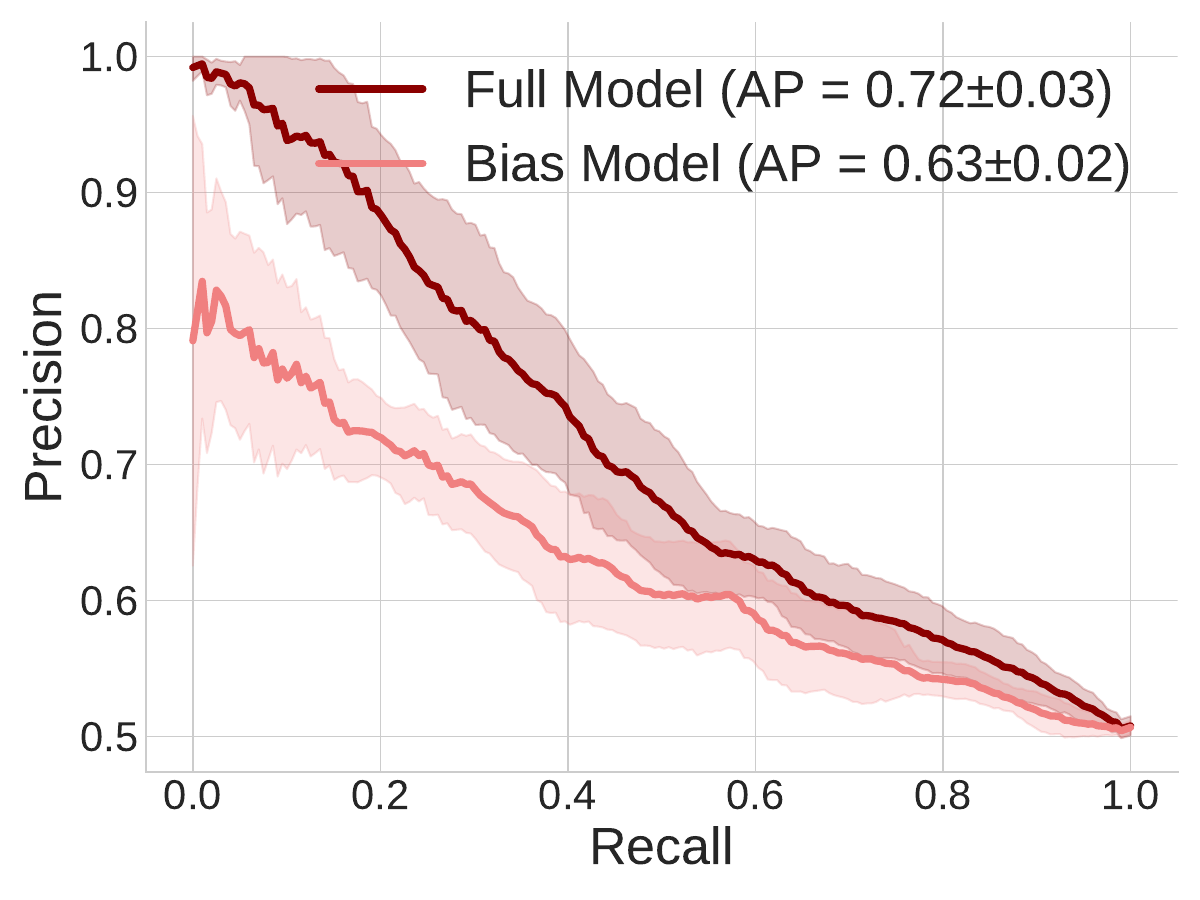}
    \caption{\small PR Economic}
  \end{subfigure}

  \caption{\textbf{Model behavior and structural interpretation on village network \# 3.} 
 Panel (a, right) shows the learned multiplex role simplex with directed ties: social (green), health (blue), and economic (red). Arrows indicate directionality; circles represent source and target embeddings $\mathbf{Z}, \mathbf{W} \in \Delta_2$. Panel (a, left) displays the same structure with ties shown separately by layer, with nodes colored by dominant role.
Panel (b) presents violin plots of source and target role strengths across layers, highlighting the dominance of the social role.
Panels (c)–(q) show the inferred multi-scale structure per layer, with hierarchy strengths $s_h^l$ quantifying each level's contribution to link formation. Adjacency matrices are reordered by dominant source and target memberships $\mathbf{u}_i^{l,h}$ and $\mathbf{v}_j^{l,h}$. Deeper hierarchies reveal increasingly fine-grained structures.
Panel (r) shows that the \textit{bias model} (capturing dependence/independence) yields strong correlations between degree and inferred biases. Panel (s) shows these correlations drop under the \textit{full model}, indicating structural signal is reallocated to interdependence.
Panels (t)–(v) display ROC curves and AUC scores comparing models via ten-fold cross-validation; panels (w)–(y) show PR curves. The \textit{full model} improves prediction, particularly in the social layer, with clearer gains visible in PR performance.}
  \label{fig:overall}
\end{figure}

\begin{figure}[h!]
\centering

\begin{subfigure}[t]{0.66\textwidth}
    \centering
    \includegraphics[width=\textwidth]{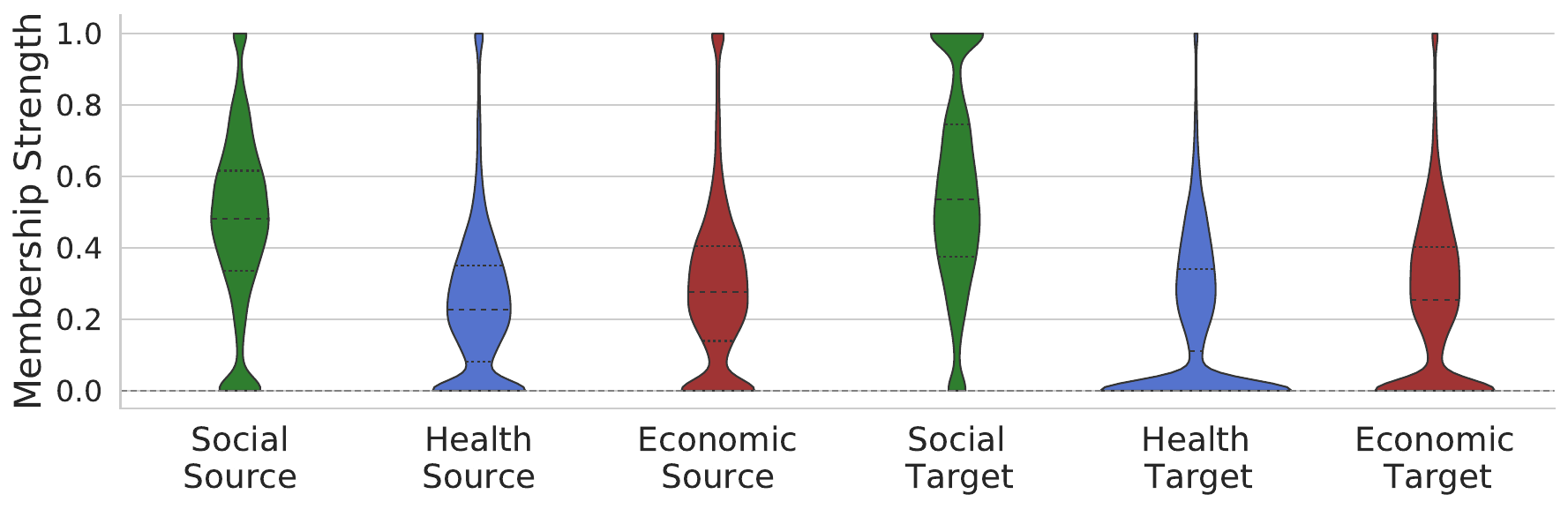}
    \caption{Distribution of node-role memberships across 176 multiplex networks, summarized by source ($\mathbf{Z}$) and target ($\mathbf{W}$) embeddings. Each violin plot shows the distribution of membership strengths for each layer, aggregated across all networks. The plots summarize how nodes distribute their roles across layers.}
    \label{fig:violin_means}
\end{subfigure}%
\hfill
\begin{subfigure}[t]{0.33\textwidth}
    \centering
    \includegraphics[width=\textwidth]{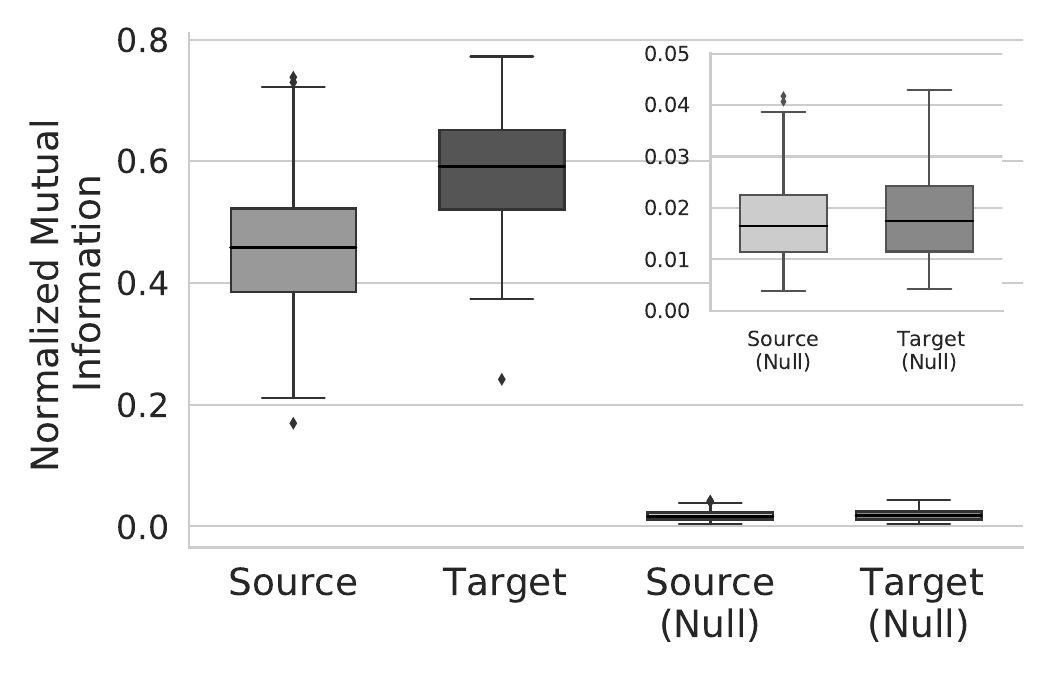}
    \caption{Normalized mutual information (NMI) across 176 multiplex networks, shown alongside the null permutation distribution (inset: zoomed view of the null).}
    \label{fig:nmi_box}
\end{subfigure}

\vspace{0.5em}
\caption{\textbf{Summary statistics across 176 multiplex networks.} Panel (a) shows the distribution of mean node-role memberships, computed from the simplex embeddings for source ($\mathbf{Z}$) and target ($\mathbf{W}$) positions. Each violin represents the across-network distribution of the average membership strength for each layer, revealing systematic differences in dominant layer expression across source and target embeddings. Higher mean values indicate that, on average, nodes in that network allocate more of their role expression to that layer. Panel (b) shows the normalized mutual information (NMI) between inferred node embeddings and degree-based layer activity vectors (alongside the null distribution obtained by permutation testing). Target embeddings ($\mathbf{W}$) exhibit slightly higher alignment with normalized in-degree than source embeddings ($\mathbf{Z}$) with out-degree, suggesting that role structure at the target level is more degree-driven. The imperfect alignment in both cases indicates that the model captures additional structural variation beyond degree, including higher-order interactions and multiplex dependencies. NMI scores under the null distribution are very close to zero (see also the inset plot at the top right), indicating no retrieval of structured signal.
}
\label{fig:summary_stats}
\end{figure}

\begin{figure}[h!]

\centering
\begin{subfigure}[t]{0.49\textwidth}
    \includegraphics[width=\textwidth]{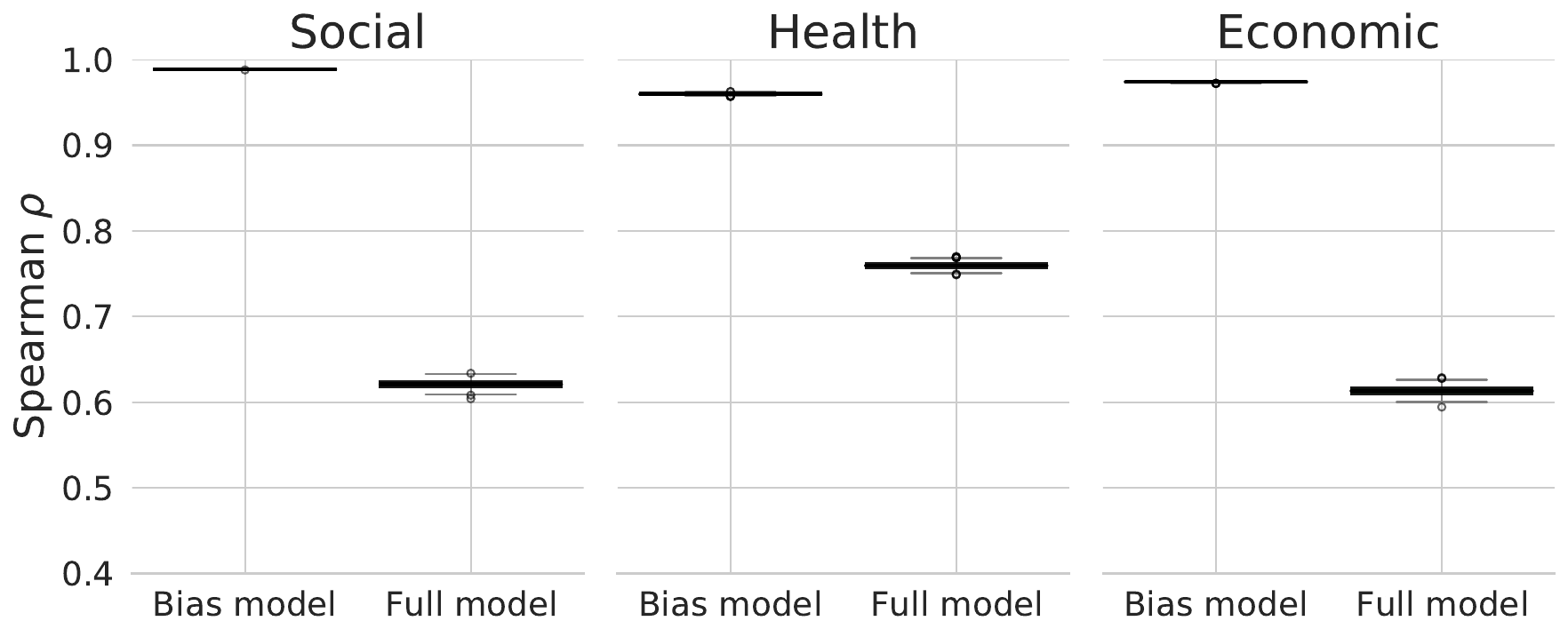}
    \caption{\small Mean bootstrapped correlations between inferred \textbf{target bias} ($\gamma_j^{l}$) and \textbf{node in-degree} in the social, health, and economic layers, comparing the \textit{bias model} with the \textit{full model} that accounts for interdependence.}
\end{subfigure}
\hfill
\begin{subfigure}[t]{0.49\textwidth}
    \includegraphics[width=\textwidth]{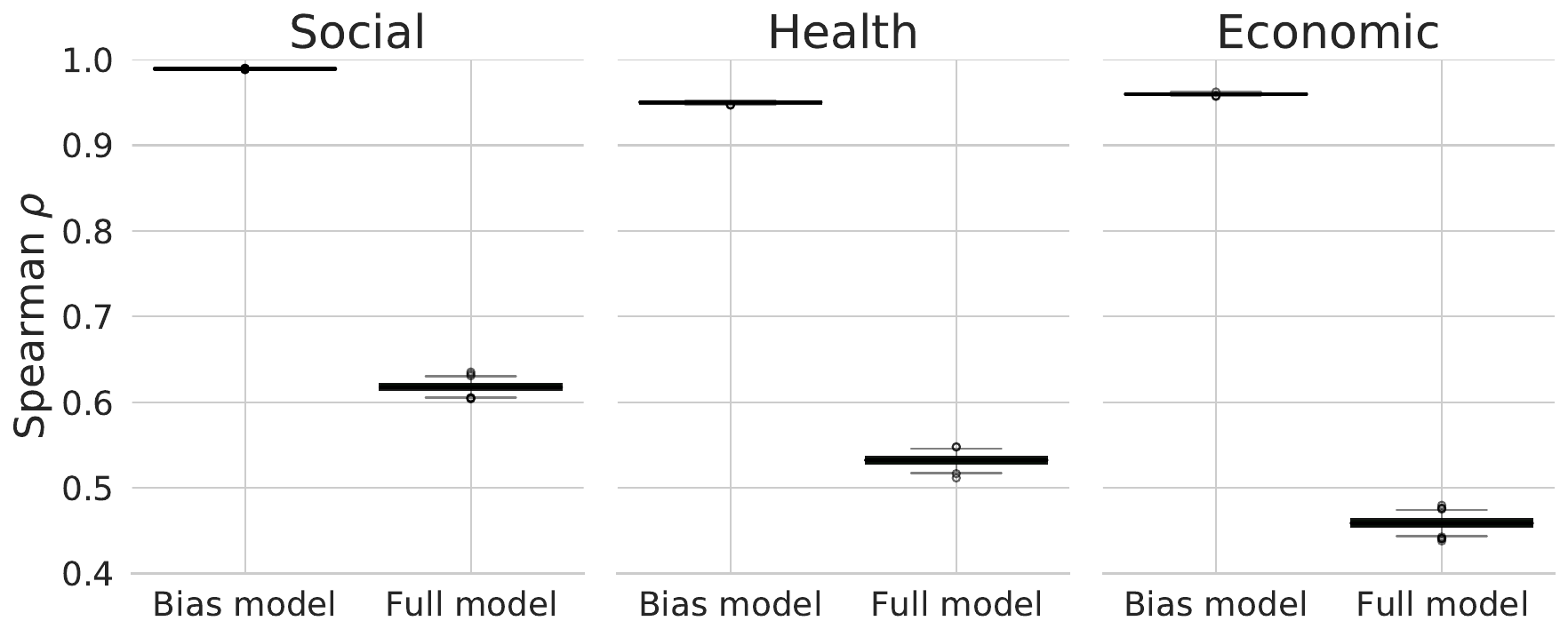}
    \caption{\small Mean bootstrapped correlations between inferred \textbf{source bias} ($\beta_i^{l}$) and \textbf{node out-degree} in the social, health, and economic layers, comparing the \textit{bias model} with the \textit{full model} that accounts for interdependence.}
\end{subfigure}
\caption{\textbf{Bootstrapped Spearman correlations ($\rho$) between the inferred target bias ($\gamma^{l}_{j}$) as well as source bias ($\beta^{l}_{j}$) and node in-/out-degree for the social, health, and economic layers.} For every bootstrap iteration we compute $\rho$ separately in each of the 176 village networks and then average these 176 values to obtain a single “network-averaged” correlation. Repeating this procedure over 1,000 bootstrap iterations yields 1,000 mean correlations; their distribution is displayed as a box-and-whisker plot. Colors contrast the bias-only model with the \textit{full model} that incorporates interdependence, showing how interdependence changes the global (across-village) link between inferred node biases and degree centrality.
}
\label{fig:RE_degrees}
\end{figure}

\begin{figure}[h!]
    \includegraphics[width=\textwidth]{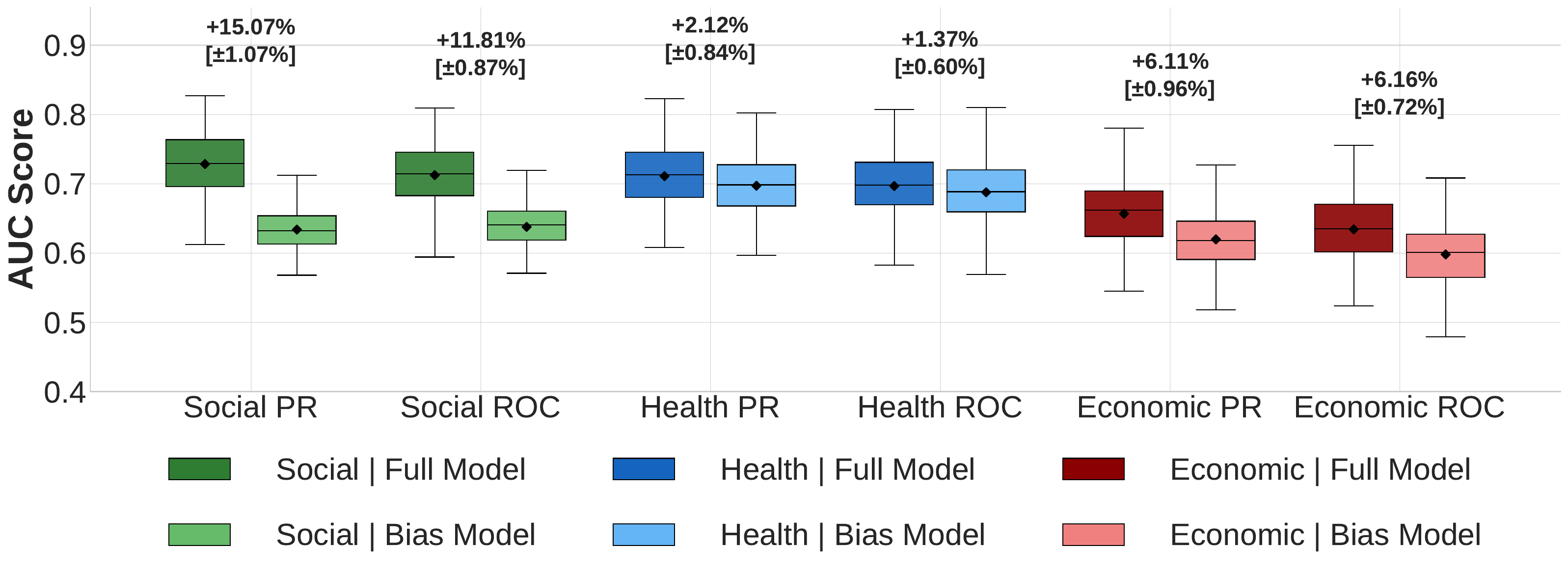}
\caption{\textbf{Quantifying Interdependence Effects Across Social, Health, and Economic Layers.} Average change in prediction performance across ten-fold cross-validation when comparing a model that captures only dependence and independence with the full \textsc{MLT} model that also includes interdependence, in accordance with Social Exchange Theory. Results are shown for all 176 village networks, using Area Under Curve of the PR (AUC-PR) and Area Under Curve of the ROC (AUC-ROC). Performance gains are shown above each pair of Full versus Bias models for each layer, along with 95\% confidence intervals based on a paired t-test of the change in AUC scores when transitioning to the \textit{full model} that accounts for interdependence.}
\label{fig:SET_all}
\end{figure}

\subsection*{Framework Illustration Using a Representative Village Network} We uncover role-based trade-offs, hierarchical structures, and patterns of interdependence in multiplex networks. Our central contribution is to show how trade-offs across relational layers give rise to emergent roles, while preserving distinct community structure within each layer. Crucially, our role-based framework enables a principled quantification of social exchange by comparing predictive performance against a null model that assumes tie formation is driven solely by individual status (i.e., sender and receiver effects). Gains from modeling interdependence beyond this baseline reveal the structural contribution of reciprocal, role-based dynamics across layers. We illustrate the full modeling pipeline using a single-village network, showcasing role discovery, trade-offs, multi-scale structure, and predictive performance.

Central to the model is the role simplex: a low-dimensional representation in which each node is positioned as a convex combination of social, health, and economic latent roles. In Figure~\ref{fig:overall}, panel (a, right) the multiplex network is directly overlayed onto this simplex, revealing how ties across layers map onto latent role structures. When ties are visualized separately by layer (panel a, left), we observe the distinct clustering along role axes, showcasing that the simplex captures not only role expression, but also the inherent trade-offs between layers describing multiplexity. The model distinguishes between source and target roles, encoded by the embeddings $\mathbf{Z}$ and $\mathbf{W}$, respectively. Panel (b) summarizes this asymmetry across all nodes in the network. Social activity dominates, both in source and target positions, with target membership strengths showing the heaviest tail reflecting nodes acting in terms of pure social roles. Membership strength distributions over the health and economic layers showcase lower membership strengths compared to the social layer having also no heavy tails indicating the relative absence of pure health and economic roles. 

Zooming into the latent community organization, panels (c)--(q) display the inferred layer-specific multi-scale structure (we further validate that the model effectively leverages the multi-scale structure and information (SI  Figure~\ref{fig:permute_h})). At each hierarchical level $h$, adjacency matrices are reordered based on dominant source ($\mathbf{u}_i^{l,h}$) and target ($\mathbf{v}_j^{l,h}$) memberships. We observe that all layers can be decomposed into increasingly fine-grained communities, revealing highly localized patterns of interdependent interactions effectively captured by the model. Corresponding hierarchy strengths $s_h^l$ quantify the contribution of each level $h$ to explaining the link structure in layer $l$, highlighting that the model relies more on deeper levels to explain the link structure in each layer.

As MLT captures structural patterns aligned with key dynamics of social exchange, to understand the contribution of the extracted interdependence, we compare two model variants: the \textit{bias model} which includes only node-level effects (dependence and independence; $r_{ij} = \beta_i^l + \gamma_j^l$), and the \textit{full model}, which adds the dyadic interdependence ($r_{ij} = \beta_i^l + \gamma_j^l + \eta_{ij}$). The \textit{bias model} essentially expresses the null assumption of status-driven tie formation. Panel (r) shows that, in the \textit{bias model} (that does not capture interdependence), inferred node biases $\beta_i^l$ and $\gamma_j^l$ simply correlate with out-degree and in-degree respectively, confirming that the model captures node status. In contrast, panel (s) shows that in the \textit{full model} (which includes interdependence), these simple correlations with degree drop significantly. This indicates that interdependence is needed to capture the structural signal. 

Finally, panels (t)--(y) evaluate predictive performance using ten-fold cross-validation.  Receiver Operating Characteristic (ROC) and Precision-Recall (PR) curves compare the bias and full models for link prediction in each layer. The \textit{full model} consistently outperforms the \textit{bias model}, in particular when considering the PR-AUC. Whereas ROC-AUC improvements are modest for health and economic layers, PR gains are more substantial, underscoring the value of modeling interdependence for accurate prediction in terms of false positive and false negative ties. Additional visualizations of villages' learned representations are provided in SI Figures~\ref{fig:overall_0}, \ref{fig:overall_50}, and \ref{fig:overall_174}.


\subsection*{Analysis Across All Village Networks}
We now extend our analysis to all 176 village networks, encompassing 22,584 individuals. We begin by examining engagement across the social, health, and economic layers to understand how individuals prioritize different exchange domains, providing a global view of trade-off structures. Next, we assess reliance on the \textit{bias model}, which represents the null assumption that interactions are solely driven by node status. Finally, we generalize the impact of interdependence across the three domains, social, health, and economic, where emerging roles reflect patterns of expected behavior, such as obligations, expectations, and norms of reciprocity that vary by layer.

\subsubsection*{Role Engagement Across Layers} In general, villages are primarily oriented toward social interactions, with health and economic layers showing comparable but secondary emphasis. Moreover, role allocations in the trade-off space are related but do not explicitly follow node degree patterns. Given the effectiveness of the simplex framework in capturing role allocation and trade-offs in the single-village example above, we examine node-role memberships across layers for all villages. We fit a model separately for each of the $176$ villages.

In Figure~\ref{fig:summary_stats} (a), violin plots illustrate the distribution of membership strengths for both source and target roles across all $176$ networks. The patterns closely resemble those observed in illustrative Village \# 3 (Figure \ref{fig:overall}), suggesting a consistent trend. We find that all villages exhibit a dominant orientation toward the social layer, with both source and target roles showing strong engagement in social interactions. The social layer displays a relative absence of values near zero, indicating broad participation, particularly as a target. Notably, the social target role also shows the heaviest tail towards one, suggesting a subset of individuals with highly specialized social roles. 

In contrast, node-role memberships in the health and economic layers show lower overall participation. Their distributions feature heavier tails near zero, especially in the target roles, most prominently in the health layer, indicating that many individuals are not targeted in these layers. This is expected, as being a target in health or economic contexts often requires specific attributes such as medical knowledge or financial capacity. 

To assess whether activity differed significantly across layers, we tested the null hypothesis that layer orderings were exchangeable. A permutation test (1,000 iterations) revealed a consistent ordering of activity: social $>$ economic $>$ health ($p = 0.0001$), indicating that social activity dominates across villages, for both source and target cases. While economic and health activities were more comparable, a statistically significant difference favored economic activity. This effect is consistent with Social Exchange Theory, which characterizes health and economic ties as predominantly instrumental and goal-oriented, in contrast to the more expressive nature of social relationships \cite{Blau1964ExchangeAP,homans}. 

We further examined the relative prominence of each layer by treating the components of each simplex as exchangeable under the null, implying no dimension has privileged status. Using 1,000 node-wise permutations (permuting components within each row), we tested for over- or under-representation of each layer. In both the source ($\mu^z_{social} = 0.463$) and target ($\mu^w_{social} = 0.560$) cases, the social layer was significantly overrepresented ($p = 0.001$). In contrast, both the health ($\mu^z_{health} = 0.245$, $\mu^w_{health} = 0.186$) and economic ($\mu^z_{economic} = 0.292$, $\mu^w_{economic} = 0.254$) layers were significantly underrepresented compared to the null ($p = 0.001$ for both). To assess the extent to which inferred role structures align with basic connectivity patterns, we compute the \textit{normalized mutual information} (NMI)~\citep{AA_fmri} between simplex embeddings and the normalized degree distribution across layers. For each node $i$, we calculate its layer-specific degree proportions as:
$
\hat{d}^{\mathrm{(in)}}_{i,l} = \frac{d^{\mathrm{(in)}}_{i,l}}{\sum_{k} d^{\mathrm{(in)}}_{i,k}}, \quad \hat{d}^{\mathrm{(out)}}_{i,l} = \frac{d^{\mathrm{(out)}}_{i,l}}{\sum_{k} d^{\mathrm{(out)}}_{i,k}},
$
where $d^{\mathrm{(in)}}_{i,l}$ and $d^{\mathrm{(out)}}_{i,l}$ denote the in-degree and out-degree of node $i$ in layer $l$. These vectors characterize how nodes distribute their ties across layers. We then compare these normalized degree profiles to the learned source ($\mathbf{Z}$) and target ($\mathbf{W}$) simplex embeddings across 176 multiplex networks. As shown in Figure~\ref{fig:summary_stats} (b), the NMI values indicate moderate alignment, with target embeddings ($\mathbf{W}$) showing slightly stronger correspondence with out-degree distributions. However, the incomplete alignment in both source and target roles suggests that the model captures more complex structural patterns beyond degree distributions, including higher-order and interdependent interactions. We also validated the NMI scores under the null hypothesis of no structural alignment between the inferred role positions and the normalized degree distributions. Figure~\ref{fig:summary_stats}(b) shows that the NMI scores under the null distribution are centered around zero, indicating no recovery of structured signal in the permuted embeddings ($p < 0.0001$).

\subsubsection*{Interdependence Reduces Bias Reliance} In models capturing only dependence and independence, node biases align almost perfectly with degree, reflecting status through activity levels. When interdependence is incorporated, these correlations weaken, as the model shifts away from being purely status-driven and instead relies on interdependence terms to account for structural regularities. 

To explore the role of layer- and node-specific biases, we analyze their inferred values across all $176$ networks in relation to status, measured to in-degree and out-degree (Figure~\ref{fig:RE_degrees}). Additional centrality measures are provided in (SI Figure~\ref{fig:RE_centr}). This analysis clarifies how node biases interact with local and global network structure. For each layer of the multiplex network, treated independently, we compute the centrality measures and calculate Spearman correlations between source ($\beta_i^l$) and target ($\gamma_i^l$) biases and each centrality metric across the 176 networks. Using 1,000 bootstrap samples, we report distributions of mean correlations across networks as box plots. In the bias-only model, source and target biases are nearly perfectly correlated with out-degree and in-degree, respectively, indicating that the model captures status almost entirely through node activity. When interdependence is included, these correlations drop substantially, though they remain significant, showing that the model shifts away from being purely status-driven. Instead, it begins to rely on interdependence terms to explain structural regularities, redistributing explanatory power from node-level effects to dyadic interactions. Similar patterns of attenuation are observed with betweenness, closeness, and Katz centralities. This shift highlights the importance of modeling interdependence to capture network patterns that cannot be explained by individual activity alone. Non-bootstrapped (observed) correlation scores are reported in SI Figure~\ref{fig:RE_core}.

\subsubsection*{Layered Effects of Interdependence} Social relationships show strong signatures of interdependence, while economic and health ties remain primarily status-driven. To quantify the effect of accounting for interdependence (\textit{full model}) as opposed to capturing only independence and dependence terms (\textit{bias model}), we evaluate predictive performance for each network layer using ten-fold cross-validation (Material and Methods \ref{sec:materials}) across all $176$ villages. Improvements in link prediction performance over the \textit{bias model} thus indicate structural interdependence beyond what can be explained by node status alone. Results for the social, health, and economic layers are shown in Figure \ref{fig:SET_all} while extended versions are provided in the SI Figures \ref{fig:SET_social}, \ref{fig:SET_heal}, and \ref{fig:SET_econ}. Specifically, each pair of Figure \ref{fig:SET_all} compares the average predictive performance of a model that accounts only for dependence and independence ($\textit{bias model}:  r_{ij} = \beta_{i}^{l} + \gamma_{j}^{l}$) with that of a total model that additionally incorporates interdependence (\textit{full model}: $ r_{ij} = \beta_{i}^{l} + \gamma_{j}^{l} + \eta_{ij}$) in terms of Precision-Recall Area Under the Curve (PR-AUC) score, and based on Receiver Operating Characteristic Area Under the Curve (ROC-AUC) score. We define the performance difference as $\text{PR-AUC Difference} = \text{AUC-PR}_{\text{Full}} - \text{PR-AUC}_{\text{Bias}}$ with the ROC-AUC difference defined analogously. Across the three layers, we observe a substantial increase in predictive performance when modeling interdependence, particularly in the \textit{social} layer. Here, the model achieves a PR-AUC gain of 15.07\% (raw increase = 0.092) and a ROC-AUC gain of 11.81\% (raw increase = 0.074). In contrast, gains in the \textit{health} layer are comparatively modest, with a PR-AUC gain of 2.12\% (raw increase = 0.014) and a ROC-AUC gain of 1.37\% (raw increase = 0.009). Similarly, in the \textit{economic} layer, the model achieves a PR-AUC gain of 6.11\% (raw increase = 0.037) and a ROC-AUC gain of 6.16\% (raw increase = 0.036). These differences highlight that interdependence modeling yields the largest improvements in the social domain, while effects in health and economic layers are more moderate. We further validate that the mean difference is statistically significant in each layer using a paired $t$-test, yielding for the social layer ($\text{PR-AUC}: t = 28.43, p < 0.0001$; $\text{ROC-AUC}: t = 27.21, p < 0.0001$), 
for the health layer ($\text{PR-AUC}: t = 4.35, p < 0.0001$; $\text{ROC-AUC}: t = 4.75, p < 0.0001$), 
and for the economic layer ($\text{PR-AUC}: t = 17.32, p < 0.0001$; $\text{ROC-AUC}: t = 12.56, p < 0.0001$). These findings suggest that interdependent ties play a more prominent role in shaping social relationships, whereas health and economic ties appear to be driven primarily by individual status, i.e., the role of an individual as a sender or receiver within the layer, rather than by mutual exchanges.


\subsection*{Role-Dependent Activity, Network Structure, and Interdependent Exchange}

We assess whether the predictive benefit of modeling interdependence relates to behavioral role engagement by correlating average source and target activations ($\Bar{\mathbf{Z}}[l]$, $\Bar{\mathbf{W}}[l]$) with the improvement in link prediction (PR- and ROC-AUC) across 176 village networks. Significant positive correlations emerge in the health ($\tilde{\rho}_{\text{health}} = 0.431$, 95\% CI: [0.286, 0.564]) and economic ($\tilde{\rho}_{\text{economic}} = 0.321$, 95\% CI: [0.169, 0.455]) layers, but not in the social layer ($\tilde{\rho}_{\text{social}} = -0.074$, 95\% CI: [–0.222, 0.086]), suggesting that instrumental interdependence is activity-dependent, while social interdependence is structurally embedded. This pattern holds across both sender and receiver roles, with health correlations significantly stronger than economic (Mann–Whitney U, p < 0.001). These results align with Social Exchange Theory: instrumental exchange requires active role performance, whereas social ties persist through generalized obligation. Full correlations and confidence intervals are in SI Figure~\ref{fig:mean_cor} and SI Figure~\ref{fig:mean_cor_sup}.

Given that reciprocity is often linked to interdependence in Social Exchange Theory, we compute average reciprocity across layers: $\text{social} = 0.311 \pm 0.045 (SD)$, $\text{economic} = 0.323 \pm 0.076 (SD)$, and $\text{health} = 0.153 \pm 0.063 (SD)$ (see SI Figure~\ref{fig:side_by_side_figures}). While the social and economic layers have comparable reciprocity, performance improvement is significantly higher in the social layer. The health layer shows both the lowest reciprocity and the smallest gains. This suggests that reciprocity may signal interdependence but does not fully explain performance differences across layers. To probe this further and generalize across structural patterns, we assess how predictive improvement from modeling interdependence relates to multiple network features; extended results are provided in the SI Appendix.

Overall, reciprocity is a strong signal across social, health, and economic networks. This suggests that mutual interactions are consistently predictive of model performance improvements, regardless of domain. This aligns with the belief that reciprocity is foundational in Social Exchange Theory, about how social obligations form and how trust and cooperation are maintained \cite{homans, Blau1964ExchangeAP}. According to Emerson \citep{Emerson1962PowerDependenceR}, reciprocal ties reflect mutual dependence, which stabilizes the relationship and fosters predictable exchange. Meanwhile, we observe the significance of transitivity and clustering in only the health and economic layers suggests that indirect exchange, norm-based cohesion, and role expectations \cite{Blau1964ExchangeAP} are more predictive in institutionalized or role-structured contexts, but less so in loosely organized social networks. 

Finally, the positive association between model performance and average network degree may suggests that denser networks, those offering more opportunities for interaction and exchange, foster more regular and predictable behavioral patterns. From the perspective of Social Exchange Theory, this supports the view that individual behavior is shaped by the structure of interdependence within which actors are embedded, with richer social contexts leading to stronger norms, expectations, and constraints that guide action \cite{Granovetter1985Economic}.

\section*{Discussion}
Here, we introduce the Multiplex Latent Trade-off (\textsc{MLT}) model , which employs simplex-constrained latent embeddings. This approach provides a comprehensive framework for understanding how interdependence manifests across core domains of life by modeling link formation in multiplex networks through the lens of social exchange. Using a collection of 176 real-world village networks, we show that interdependence the mutual contribution of both sender and receiver to tie formation is not uniformly distributed across relational domains, but is instead domain-specific and structurally nuanced. Our results demonstrate that the \textit{social layer} exhibits the most substantial gains in predictive performance when interdependence is explicitly modeled. These gains are robust across networks and persist regardless of the level of activity in the layer. This suggests that interdependence in social ties is \textit{structurally embedded}, aligning with Blau's notion of generalized exchange \cite{Blau1964ExchangeAP}, where reciprocity and mutuality are supported by social norms rather than individual behavioral strategies. Moreover, we find that predictive gains in the social layer are not significantly correlated with network statistics such as clustering or transitivity, nor with node-level engagement. These patterns reinforce the idea that social interdependence arises from diffuse structural mechanisms rather than local or behavioral drivers. In contrast, the \textit{health} and \textit{economic} layers show weaker average performance improvements when modeling interdependence. However, in these instrumental domains, we observe significant positive correlations between predictive gain and average node activity or layer engagement. 

Importantly, our model incorporates trade-offs in relational investments by learning node-specific activity profiles across layers. These profiles capture the relative prioritization of each domain for every node and reveal how resource allocation influences the emergence of interdependence. While nodes tend to allocate most of their activity to the social layer, only in the health and economic domains do we observe a behavioral-performance link, suggesting that interdependence in these layers is contingent upon active engagement.

We further analyze the role of node-level biases, comparing their alignment with network centralities across model variants. In the bias-only model, sender and receiver biases correlate nearly perfectly with degree, indicating that structure is explained solely through node activity. In the \textit{full model}, these correlations are decreased suggesting that the model reallocates explanatory weight to dyadic relationships. This shift confirms that the \textit{full model} captures interdependence as a distinct structural signal not reducible to degree alone.

Finally, our model provides a principled framework for analyzing trade-offs in multiplex networks while capturing key dimensions of social exchange, offering a deeper understanding of how individuals navigate across various relational domains. This advance is important for bridging the gap between network theory and real-world social behavior, shedding light on the strategic allocation of relationships across social, health, and economic activities. 

While we focused here on social networks, our model can be extended to any type of network where trade-offs are present, highlighting the generalizability of the approach. By endowing latent space modeling with constraints that observations reside on simplices, we propose a novel model to address social exchange, roles, and trade-offs in multiplex networks, while also being highly interpretable. Uncovering such structures in multiplex networks can also facilitate further analysis of the driving forces shaping structures, such as the search for ``relational phenotypes'' and profiles that align with specific distinct roles, and enables meta-data analysis to understand why these structures emerge. Importantly, the modeling components enable a comprehensive understanding of the diverse mechanisms through which ties emerge across layers. They quantify the respective contributions of independence, dependence, and interdependence in the formation of multiplex social relationships. While MLT provides a probabilistic framework for uncovering structural patterns of independence, dependence, and interdependence in multiplex networks, it does not directly model the mechanistic processes of tie formation. Instead, it offers a statistical lens for examining trade-offs and the emergence of roles across relational domains, which may reflect, but do not formally encode, underlying social exchange dynamics.

\section*{Materials and Methods}\label{sec:materials}

Let \(\mathcal{G}=(V, \mathcal{E})\) be a directed multiplex network, where \(N := |V|\) is the number of nodes, and \(\mathcal{E} := \{E^{(1)}, E^{(2)}, \dots, E^{(L)}\}\) represents the set of edge sets, with each \(E^{(l)}\) corresponding to the directed edges in layer \(l \in \{1, 2, \dots, L\}\). The adjacency tensor \(\mathcal{Y}_{N \times N \times L} = \left(y_{i,j}^{(l)}\right)\) encodes the presence of directed edges, where \(y_{i,j}^{(l)} = 1\) if \((i,j) \in E^{(l)}\), and \(y_{i,j}^{(l)} = 0\) otherwise, for all \(1 \leq i, j \leq N\) and \(1 \leq l \leq L\). 

In this study, we focus on multiplex networks that describe human behavior, with layers representing core needs for navigating life, such as social, health, and economic relationships. We note here though, that our approach generalizes to any type of network that can be characterized by trade-offs in multiplexity. We aim at constructing a model accounting for trade-offs in multiplexity while characterizing the three interaction types of \textit{independence}, \textit{dependence}, and \textit{interdependence} under Social Exchange Theory. 

Given a directed multiplex network, a straightforward approach for modeling \textit{independence} and \textit{dependence} driven interactions under a Bernoulli likelihood would define the log-odds as:
  \begin{equation}
  r^l_{ij}=\Big(\beta_{i}^l + \gamma_{j}^l\Big),
  \label{eq:log_odds_bias}
   \end{equation}
where $\beta_i^l, \gamma_j^l \in \mathbb{R}$ denote the node- and layer-specific biases for the sender and receiver, respectively. These biases capture important structural properties of the network, such as degree heterogeneity across both nodes and layers. Given a node pair $\{i, j\}$ and a directed tie formation $i \rightarrow j$, \textit{independence} is captured via the sender bias $\beta_i^l$, where higher values of $\beta_i^l$ indicate that tie formation depends primarily on node $i$'s own initiative. A high $\beta_i^l$ reflects greater independent activity by node $i$ in layer $l$, increasing the likelihood that it initiates ties regardless of the attributes or status of potential recipients $j$. On the other hand, \textit{dependence} is captured by the receiver bias $\gamma_j^l$, where higher values indicate that tie formation from node $i$ to node $j$ is more influenced by the attributes or position of the receiver. In this case, node $j$ is more likely to receive connections due to its high receptivity or prominence, regardless of the status of the sender $i$. Such a model represents a null assumption in which network structure can be predicted solely from individual node-level tendencies, such as activity or popularity, without accounting for dyadic or relational dependencies.

To further account for trade-offs and roles in multiplexity, we can extend the log-odds expression for layer $l$ as:

  \begin{equation}
  r^l_{ij}=\Big(\beta_{i}^l + \gamma_{j}^l +\mathbf{z}_i[l]\cdot s \cdot \mathbf{w}_j[l]\Big),
  \label{eq:log_odds}
   \end{equation}
where the node embeddings $\mathbf{z_i},\mathbf{w_i} \in [0,1]^{L}$ and $\sum_{l=1}^{L} z_{il}=1, \: \sum_{l=1}^{L} w_{il}=1$ each of them belonging to the $(L-1)-\text{simplex}$ or else $\bm{\Delta}_{L-1}$, $s \in \mathbb{R}_+$ is the relative strength of the trade-offs term. In Equation \ref{eq:log_odds}, the node embeddings $\mathbf{Z}$, and $\mathbf{W}$ are constrained to the $(L-1)-\text{simplex}$. In addition, the log-odds expression for layer $l$ acts only in the $l$'th dimension of the embeddings. Essentially, this translates to positioning nodes closer to the $l$'th corner of the $(L-1)-\text{simplex}$ the higher the activity one node has in the specific layer. This leads to the successful modeling of trade-offs of the node activity across each layer. In addition, nodes placed solely in the corner of the simplex can be considered that specific layers' unique role. Furthermore, we extend the characterization of trade-offs and distinct profiles to directed networks, distinguishing between source $\mathbf{z_i}$ and target $\mathbf{w_i}$ node representations, as well as their roles within the multiplex network, thereby satisfying such a decoupling of roles. An overview of the modeling approach is given in Figure \ref{fig:model_overview} panel (a). The dyadic term $\eta_{ij} = \mathbf{z}_i[l] \cdot s \cdot \mathbf{w}_j[l]$ captures interdependence in tie formation, on top of trade-offs in multiplexity. Specifically, the directed connection $i \rightarrow j$ is influenced by the joint characteristics, efforts, and activity of both nodes, and how these interact in a layer-specific manner. Importantly, interdependence differs from reciprocity, as $\eta_{ij} \neq \eta_{ji}$; it models the mutual contribution to tie formation while preserving the directionality of the relationship.

The model described by Equation \ref{eq:log_odds} so far successfully accounts for trade-offs across layers, effectively characterizing the global structure of the multiplex network. Accounting for characterization of the type of tie emergence, under social exchange. However, it does not account for detailed layer-specific structures. The existence of layer-specific multi-scale structures, can potentially affect or bias the emergence of trade-offs in the multiplex networks. To address this limitation, we extend the model's log-odds expression as follows:

 \begin{equation}
r^l_{ij} = \beta_i^l + \gamma_j^l +  \mathbf{z}_i[l] \cdot \Big[\sum_{h=0}^{H}s^l_h\big(\bm{u}_i^{l,h} \times \bm{v}_j^{l,h^T}\big)\Big] \cdot \mathbf{w}_j[l].
  \label{eq:log_odds2}
\end{equation}
We thereby introduce layer-specific hierarchical embeddings $\bm{v}_i^{l,h}, \bm{u}_j^{l,h} \in [0,1]^{D_h}$, each constrained to lie on the $(D_h{-}1)$-simplex or $\bm{\Delta}_{D_h{-}1}$, when $h>1$; that is, $\sum_{d_h=1}^{D_h} v_{i,d_h}^{l,h} = 1$ and $\sum_{d_h=1}^{D_h} u_{j,d_h}^{l,h} = 1$, where $h$ denotes the height of the hierarchy (i.e., the resolution level of the multi-scale structure). Formally, we define the hierarchical membership embeddings $\bm{u}_i^{l,h}, \bm{v}_j^{l,h} \in \Delta_{2^h - 1}$ as the Kronecker product:
\begin{center}
\begin{minipage}{0.45\textwidth}
\begin{equation}
\bm{u}_i^{l,h} = \bigotimes_{h'=0}^{h-1} \bm{b}_i^{l,h'},
\end{equation}
\end{minipage}
\hfill
\begin{minipage}{0.45\textwidth}
\begin{equation}
\bm{v}_j^{l,h} = \bigotimes_{h'=0}^{h-1} \bm{g}_j^{l,h'},
\end{equation}
\end{minipage}
\end{center}

where each $\bm{b}_i^{l,h'},\bm{g}_j^{l,h'} \in \Delta_1$ represents binary branching proportions at hierarchy level $h'$. This construction ensures that $\bm{u}_i^{l,h}$, $\bm{v}_j^{l,h}$ captures all $2^h$ possible paths through the hierarchy, assigning to each a probability given by the product of branching weights along that path.

These embeddings allow the model to capture community structure or local relational patterns within each layer at varying levels of resolution, thereby increasing expressiveness. To account for the fact that different layers may exhibit structure at different scales, we introduce the parameter $s^l_h\geq 0$ describing the strength in which links are being generated at the h'th level of the hierarchy. This parameter models the responsibility of each hierarchical level $h$ for explaining the observed link structure.  We reparameterize this as $s^l_h = s \cdot \pi^l_h$, where $\sum_h \pi^l_h = 1$ and $\pi^l_h \geq 0$. Here, $s \geq 0$ is a global strength parameter that controls the overall influence of the hierarchical structure across the model, while the $\pi^l_h$ values determine how this strength is distributed across hierarchy levels within each layer. This reparameterization shares a single global strength $s$ across all layers, thereby constraining each layer to have the same overall capacity for hierarchical expression. It enables trade-offs in how different layers allocate representational emphasis across hierarchy levels, while maintaining a globally consistent capacity. Notably, we include the $h=0$ level (i.e., the root of the tree), which corresponds to the absence of detailed community structure in a layer. This ultimately leads to the log-odds expression in Equation~\ref{eq:log_odds2}. Without loss of generality, we assume binary splits in the hierarchy such that $D_0 = 1$, $D_1 = 2$, $D_2 = 4$, and in general $D_H = 2^H$. To model the correct hierarchical ordering in practice, we define the embeddings at the finest level $D_H$ as parameters, and construct coarser levels by iterative summation: every pair of adjacent dimensions at level $h$ is summed to form a dimension at level $h{-}1$, thereby preserving the nested multi-resolution structure. Without loss of generality, we define the number of binary splits of the hierarchical tree as the natural logarithm $H = \log N$, where $N$ is the number of nodes in the network. An overview of this modeling characteristic is given in Figure \ref{fig:model_overview} panel (b). 

To optimize this model, we define the normalized Bernoulli negative log-likelihood as our loss function to be minimized:

 \begin{equation}
 \mathcal{L}(\mathcal{H})=-\log P(\mathcal{Y}|\mathcal{H}))= -\frac{1}{(N(N-1))}
\sum_{l=1}^L\Big(\sum_{y_{ij}^{(l)}=1}r^{l}_{ij}\;-  \sum_{i\neq j}
\log(1 + \exp{r^{l}_{ij}})\Big ) ,
\end{equation}
where we utilize the $\text{softplus}(x) = \log(1 + e^x)$, and the log-odds tensor \(\bm{R}_{N \times N \times L} = \left(r_{i,j}^{l}\right)\).

\subsection*{Complexity analysis}
The computational complexity of the proposed \textsc{MLT} model scales as $\mathcal{O}(LN^2)$, where $L$ denotes the number of layers in the multiplex network and $N$ the number of nodes. This is due to the need to compute the pairwise Gram matrix between nodes. Such complexity poses challenges for analyzing large-scale networks. To overcome this limitation, we introduce an unbiased estimator of the log-likelihood via random sampling \citep{nakis2023characterizing,nakis2024signed}. Specifically, gradient updates are computed based on the log-likelihood of a block corresponding to a randomly sampled subset $S$ of network nodes (sampled with replacement at each iteration). This strategy enhances scalability by reducing both time and space complexity to $\mathcal{O}(LS^2)$, thereby enabling efficient analysis of large-scale graphs.

\subsection*{Data}
\textbf{Real-world networks:}\label{sec:real_nets}
Our data come from a sociocentric network study of $22,584$ people in $176$ geographically isolated villages in western Honduras \cite{honduras_vils}. This research was approved by the Yale IRB and by the Honduran Ministry of Health (Protocol \# 1506016012), and all participants provided informed consent upon enrolling in the study. Using this empirical dataset, we construct $176$ binary-directed multiplex networks with no self-loops or multiple ties within the same layer. Each network contains 3-layers, describing the social, health, and economic ties. Throughout this analysis, we denote with $l=1$ the social layer, with $l=2$ the health layer, and with $l=3$ the economic layer. For the construction of the social layer we define a connection based on three name generators (s1: “Who do you spend your free time with?” s2: “Who is your closest friend?” and s3: “Who do you discuss personal matters with?”). For the health layer we make use of two name generators (h1: “Who would you ask for advice about health related matters?” and h2: “Who comes to you for health advice?”). Finally, for the economic layer we make use of two name generators (e1: “Who would you feel comfortable asking to borrow $200$ lempiras from if you needed them for the day?” and e2: “Who do you think would be comfortable asking you to borrow $200$ lempiras for the day?”). For the name generators h2 and e2, we reverse the direction of nominations so that individuals providing the service are treated as targets in the directed ties within those layers. We consider the nodes that define the strongly connected component of the collapsed (aggregated) multiplex directed network to make sure that information flows in every node in the network during learning.

\subsection*{Training details}
All experiments were conducted on a high-performance computing node equipped with two Intel(R) Xeon(R) Gold 6346 CPUs running at $3.10$GHz. We adopt the AdamW optimizer \cite{ADAMW} with a weight decay of $0.01$ and an initial learning rate of $0.1$, combined with a learning rate scheduler that reduces the rate by a factor of $0.5$ if the training likelihood plateaus for more than ten optimization steps. We fit a model separately for each of the 176 villages. Training is terminated when the learning rate falls below $10^{-7}$, indicating convergence to a local minimum. All model parameters are randomly initialized. We employ a sequential learning strategy: for the first 2000 optimization steps, we optimize only the node-specific biases $\beta_{i}^l$ and $\gamma_{j}^l$, after which the remaining model parameters are jointly learned. To mitigate the effect of local minima, we run the model five times with different initializations and select the configuration that achieves the lowest training log-likelihood.

We assessed link‐prediction performance via ten‐fold cross‐validation. In each fold, we generated 100 independent negative, edge sets, each set containing as many non‐edges as there were positives in that fold’s test set, and computed the metric of interest on each set. We then averaged these 100 values to obtain a stable performance estimate for the fold. Finally, we averaged the ten fold‐level estimates to yield the overall predictive performance for each network. For correlation analyses, we report Spearman coefficients (Pearson coefficients are provided in the SI). To ensure robustness across varying test set configurations, we first compute the relevant metric (e.g., performance gain or node-level quantity) within each fold and then average the values across all ten folds before calculating the correlation. This averaging procedure reduces variance due to fold-specific test sets and enables more stable and interpretable comparisons across networks and layers. Correlations pertaining to compositional data were also computed using a centered log-ratio (CLR) transformation, which had minimal influence on the results, confirming that closure bias effects \cite{aitch} were negligible.

\bibliographystyle{plainnat}  
\bibliography{references}

\section*{Acknowledgements}
This work was supported by the Bill and Melinda Gates Foundation
NIH grant R01-AG081814 from the National Institute on Aging.

\section*{Author Contributions}
N.N., S.L., N.A.C., and M.M. designed research; N.N., S.L., N.A.C., and M.M. performed research; N.N. analyzed data; N.A.C. supervised data collection; and N.N., S.L., N.A.C., and M.M. wrote the paper.

\section*{Ethics Declaration}
The authors declare no competing interest.

\clearpage
\setcounter{figure}{0}
\setcounter{table}{0}
\renewcommand{\thefigure}{\arabic{figure}}
\renewcommand{\thetable}{\arabic{table}}

\captionsetup[figure]{name=SI Figure}
\captionsetup[table]{name=SI Table}

\onecolumn
\section*{Supporting Information}

\subsection*{ Additional networks behavioral and structural model interpretation; village networks \# 1, \# 51, and \# 175.}

\begin{figure*}[b!]

\centering
  \centering
  \adjustbox{valign=b}{%
    \begin{minipage}[t]{0.62\textwidth}
      \begin{subfigure}[t]{\linewidth}
        \includegraphics[width=1\linewidth]{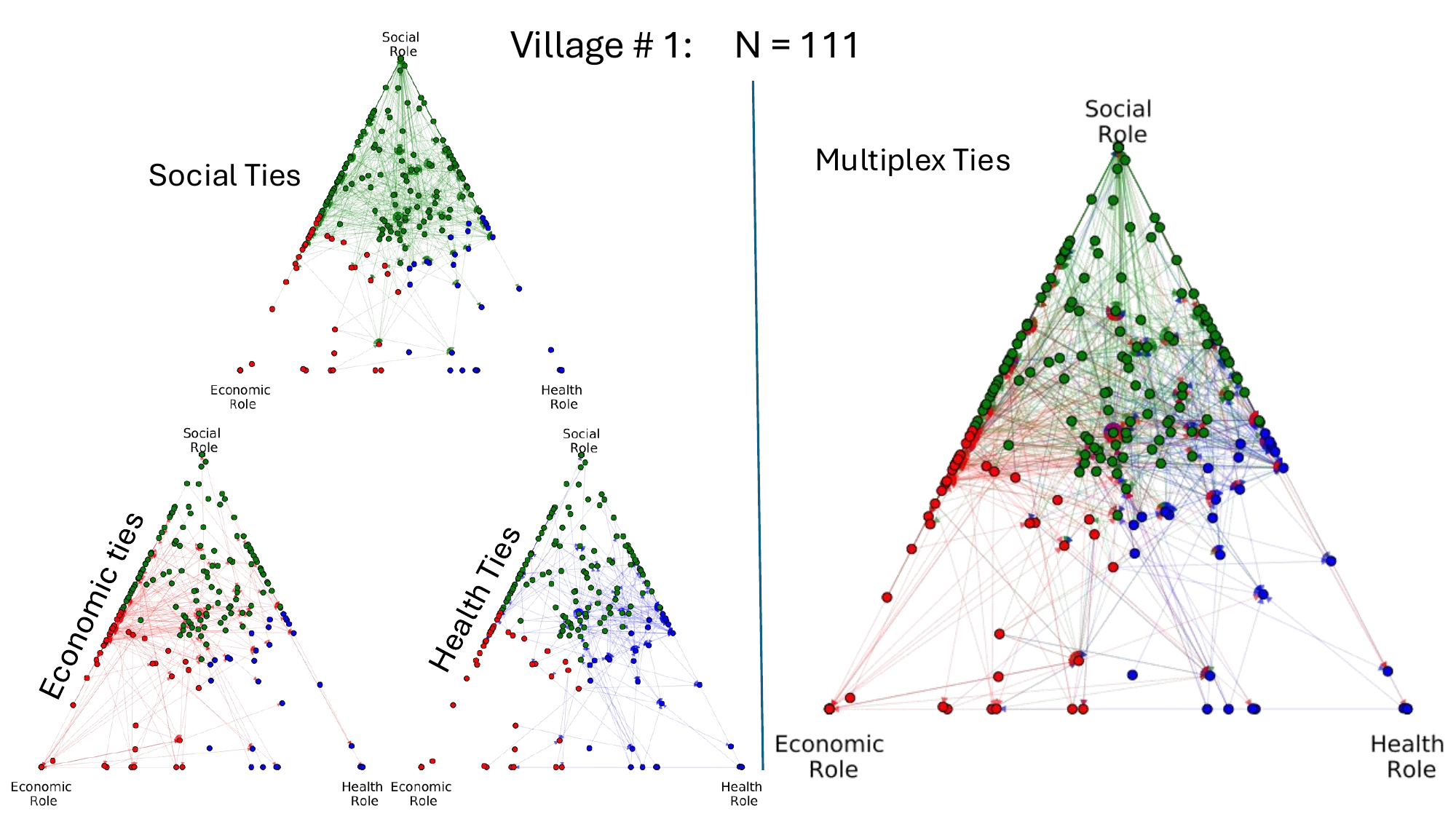}
        \caption{\small Role simplex visualizations for village $\# 1$. Each point represents a node, colored by dominant role. Left: tie-specific roles. Right: multiplex network with all ties overlaid. }
      \end{subfigure}
      \vspace{1em}

      \begin{subfigure}[t]{\linewidth}
        \includegraphics[width=1\linewidth]{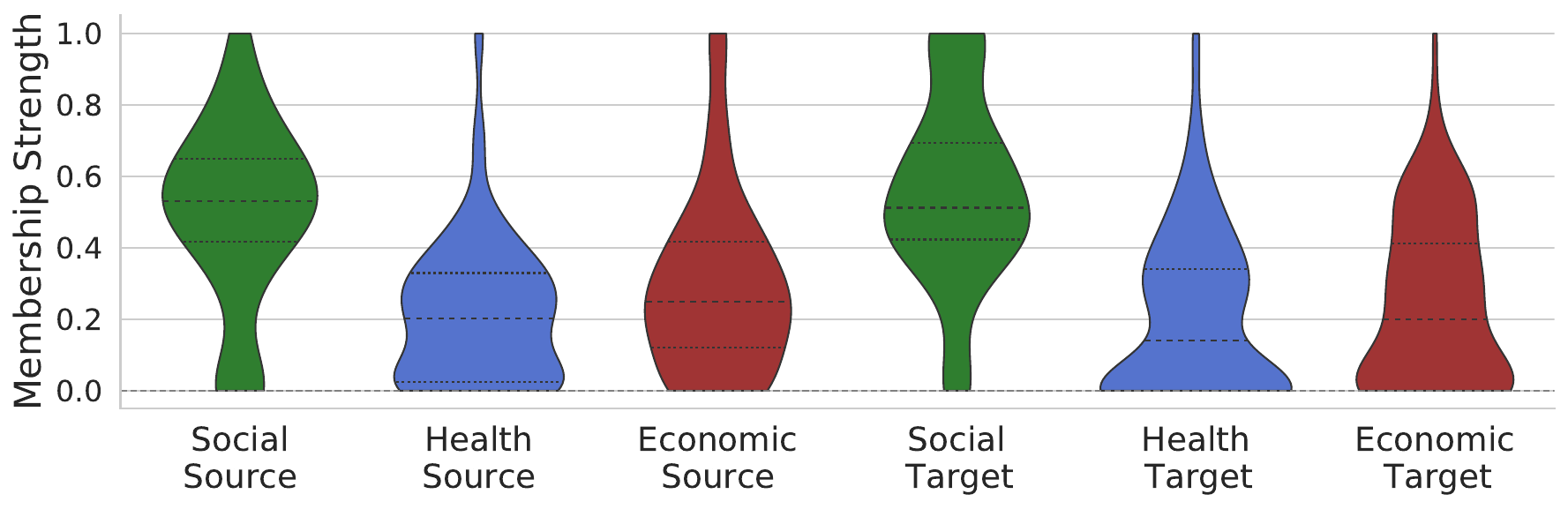}
        \caption{\small Violin plots of node-role membership strengths for source and target positions across domains. }
      \end{subfigure}
    \end{minipage}
  }\hfill
  \adjustbox{valign=b}{%
    \begin{minipage}[t]{0.36\textwidth}

\makebox[1\linewidth]{\centering \textbf{Multi-scale Layer Structure:}}
\\[1ex]
 \makebox[0.32\linewidth]{\centering \textbf{Social}} \hfill
  \makebox[0.32\linewidth]{\centering \textbf{Health}} \hfill
  \makebox[0.32\linewidth]{\centering \textbf{Economic}} \\[-1ex]
    
\centering

  \begin{minipage}[c]{\linewidth}
    \begin{minipage}[c]{0.04\linewidth}
      \centering
      \rotatebox{90}{\small \textbf{$D_{h=1}=2$}}
    \end{minipage}%
    \hfill
     \captionsetup[subfigure]{skip=-5pt}
    \begin{subfigure}[c]{0.3\linewidth}
\includegraphics[width=\linewidth]{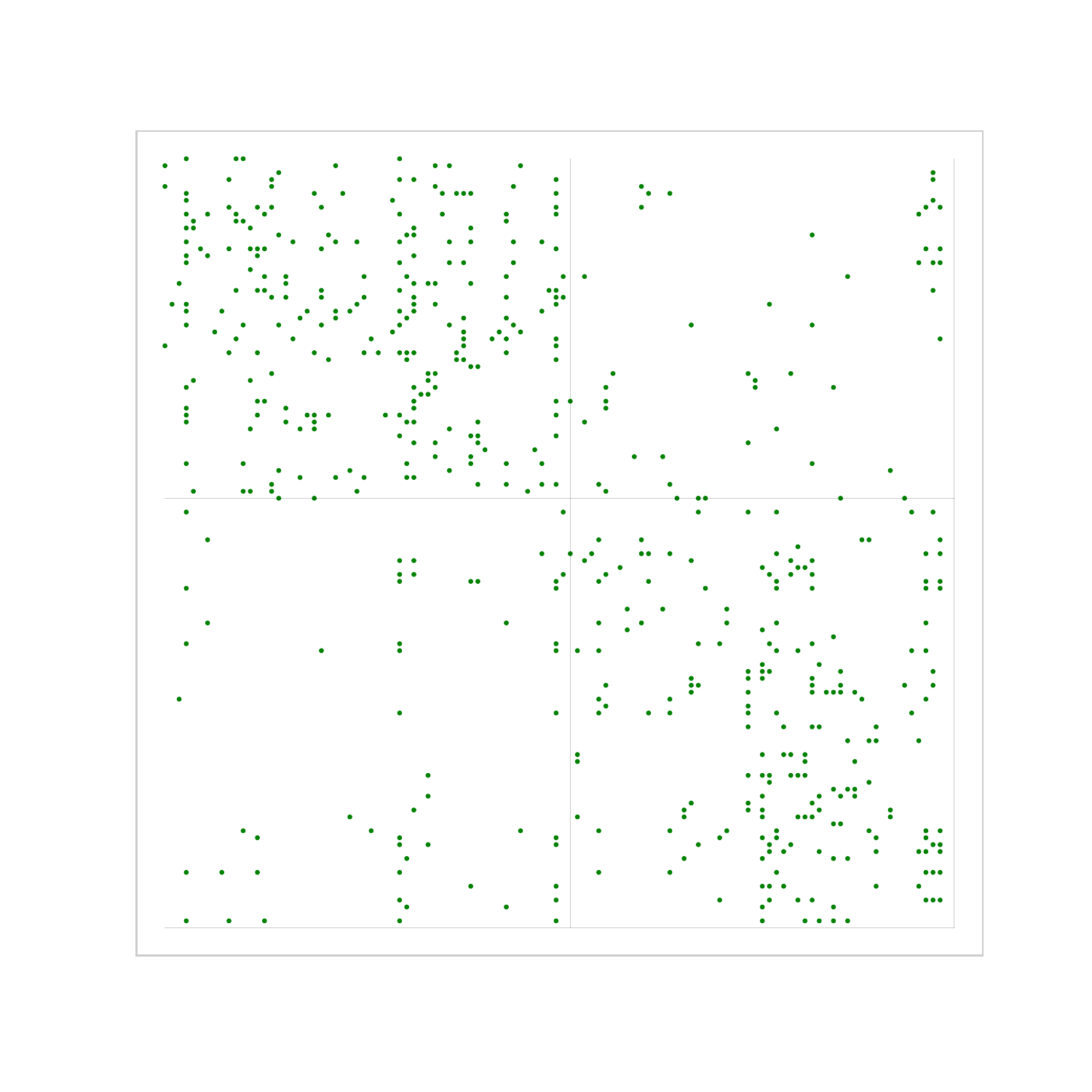} \caption{\tiny\\$s_1^1=0.001$}
    \end{subfigure}\hfill
     \captionsetup[subfigure]{skip=-5pt}
    \begin{subfigure}[c]{0.3\linewidth}
\includegraphics[width=\linewidth]{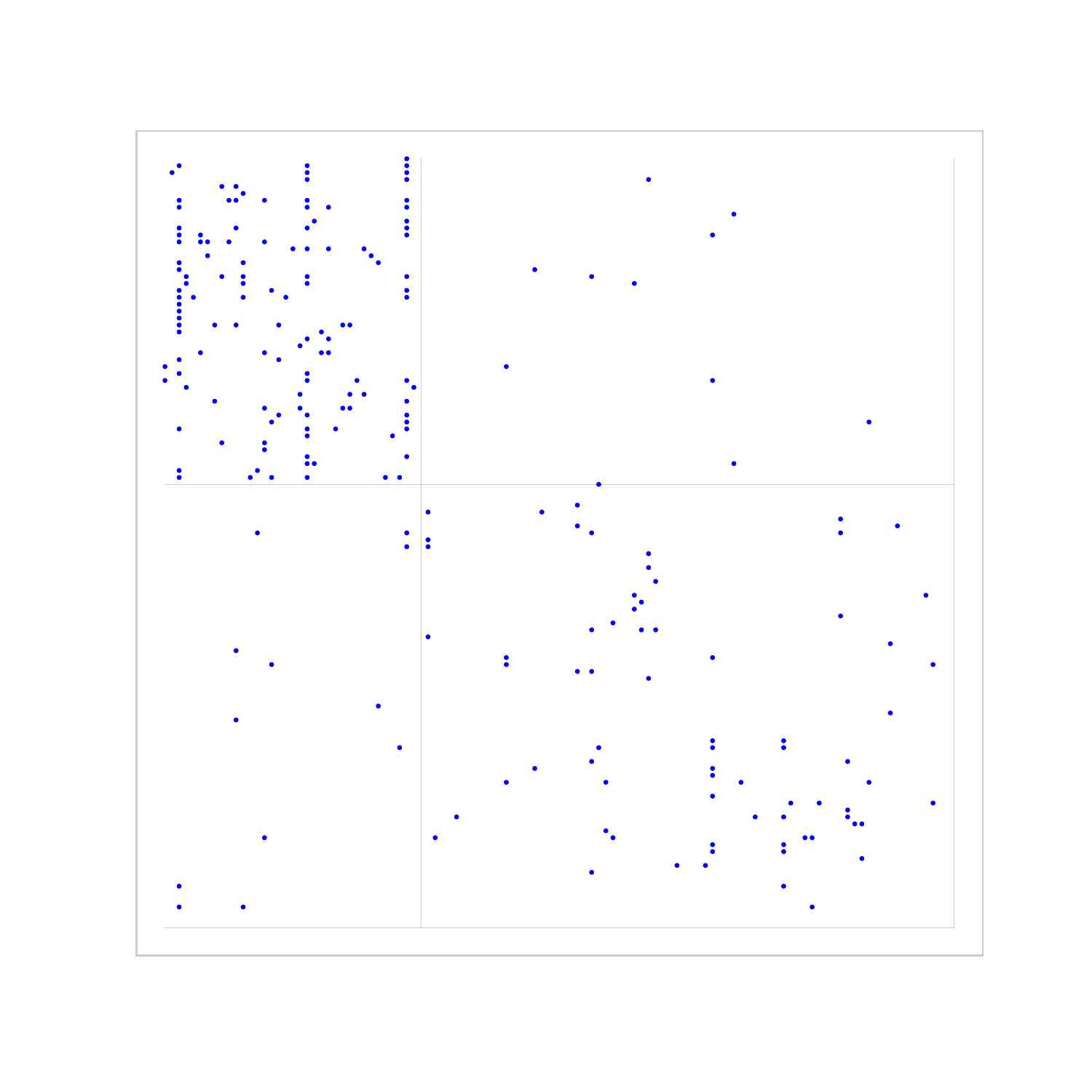} \caption{\tiny\\$s_1^2=0.005$}
    \end{subfigure}\hfill
    \captionsetup[subfigure]{skip=-5pt}
    \begin{subfigure}[c]{0.3\linewidth}
\includegraphics[width=\linewidth]{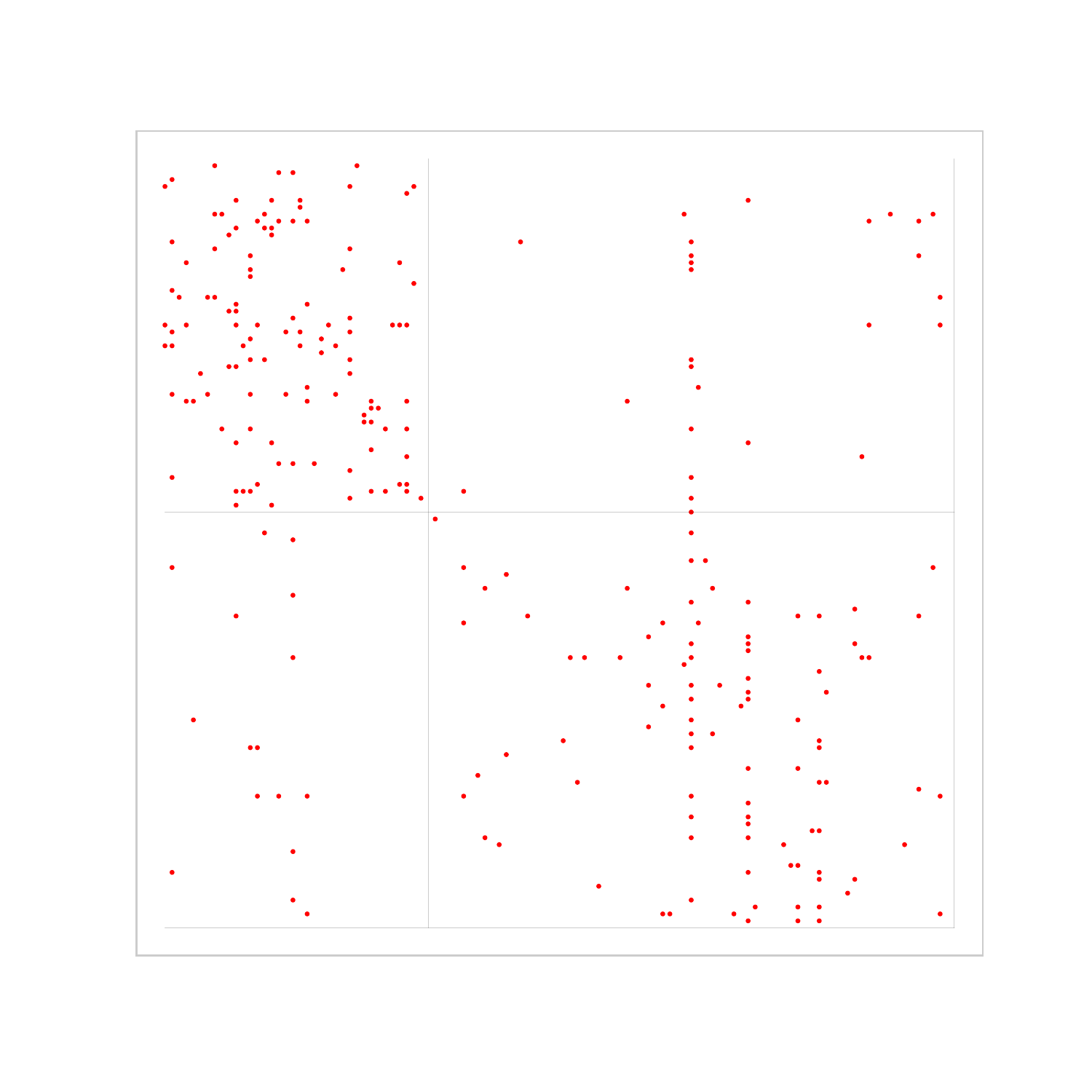}
      \caption{\tiny\\$s_1^3=0.001$}
    \end{subfigure}
  \end{minipage}

  \begin{minipage}[c]{\linewidth}
    \begin{minipage}[c]{0.04\linewidth}
      \centering
      \rotatebox{90}{\small \textbf{$D_{h=2}=4$}}
    \end{minipage}%
    \hfill
     \captionsetup[subfigure]{skip=-5pt}
    \begin{subfigure}[c]{0.3\linewidth}
      \includegraphics[width=\linewidth]{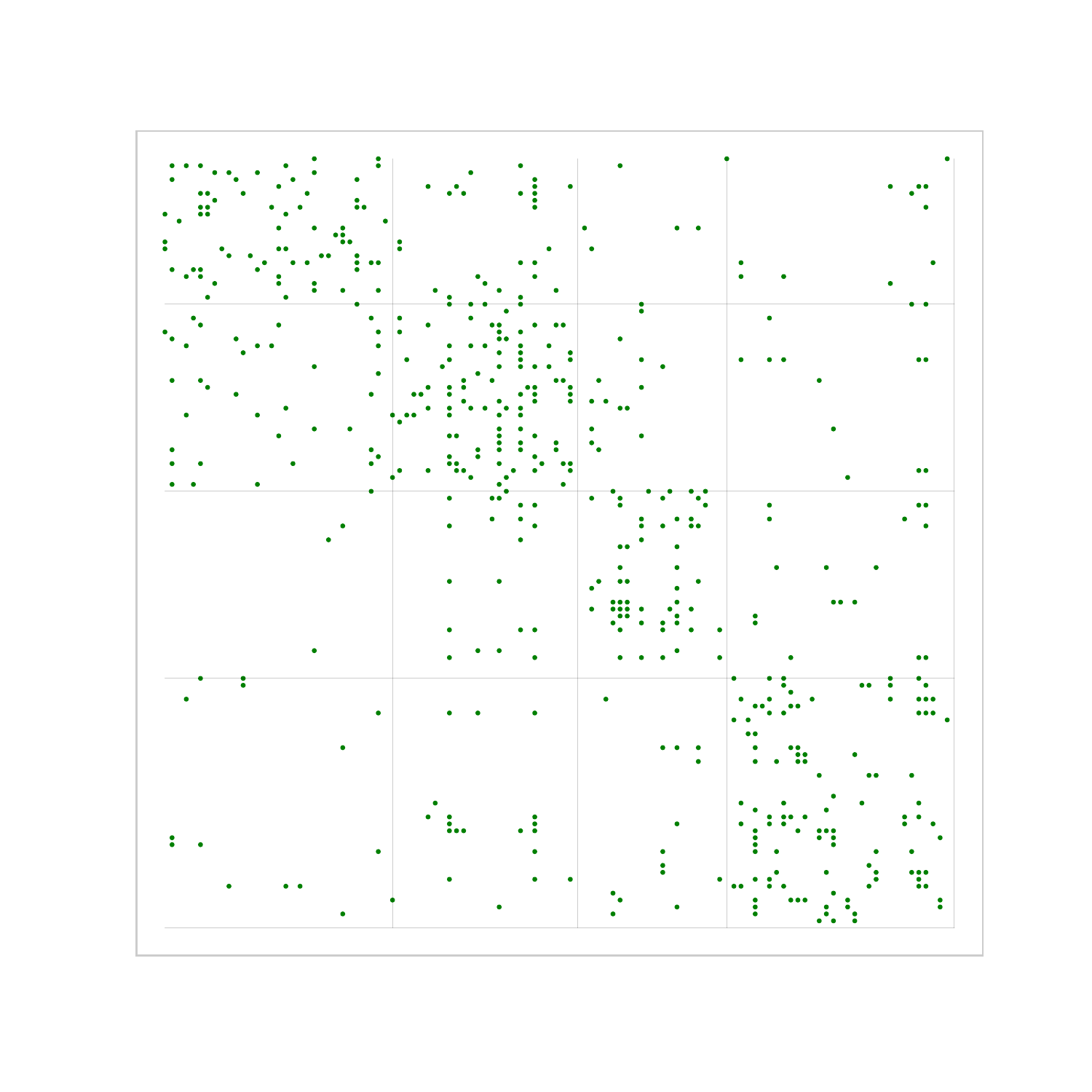}
       \caption{\tiny\\$s_2^1=0.030$}
    \end{subfigure}\hfill
     \captionsetup[subfigure]{skip=-5pt}
    \begin{subfigure}[c]{0.3\linewidth}
      \includegraphics[width=\linewidth]{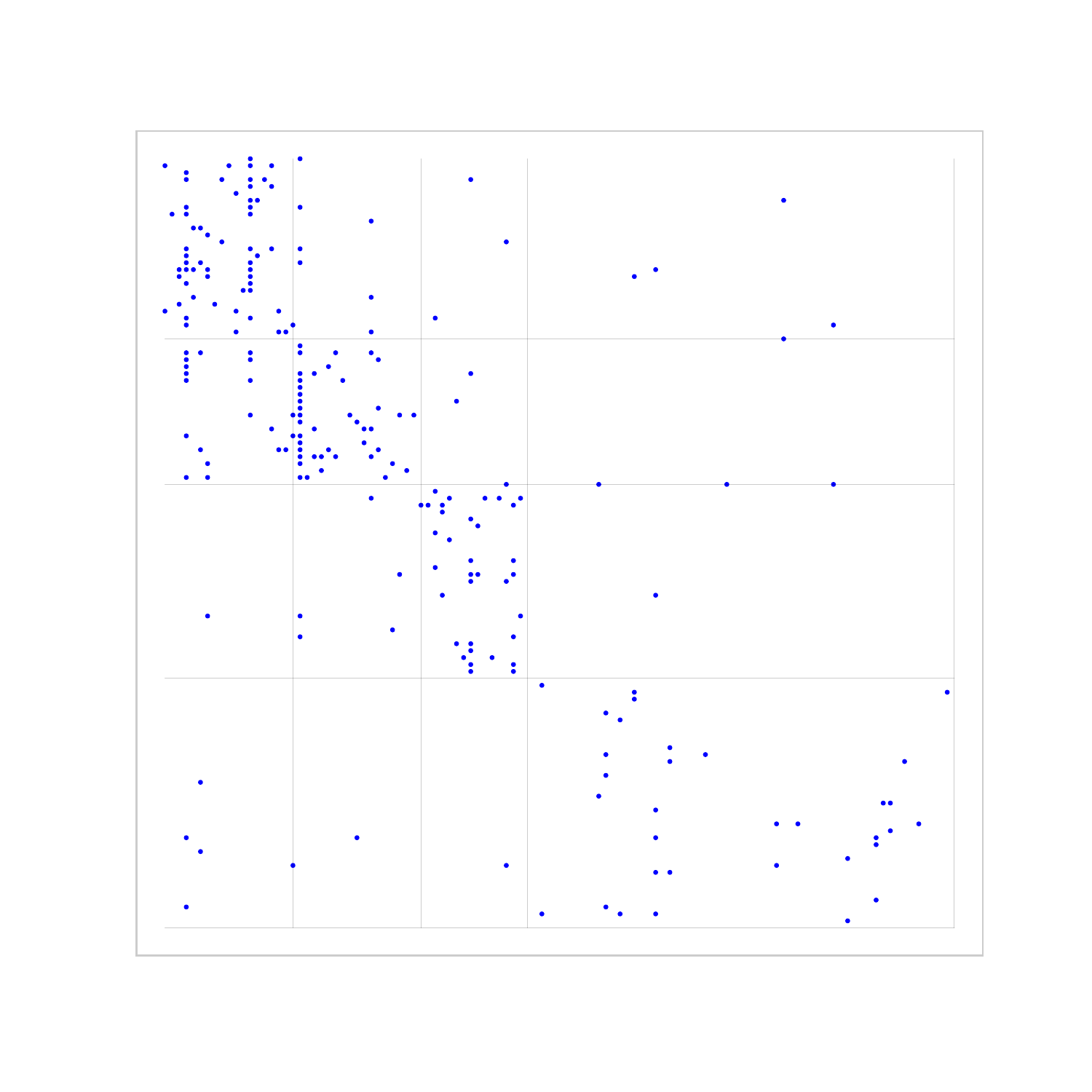}
       \caption{\tiny\\$s_2^2=2.925$}
    \end{subfigure}\hfill
     \captionsetup[subfigure]{skip=-5pt}
    \begin{subfigure}[c]{0.3\linewidth}
      \includegraphics[width=\linewidth]{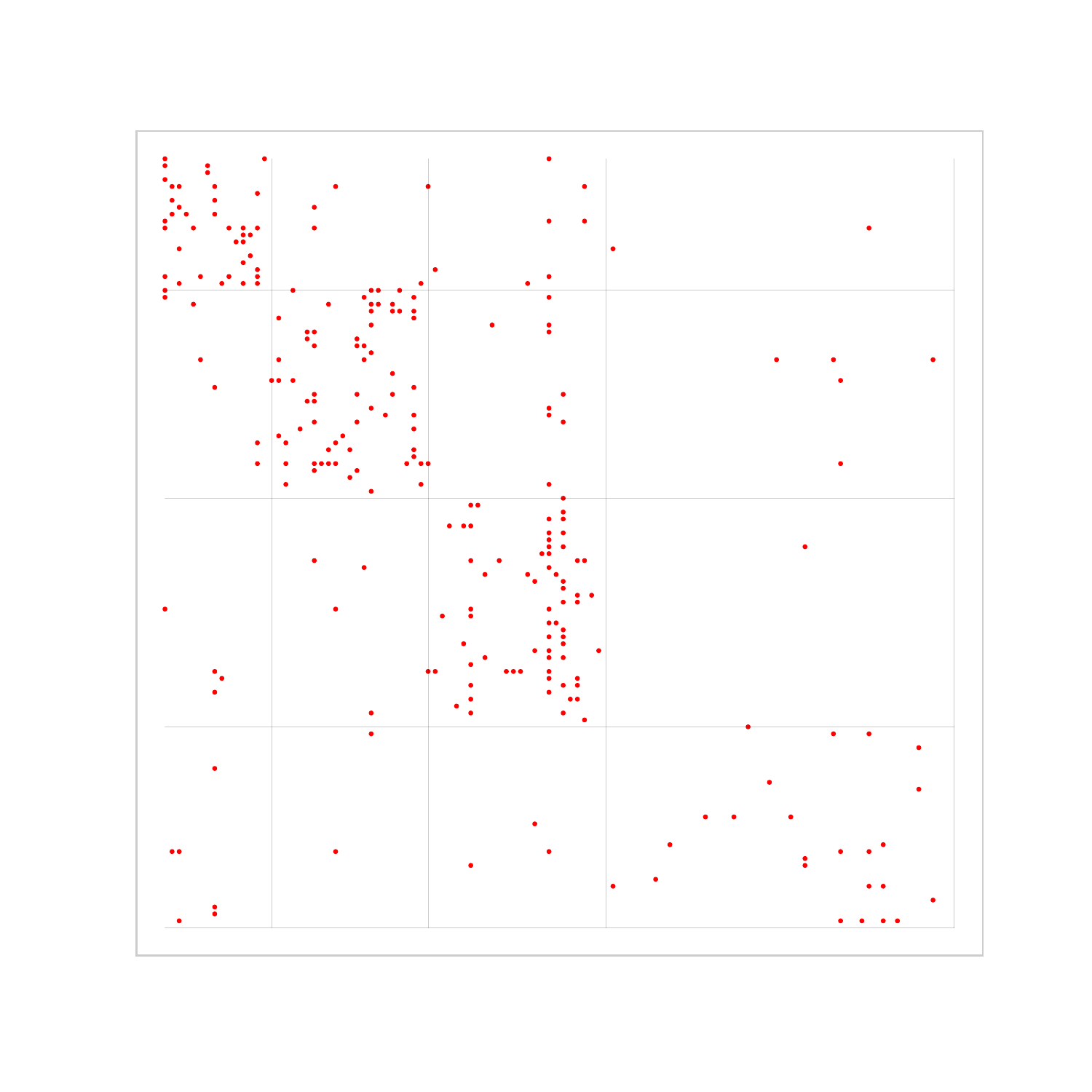}
       \caption{\tiny\\$s_2^3=0.119$}
    \end{subfigure}
  \end{minipage}
  
  \begin{minipage}[c]{\linewidth}
    \begin{minipage}[c]{0.04\linewidth}
      \centering
      \rotatebox{90}{\small \textbf{$D_{h=3}=8$}}
    \end{minipage}%
    \hfill
     \captionsetup[subfigure]{skip=-5pt}
    \begin{subfigure}[c]{0.3\linewidth}
      \includegraphics[width=\linewidth]{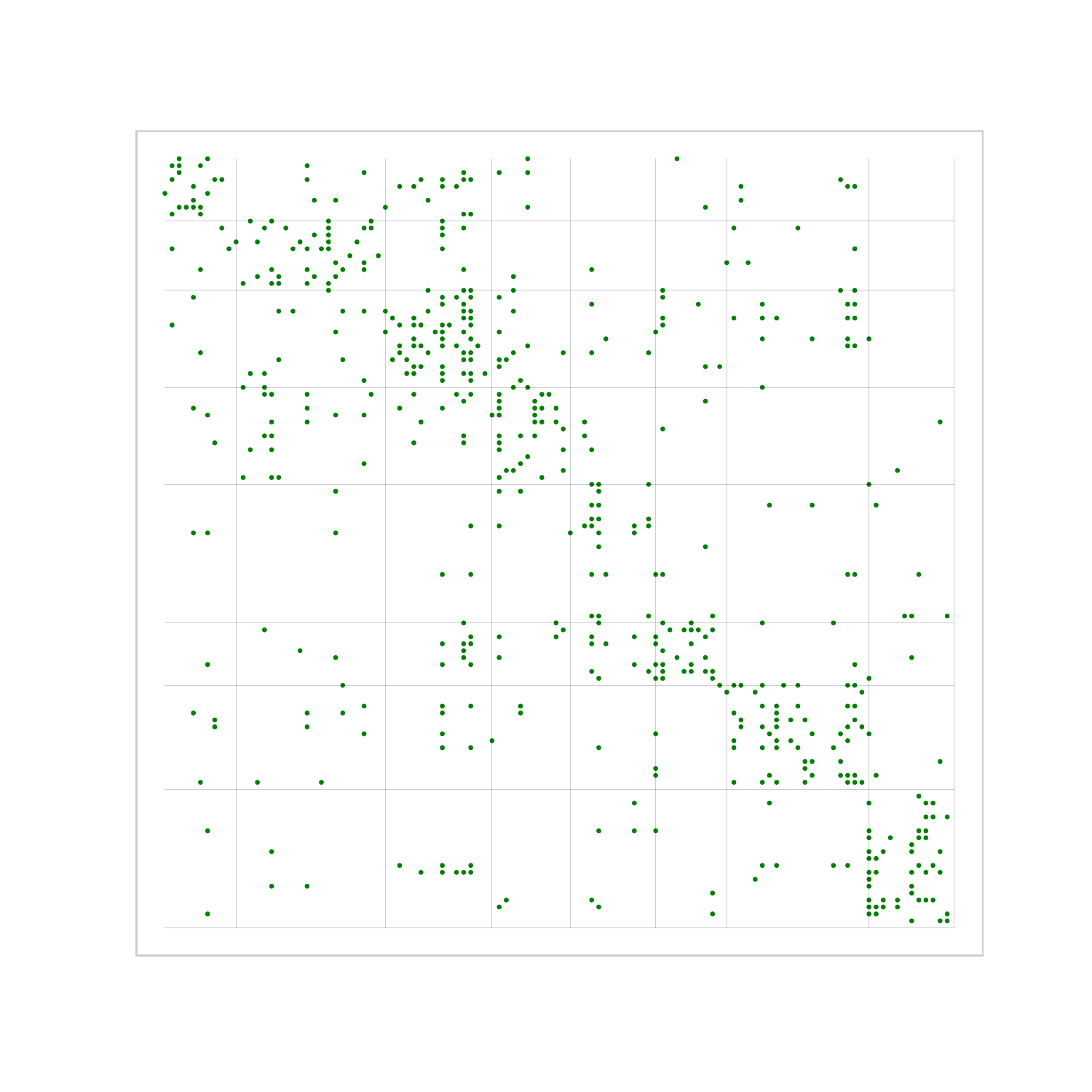}
       \caption{\tiny\\$s_3^1=9.311$}
    \end{subfigure}\hfill
     \captionsetup[subfigure]{skip=-5pt}
    \begin{subfigure}[c]{0.3\linewidth}
      \includegraphics[width=\linewidth]{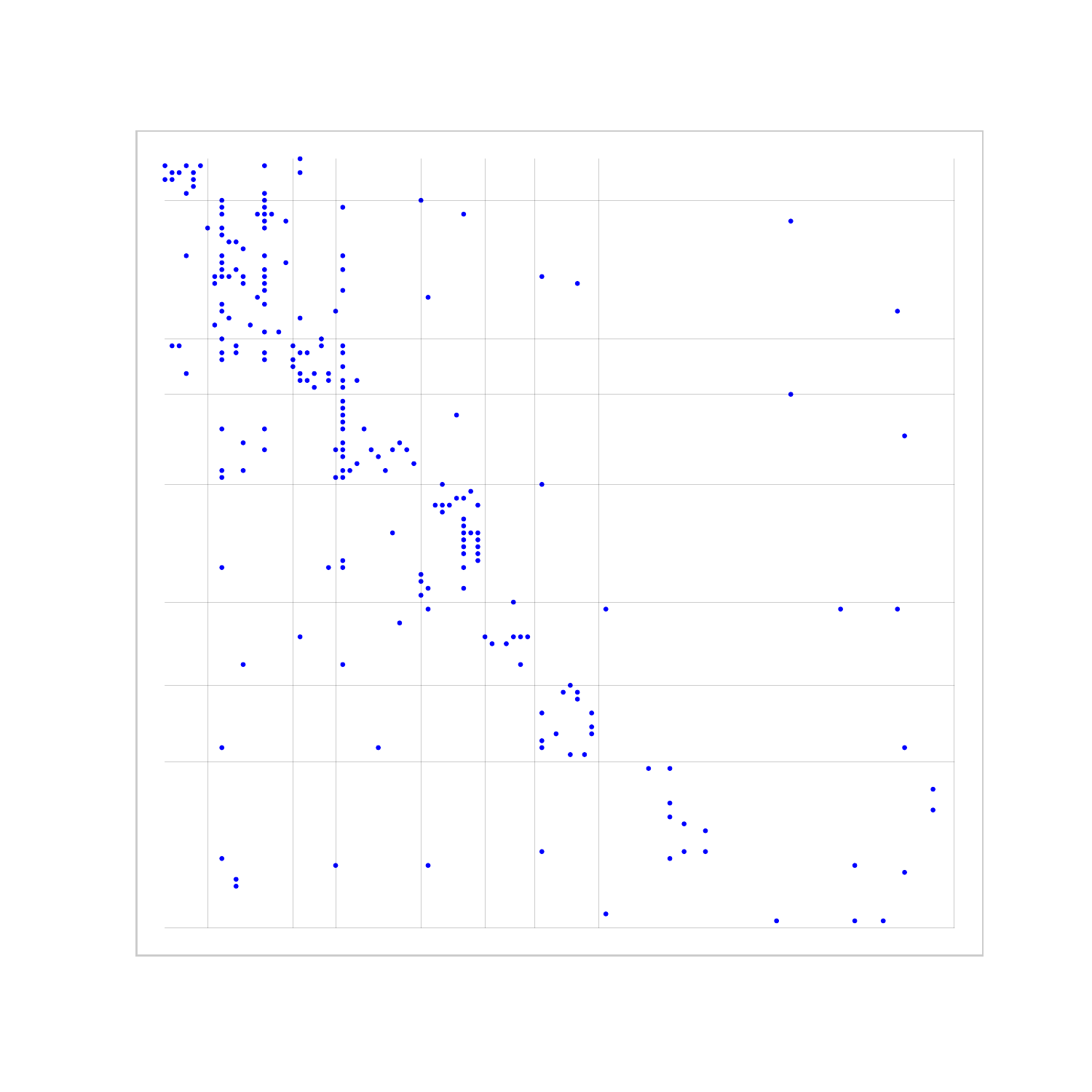}
       \caption{\tiny\\$s_3^2=1.899$}
    \end{subfigure}\hfill
     \captionsetup[subfigure]{skip=-5pt}
    \begin{subfigure}[c]{0.3\linewidth}
      \includegraphics[width=\linewidth]{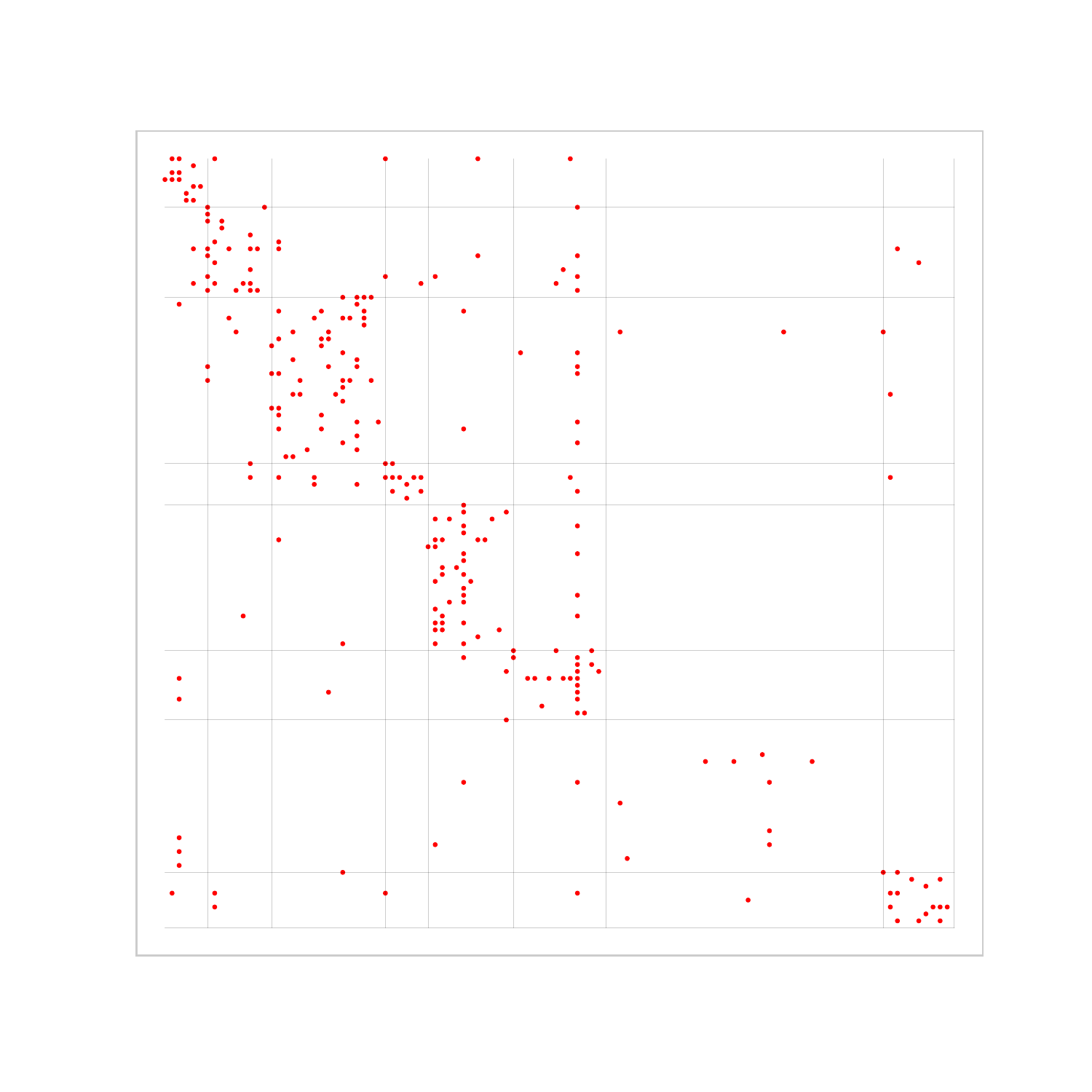}
       \caption{\tiny\\$s_3^3=4.846$}
    \end{subfigure}
  \end{minipage}
  
  \begin{minipage}[c]{\linewidth}
    \begin{minipage}[c]{0.04\linewidth}
      \centering
      \rotatebox{90}{\small \textbf{$D_{h=4}=16$}}
    \end{minipage}%
    \hfill \captionsetup[subfigure]{skip=-5pt}
    \begin{subfigure}[c]{0.3\linewidth}
      \includegraphics[width=\linewidth]{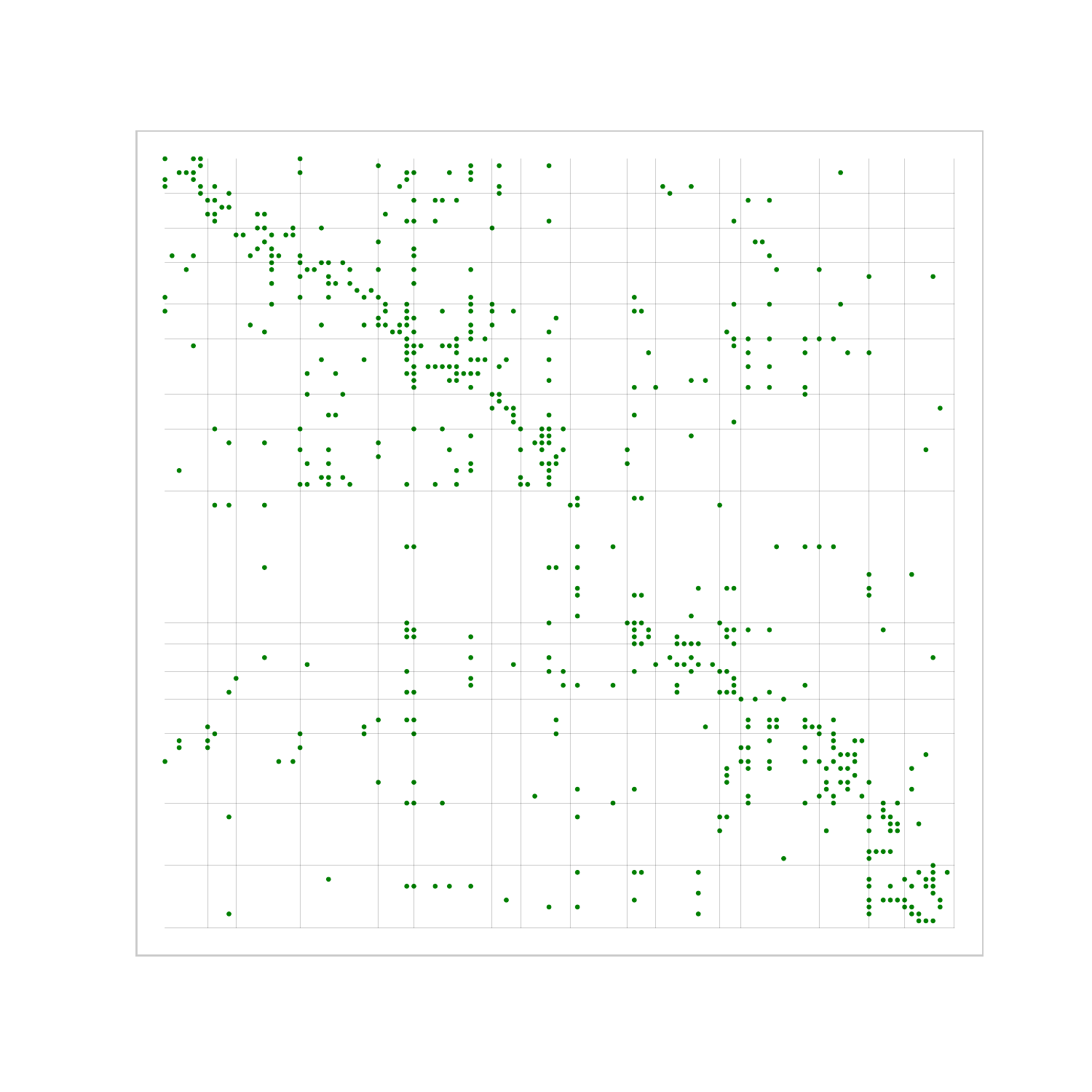} \caption{\tiny\\$s_4^1=0.157$}
    \end{subfigure}\hfill
     \captionsetup[subfigure]{skip=-5pt}
    \begin{subfigure}[c]{0.3\linewidth}
      \includegraphics[width=\linewidth]{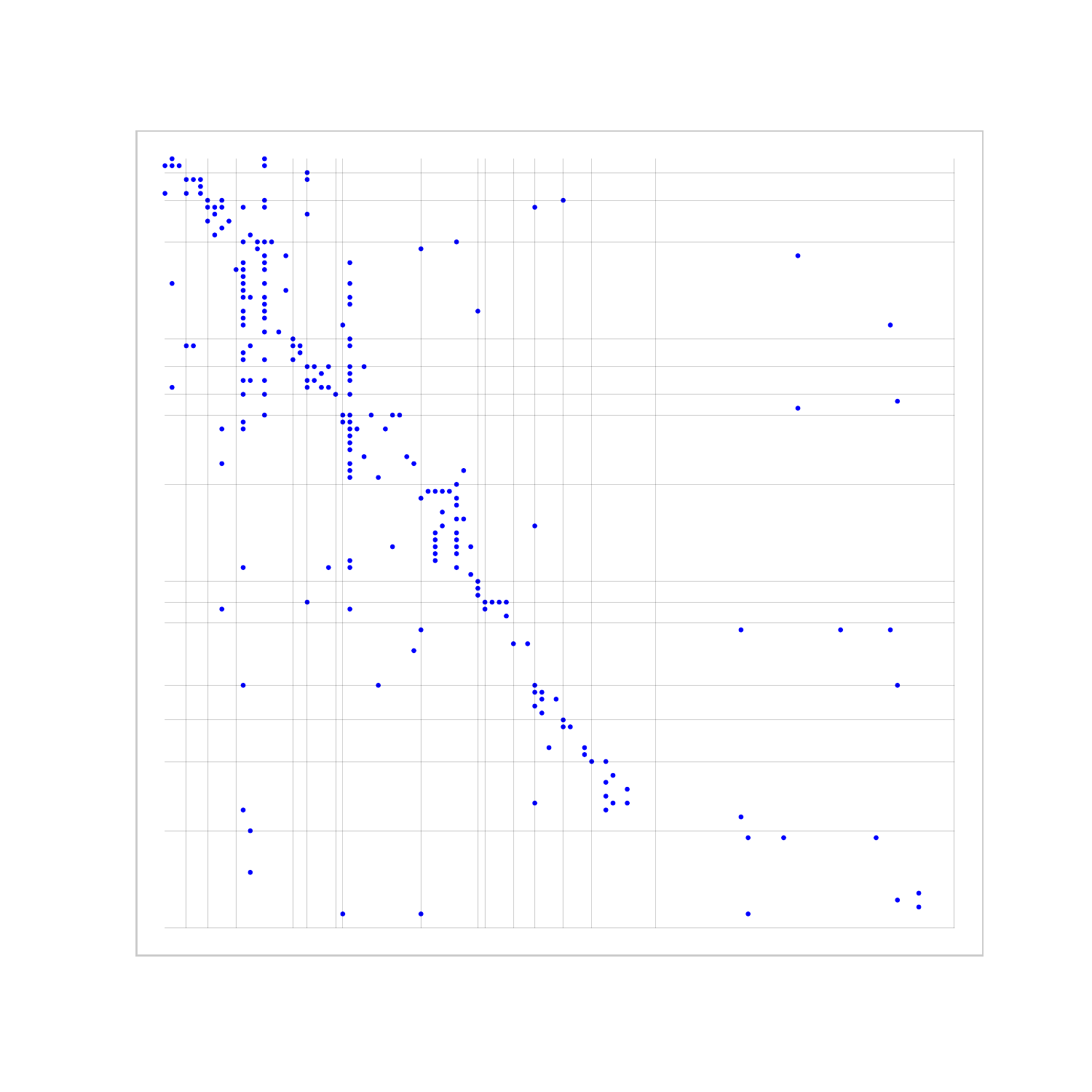} \caption{\tiny\\$s_4^2=16.332$}
    \end{subfigure}\hfill
     \captionsetup[subfigure]{skip=-5pt}
    \begin{subfigure}[c]{0.3\linewidth}
      \includegraphics[width=\linewidth]{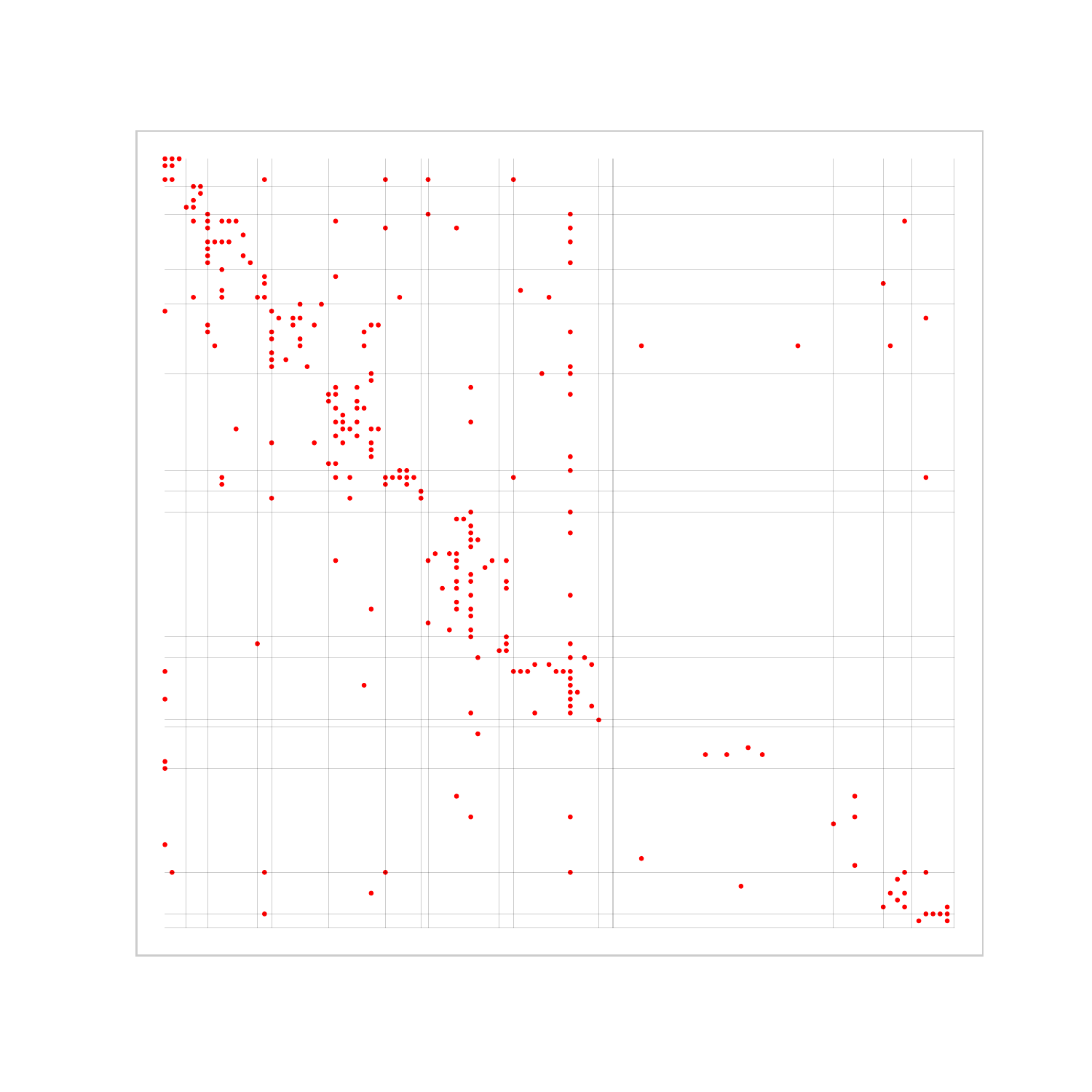} \caption{\tiny\\$s_4^3=5.939$}
    \end{subfigure}
  \end{minipage}

  \begin{minipage}[c]{\linewidth}
    \begin{minipage}[c]{0.04\linewidth}
      \centering
      \rotatebox{90}{\small \textbf{$D_{h=5}=32$}}
    \end{minipage}%
    \hfill  \captionsetup[subfigure]{skip=-5pt}
    \begin{subfigure}[c]{0.3\linewidth}
      \includegraphics[width=\linewidth]{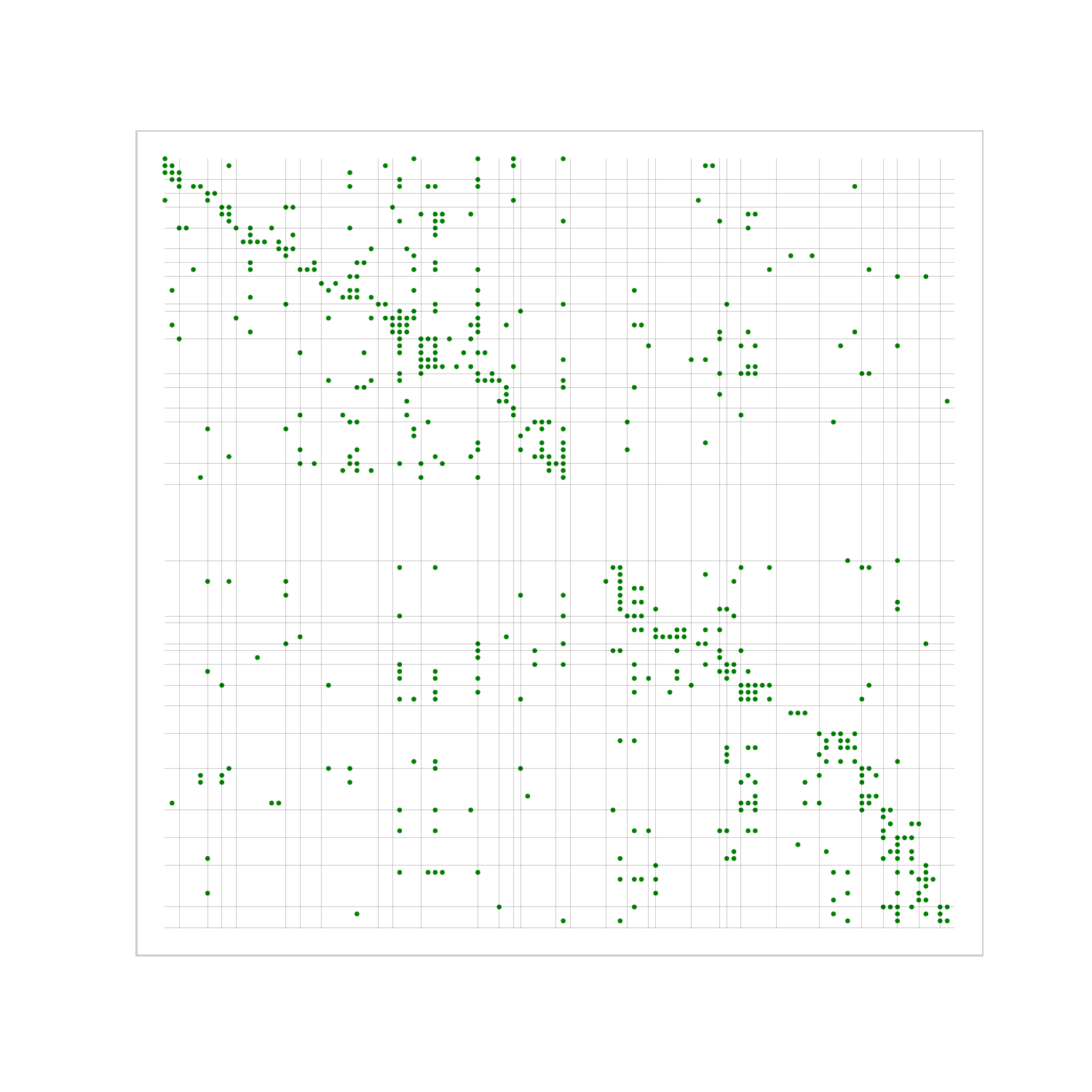}
      \caption{\tiny\\$s_5^1=81.420$}
    \end{subfigure}\hfill  \captionsetup[subfigure]{skip=-5pt}
    \begin{subfigure}[c]{0.3\linewidth}
      \includegraphics[width=\linewidth]{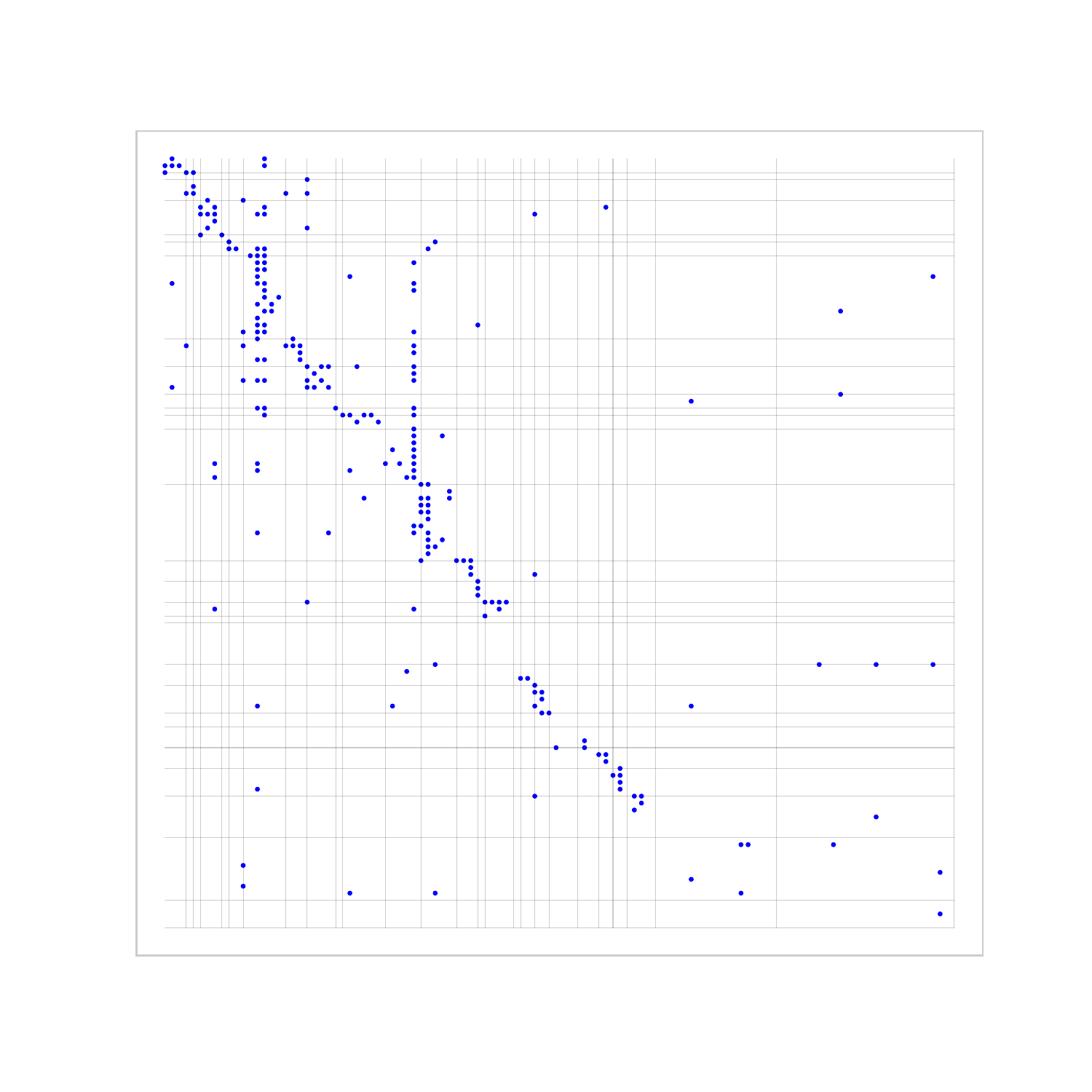}
      \caption{\tiny\\$s_5^2=67.544$}
    \end{subfigure}\hfill  \captionsetup[subfigure]{skip=-5pt}
    \begin{subfigure}[c]{0.3\linewidth}
      \includegraphics[width=\linewidth]{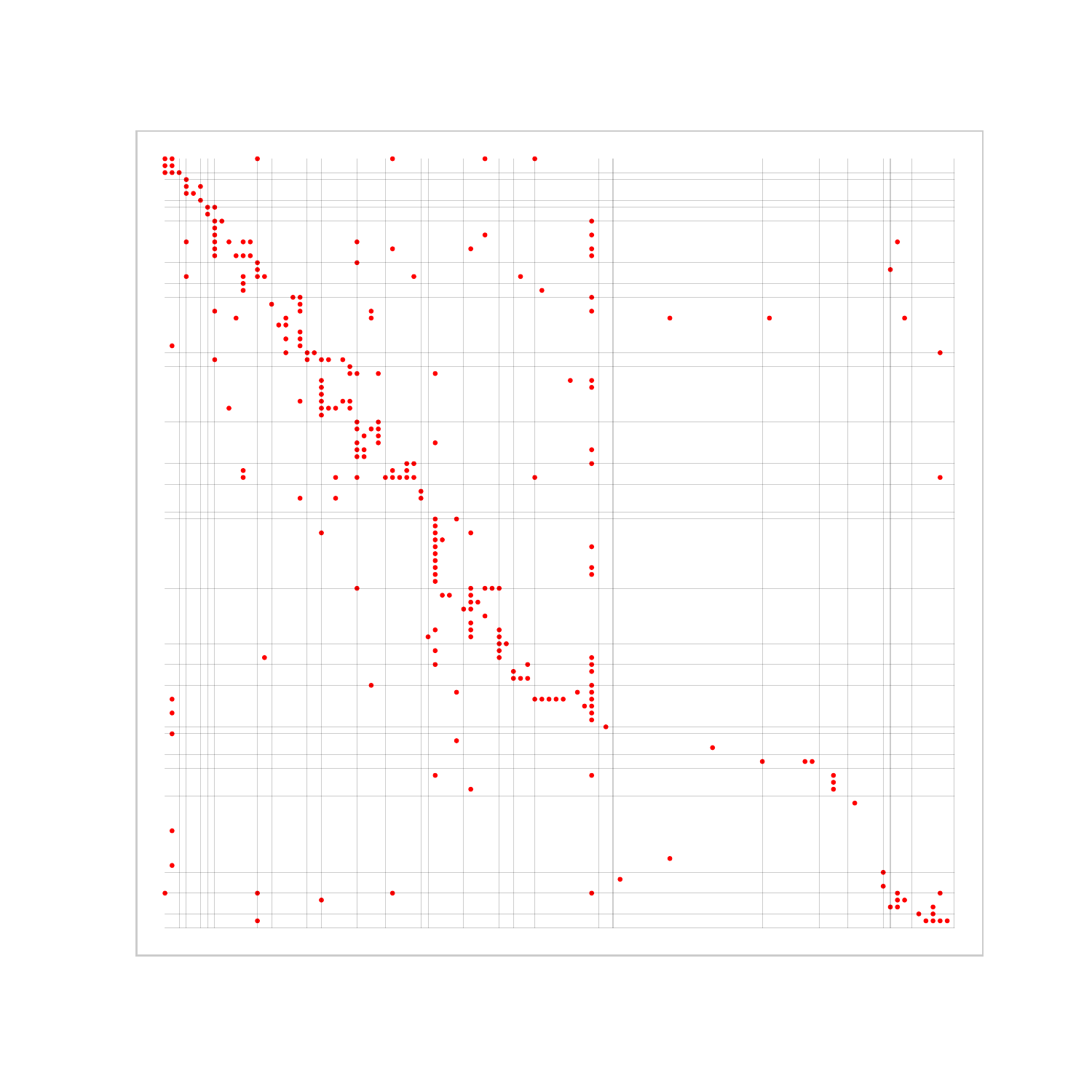}
      \caption{\tiny\\$s_5^3=80.027$}
    \end{subfigure}
  \end{minipage}

      \hfill 
    \end{minipage}
  }

  \centering
  \begin{subfigure}[t]{0.49\textwidth}
    \includegraphics[width=1\linewidth]{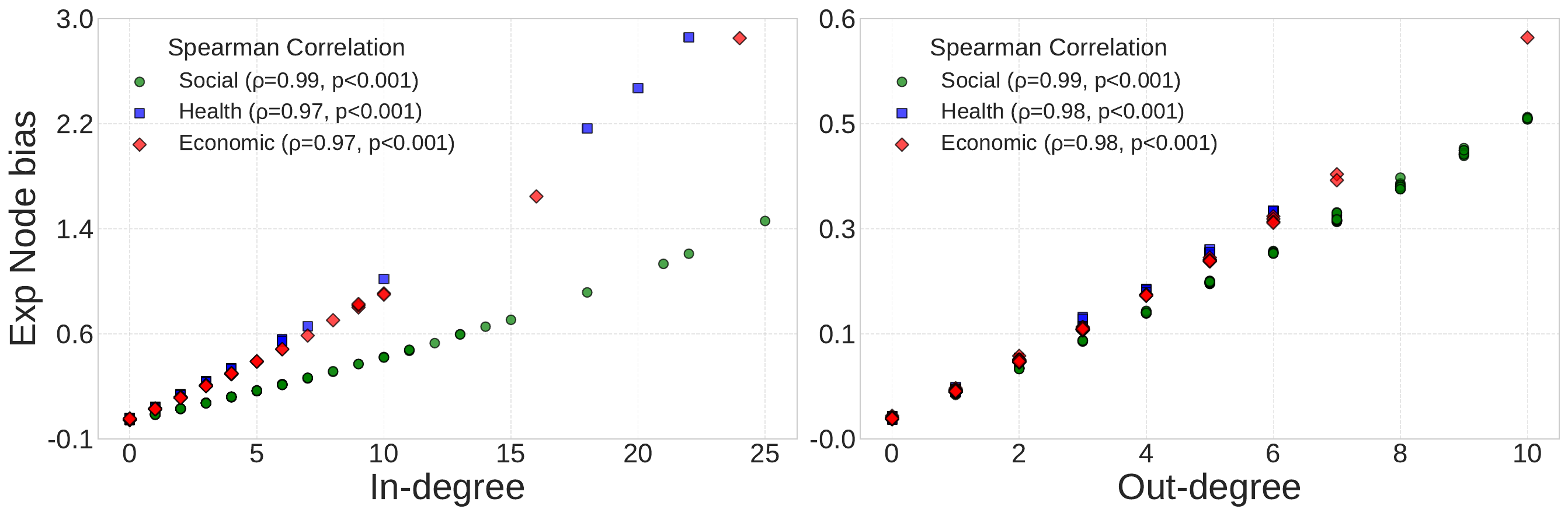}
    \caption{\small Bias model (dependence and independence): inferred node biases vs. in- and out-degree across layers.  }
  \end{subfigure}\hfill
  \begin{subfigure}[t]{0.49\textwidth}
    \includegraphics[width=1\linewidth]{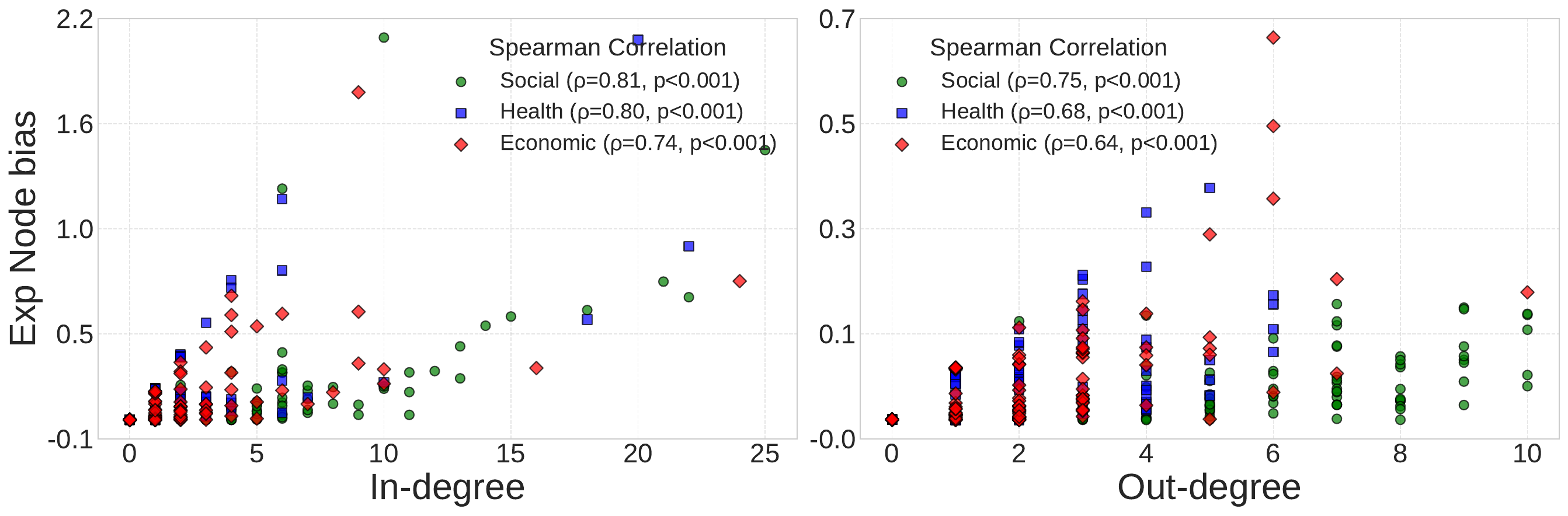}
    \caption{\small Full model (adds interdependence): inferred node biases vs. in- and out-degree across layers. }
  \end{subfigure}
  \begin{subfigure}[t]{0.16\textwidth}
    \includegraphics[width=\linewidth]{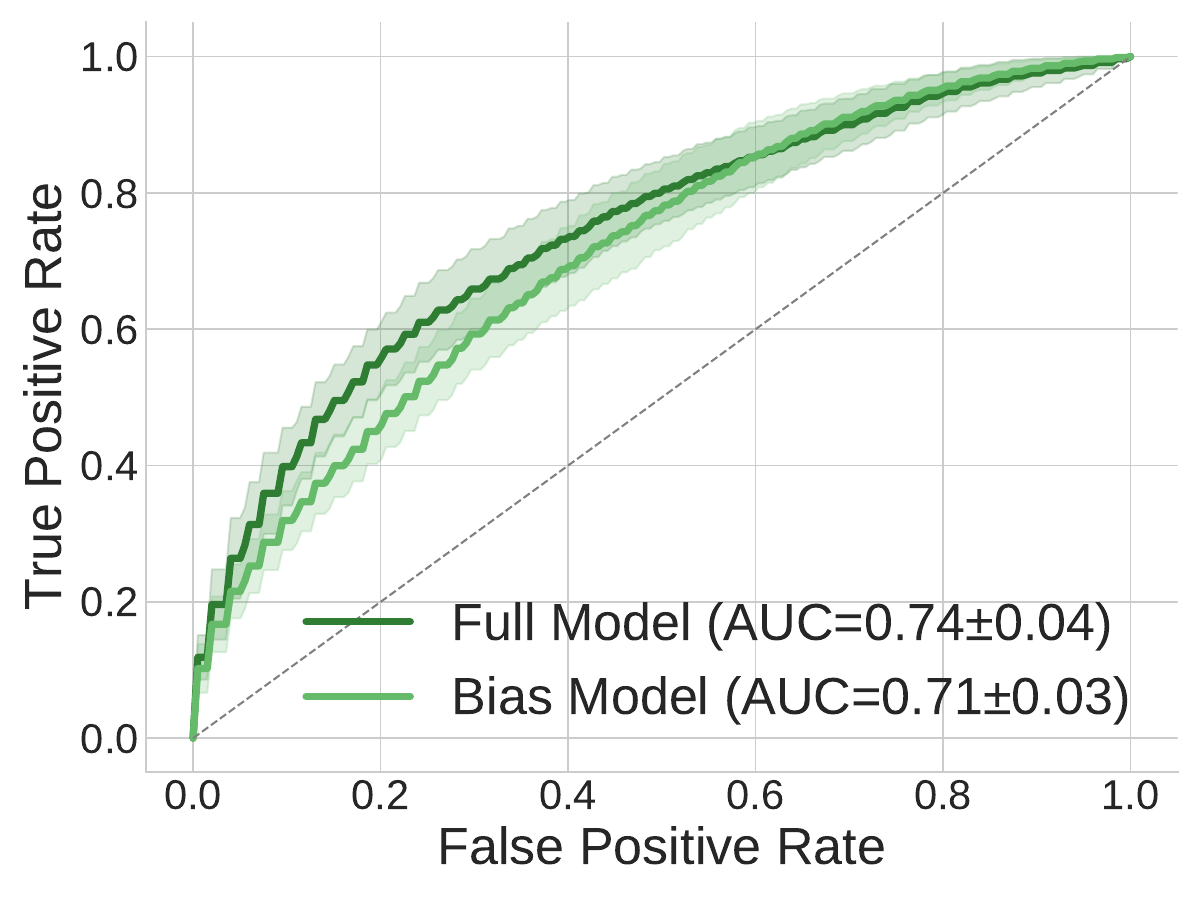}
    \caption{\small ROC Social}
  \end{subfigure}\hfill
  \begin{subfigure}[t]{0.16\textwidth}
    \includegraphics[width=\linewidth]{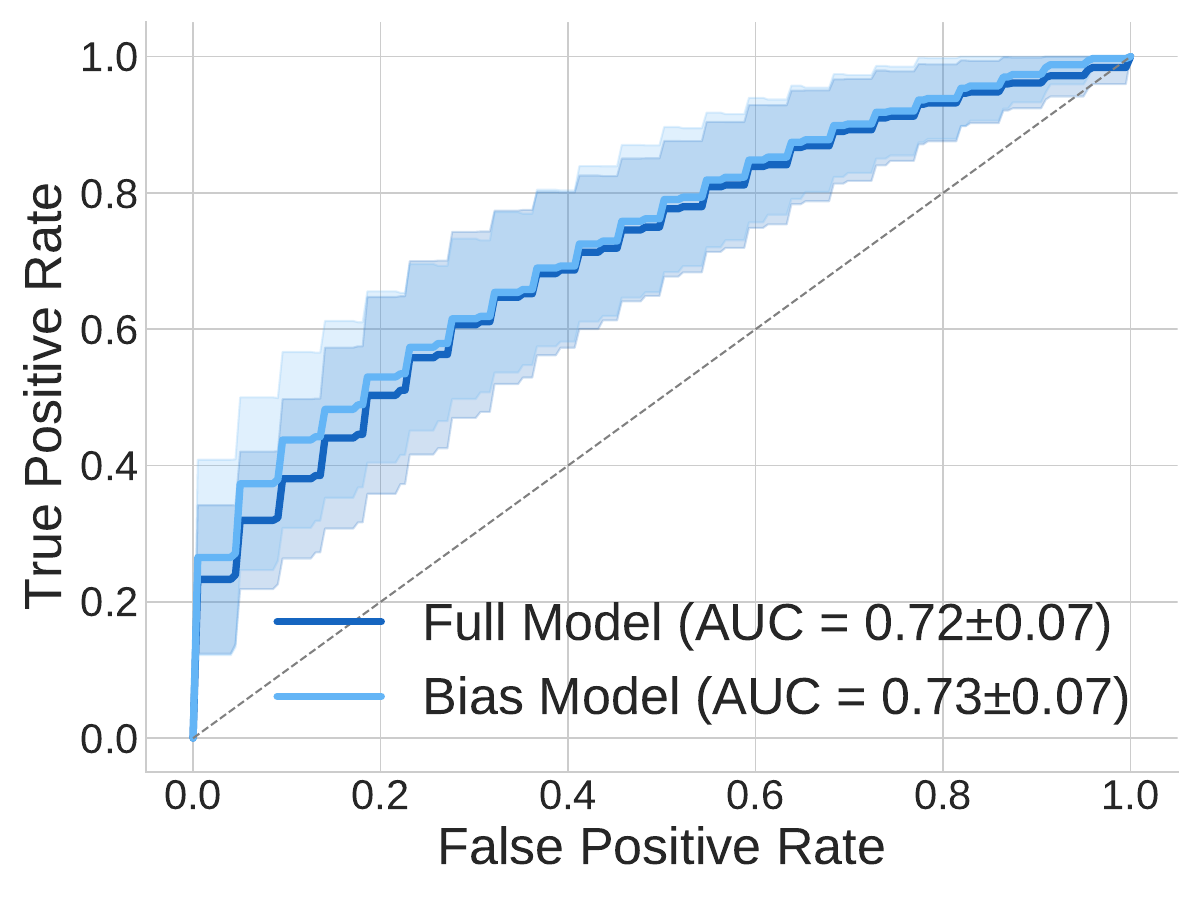}
    \caption{\small ROC Health }
  \end{subfigure}\hfill
  \begin{subfigure}[t]{0.17\textwidth}
    \includegraphics[width=\linewidth]{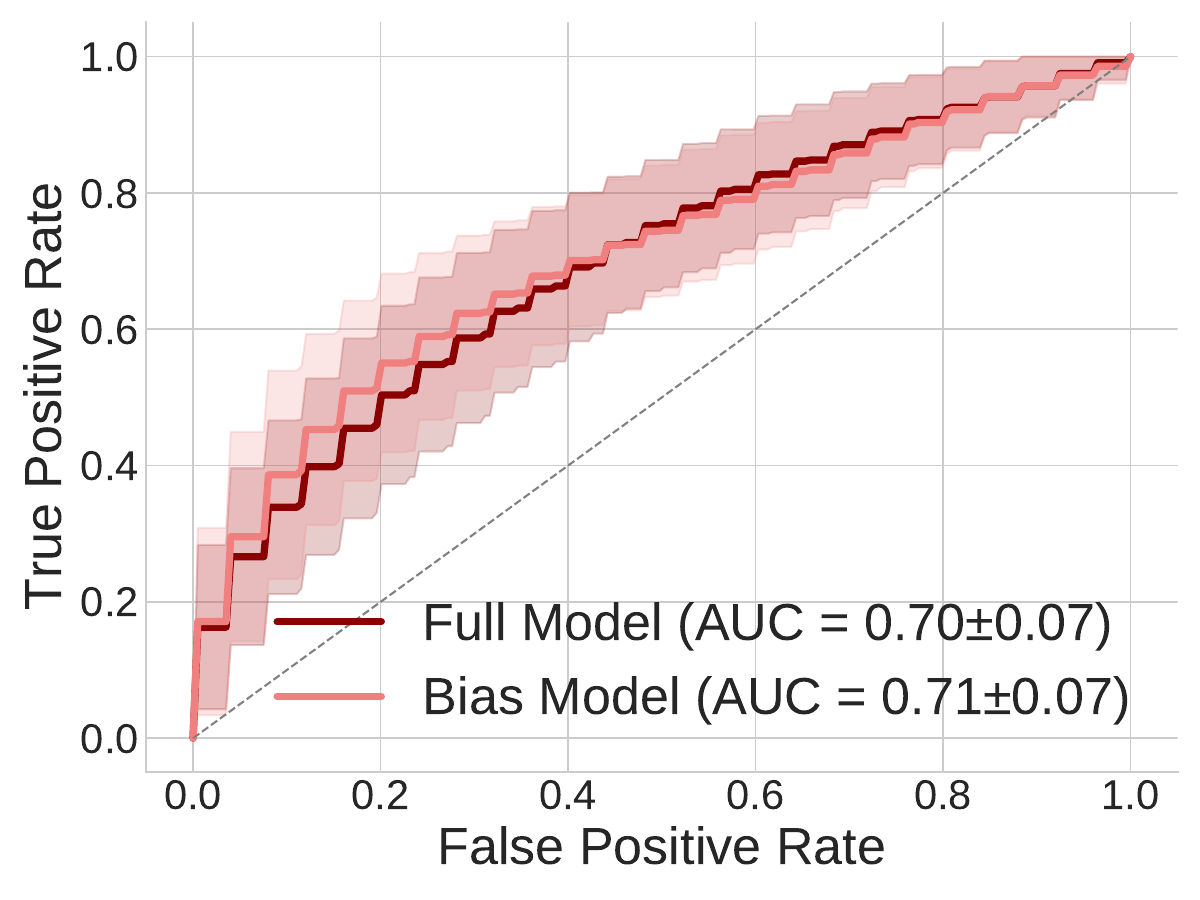}
    \caption{\small ROC Economic}
  \end{subfigure}
  \begin{subfigure}[t]{0.16\textwidth}
    \includegraphics[width=\linewidth]{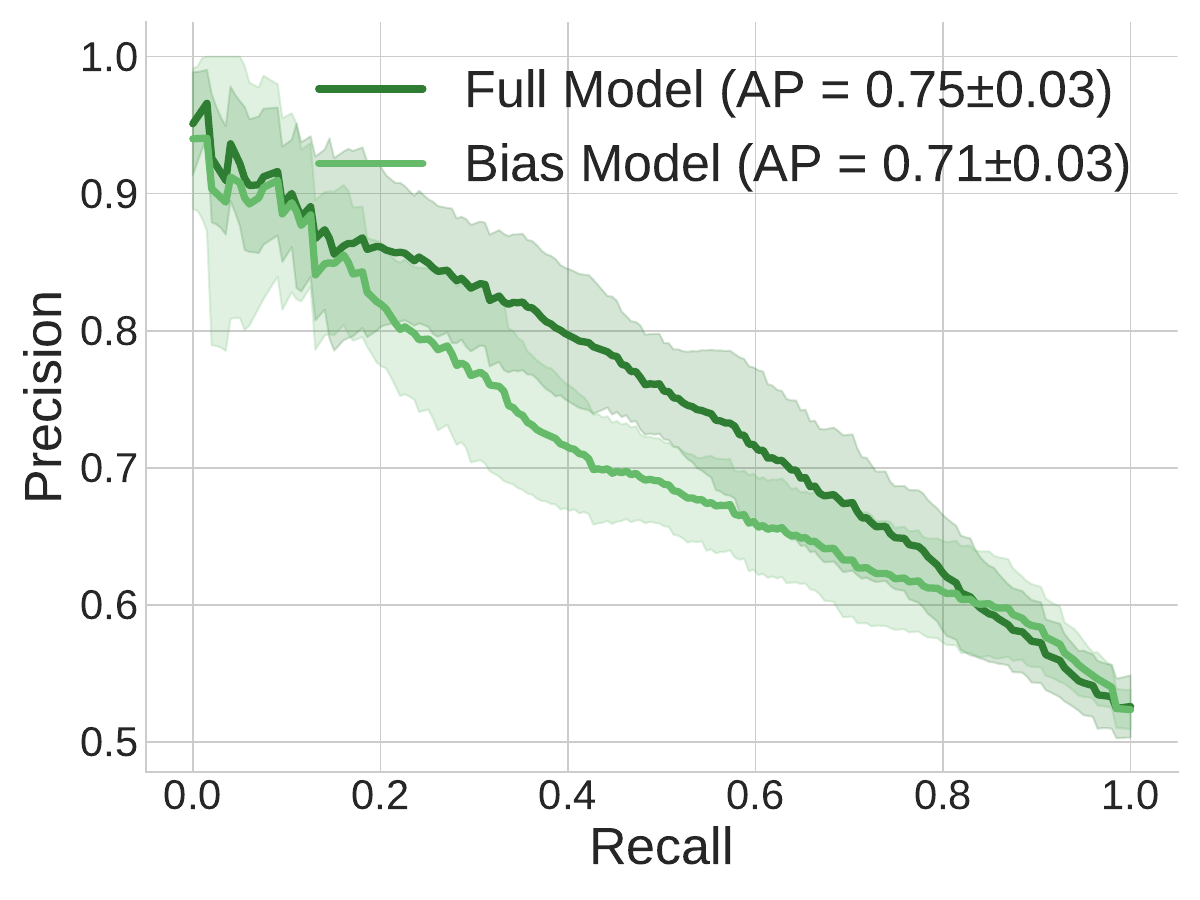}
    \caption{\small PR Social}
  \end{subfigure}\hfill
  \begin{subfigure}[t]{0.16\textwidth}
    \includegraphics[width=\linewidth]{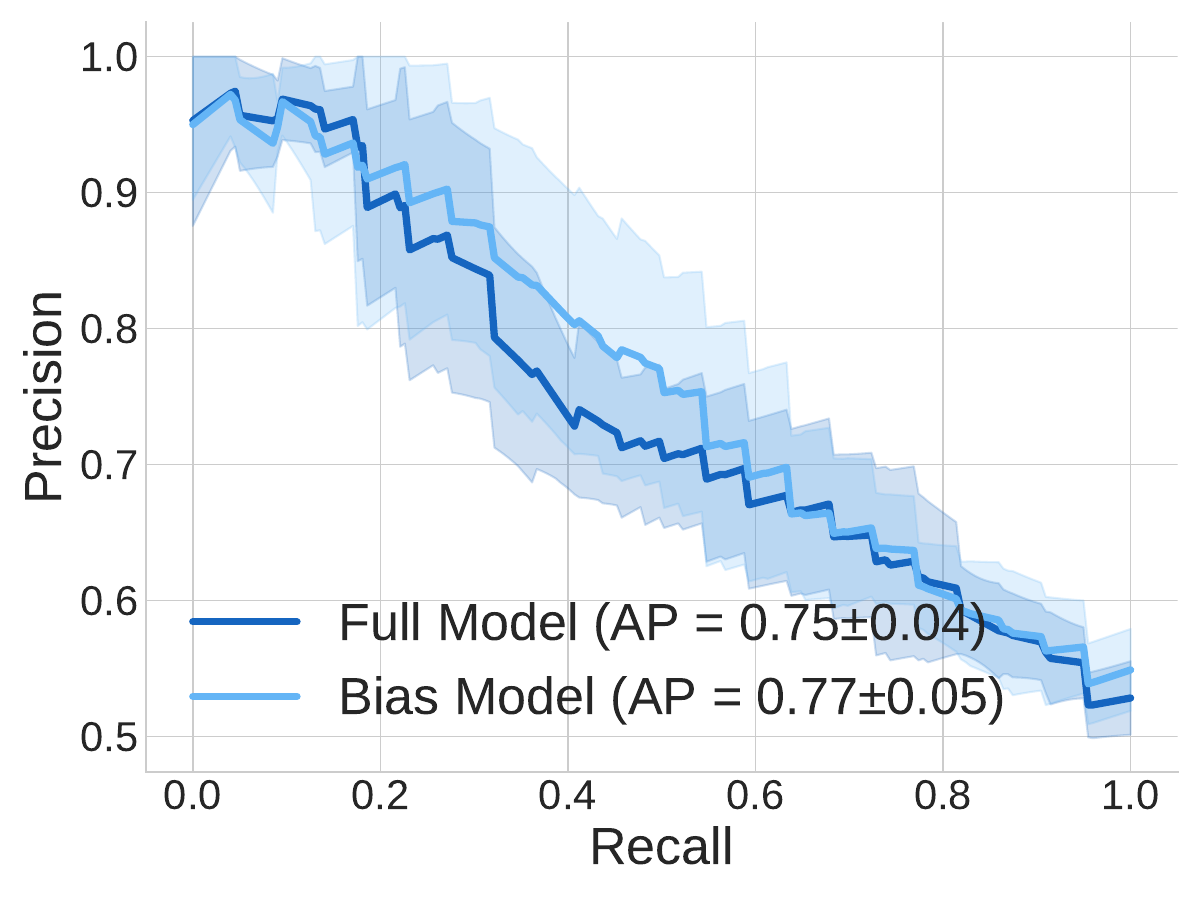}
    \caption{\small PR Health}
  \end{subfigure}\hfill
  \begin{subfigure}[t]{0.16\textwidth}
    \includegraphics[width=\linewidth]{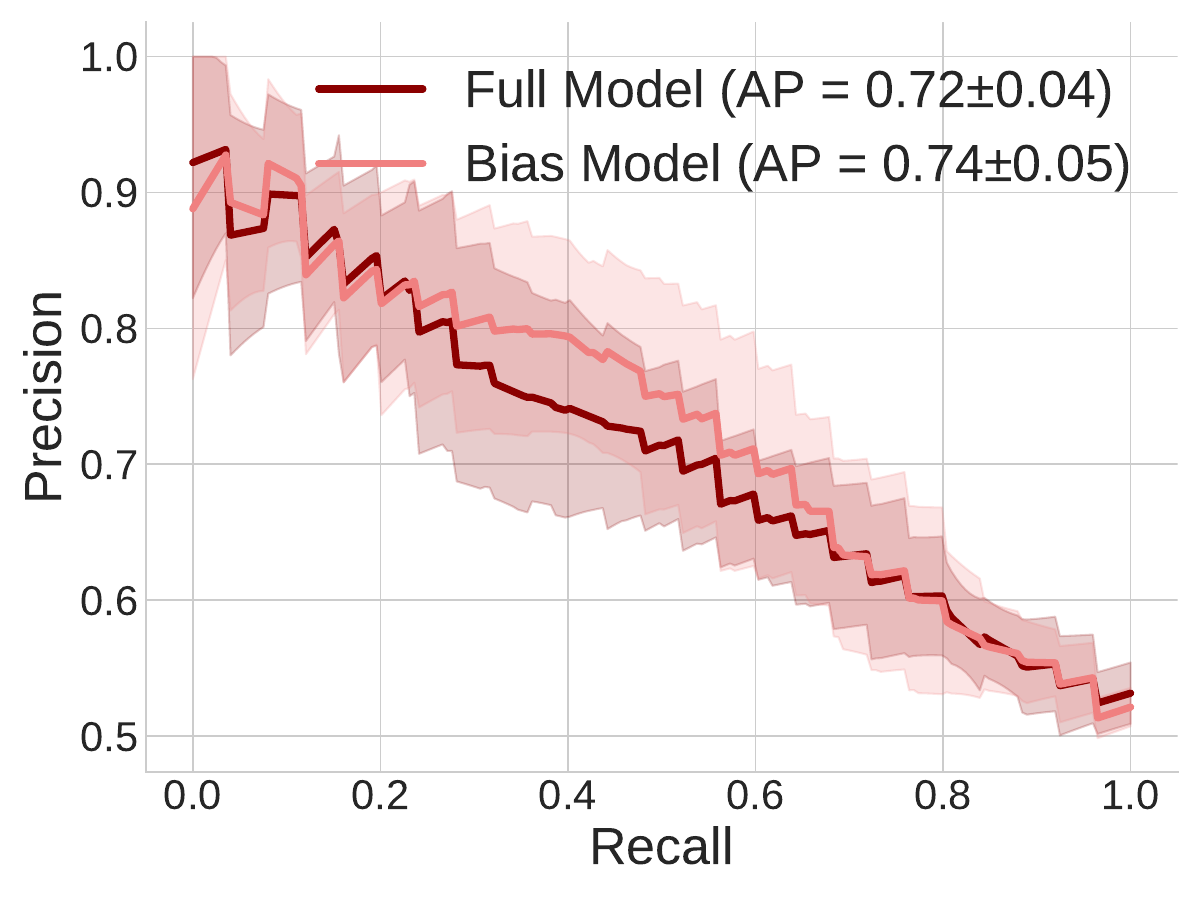}
    \caption{\small PR Economic}
  \end{subfigure}

  \caption{\textbf{Model behavior and structural interpretation on village network \# 1.}Panel (a) (right) displays the learned multiplex role simplex with directed social (green), health (blue), and economic (red) ties; arrows indicate direction, and circles represent source/target embeddings ($\mathbf{Z}, \mathbf{W} \in \Delta_2$). Panel (a) (left) shows the same structure with ties separated by layer; node colors reflect dominant roles. Panel (b) presents violin plots of node membership strengths across layers for source ($\mathbf{Z}$) and target ($\mathbf{W}$) roles. Panels (c)–(q) show the inferred multi-scale structure per layer, with hierarchy strengths $s_h^l$ quantifying the contribution of each level $h$ to link formation. Adjacency matrices are reordered by dominant memberships $\mathbf{u}_i^{l,h}$ (source) and $\mathbf{v}_j^{l,h}$ (target). Deeper rows reveal finer-grained structure. Panel (r) shows bias model correlations between degree and inferred biases: in-degree vs. $\exp{\gamma_j^l}$ (left) and out-degree vs. $\exp{\beta_i^l}$ (right), indicating network status effects. Panel (s) shows reduced correlations in the full model, reflecting shifted signal to interdependence. Panels (t)–(v) compare ROC curves and AUC scores for the full vs. bias model using 10-fold cross-validation (shaded areas = uncertainty). Panels (w)–(y) show corresponding PR curves.}
  \label{fig:overall_0}
\end{figure*}

\begin{figure*}[b!]

\centering
  \centering
  \adjustbox{valign=b}{%
    \begin{minipage}[t]{0.62\textwidth}
      \begin{subfigure}[t]{\linewidth}
        \includegraphics[width=1\linewidth]{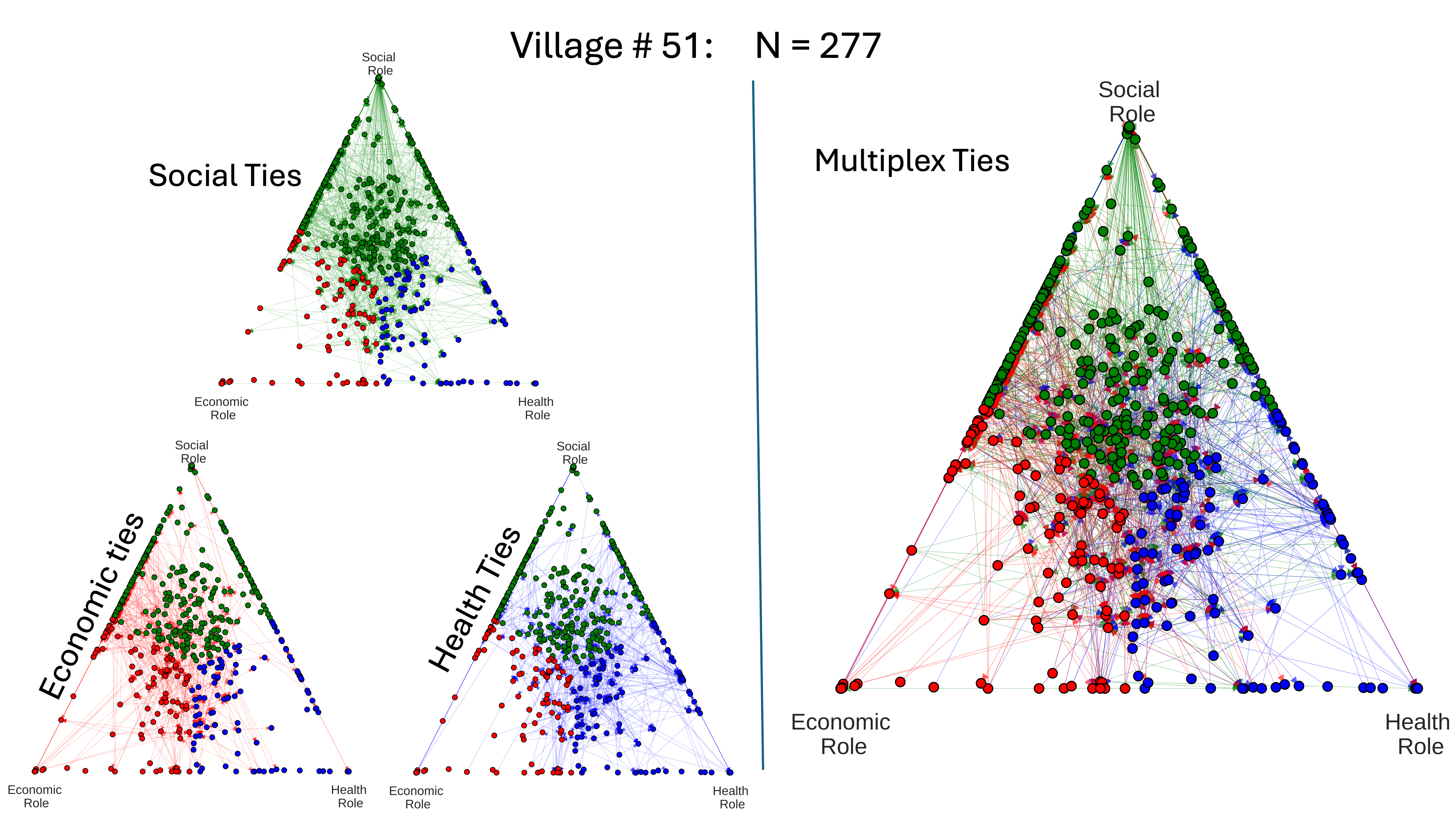}
        \caption{\small Role simplex visualizations for village $\# 51$. Each point represents a node, colored by dominant role. Left: tie-specific roles. Right: multiplex network with all ties overlaid. }
      \end{subfigure}
      \vspace{1em}

      \begin{subfigure}[t]{\linewidth}
        \includegraphics[width=1\linewidth]{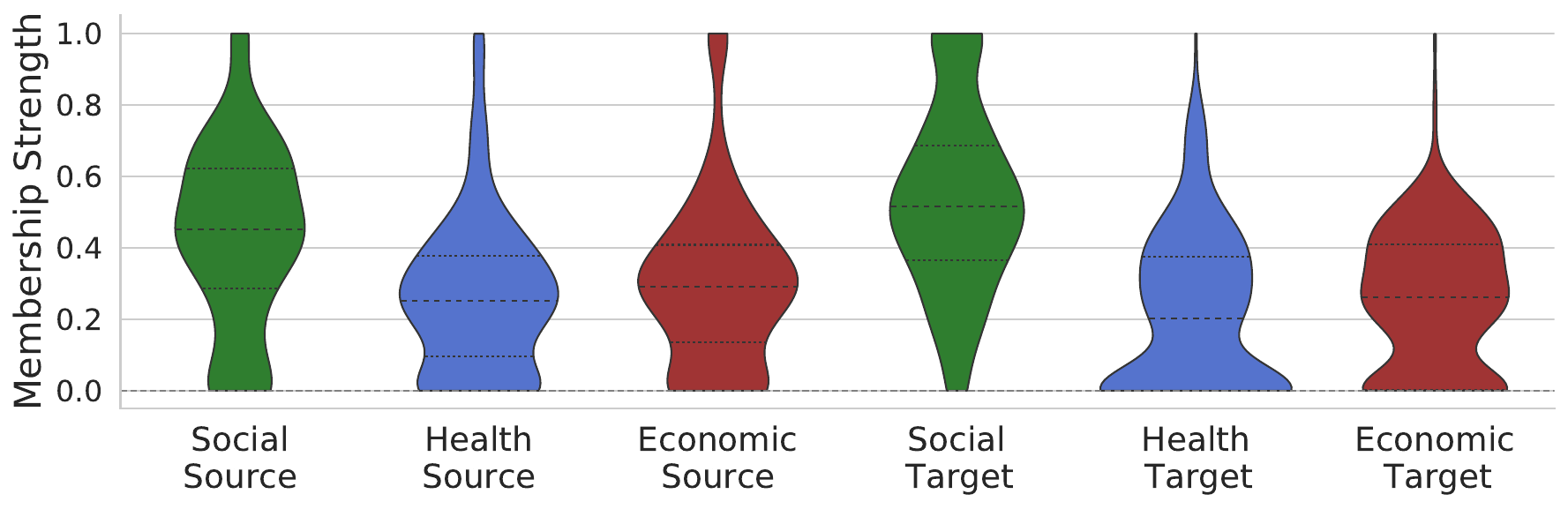}
        \caption{\small Violin plots of node-role membership strengths for source and target positions across domains. }
      \end{subfigure}
    \end{minipage}
  }\hfill
  \adjustbox{valign=b}{%
    \begin{minipage}[t]{0.36\textwidth}

\makebox[1\linewidth]{\centering \textbf{Multi-scale Layer Structure:}}
\\[1ex]
 \makebox[0.32\linewidth]{\centering \textbf{Social}} \hfill
  \makebox[0.32\linewidth]{\centering \textbf{Health}} \hfill
  \makebox[0.32\linewidth]{\centering \textbf{Economic}} \\[-1ex]
    
\centering

  \begin{minipage}[c]{\linewidth}
    \begin{minipage}[c]{0.04\linewidth}
      \centering
      \rotatebox{90}{\small \textbf{$D_{h=1}=2$}}
    \end{minipage}%
    \hfill
     \captionsetup[subfigure]{skip=-5pt}
    \begin{subfigure}[c]{0.3\linewidth}
\includegraphics[width=\linewidth]{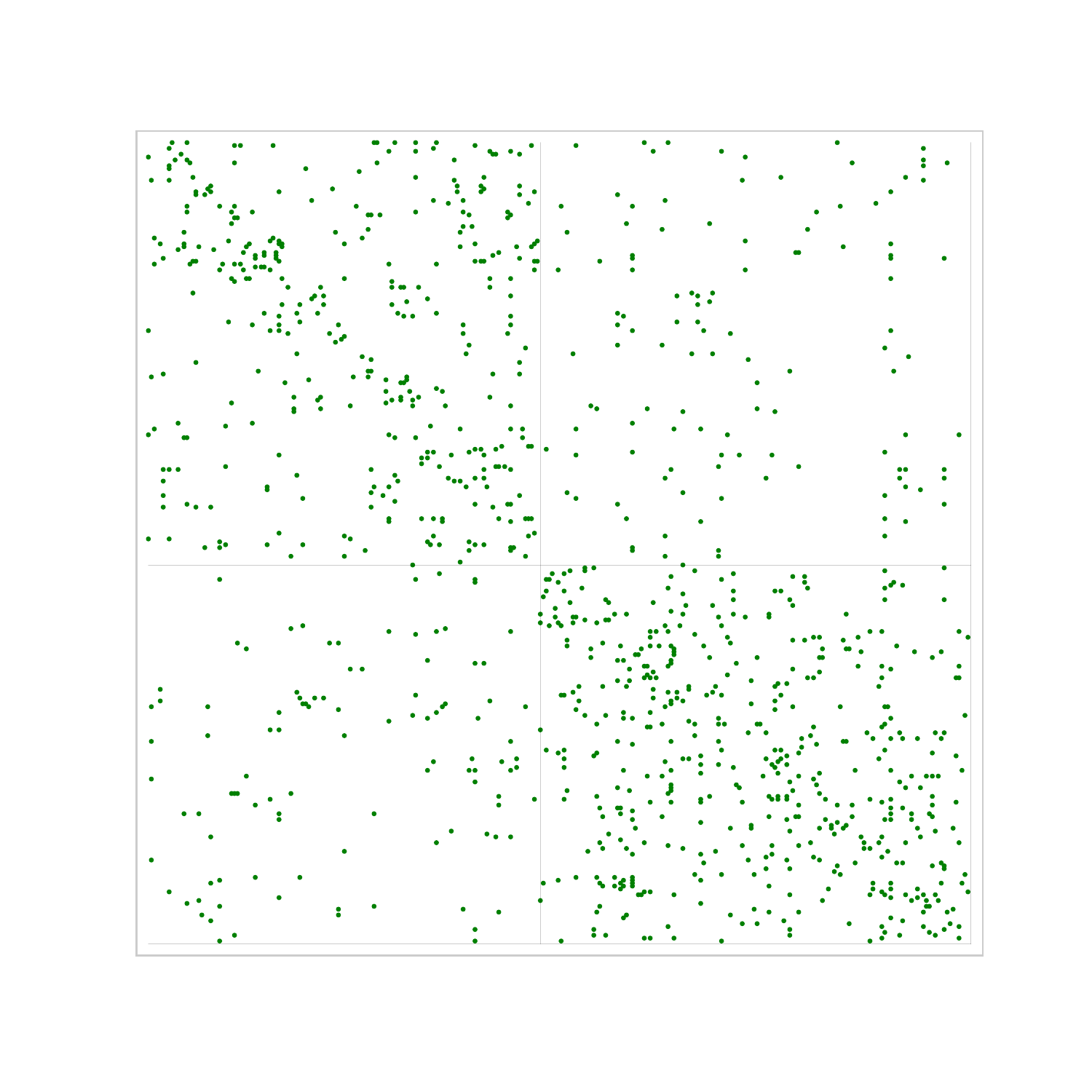} \caption{\tiny\\$s_1^1=0.001$}
    \end{subfigure}\hfill
     \captionsetup[subfigure]{skip=-5pt}
    \begin{subfigure}[c]{0.3\linewidth}
\includegraphics[width=\linewidth]{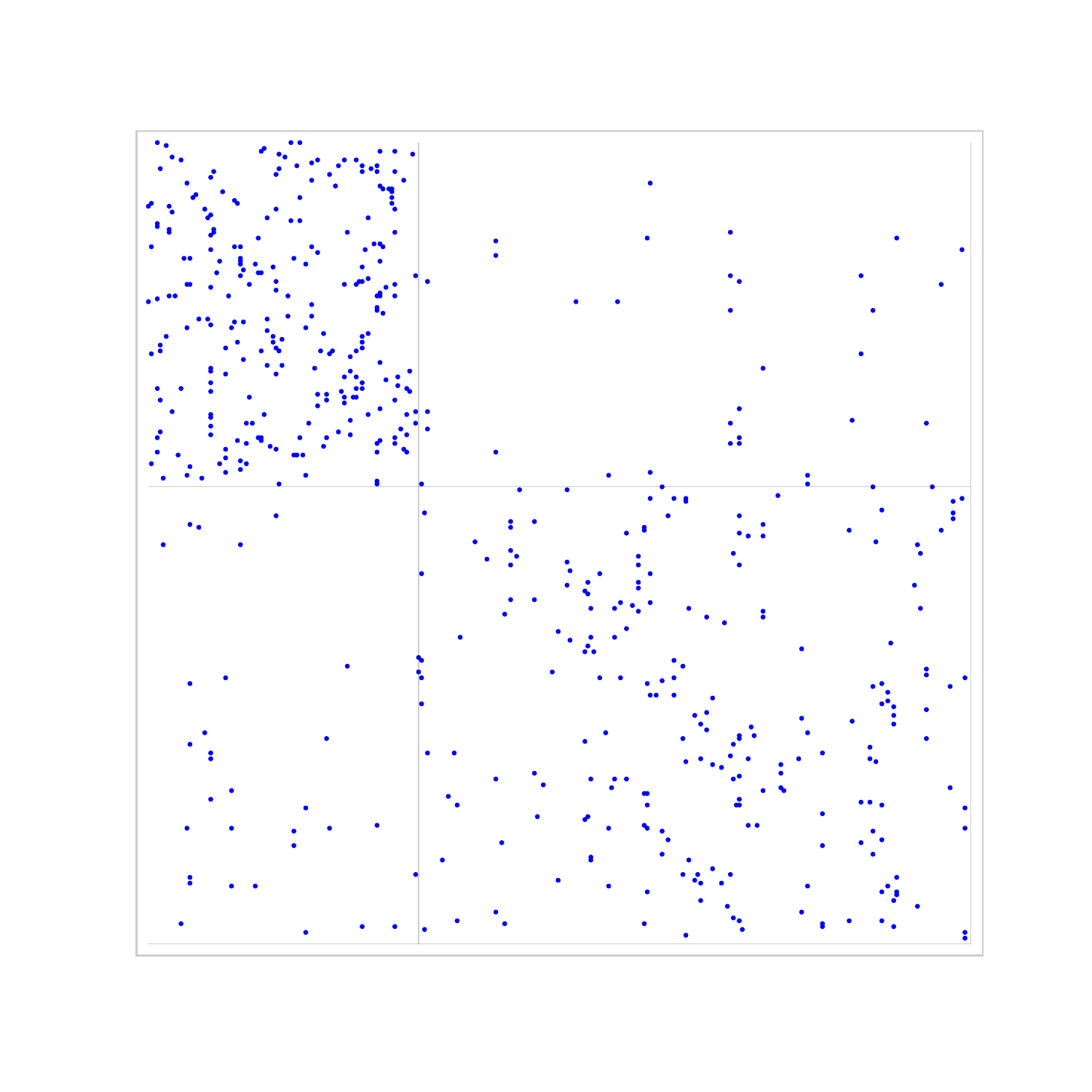} \caption{\tiny\\$s_1^2=0.004$}
    \end{subfigure}\hfill
    \captionsetup[subfigure]{skip=-5pt}
    \begin{subfigure}[c]{0.3\linewidth}
\includegraphics[width=\linewidth]{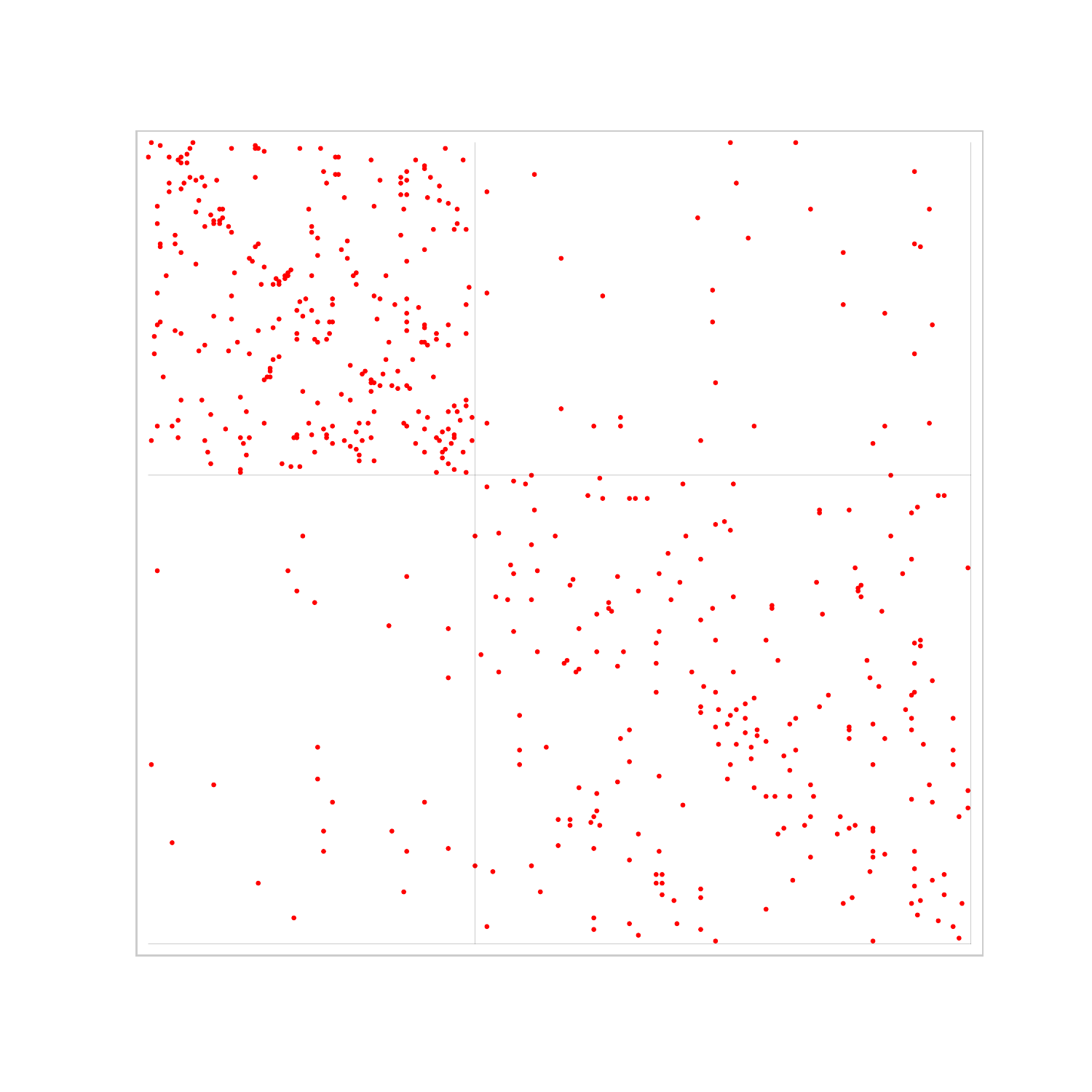}
      \caption{\tiny\\$s_1^3=0.002$}
    \end{subfigure}
  \end{minipage}

  \begin{minipage}[c]{\linewidth}
    \begin{minipage}[c]{0.04\linewidth}
      \centering
      \rotatebox{90}{\small \textbf{$D_{h=2}=4$}}
    \end{minipage}%
    \hfill
     \captionsetup[subfigure]{skip=-5pt}
    \begin{subfigure}[c]{0.3\linewidth}
      \includegraphics[width=\linewidth]{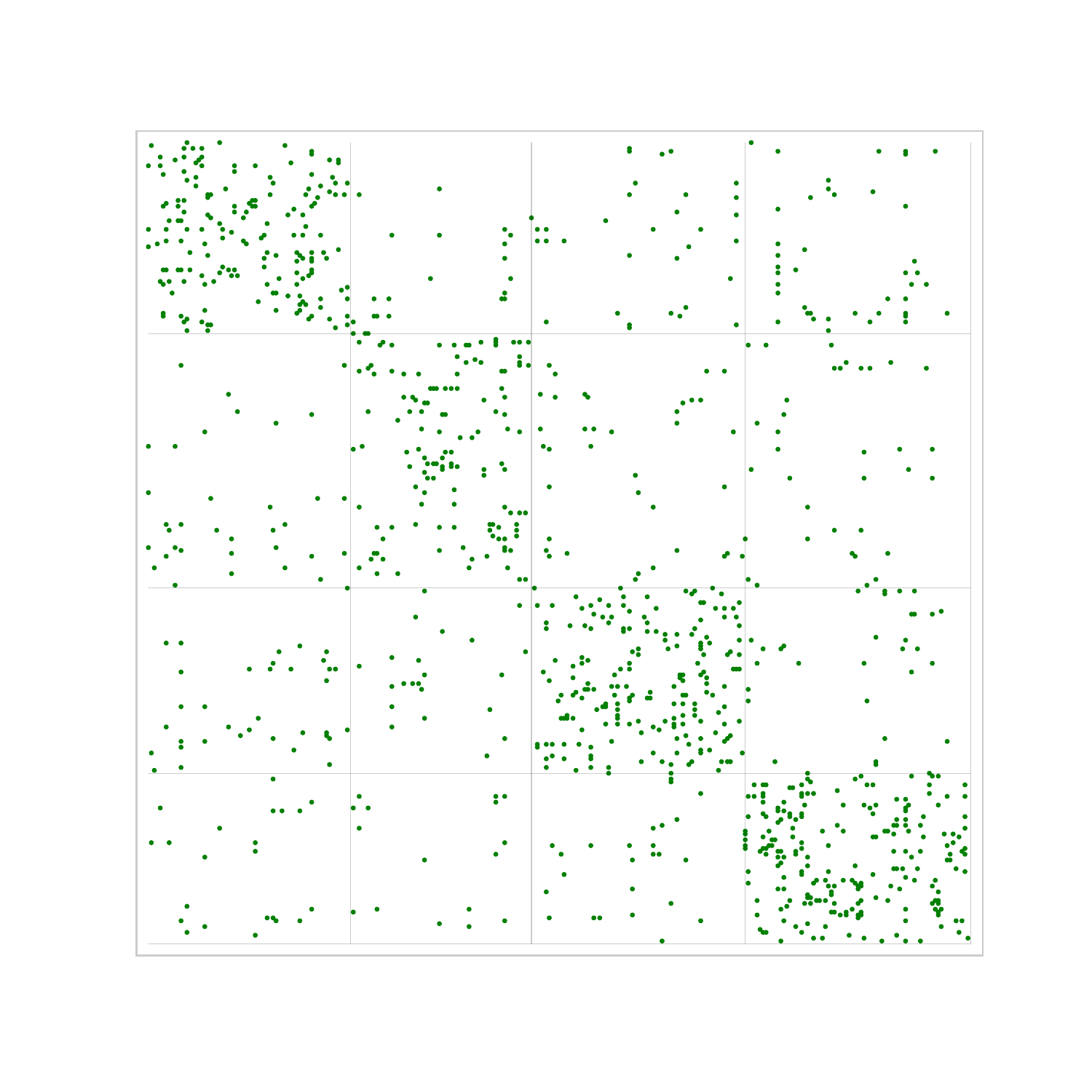}
       \caption{\tiny\\$s_2^1=0.001$}
    \end{subfigure}\hfill
     \captionsetup[subfigure]{skip=-5pt}
    \begin{subfigure}[c]{0.3\linewidth}
      \includegraphics[width=\linewidth]{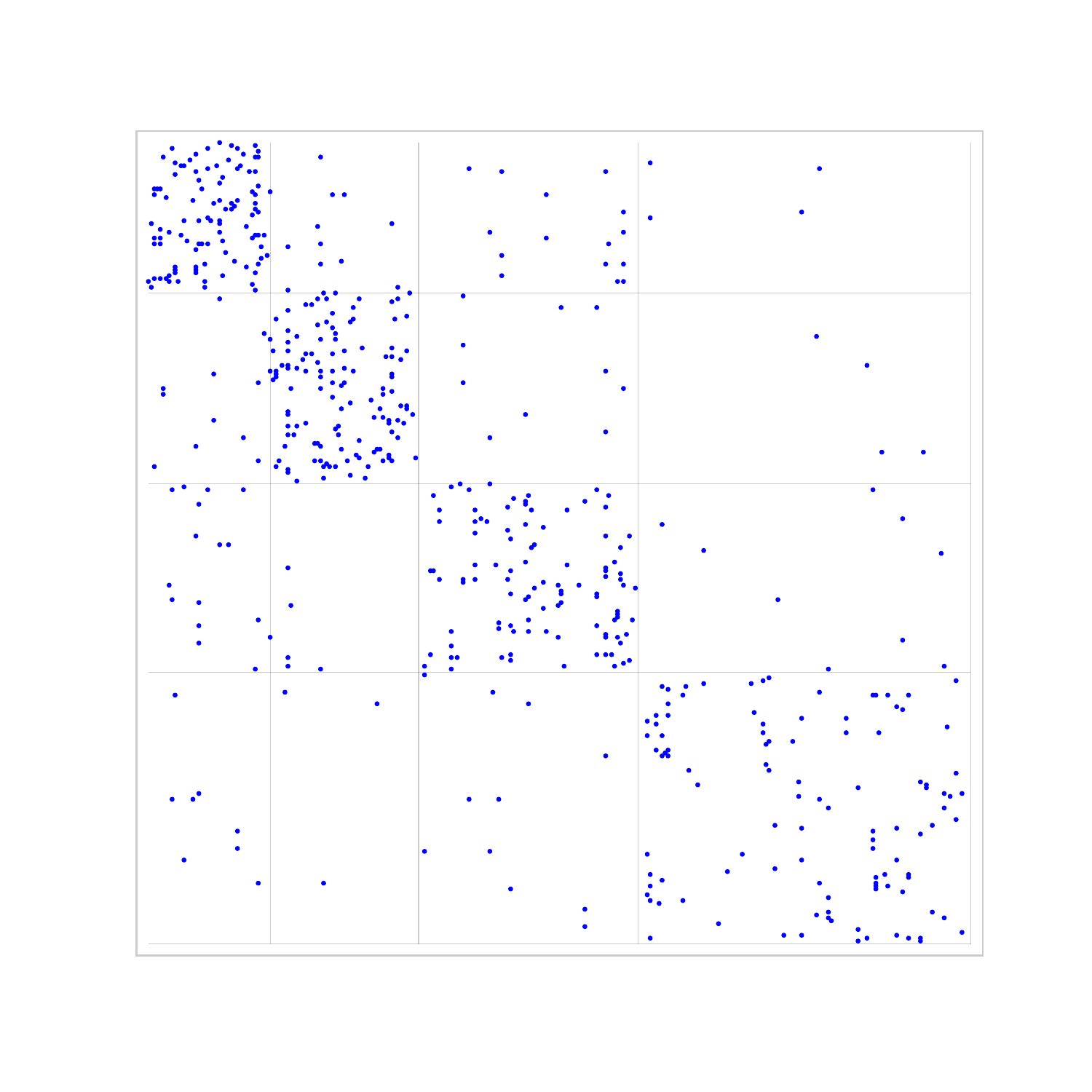}
       \caption{\tiny\\$s_2^2=0.019$}
    \end{subfigure}\hfill
     \captionsetup[subfigure]{skip=-5pt}
    \begin{subfigure}[c]{0.3\linewidth}
      \includegraphics[width=\linewidth]{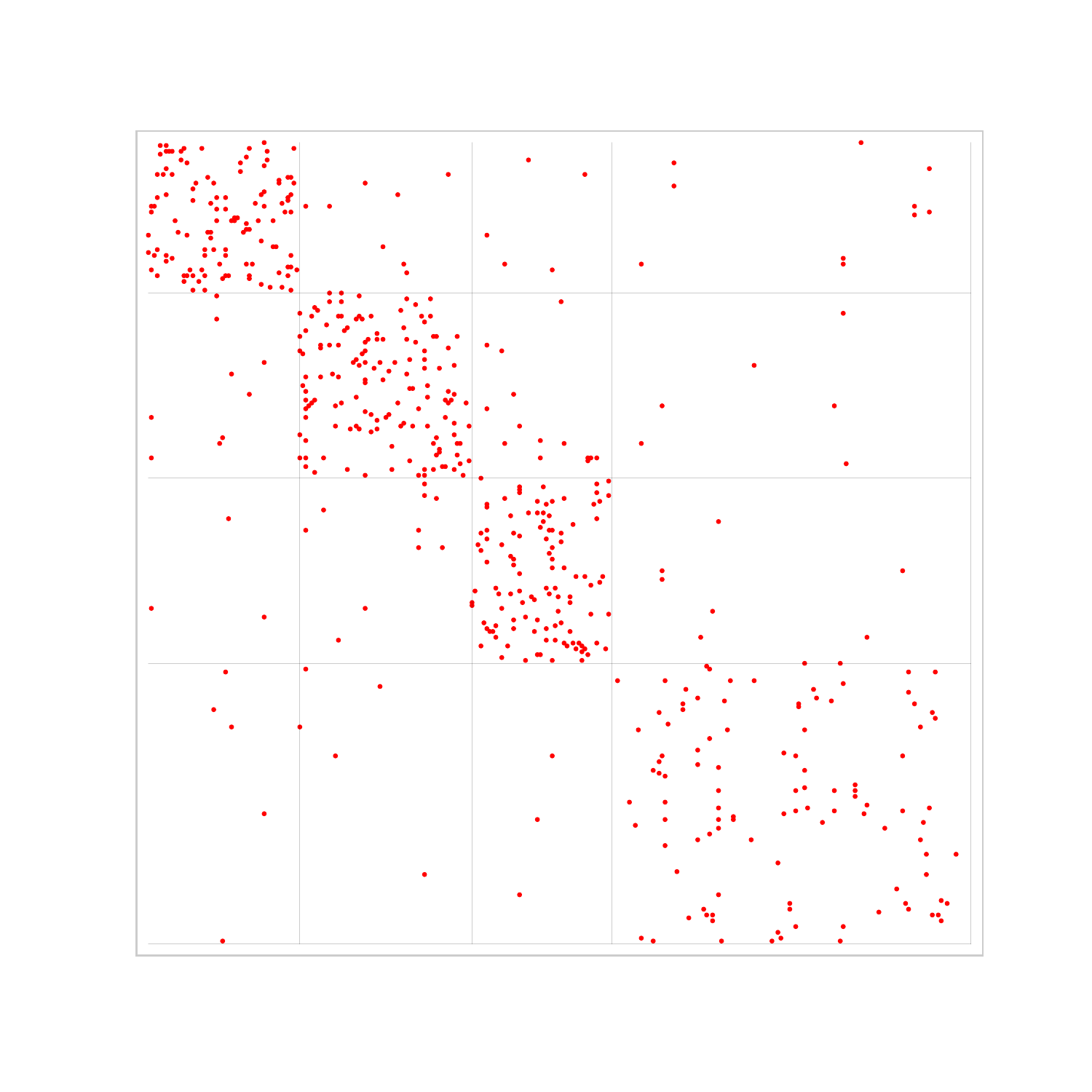}
       \caption{\tiny\\$s_2^3=0.003$}
    \end{subfigure}
  \end{minipage}
  
  \begin{minipage}[c]{\linewidth}
    \begin{minipage}[c]{0.04\linewidth}
      \centering
      \rotatebox{90}{\small \textbf{$D_{h=3}=8$}}
    \end{minipage}%
    \hfill
     \captionsetup[subfigure]{skip=-5pt}
    \begin{subfigure}[c]{0.3\linewidth}
      \includegraphics[width=\linewidth]{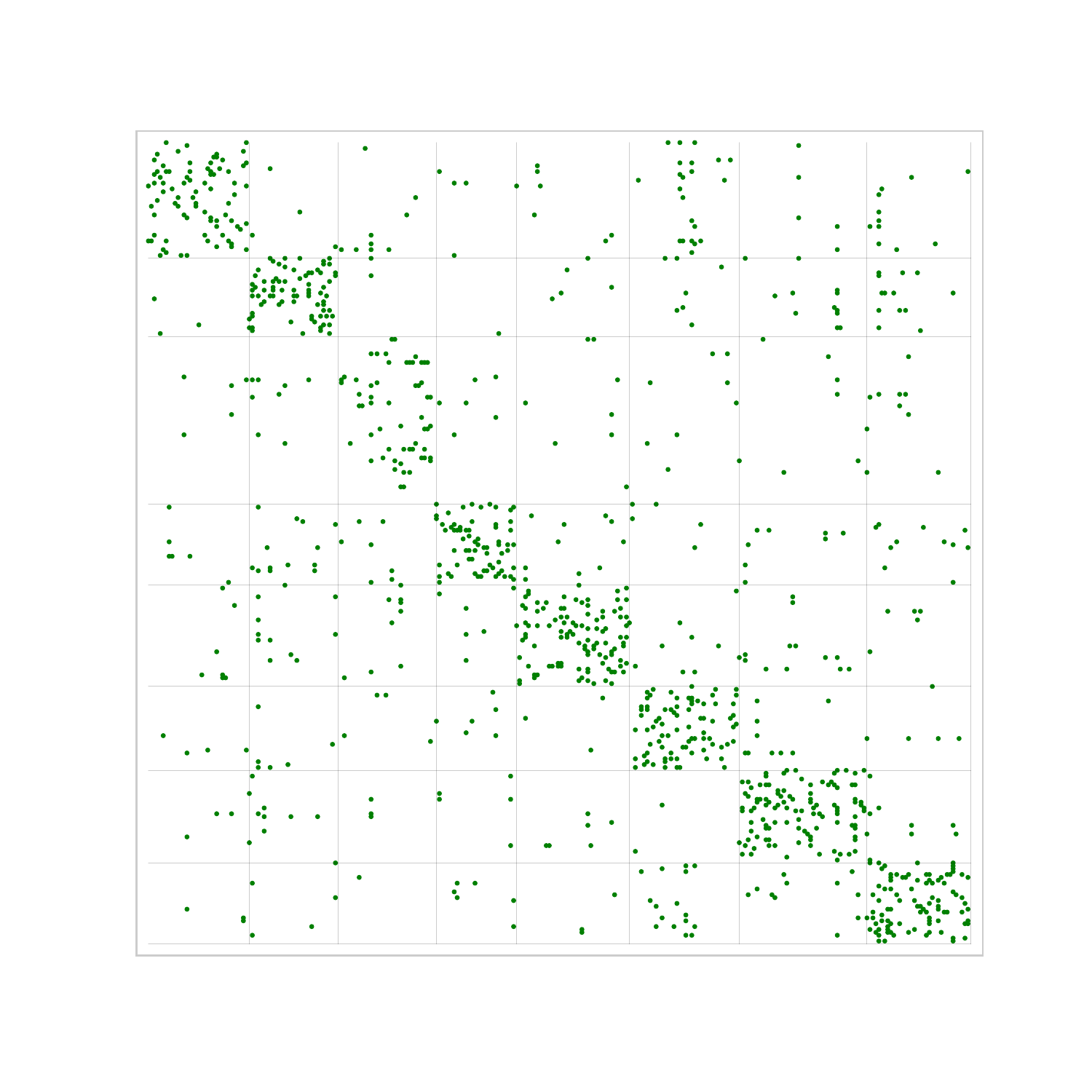}
       \caption{\tiny\\$s_3^1=9.967$}
    \end{subfigure}\hfill
     \captionsetup[subfigure]{skip=-5pt}
    \begin{subfigure}[c]{0.3\linewidth}
      \includegraphics[width=\linewidth]{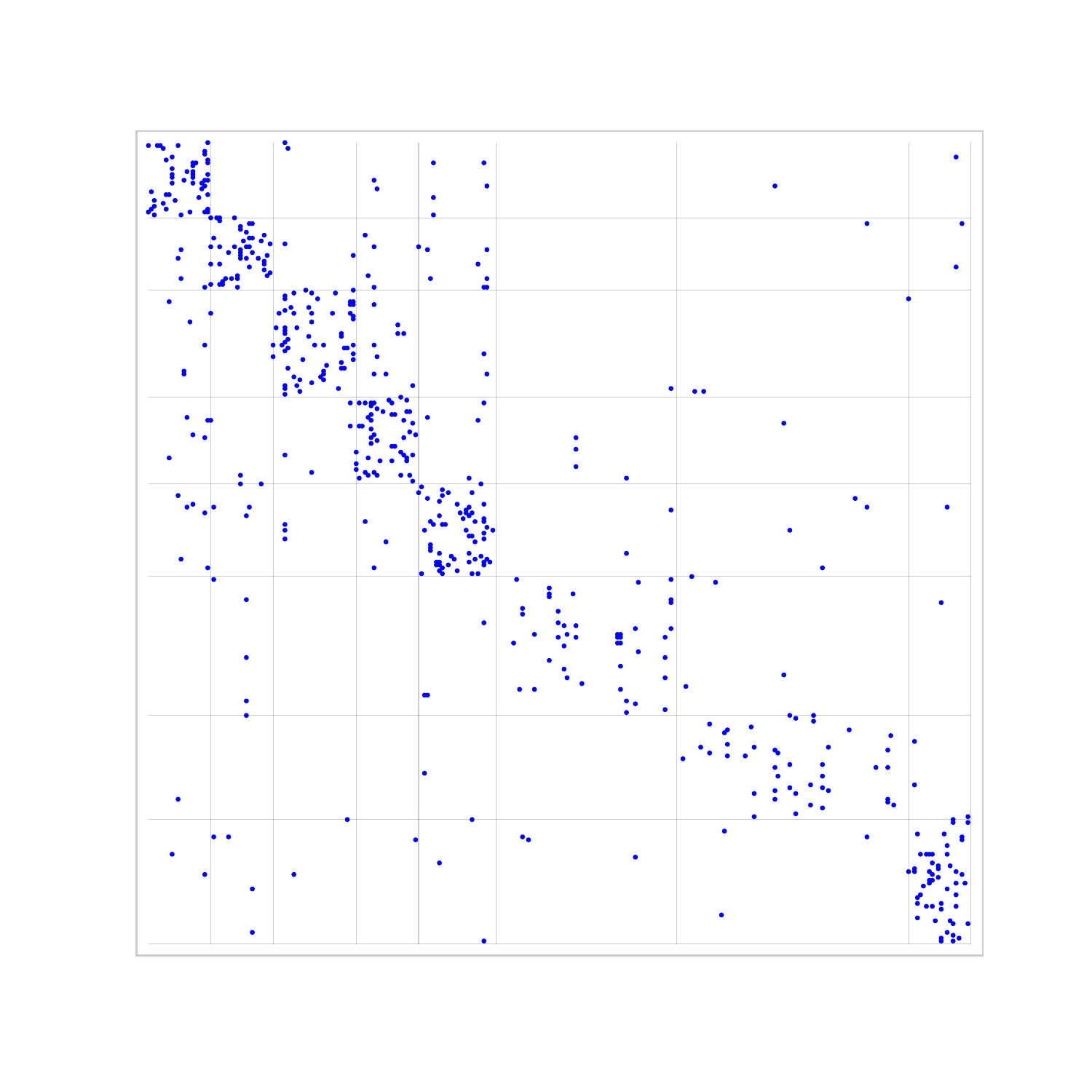}
       \caption{\tiny\\$s_3^2=6.818$}
    \end{subfigure}\hfill
     \captionsetup[subfigure]{skip=-5pt}
    \begin{subfigure}[c]{0.3\linewidth}
      \includegraphics[width=\linewidth]{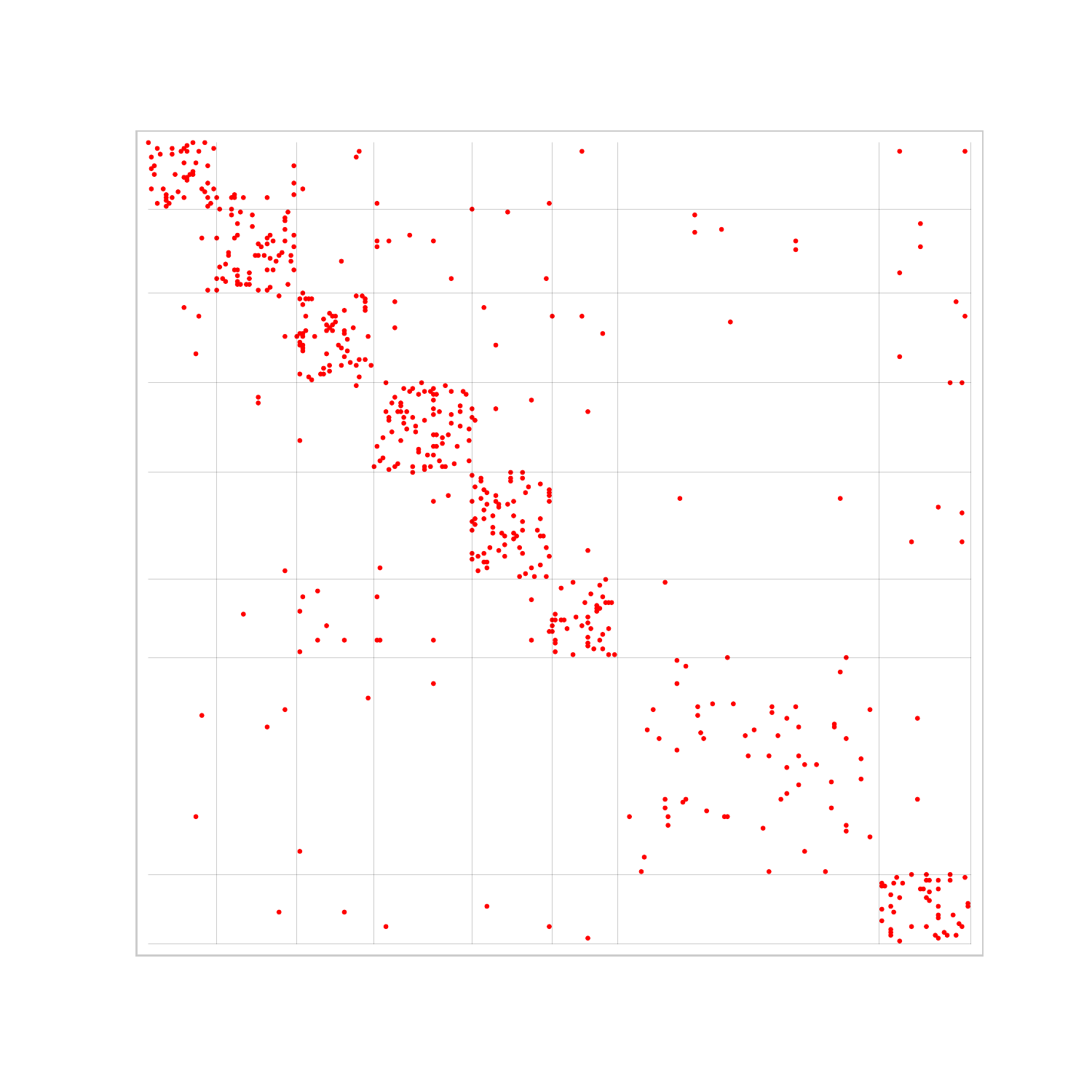}
       \caption{\tiny\\$s_3^3=6.910$}
    \end{subfigure}
  \end{minipage}
  
  \begin{minipage}[c]{\linewidth}
    \begin{minipage}[c]{0.04\linewidth}
      \centering
      \rotatebox{90}{\small \textbf{$D_{h=4}=16$}}
    \end{minipage}%
    \hfill \captionsetup[subfigure]{skip=-5pt}
    \begin{subfigure}[c]{0.3\linewidth}
      \includegraphics[width=\linewidth]{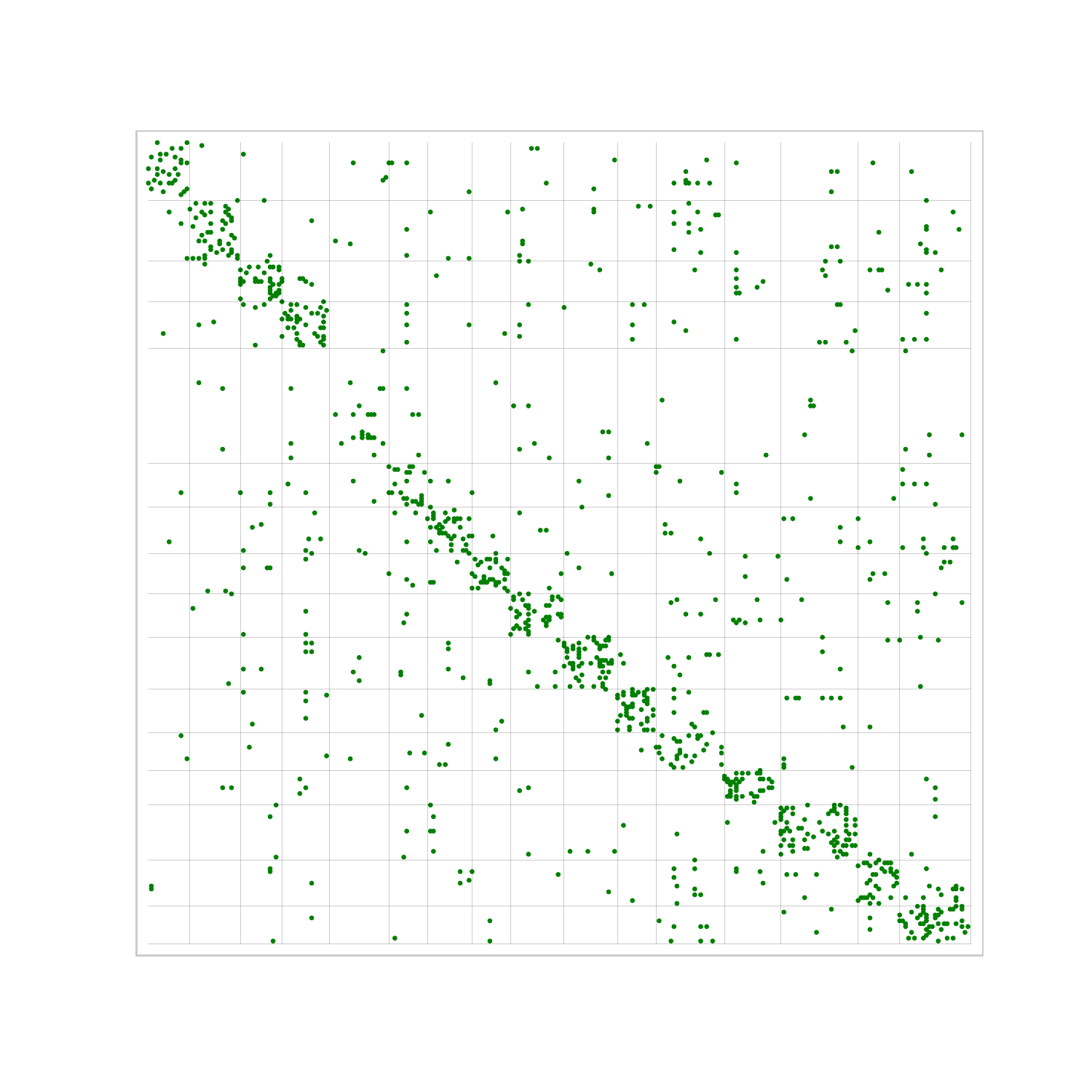} \caption{\tiny\\$s_4^1=85.048$}
    \end{subfigure}\hfill
     \captionsetup[subfigure]{skip=-5pt}
    \begin{subfigure}[c]{0.3\linewidth}
      \includegraphics[width=\linewidth]{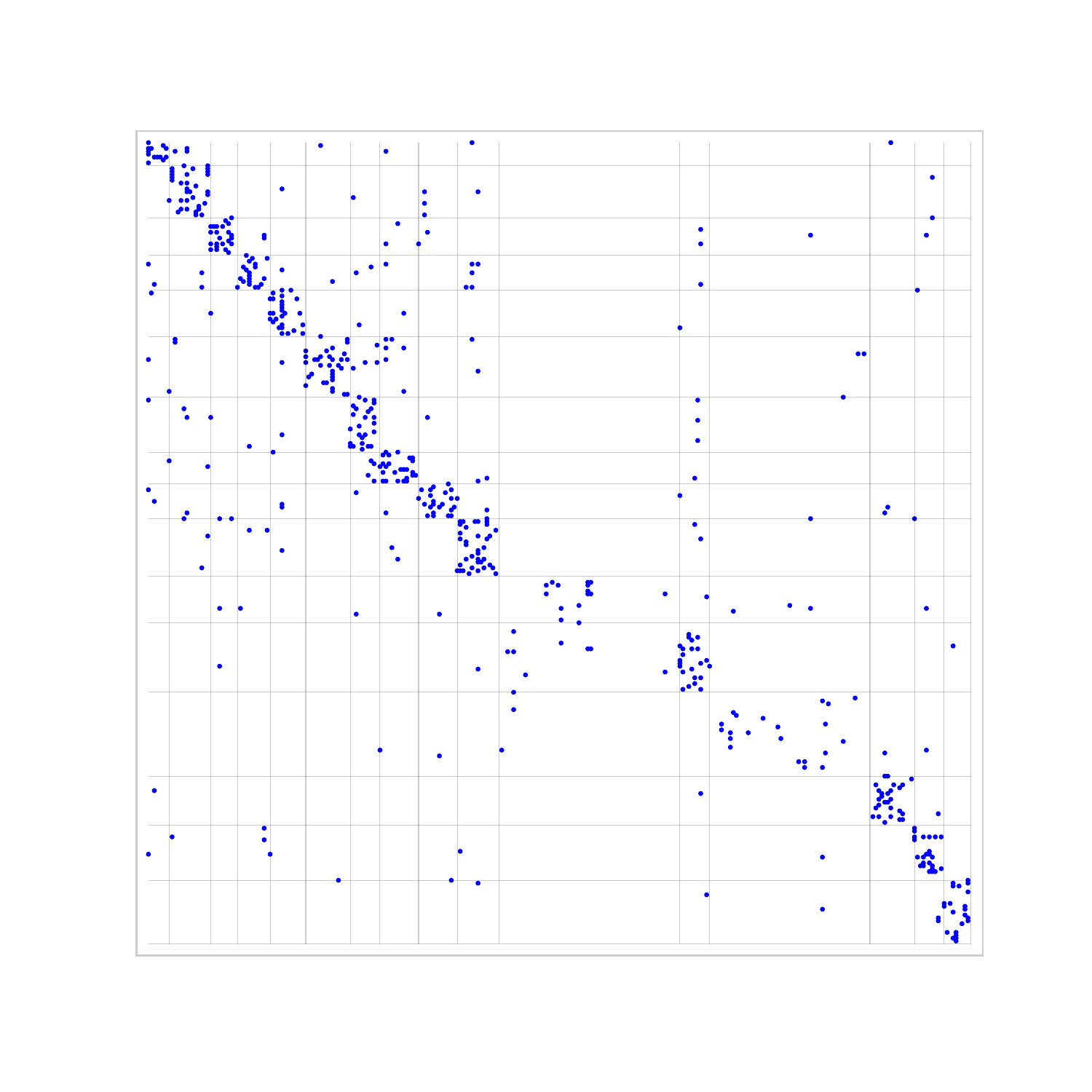} \caption{\tiny\\$s_4^2=6.516$}
    \end{subfigure}\hfill
     \captionsetup[subfigure]{skip=-5pt}
    \begin{subfigure}[c]{0.3\linewidth}
      \includegraphics[width=\linewidth]{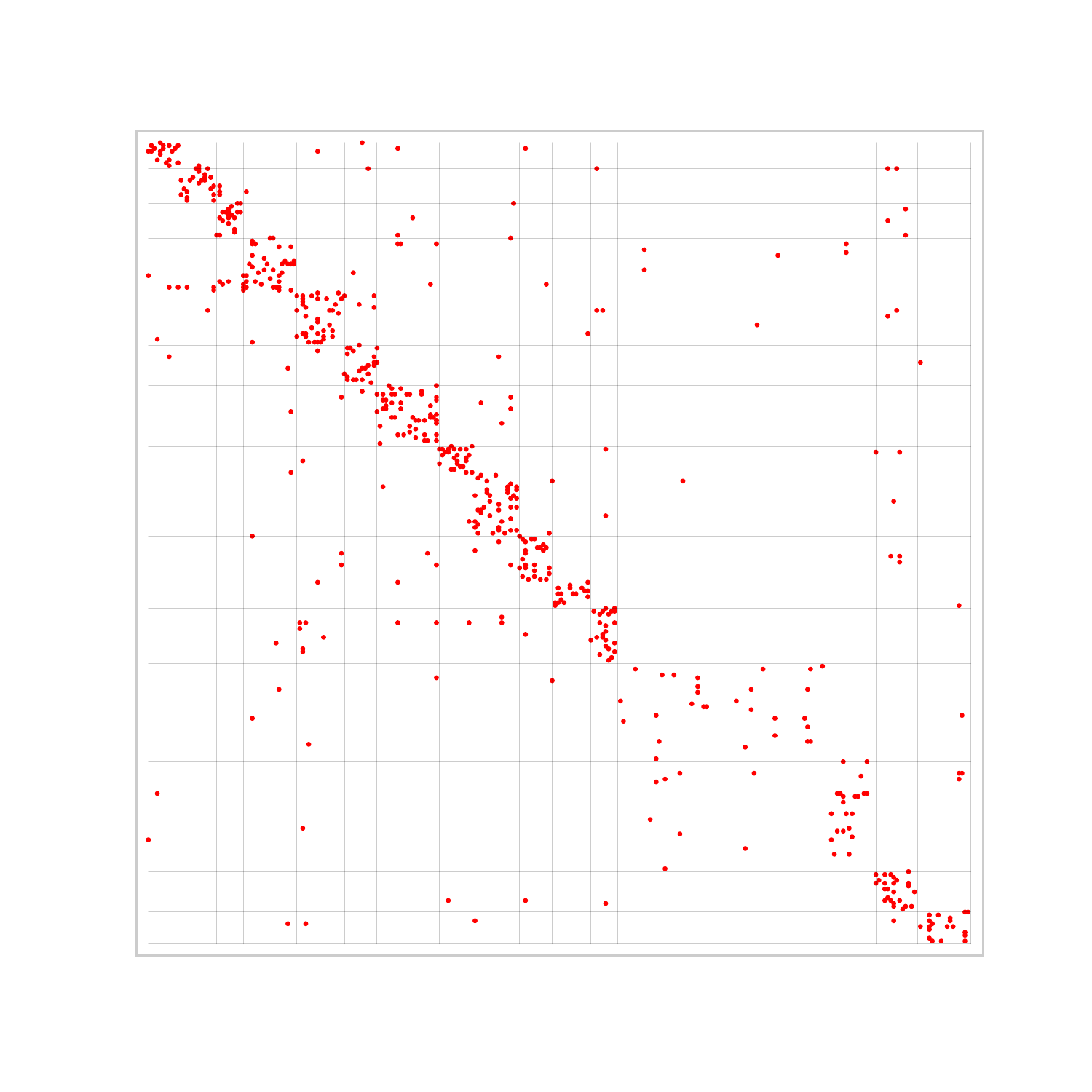} \caption{\tiny\\$s_4^3=8.162$}
    \end{subfigure}
  \end{minipage}

  \begin{minipage}[c]{\linewidth}
    \begin{minipage}[c]{0.04\linewidth}
      \centering
      \rotatebox{90}{\small \textbf{$D_{h=5}=32$}}
    \end{minipage}%
    \hfill  \captionsetup[subfigure]{skip=-5pt}
    \begin{subfigure}[c]{0.3\linewidth}
      \includegraphics[width=\linewidth]{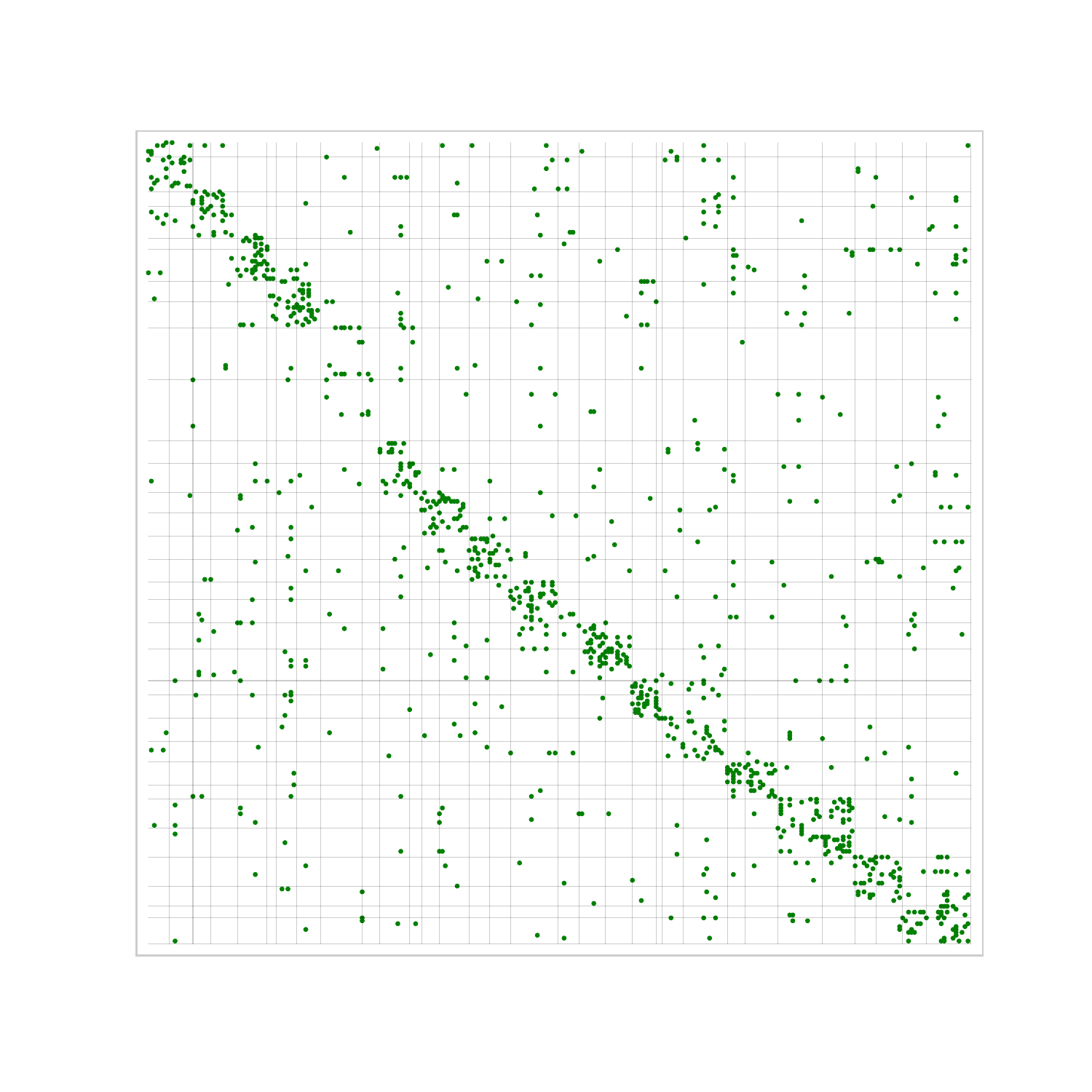}
      \caption{\tiny\\$s_5^1=0.001$}
    \end{subfigure}\hfill  \captionsetup[subfigure]{skip=-5pt}
    \begin{subfigure}[c]{0.3\linewidth}
      \includegraphics[width=\linewidth]{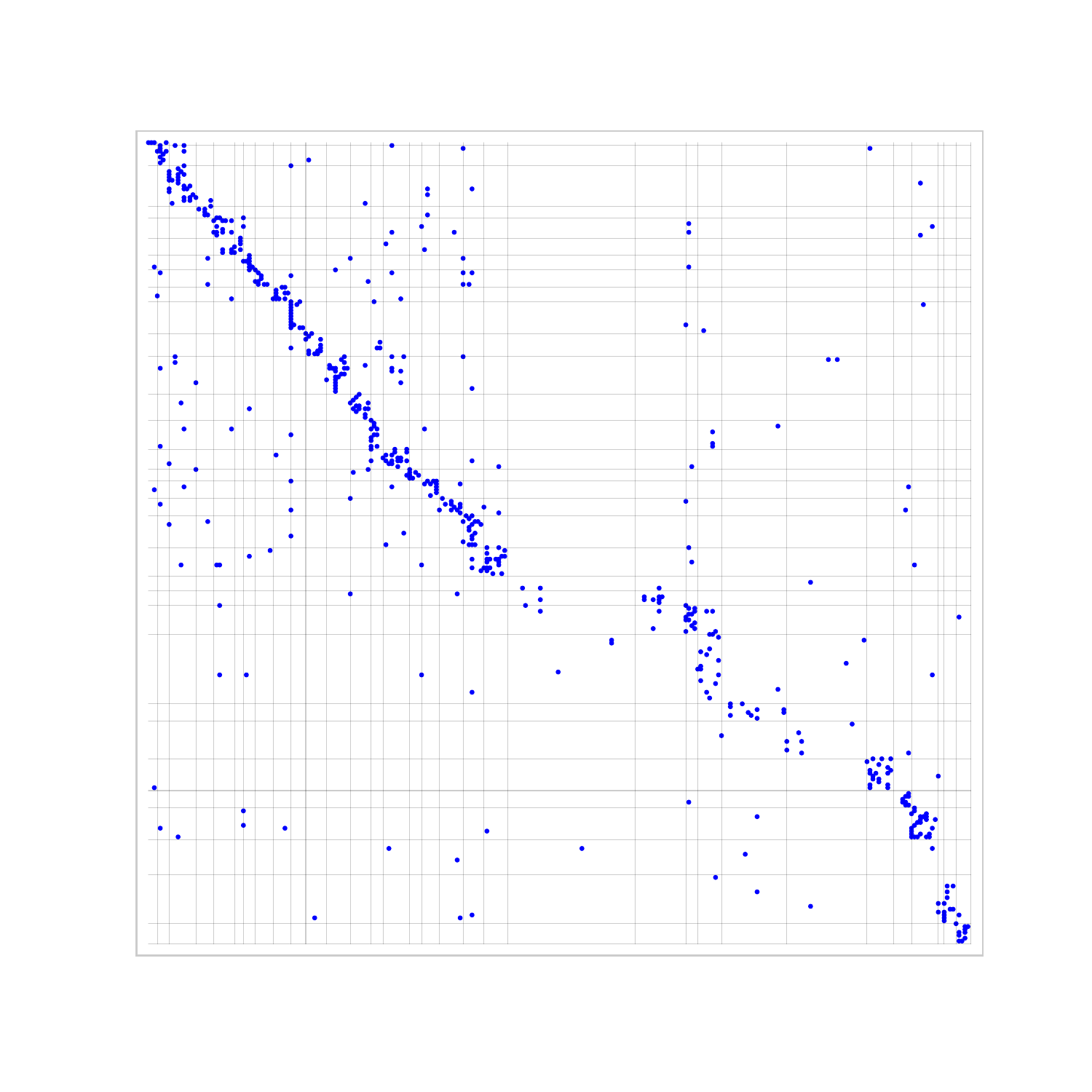}
      \caption{\tiny\\$s_5^2=19.156$}
    \end{subfigure}\hfill  \captionsetup[subfigure]{skip=-5pt}
    \begin{subfigure}[c]{0.3\linewidth}
      \includegraphics[width=\linewidth]{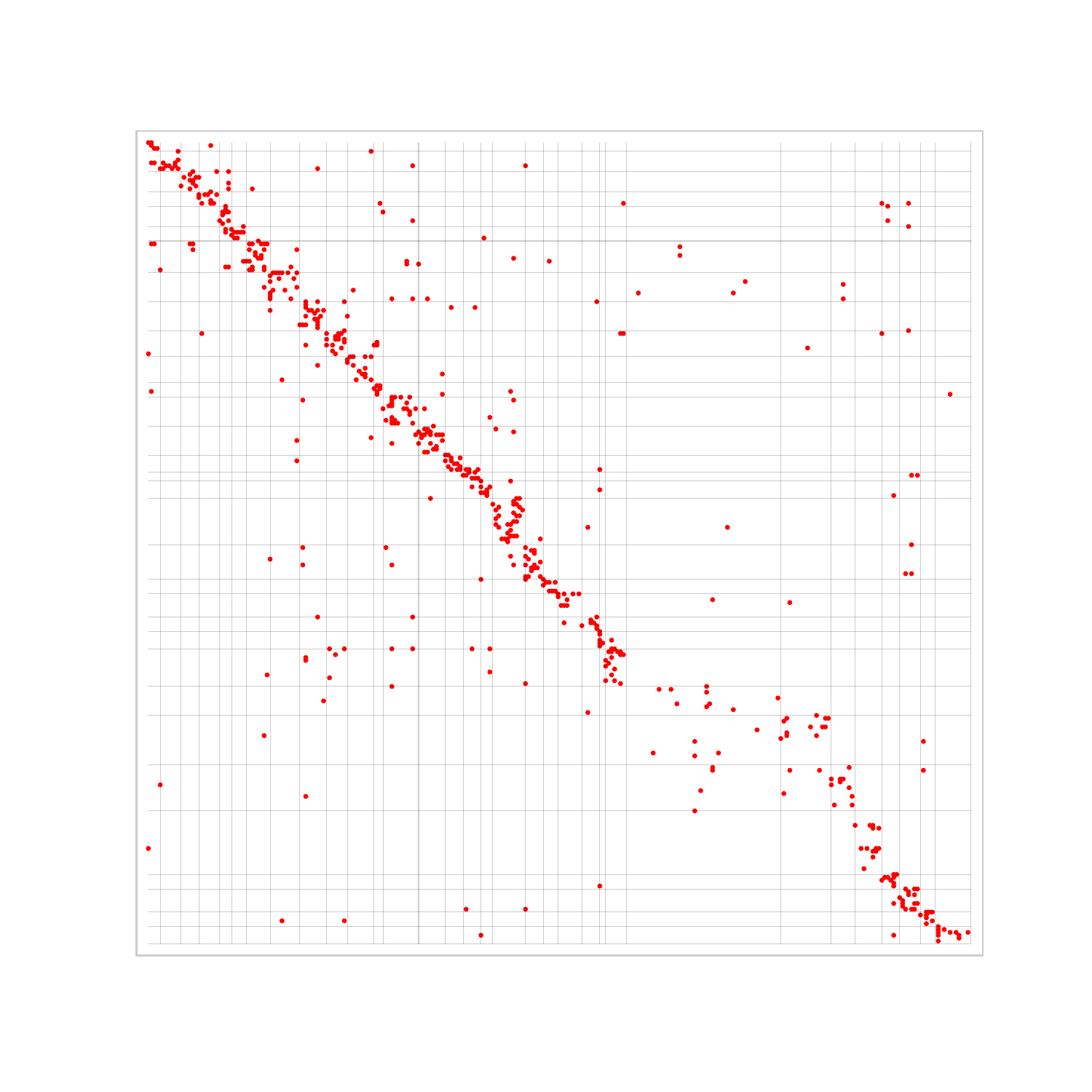}
      \caption{\tiny\\$s_5^3=12.171$}
    \end{subfigure}
  \end{minipage}

      \hfill 
    \end{minipage}
  }

  \centering
  \begin{subfigure}[t]{0.49\textwidth}
    \includegraphics[width=1\linewidth]{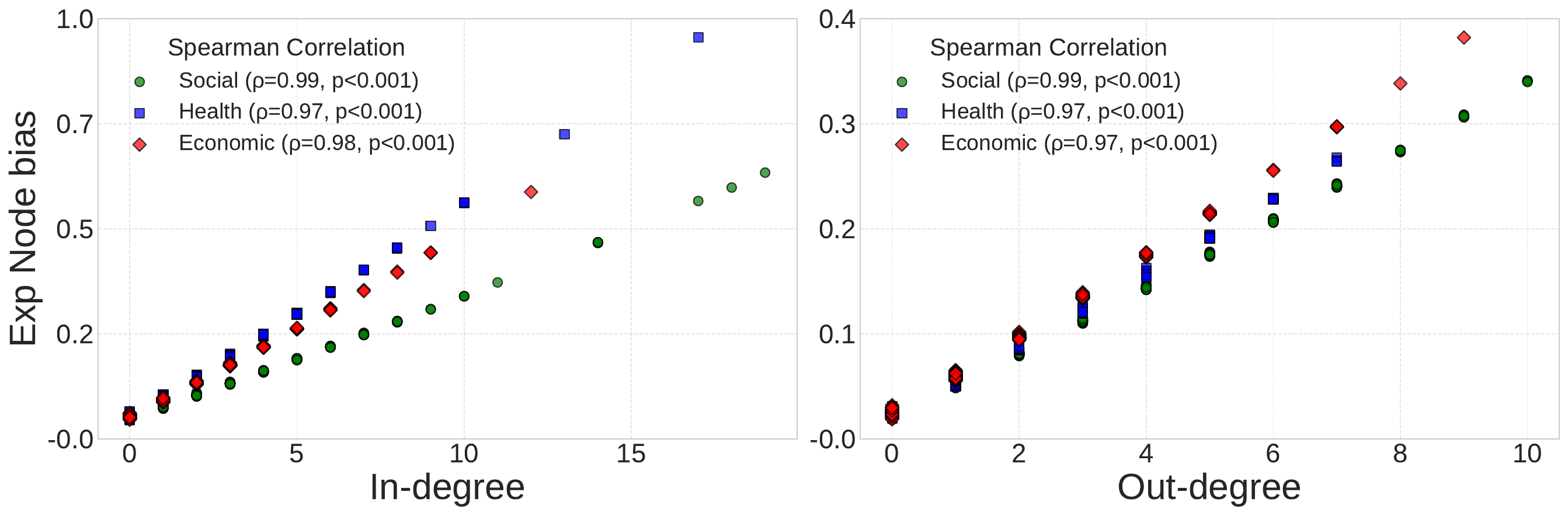}
    \caption{\small Bias model (dependence and independence): inferred node biases vs. in- and out-degree across layers.  }
  \end{subfigure}\hfill
  \begin{subfigure}[t]{0.49\textwidth}
    \includegraphics[width=1\linewidth]{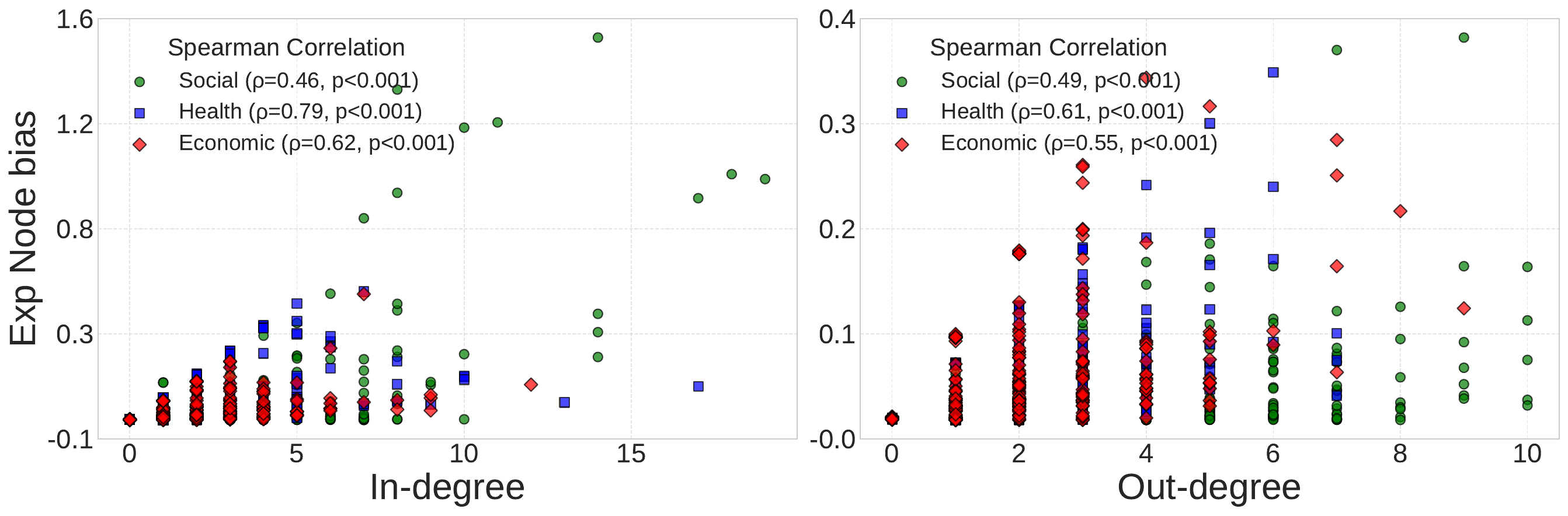}
    \caption{\small Full model (adds interdependence): inferred node biases vs. in- and out-degree across layers. }
  \end{subfigure}
  \begin{subfigure}[t]{0.16\textwidth}
    \includegraphics[width=\linewidth]{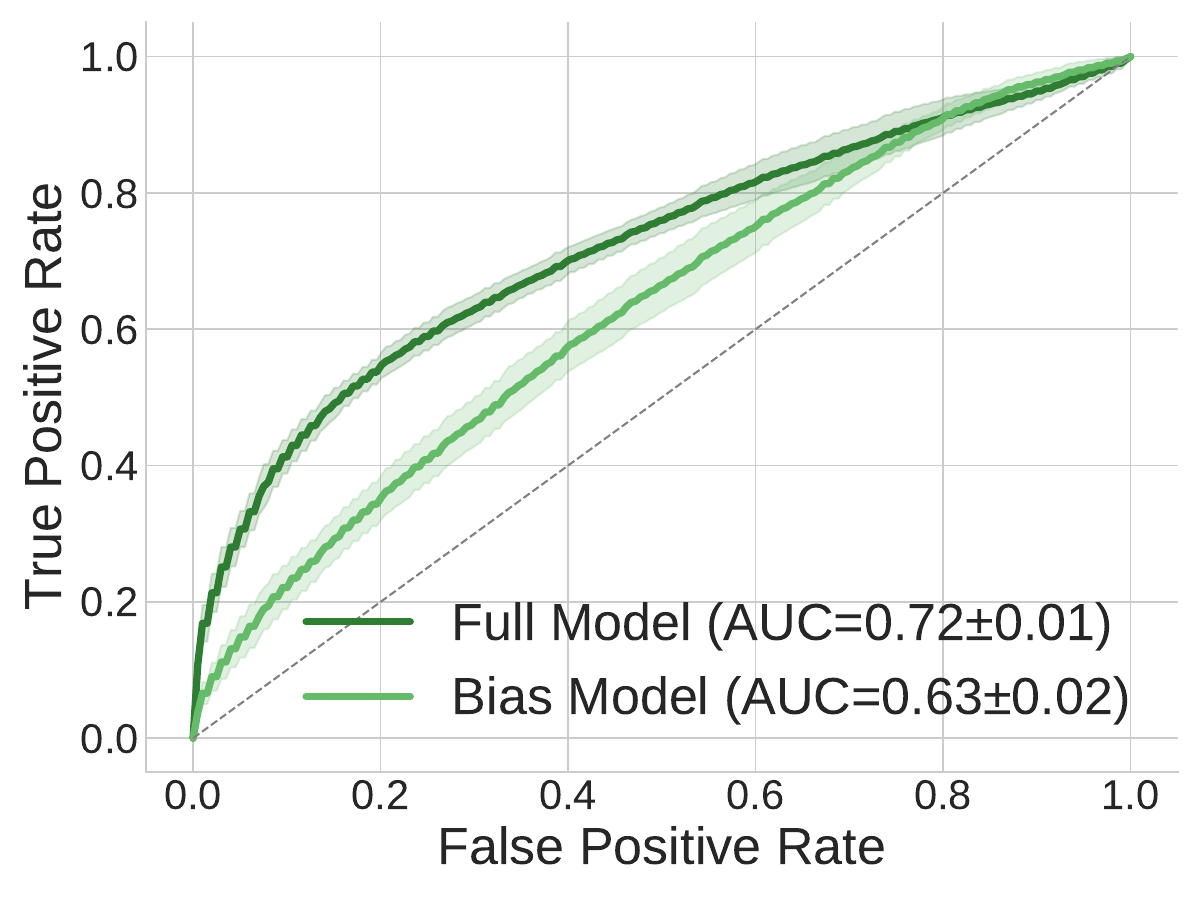}
    \caption{\small ROC Social}
  \end{subfigure}\hfill
  \begin{subfigure}[t]{0.16\textwidth}
    \includegraphics[width=\linewidth]{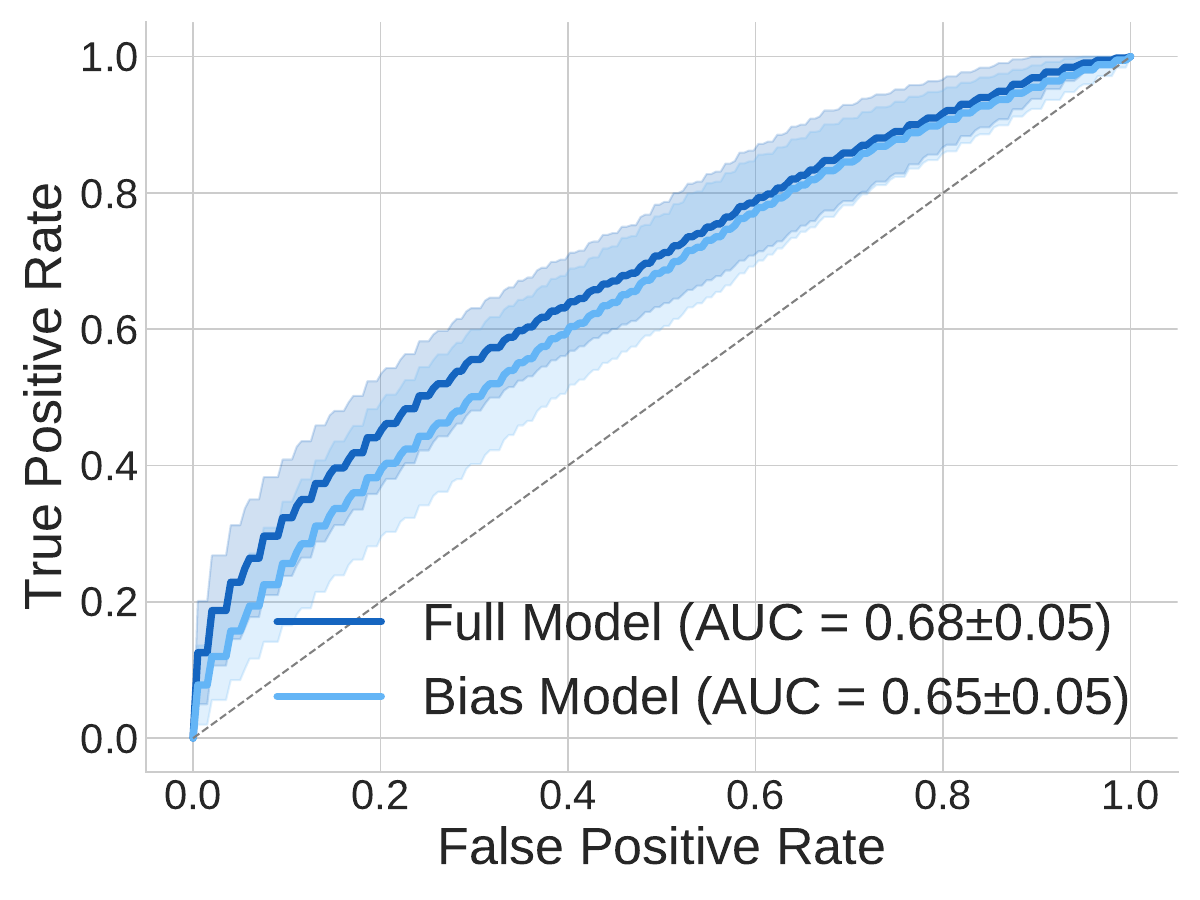}
    \caption{\small ROC Health}
  \end{subfigure}\hfill
  \begin{subfigure}[t]{0.17\textwidth}
    \includegraphics[width=\linewidth]{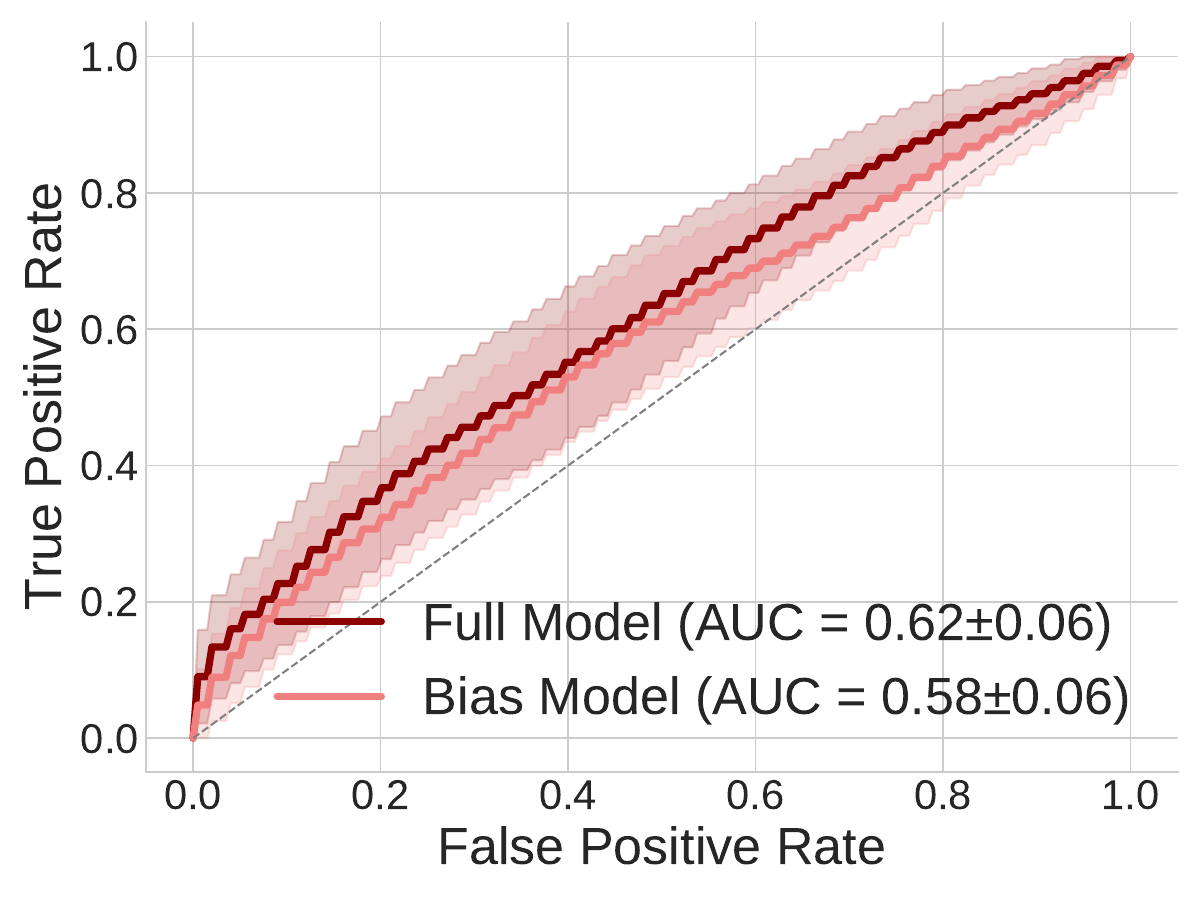}
    \caption{\small ROC Economic}
  \end{subfigure}
  \begin{subfigure}[t]{0.16\textwidth}
    \includegraphics[width=\linewidth]{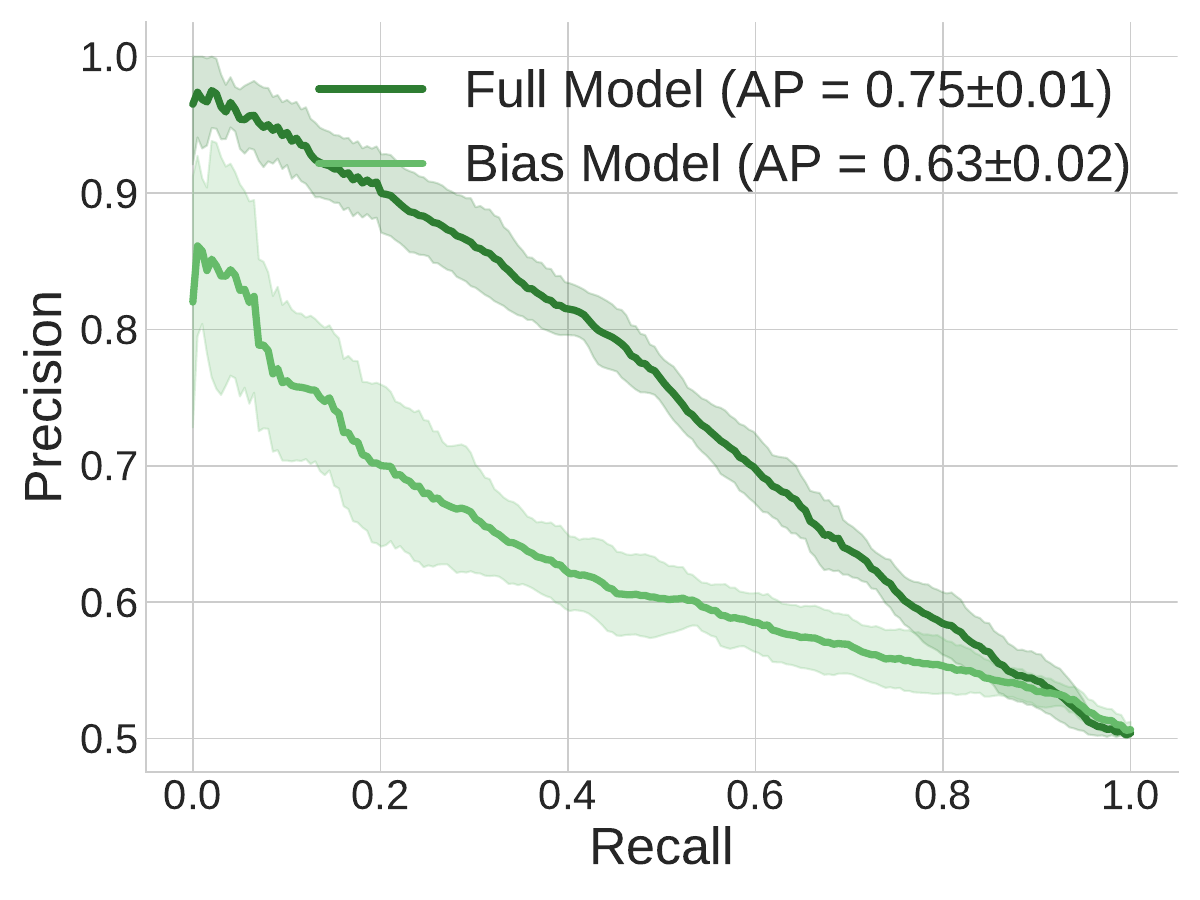}
    \caption{\small PR Social}
  \end{subfigure}\hfill
  \begin{subfigure}[t]{0.16\textwidth}
    \includegraphics[width=\linewidth]{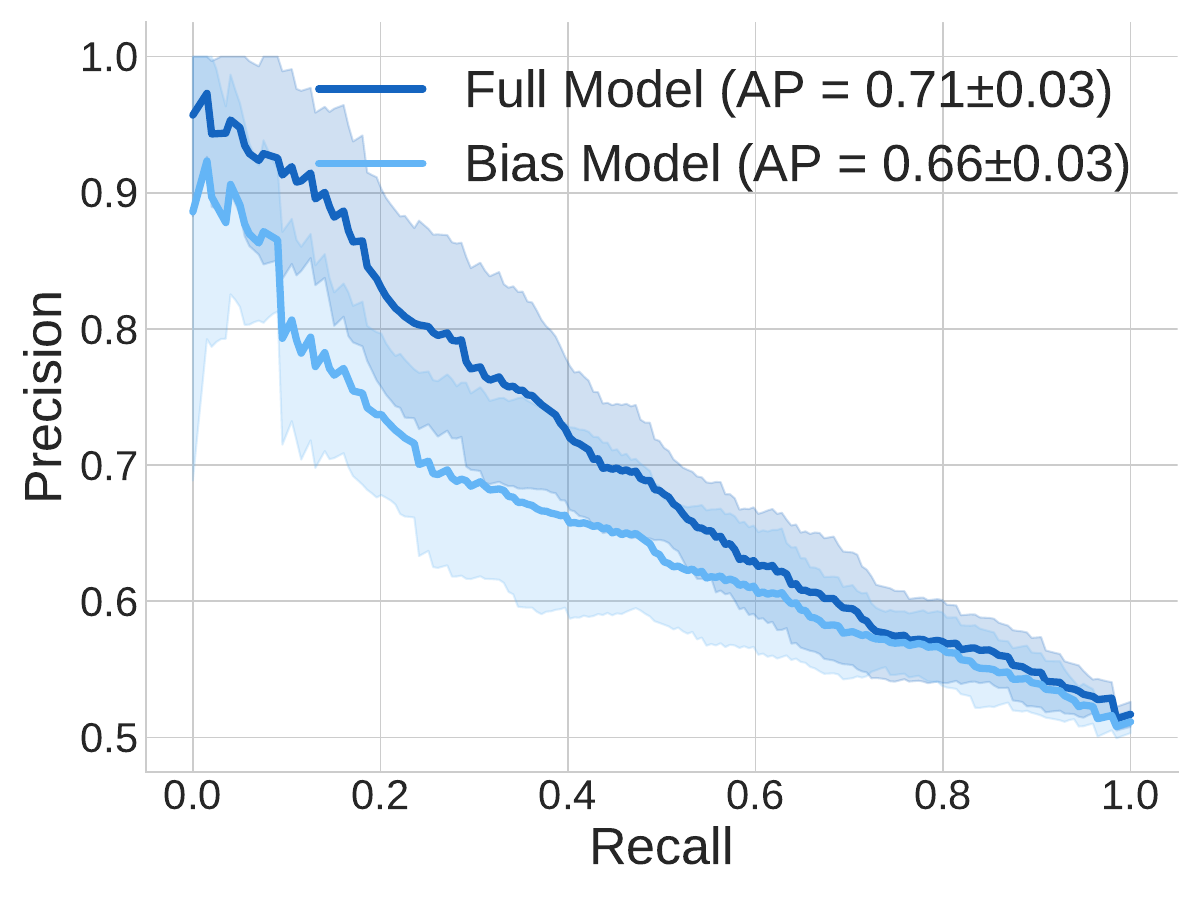}
    \caption{\small PR Health}
  \end{subfigure}\hfill
  \begin{subfigure}[t]{0.16\textwidth}
    \includegraphics[width=\linewidth]{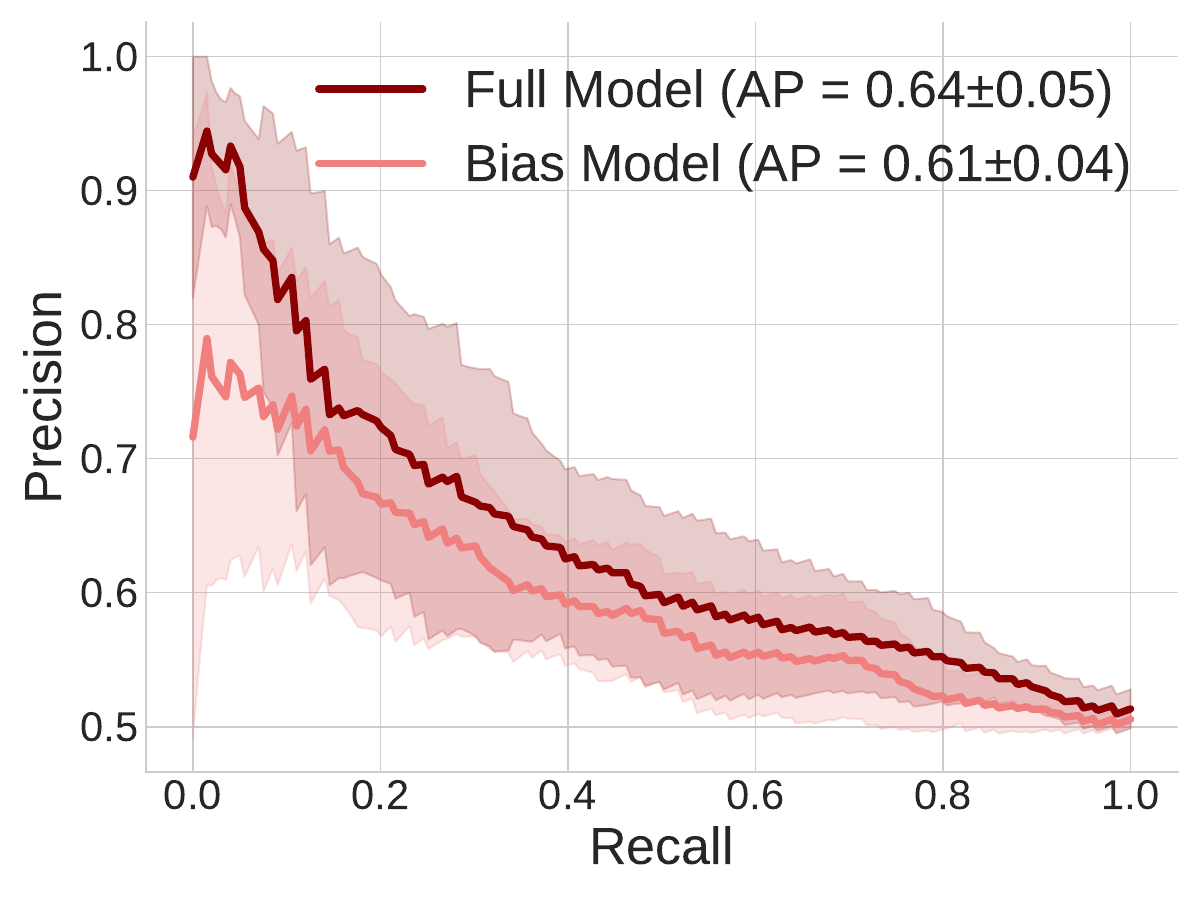}
    \caption{\small PR Economic}
  \end{subfigure}

  \caption{\textbf{Model behavior and structural interpretation on village network \# 51.} Panel (a, right) shows the learned role simplex with directed social (green), health (blue), and economic (red) ties; arrows indicate direction, and circles represent source/target embeddings ($\mathbf{Z}, \mathbf{W} \in \Delta_2$). Panel (a, left) separates ties by layer, with node colors denoting dominant roles. Panel (b) shows violin plots of node membership strengths across layers for source ($\mathbf{Z}$) and target ($\mathbf{W}$) roles. Panels (c)–(q) display the inferred multi-scale structure per layer. Adjacency matrices are reordered by dominant memberships $\mathbf{u}_i^{l,h}$ and $\mathbf{v}_j^{l,h}$; hierarchy strength $s_h^l$ quantifies the contribution of each level $h$. Deeper rows reveal finer group structure. Panel (r) shows bias model correlations between in-/out-degree and inferred biases ($\exp{\gamma_j^l}$, $\exp{\beta_i^l}$), highlighting network status effects. Panel (s) shows reduced correlations in the full model, indicating structural signal is captured via interdependence. Panels (t)–(v) compare ROC curves and AUC scores for full vs. bias models using 10-fold cross-validation (shaded = uncertainty). Panels (w)–(y) show the corresponding PR curves.}
  \label{fig:overall_50}
\end{figure*}

\begin{figure*}[b!]

\centering
  \centering
  \adjustbox{valign=b}{%
    \begin{minipage}[t]{0.62\textwidth}
      \begin{subfigure}[t]{\linewidth}
        \includegraphics[width=1\linewidth]{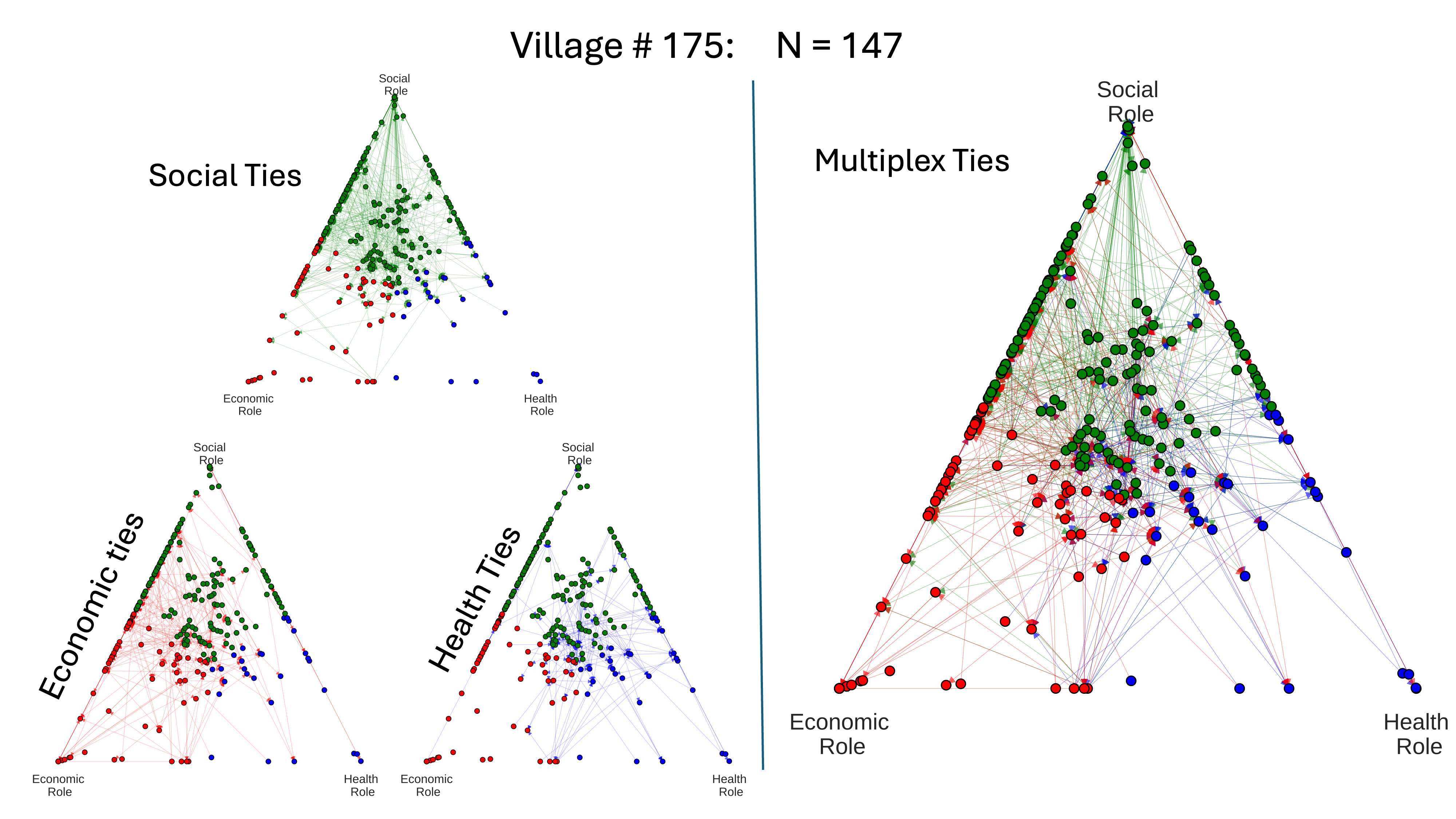}
        \caption{\small Role simplex visualizations for village $\# 175$. Each point represents a node, colored by dominant role. Left: tie-specific roles. Right: multiplex network with all ties overlaid. }
      \end{subfigure}
      \vspace{1em}

      \begin{subfigure}[t]{\linewidth}
        \includegraphics[width=1\linewidth]{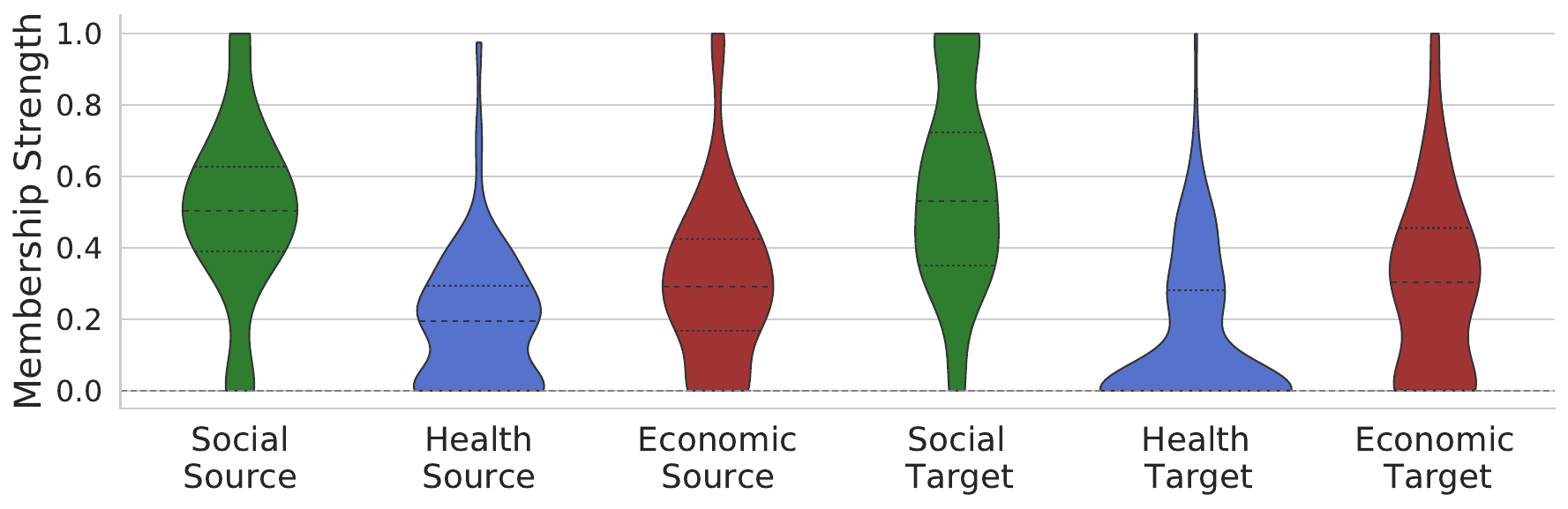}
        \caption{\small Violin plots of node-role membership strengths for source and target positions across domains. }
      \end{subfigure}
    \end{minipage}
  }\hfill
  \adjustbox{valign=b}{%
    \begin{minipage}[t]{0.36\textwidth}

\makebox[1\linewidth]{\centering \textbf{Multi-scale Layer Structure:}}
\\[1ex]
 \makebox[0.32\linewidth]{\centering \textbf{Social}} \hfill
  \makebox[0.32\linewidth]{\centering \textbf{Health}} \hfill
  \makebox[0.32\linewidth]{\centering \textbf{Economic}} \\[-1ex]
    
\centering

  \begin{minipage}[c]{\linewidth}
    \begin{minipage}[c]{0.04\linewidth}
      \centering
      \rotatebox{90}{\small \textbf{$D_{h=1}=2$}}
    \end{minipage}%
    \hfill
     \captionsetup[subfigure]{skip=-5pt}
    \begin{subfigure}[c]{0.3\linewidth}
\includegraphics[width=\linewidth]{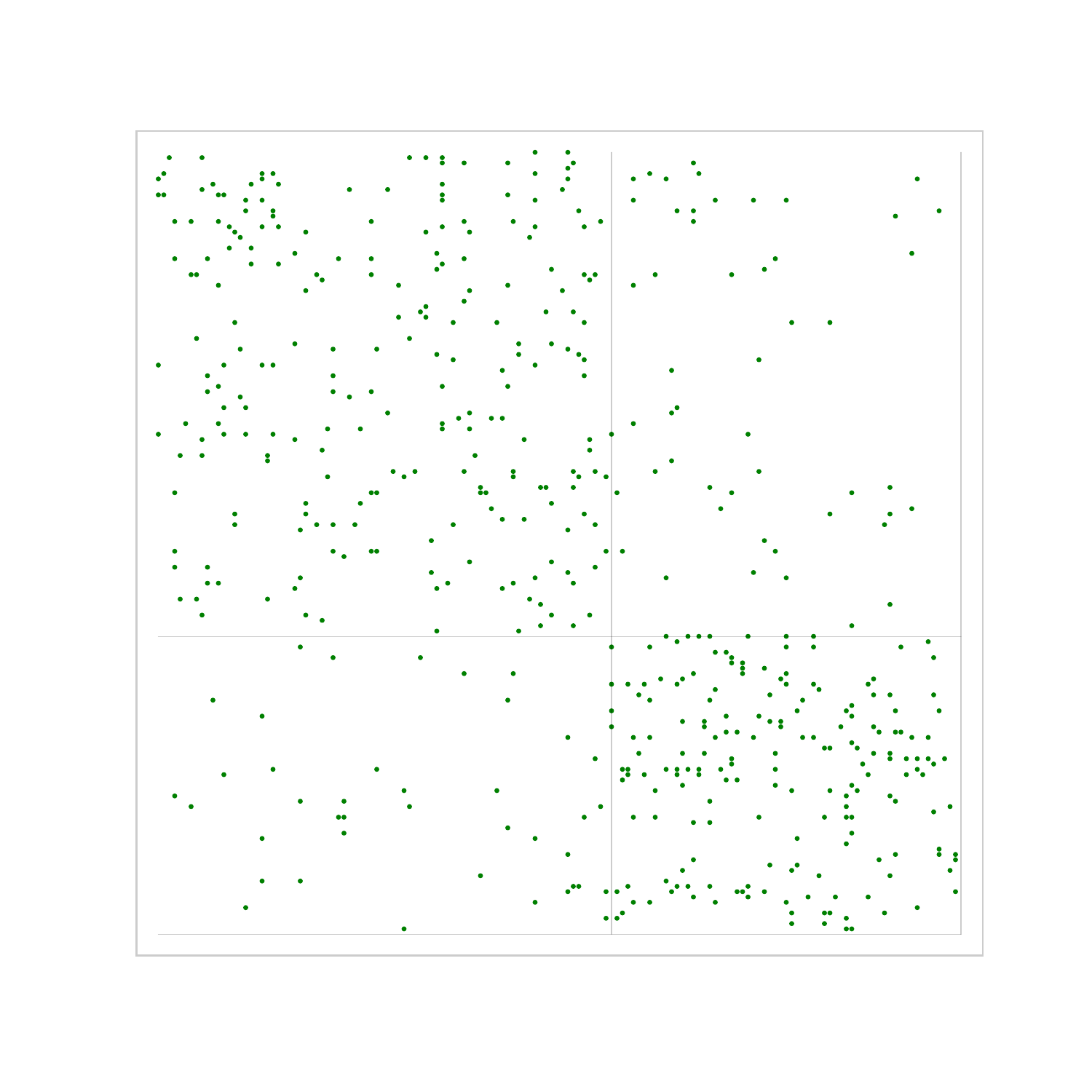} \caption{\tiny\\$s_1^1=0.001$}
    \end{subfigure}\hfill
     \captionsetup[subfigure]{skip=-5pt}
    \begin{subfigure}[c]{0.3\linewidth}
\includegraphics[width=\linewidth]{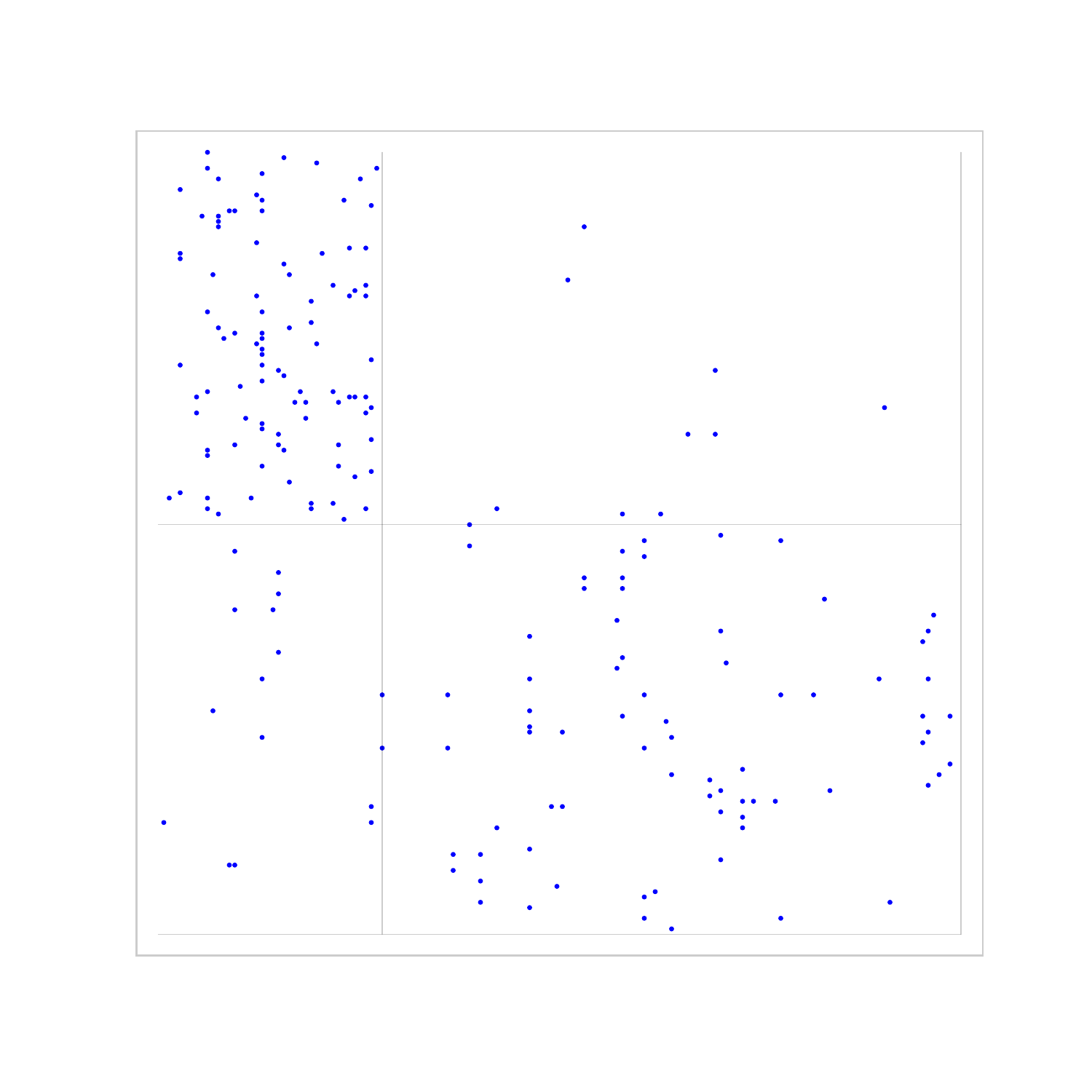} \caption{\tiny\\$s_1^2=0.004$}
    \end{subfigure}\hfill
    \captionsetup[subfigure]{skip=-5pt}
    \begin{subfigure}[c]{0.3\linewidth}
\includegraphics[width=\linewidth]{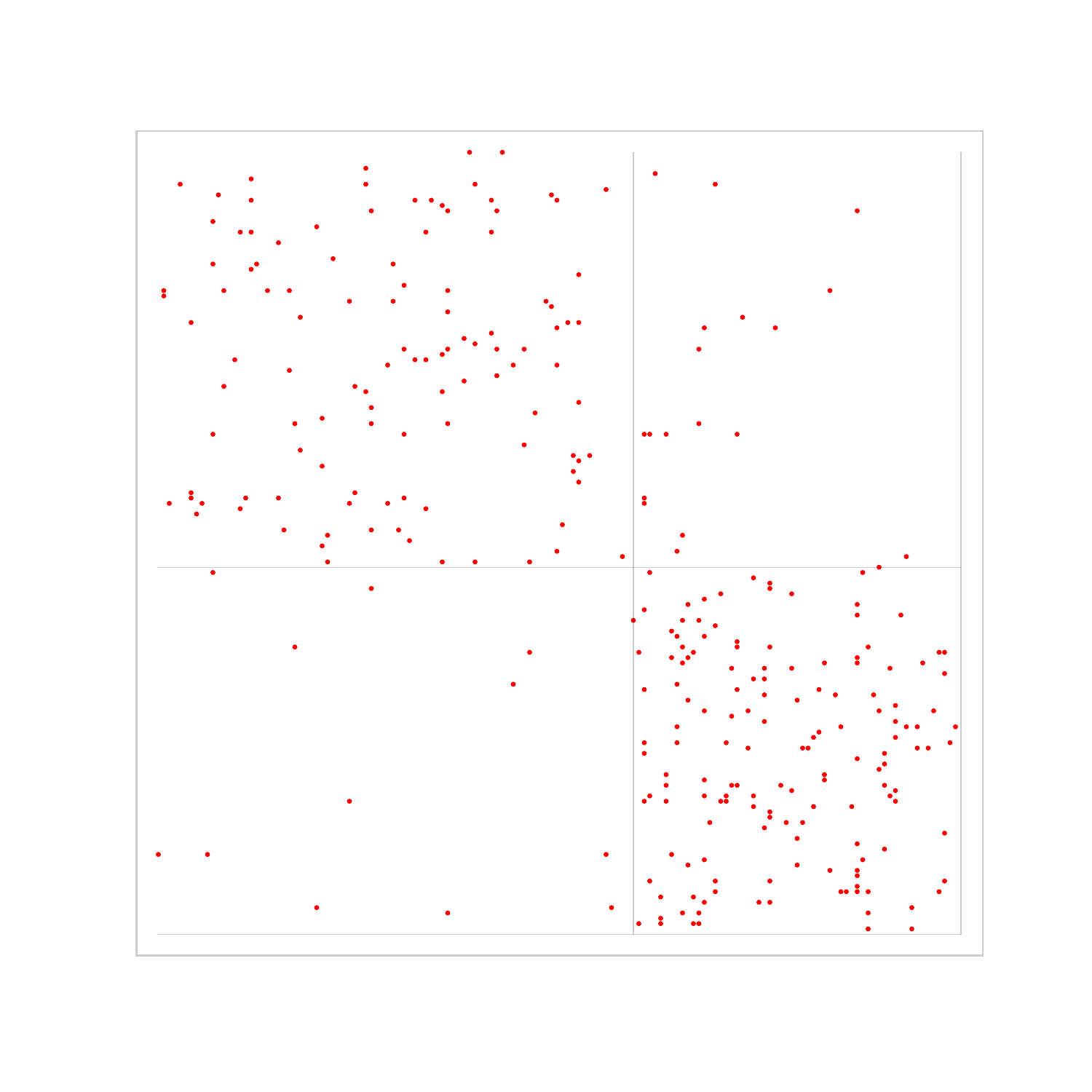}
      \caption{\tiny\\$s_1^3=0.004$}
    \end{subfigure}
  \end{minipage}

  \begin{minipage}[c]{\linewidth}
    \begin{minipage}[c]{0.04\linewidth}
      \centering
      \rotatebox{90}{\small \textbf{$D_{h=2}=4$}}
    \end{minipage}%
    \hfill
     \captionsetup[subfigure]{skip=-5pt}
    \begin{subfigure}[c]{0.3\linewidth}
      \includegraphics[width=\linewidth]{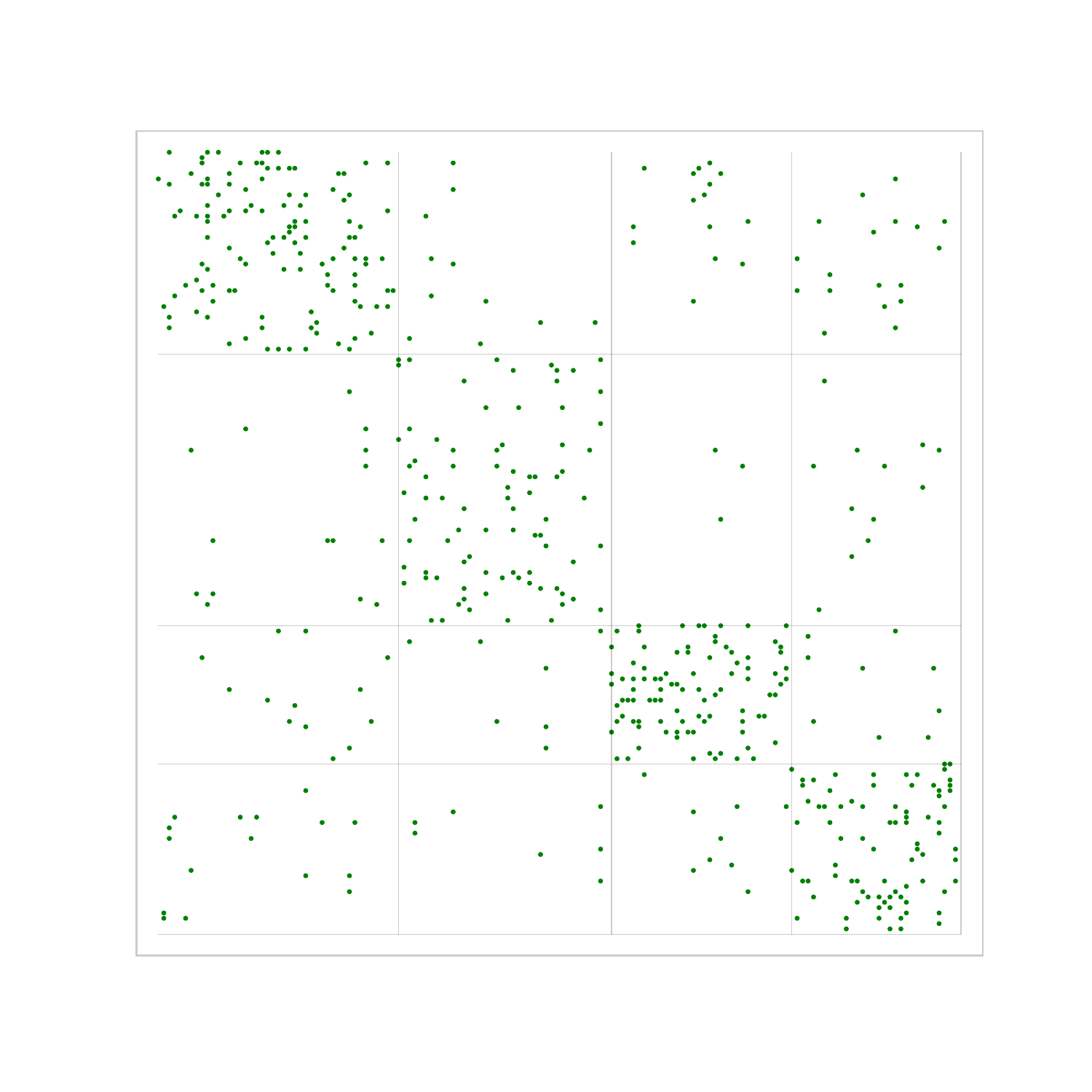}
       \caption{\tiny\\$s_2^1=0.015$}
    \end{subfigure}\hfill
     \captionsetup[subfigure]{skip=-5pt}
    \begin{subfigure}[c]{0.3\linewidth}
      \includegraphics[width=\linewidth]{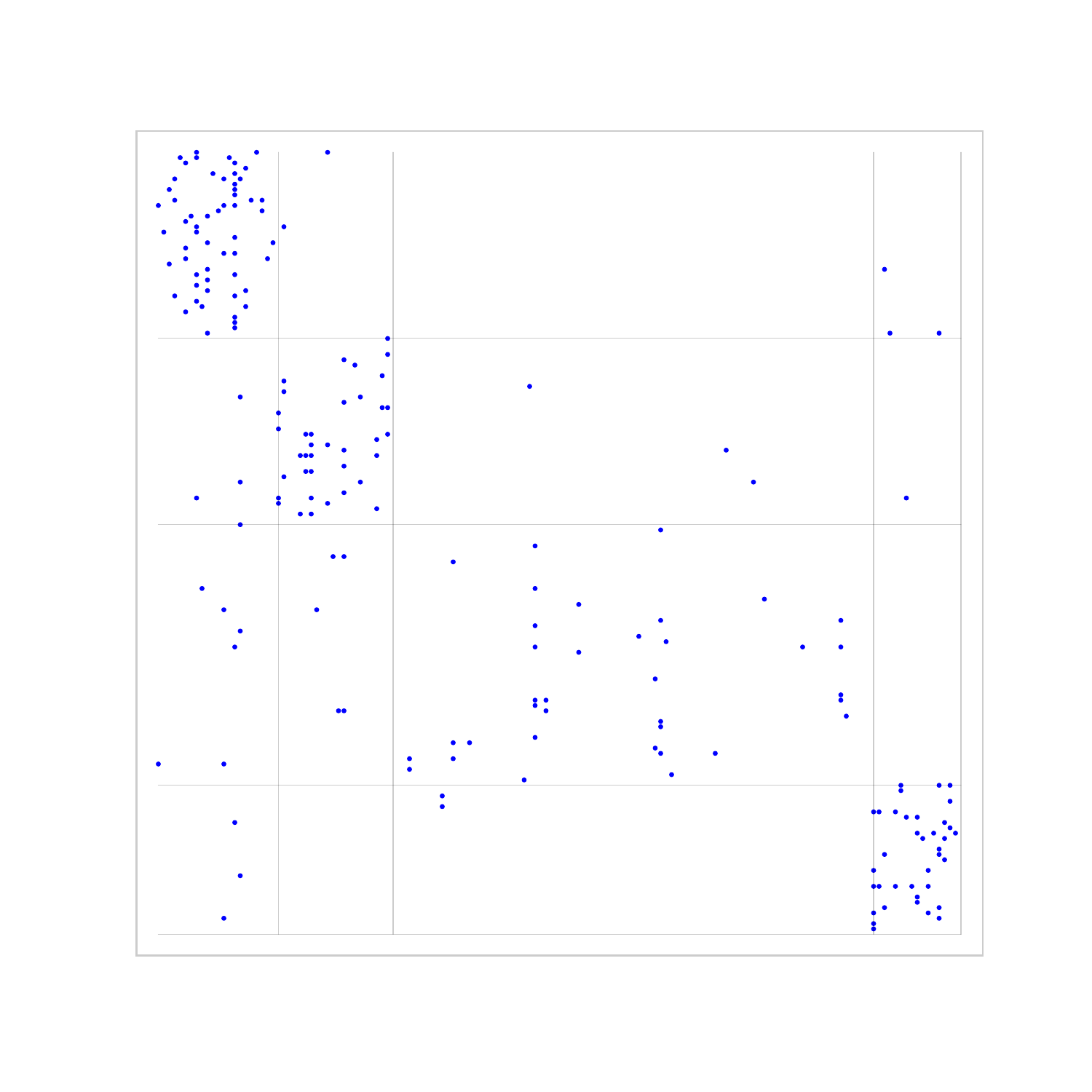}
       \caption{\tiny\\$s_2^2=3.627$}
    \end{subfigure}\hfill
     \captionsetup[subfigure]{skip=-5pt}
    \begin{subfigure}[c]{0.3\linewidth}
      \includegraphics[width=\linewidth]{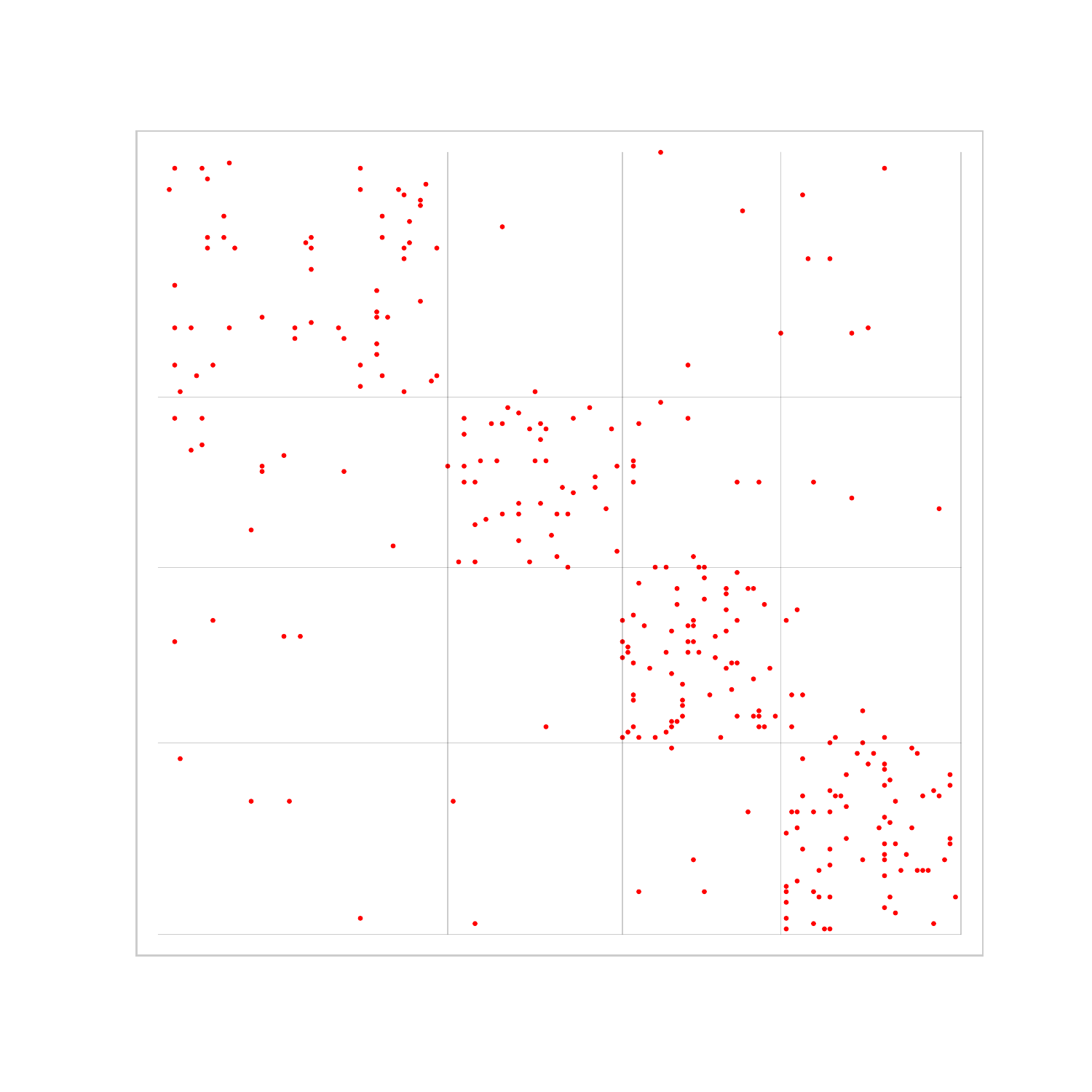}
       \caption{\tiny\\$s_2^3=1.062$}
    \end{subfigure}
  \end{minipage}
  
  \begin{minipage}[c]{\linewidth}
    \begin{minipage}[c]{0.04\linewidth}
      \centering
      \rotatebox{90}{\small \textbf{$D_{h=3}=8$}}
    \end{minipage}%
    \hfill
     \captionsetup[subfigure]{skip=-5pt}
    \begin{subfigure}[c]{0.3\linewidth}
      \includegraphics[width=\linewidth]{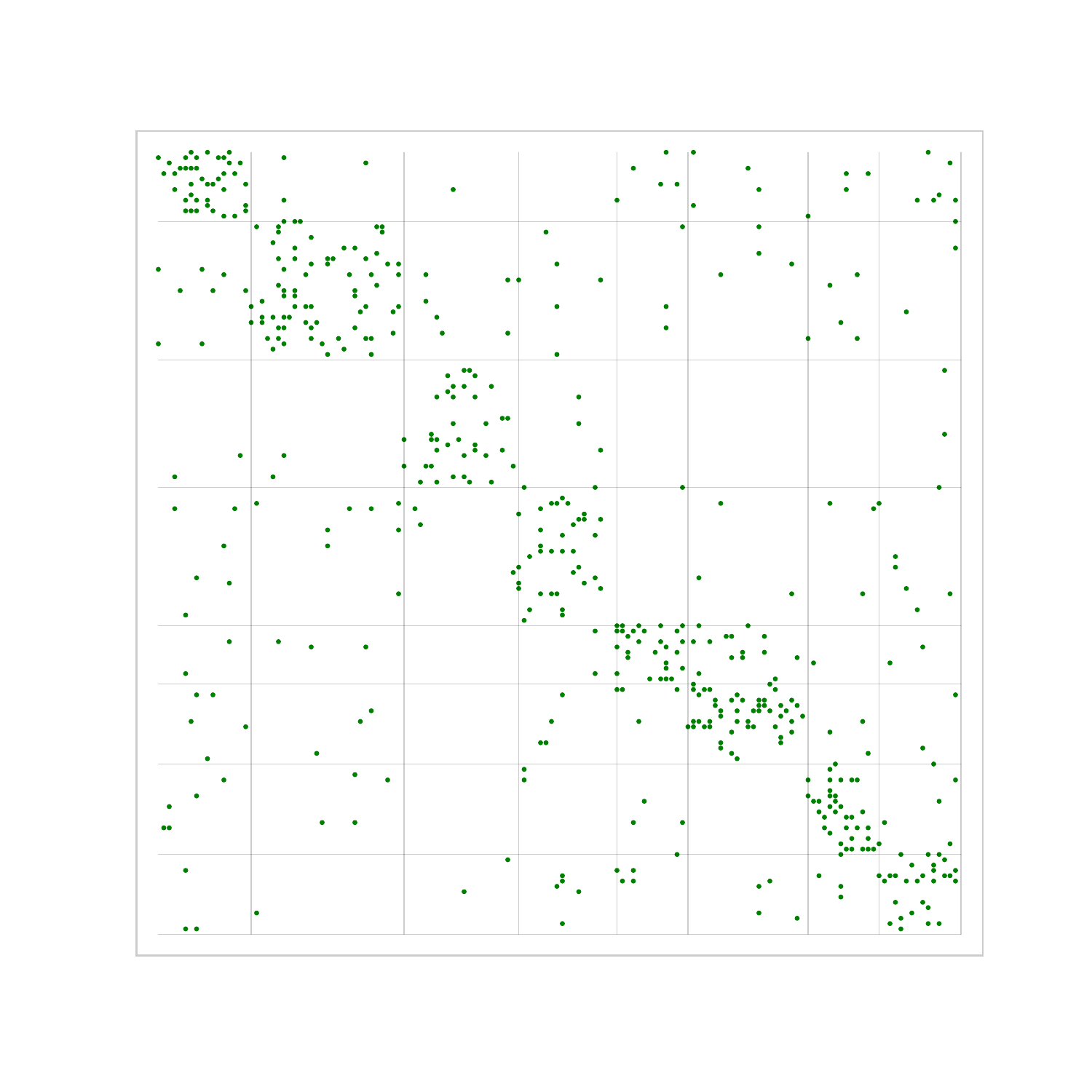}
       \caption{\tiny\\$s_3^1=0.301$}
    \end{subfigure}\hfill
     \captionsetup[subfigure]{skip=-5pt}
    \begin{subfigure}[c]{0.3\linewidth}
      \includegraphics[width=\linewidth]{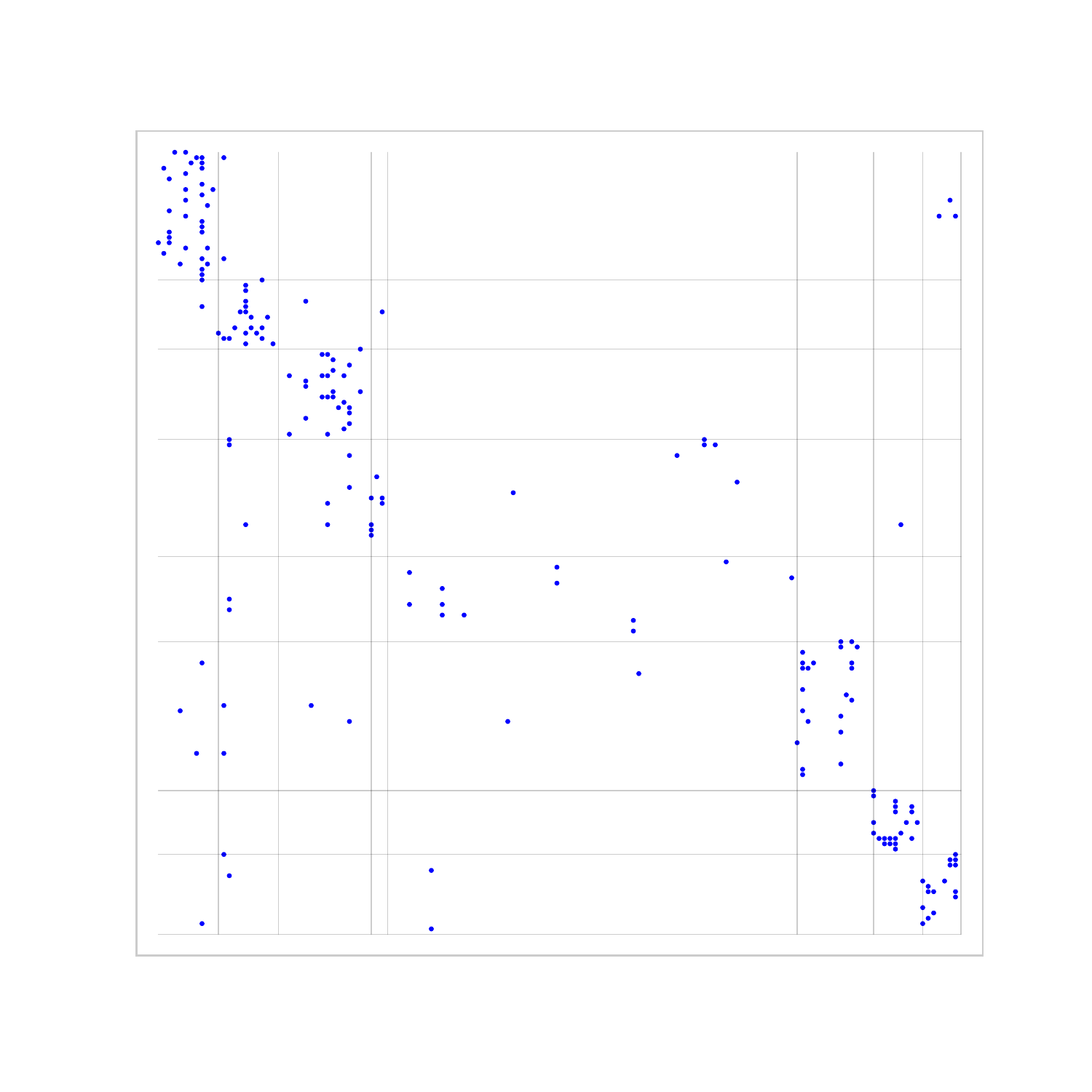}
       \caption{\tiny\\$s_3^2=4.446$}
    \end{subfigure}\hfill
     \captionsetup[subfigure]{skip=-5pt}
    \begin{subfigure}[c]{0.3\linewidth}
      \includegraphics[width=\linewidth]{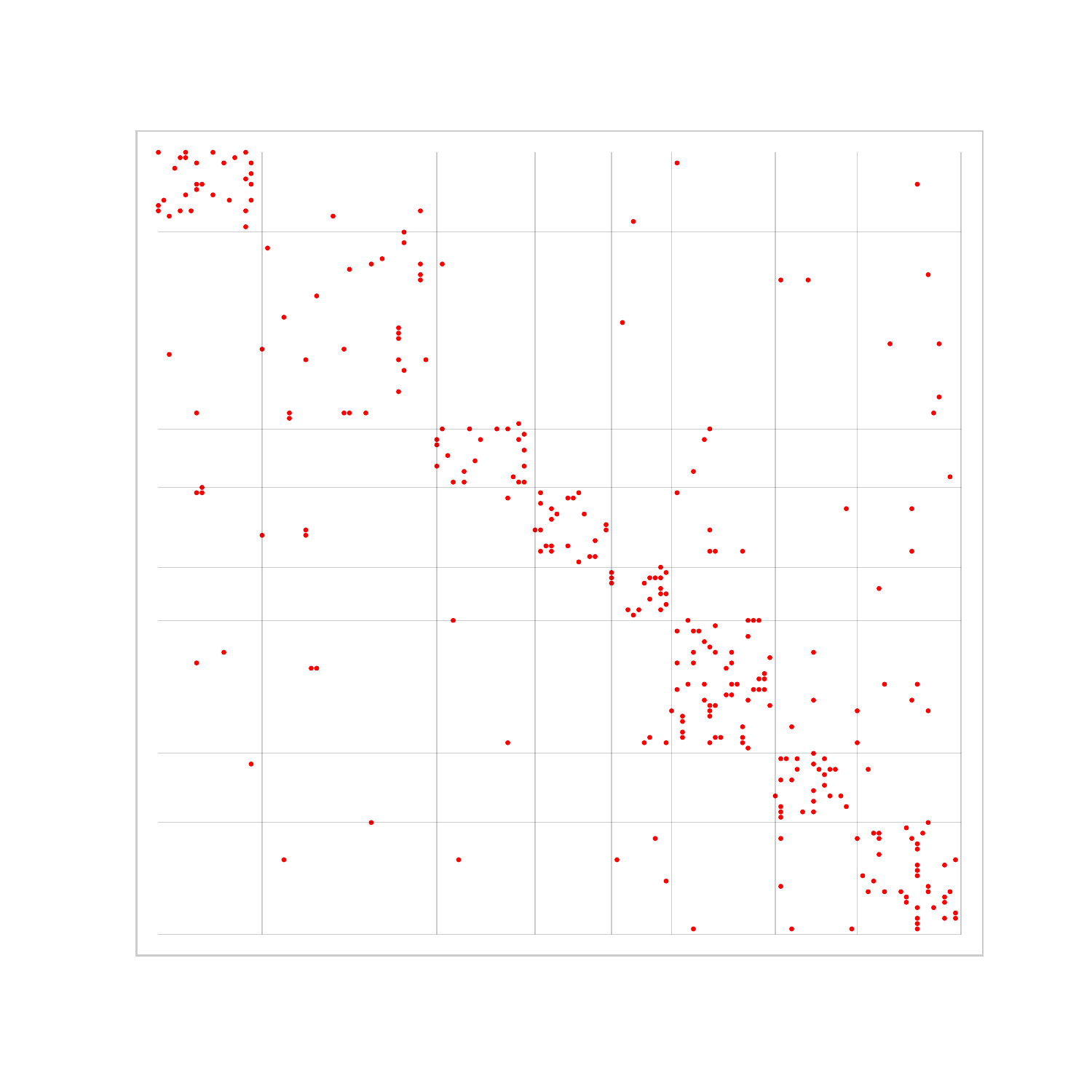}
       \caption{\tiny\\$s_3^3=8.690$}
    \end{subfigure}
  \end{minipage}
  
  \begin{minipage}[c]{\linewidth}
    \begin{minipage}[c]{0.04\linewidth}
      \centering
      \rotatebox{90}{\small \textbf{$D_{h=4}=16$}}
    \end{minipage}%
    \hfill \captionsetup[subfigure]{skip=-5pt}
    \begin{subfigure}[c]{0.3\linewidth}
      \includegraphics[width=\linewidth]{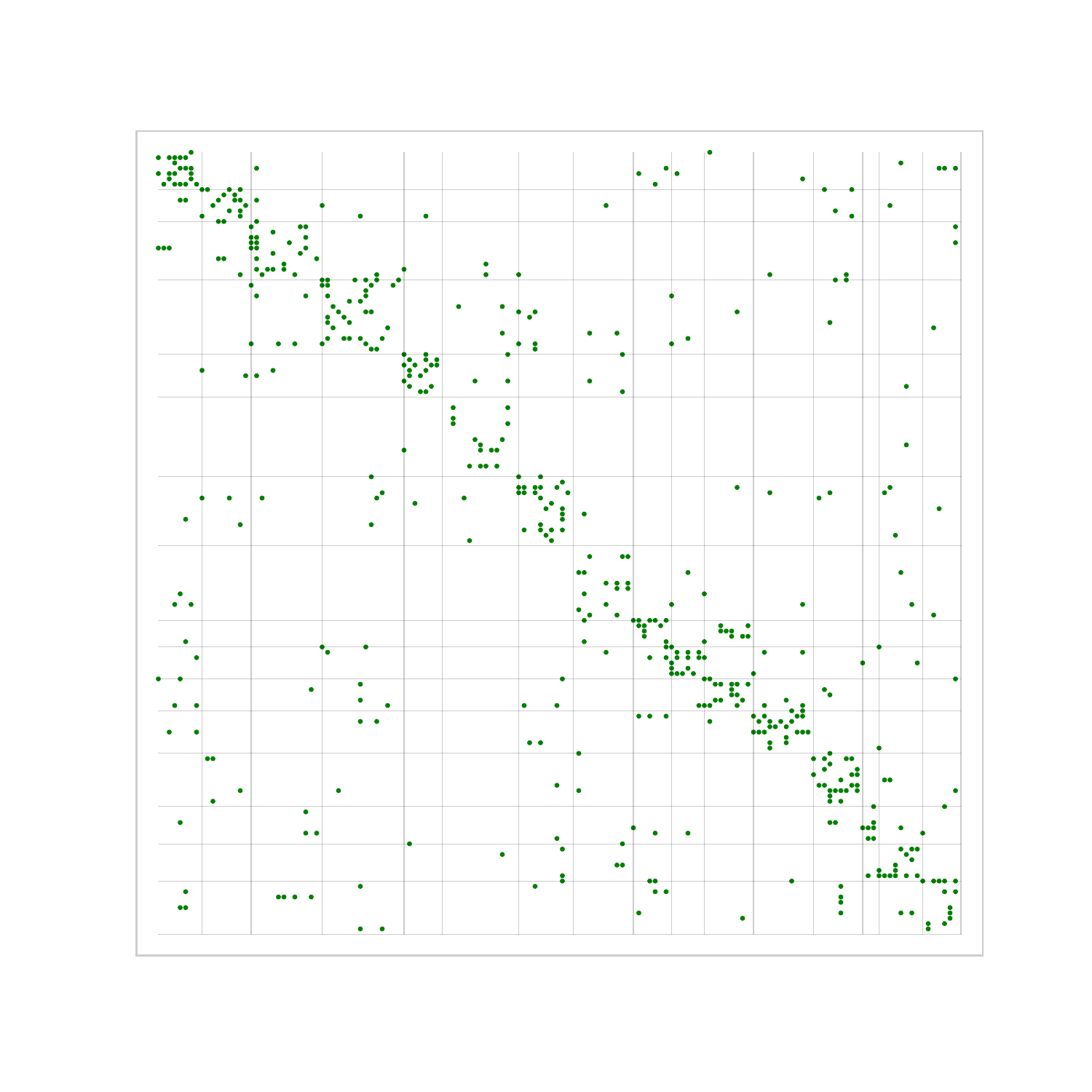} \caption{\tiny\\$s_4^1=6.741$}
    \end{subfigure}\hfill
     \captionsetup[subfigure]{skip=-5pt}
    \begin{subfigure}[c]{0.3\linewidth}
      \includegraphics[width=\linewidth]{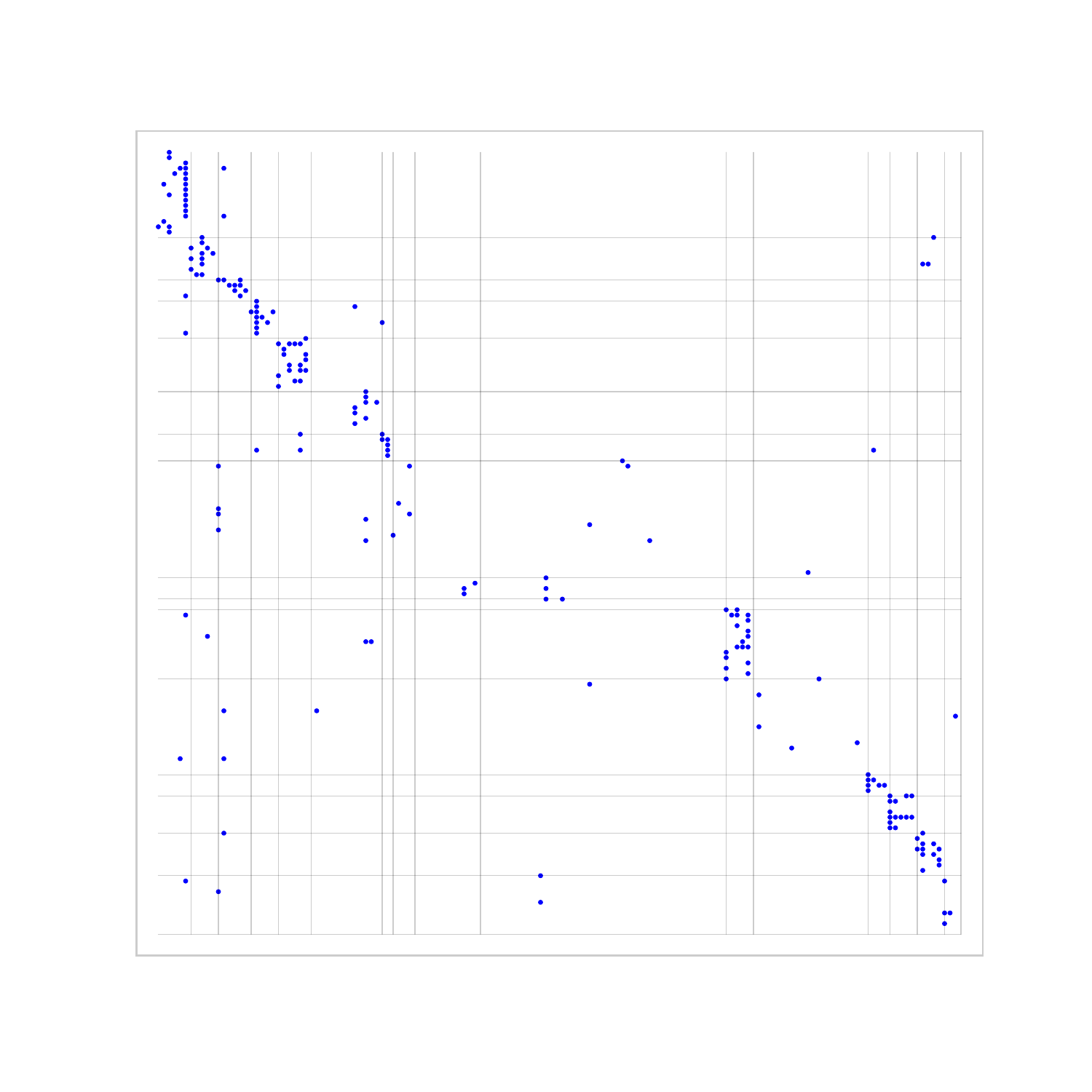} \caption{\tiny\\$s_4^2=32.855$}
    \end{subfigure}\hfill
     \captionsetup[subfigure]{skip=-5pt}
    \begin{subfigure}[c]{0.3\linewidth}
      \includegraphics[width=\linewidth]{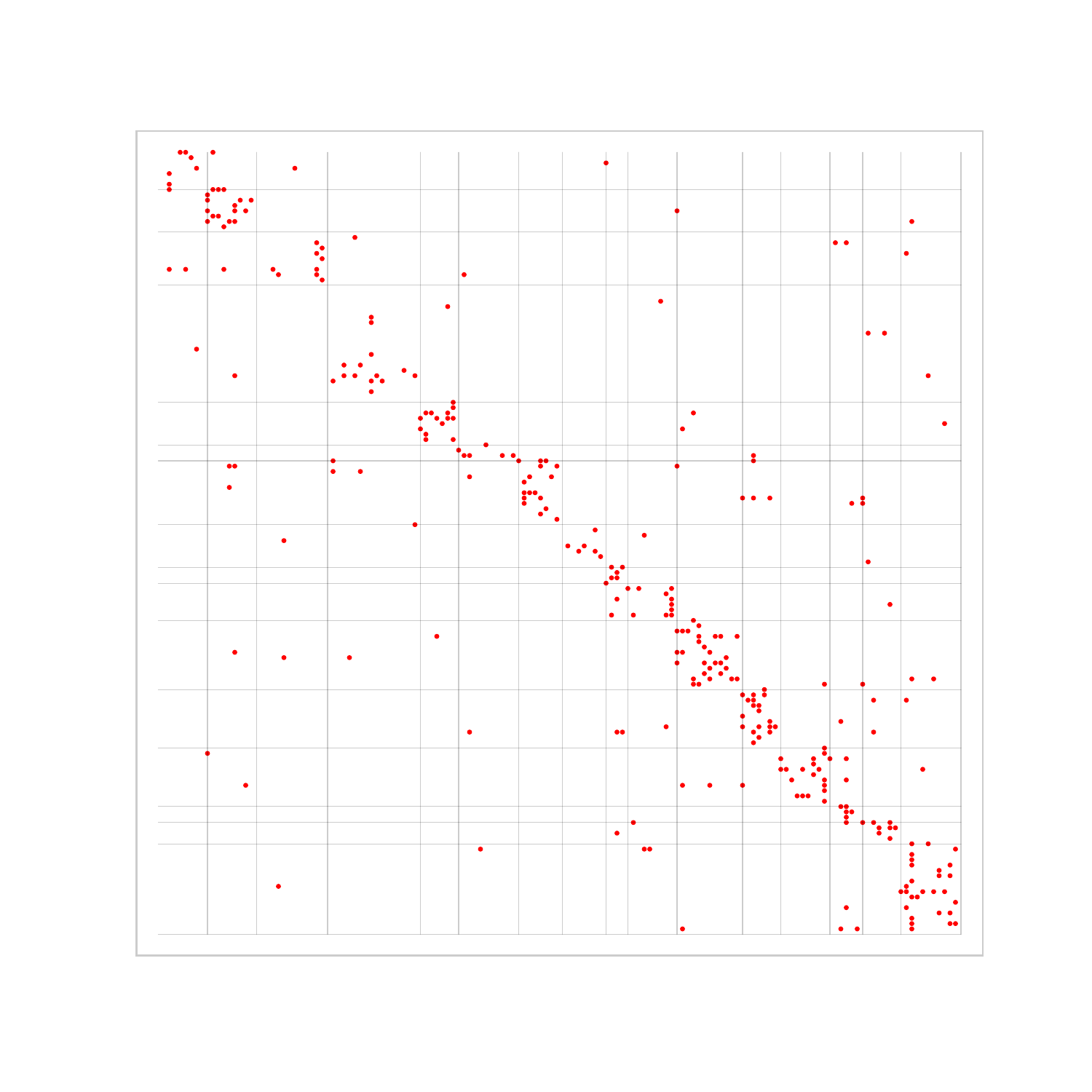} \caption{\tiny\\$s_4^3=8.496$}
    \end{subfigure}
  \end{minipage}

  \begin{minipage}[c]{\linewidth}
    \begin{minipage}[c]{0.04\linewidth}
      \centering
      \rotatebox{90}{\small \textbf{$D_{h=5}=32$}}
    \end{minipage}%
    \hfill  \captionsetup[subfigure]{skip=-5pt}
    \begin{subfigure}[c]{0.3\linewidth}
      \includegraphics[width=\linewidth]{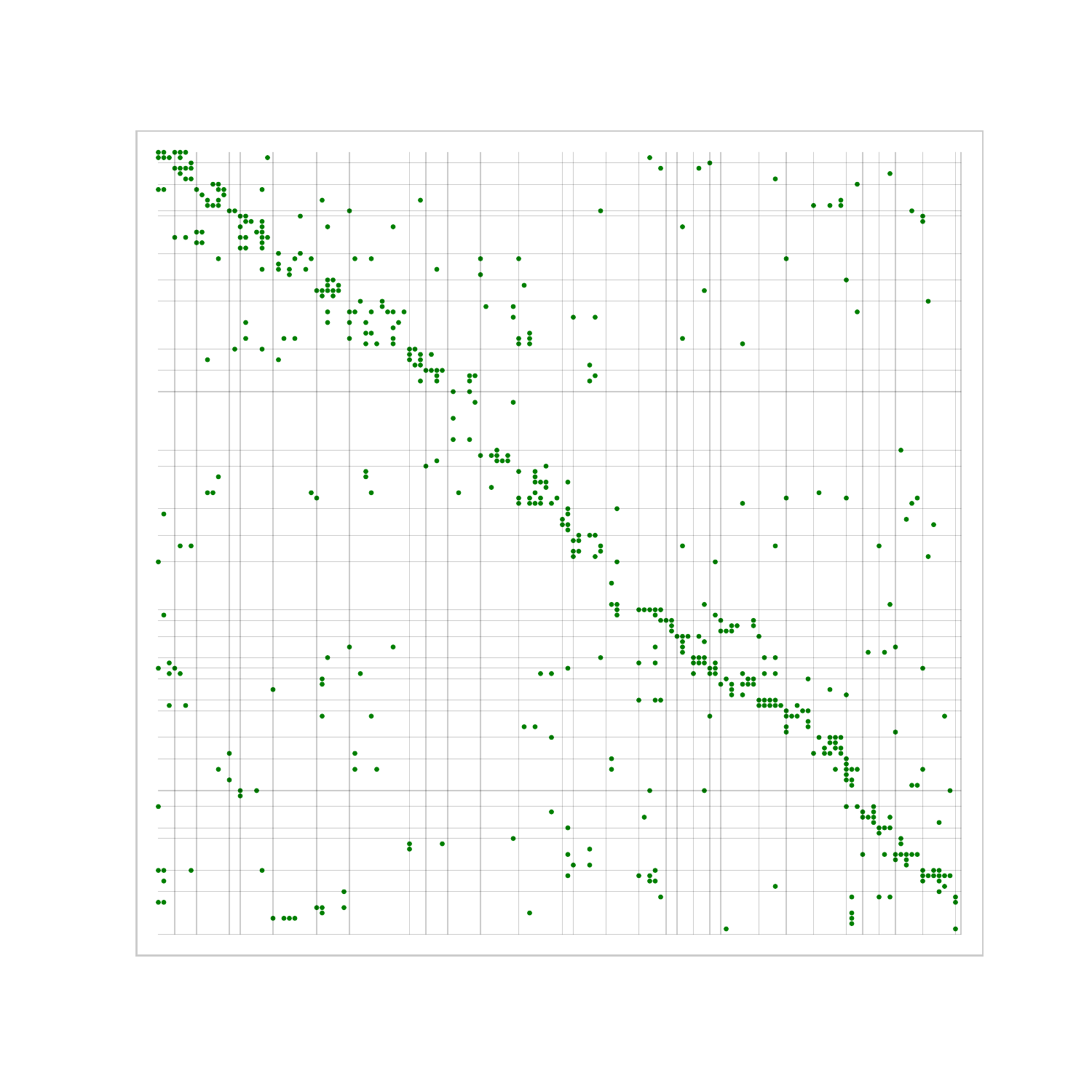}
      \caption{\tiny\\$s_5^1=84.742$}
    \end{subfigure}\hfill  \captionsetup[subfigure]{skip=-5pt}
    \begin{subfigure}[c]{0.3\linewidth}
      \includegraphics[width=\linewidth]{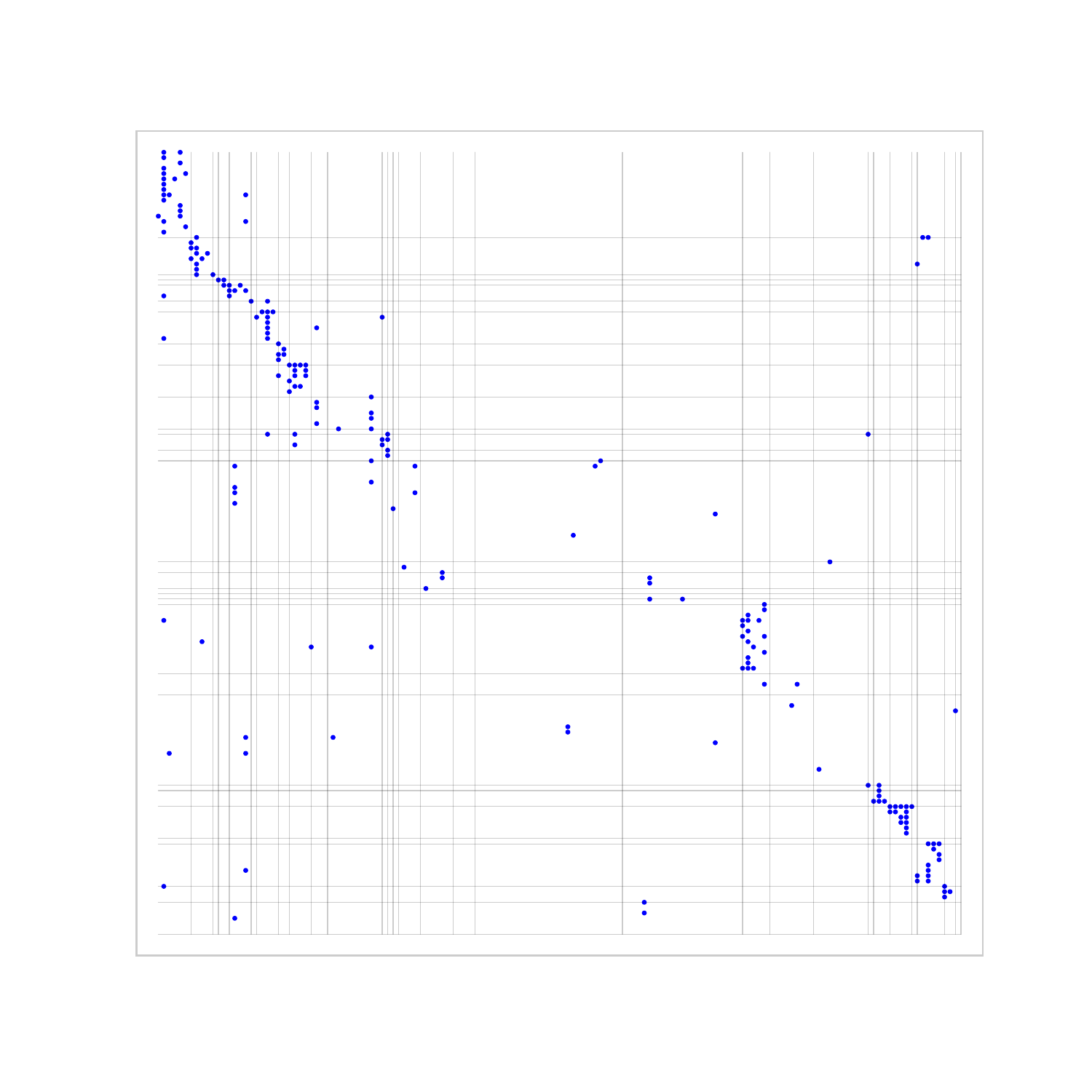}
      \caption{\tiny\\$s_5^2=50.865$}
    \end{subfigure}\hfill  \captionsetup[subfigure]{skip=-5pt}
    \begin{subfigure}[c]{0.3\linewidth}
      \includegraphics[width=\linewidth]{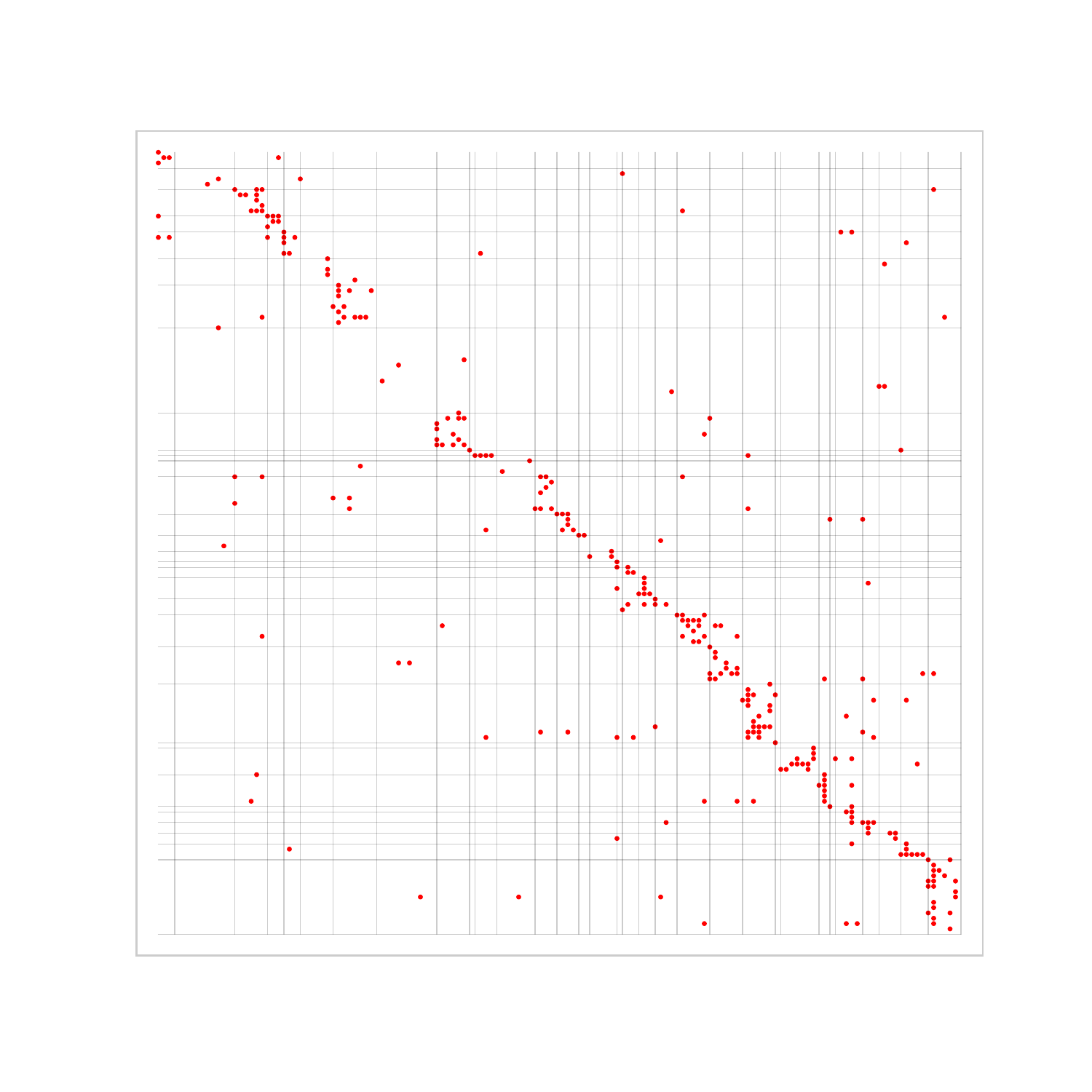}
      \caption{\tiny\\$s_5^3=73.547$}
    \end{subfigure}
  \end{minipage}

      \hfill 
    \end{minipage}
  }

  \centering
  \begin{subfigure}[t]{0.49\textwidth}
    \includegraphics[width=1\linewidth]{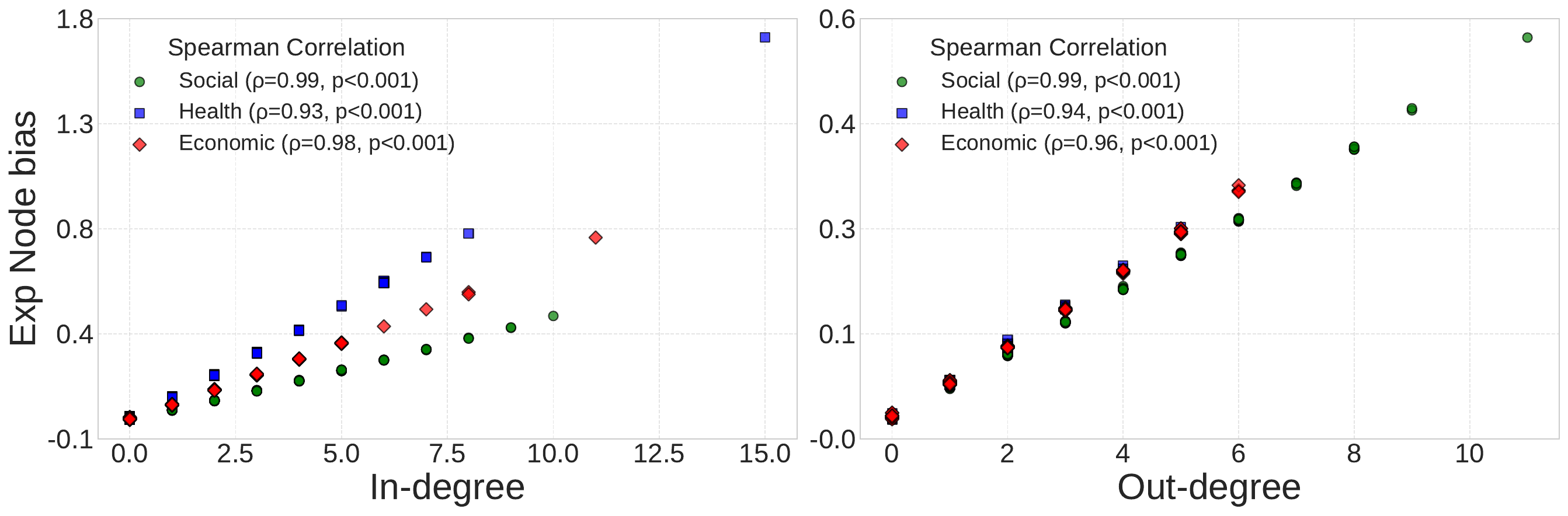}
    \caption{\small Bias model (dependence and independence): inferred node biases vs. in- and out-degree across layers.  }
  \end{subfigure}\hfill
  \begin{subfigure}[t]{0.49\textwidth}
    \includegraphics[width=1\linewidth]{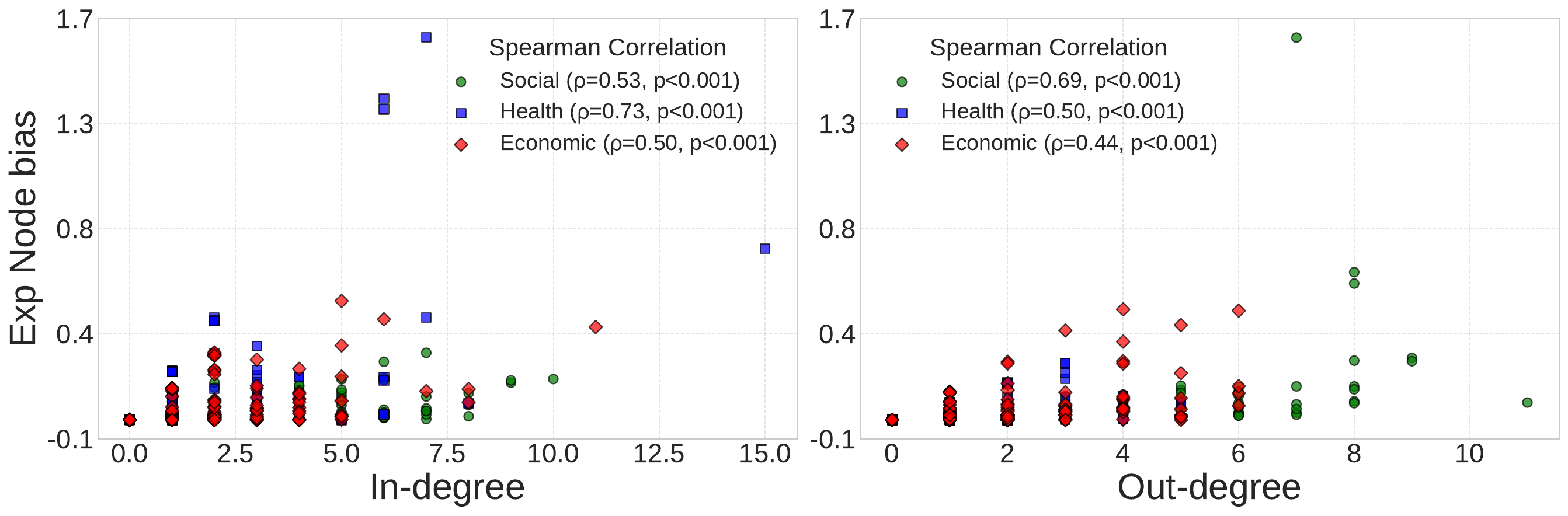}
    \caption{\small Full model (adds interdependence): inferred node biases vs. in- and out-degree across layers. }
  \end{subfigure}
  \begin{subfigure}[t]{0.16\textwidth}
    \includegraphics[width=\linewidth]{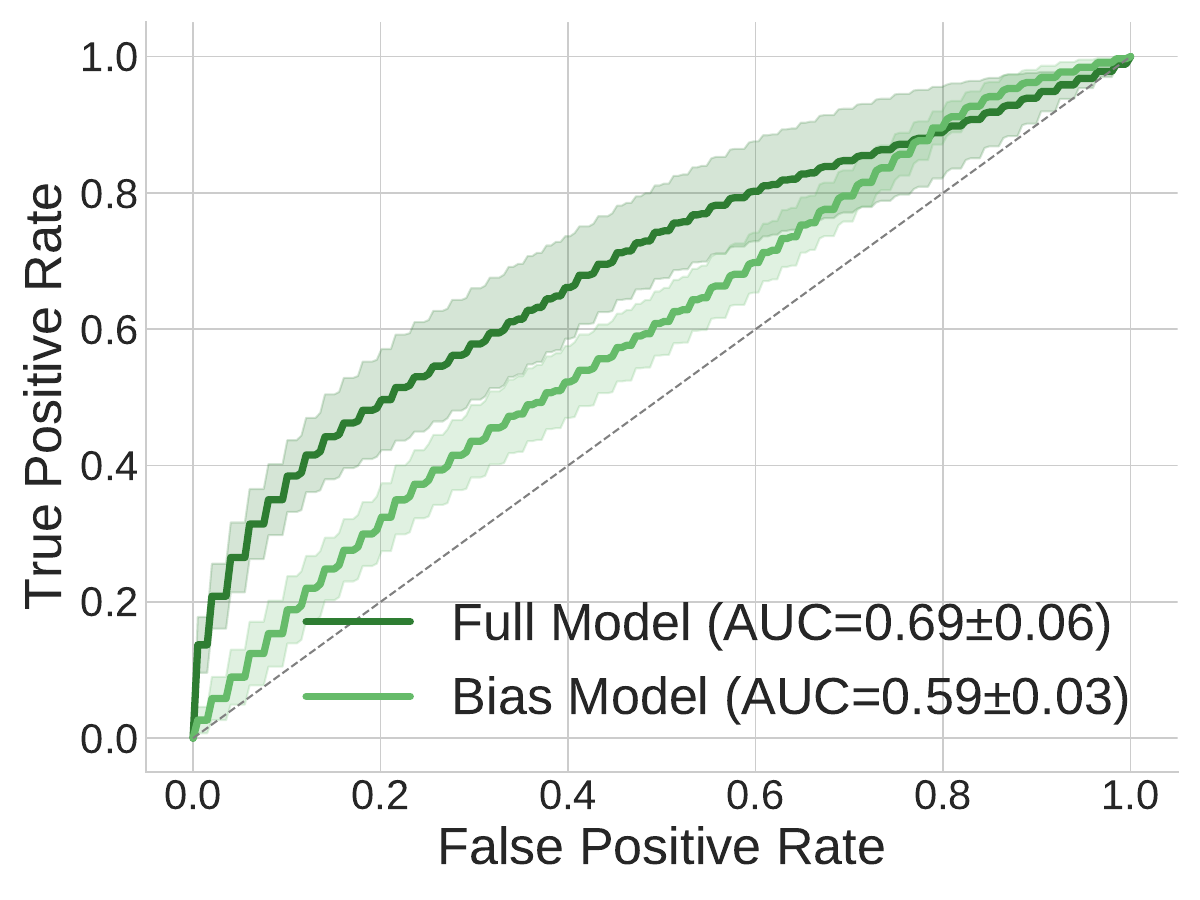}
    \caption{\small ROC Social}
  \end{subfigure}\hfill
  \begin{subfigure}[t]{0.16\textwidth}
    \includegraphics[width=\linewidth]{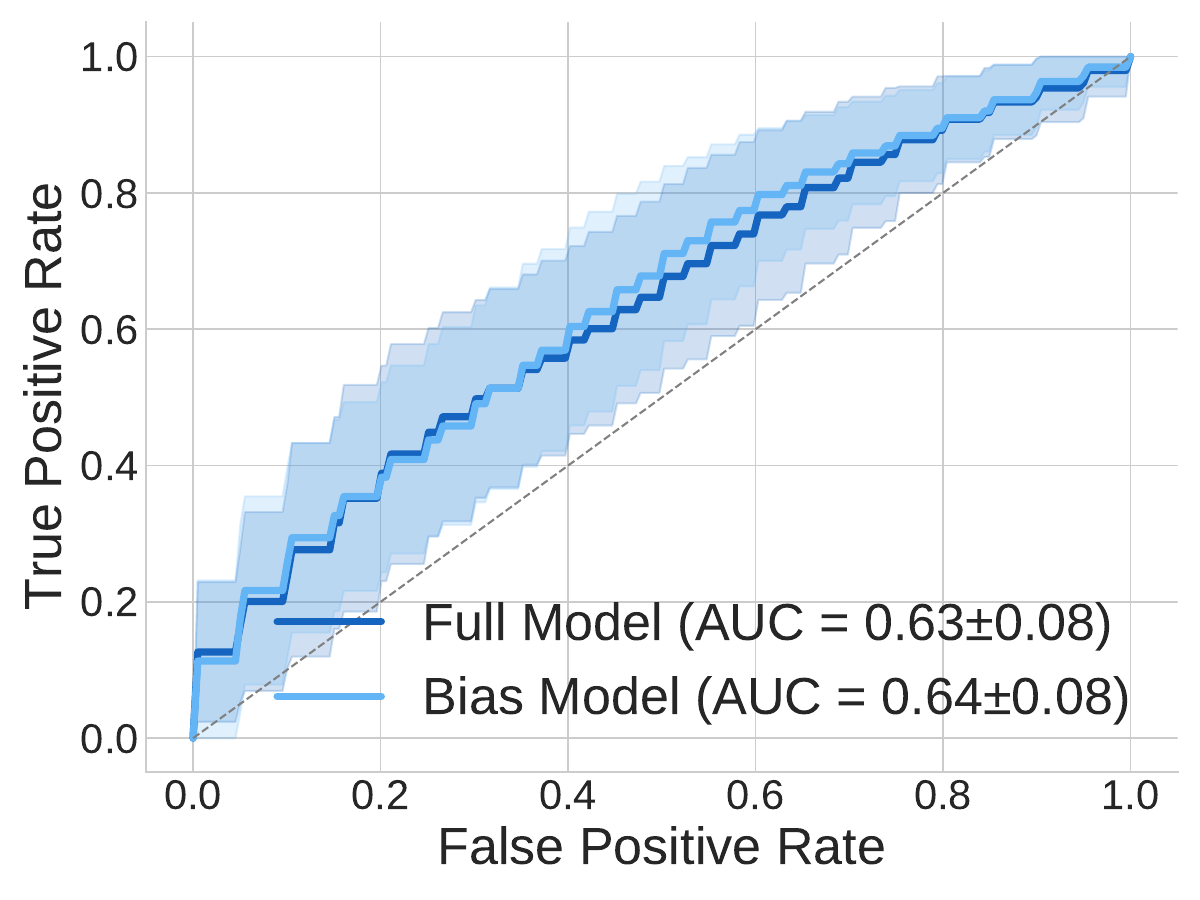}
    \caption{\small ROC Health}
  \end{subfigure}\hfill
  \begin{subfigure}[t]{0.17\textwidth}
    \includegraphics[width=\linewidth]{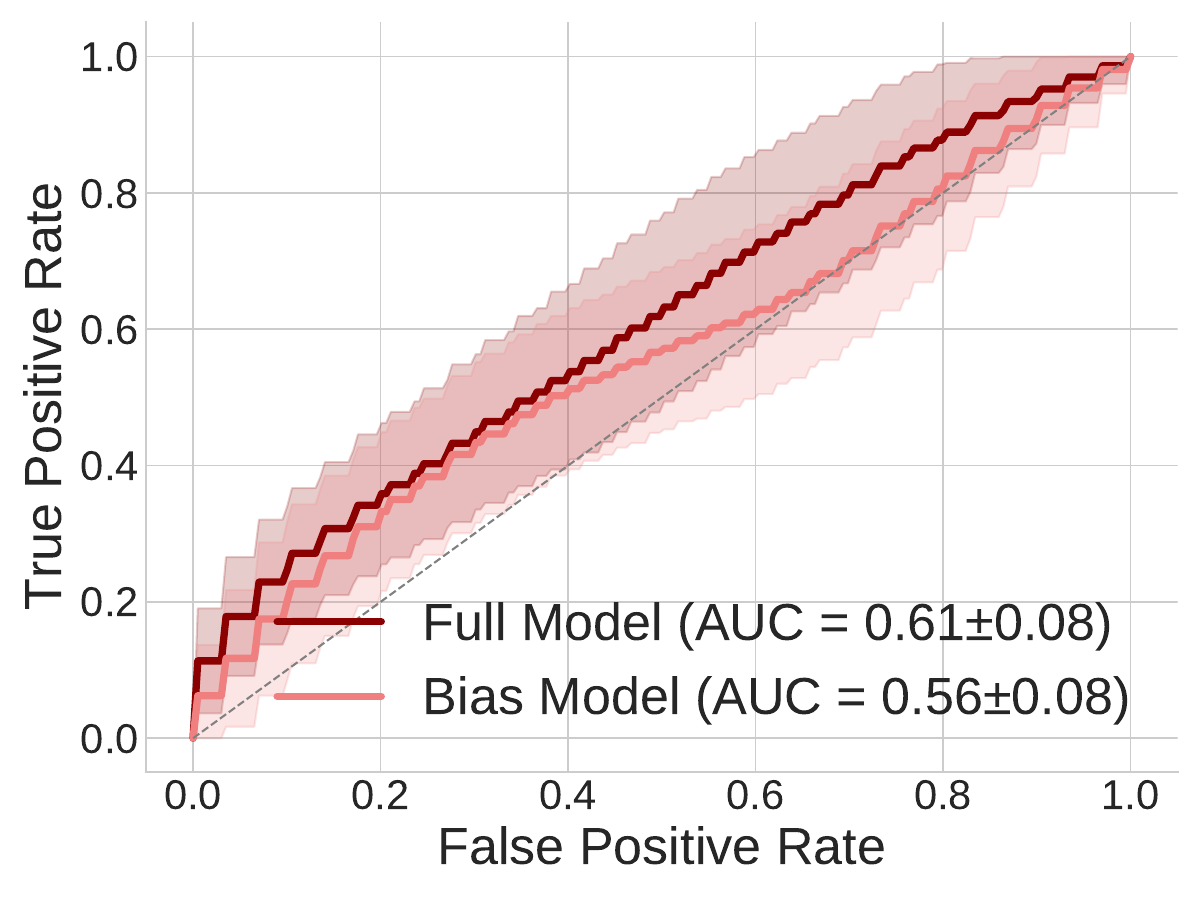}
    \caption{\small ROC Economic}
  \end{subfigure}
  \begin{subfigure}[t]{0.16\textwidth}
    \includegraphics[width=\linewidth]{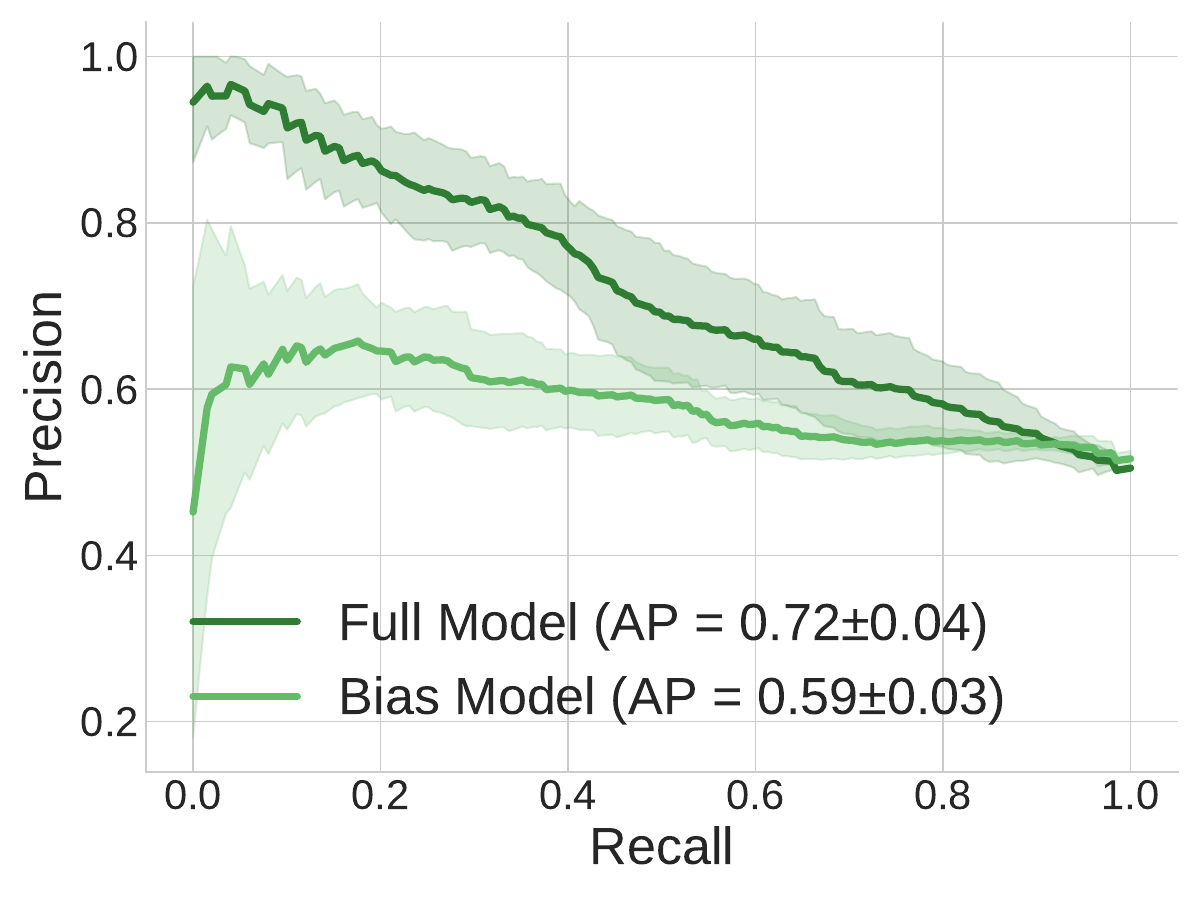}
    \caption{\small PR Social}
  \end{subfigure}\hfill
  \begin{subfigure}[t]{0.16\textwidth}
    \includegraphics[width=\linewidth]{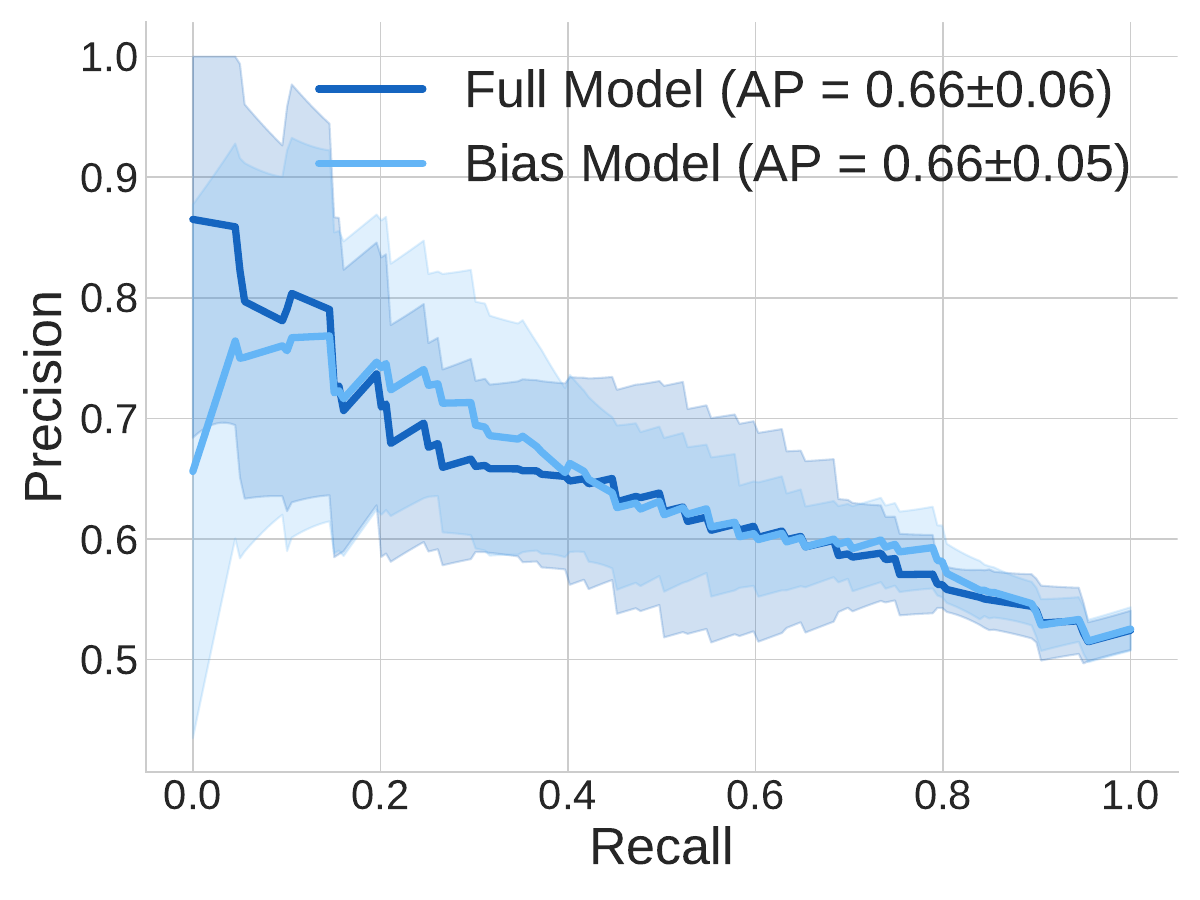}
    \caption{\small PR Health}
  \end{subfigure}\hfill
  \begin{subfigure}[t]{0.16\textwidth}
    \includegraphics[width=\linewidth]{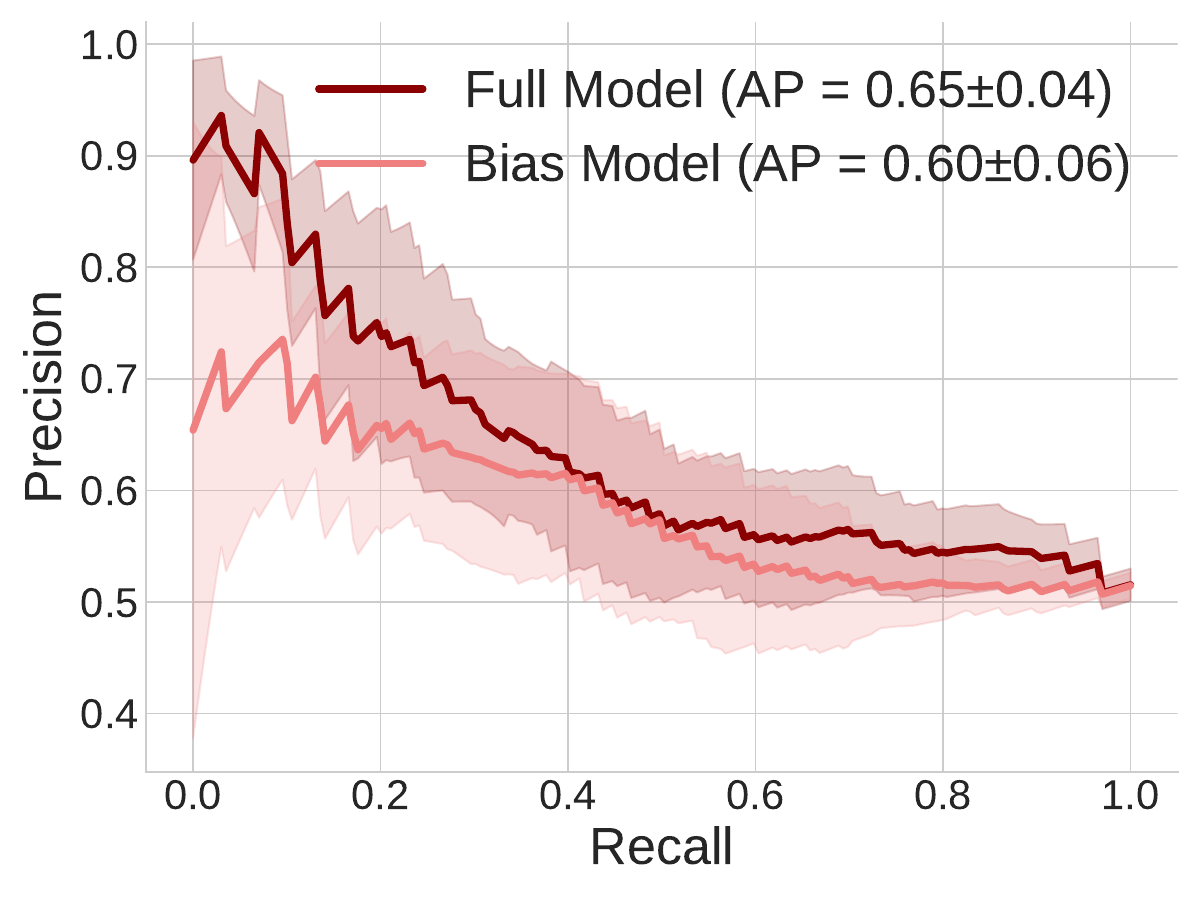}
    \caption{\small PR Economic}
  \end{subfigure}

  \caption{\textbf{Model behavior and structural interpretation on village network \# 175.} Panel (a, right) shows the learned role simplex with directed social (green), health (blue), and economic (red) ties; arrows indicate direction, and circles represent source/target embeddings ($\mathbf{Z}, \mathbf{W} \in \Delta_2$). Panel (a, left) separates ties by layer, with nodes colored by dominant role. Panel (b) shows violin plots of source and target membership strengths ($\mathbf{Z}, \mathbf{W}$) across layers. Panels (c)–(q) illustrate layer-specific multi-scale structures, with hierarchy strengths $s_h^l$ indicating each level’s contribution. Adjacency matrices are reordered by dominant memberships ($\mathbf{u}_i^{l,h}$, $\mathbf{v}_j^{l,h}$). Lower rows reveal finer-grained groups. Panel (r) shows bias model correlations: in-/out-degree vs. inferred biases ($\exp{\gamma_j^l}$, $\exp{\beta_i^l}$), capturing network status. Panel (s) shows weaker correlations under the full model, reflecting a shift toward interdependence. Panels (t)–(v) report ROC curves and AUC scores comparing full and bias models via 10-fold cross-validation (shaded = uncertainty). Panels (w)–(y) show corresponding PR curves.}
  \label{fig:overall_174}
\end{figure*}

To assess the robustness of our findings across networks, we replicate the full analysis presented in Figure~\ref{fig:overall} on three additional village networks. SI Figures~\ref{fig:overall_0}, \ref{fig:overall_50}, and \ref{fig:overall_174} show the learned role structures, membership distributions, multi-scale layer organization, degree–bias correlations, and predictive performance comparisons for Villages~\#1, \#51, and \#175. Results are consistent with those from Village~\#3, with the social layer again showing the strongest interdependence signal and predictive improvement, while gains in the economic and health layers remain modest or even absent.

\begin{figure*}[h]
   \centering
\begin{subfigure}[t]{0.32\textwidth}
    \includegraphics[width=\textwidth]{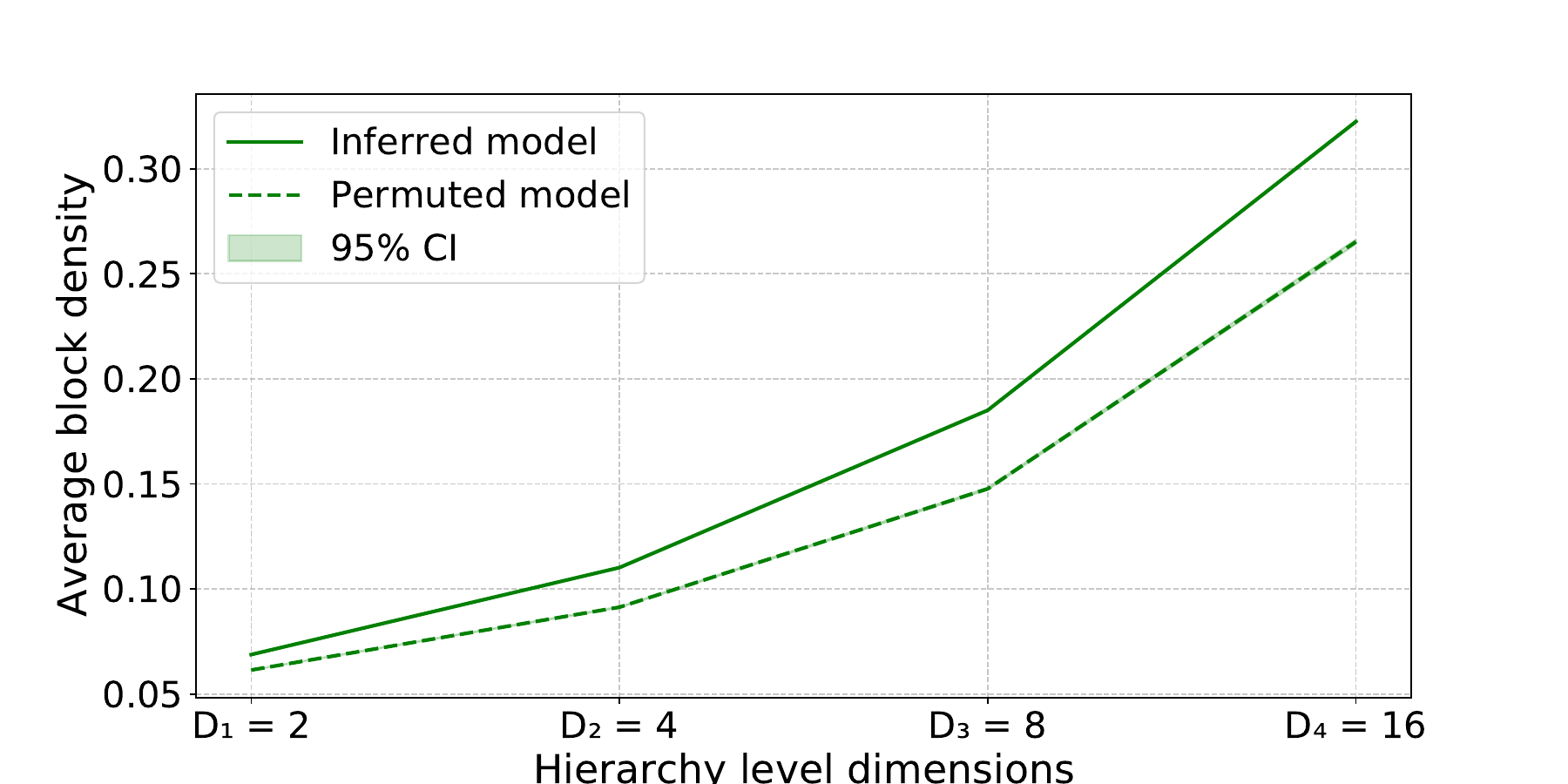}
    \caption{\small Social layer Village \# 1}
\end{subfigure}
\hfill
\begin{subfigure}[t]{0.32\textwidth}
    \includegraphics[width=\textwidth]{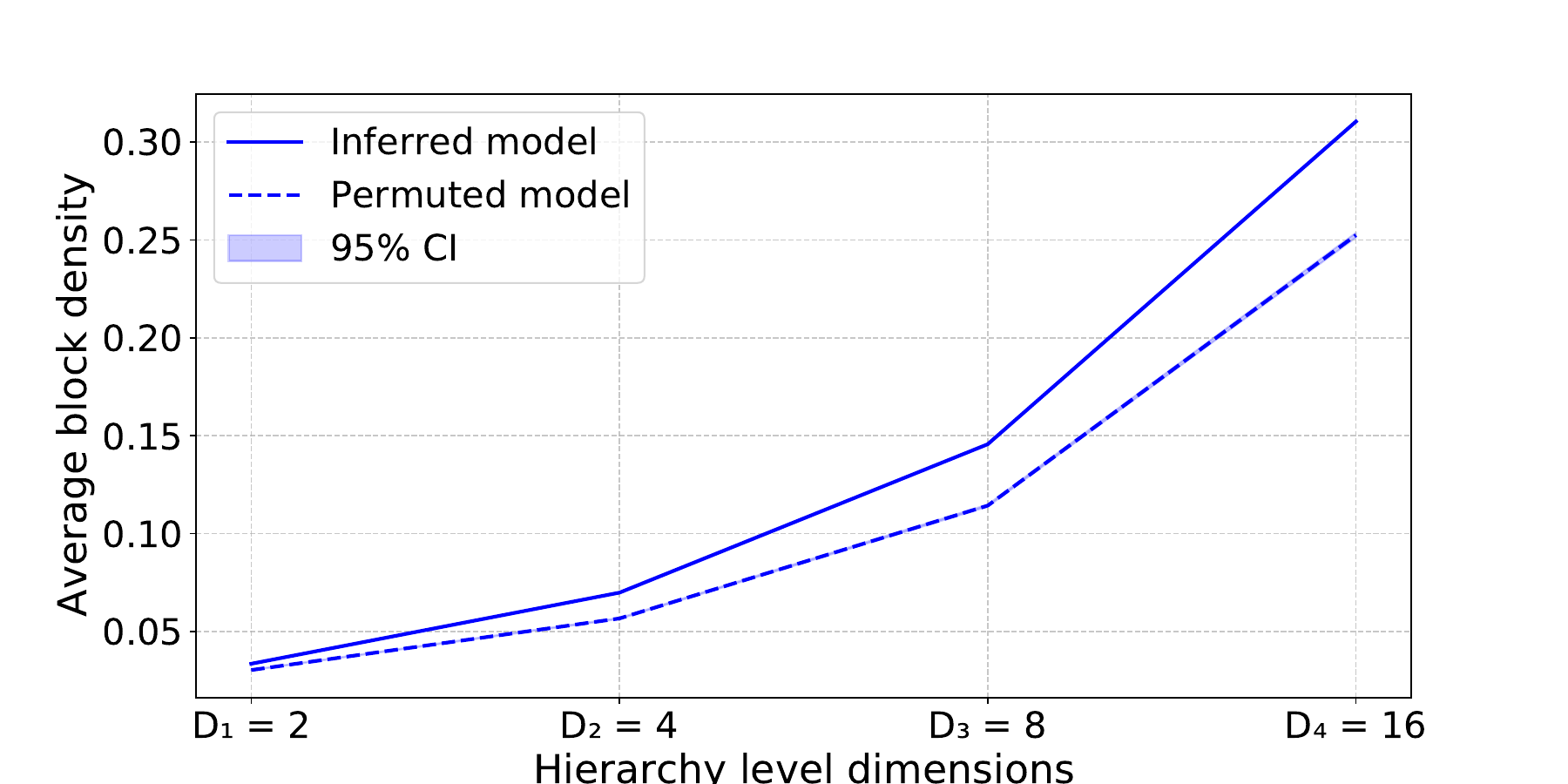}
    \caption{\small Health layer Village \# 1}
\end{subfigure}
\hfill
\begin{subfigure}[t]{0.32\textwidth}
    \includegraphics[width=\textwidth]{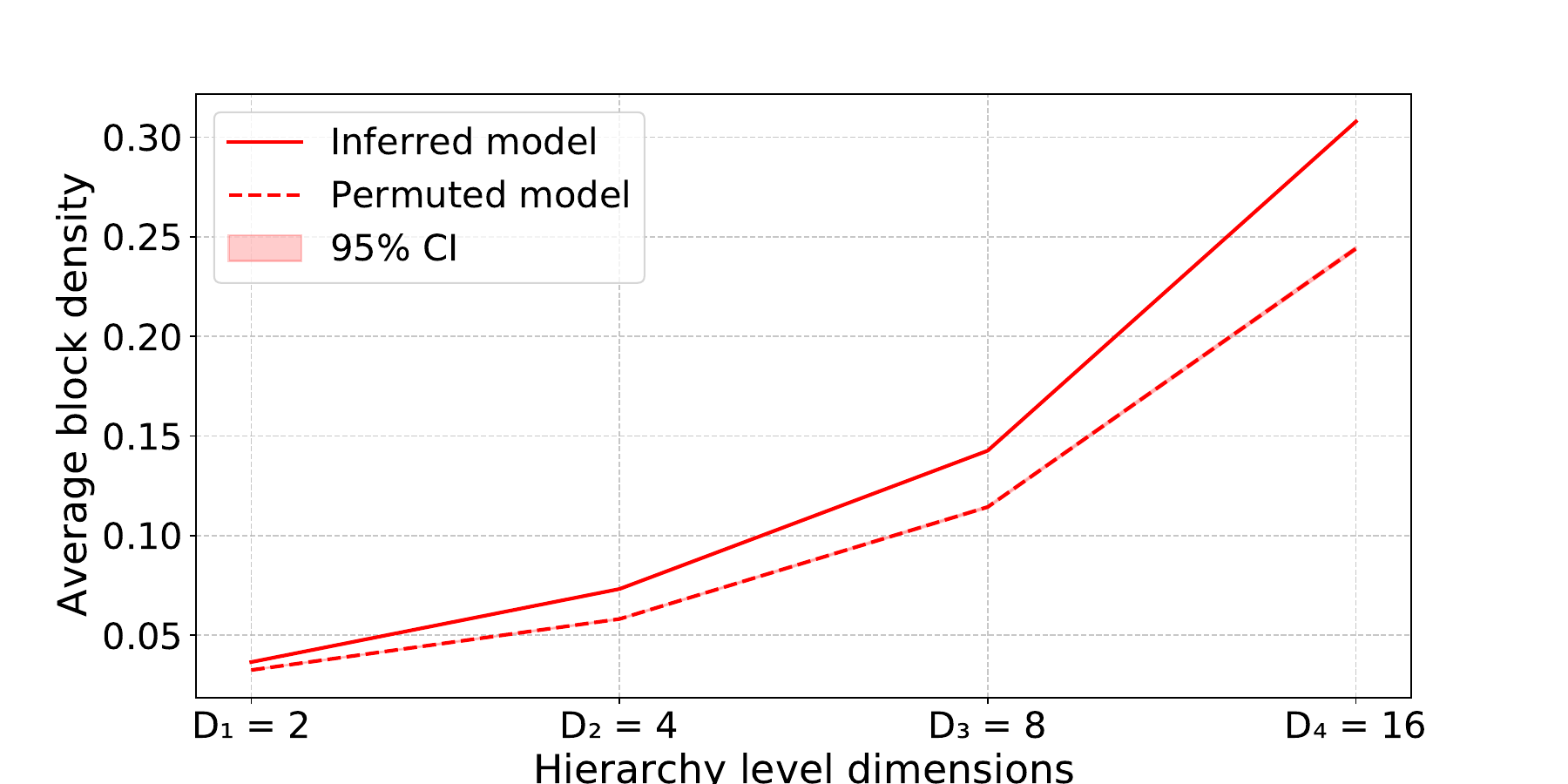}
    \caption{\small Economic layer Village \# 1}
\end{subfigure}
    \centering
\begin{subfigure}[t]{0.32\textwidth}
    \includegraphics[width=\textwidth]{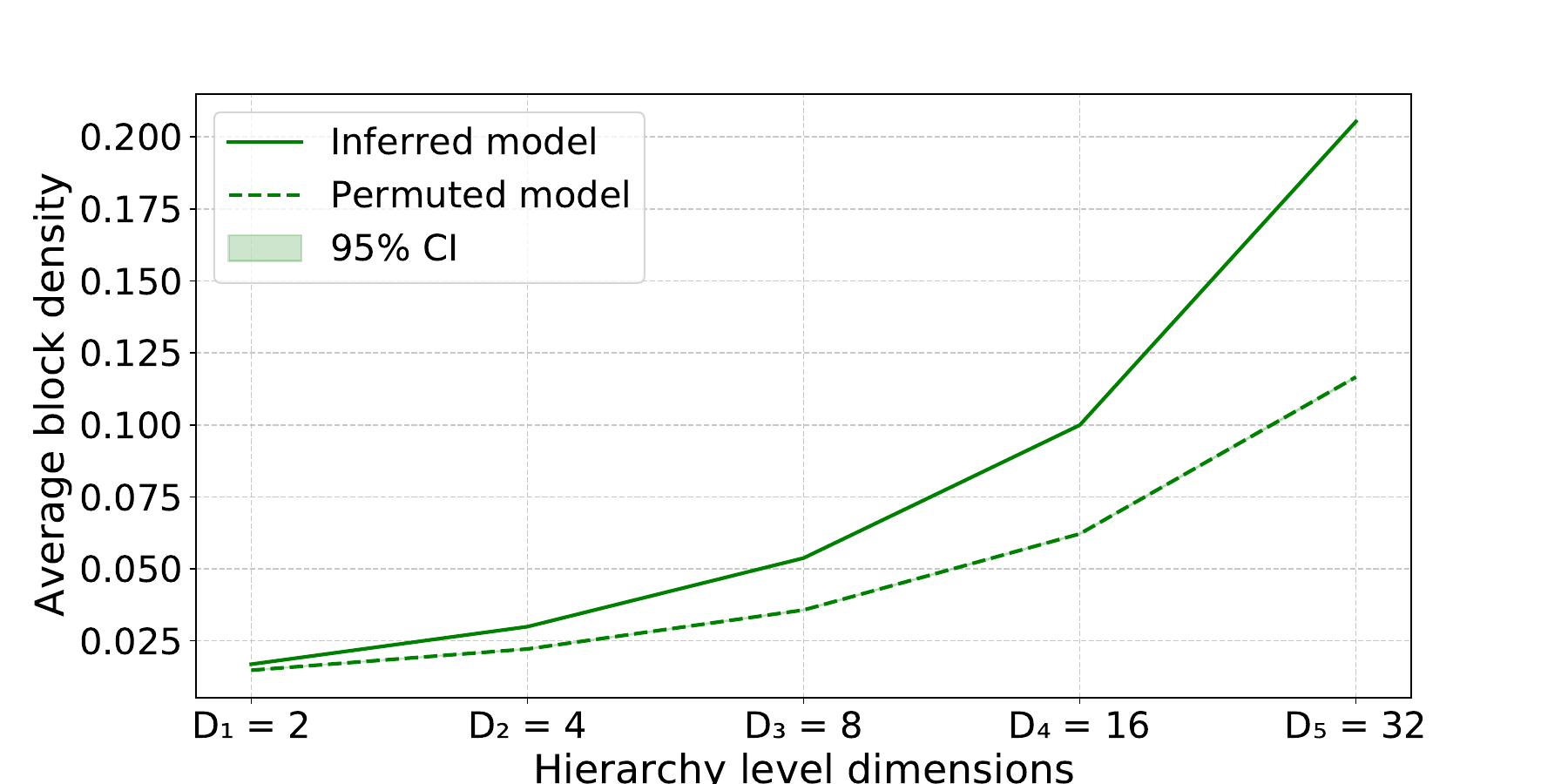}
    \caption{\small Social layer Village \# 3}
\end{subfigure}
\hfill
\begin{subfigure}[t]{0.32\textwidth}
    \includegraphics[width=\textwidth]{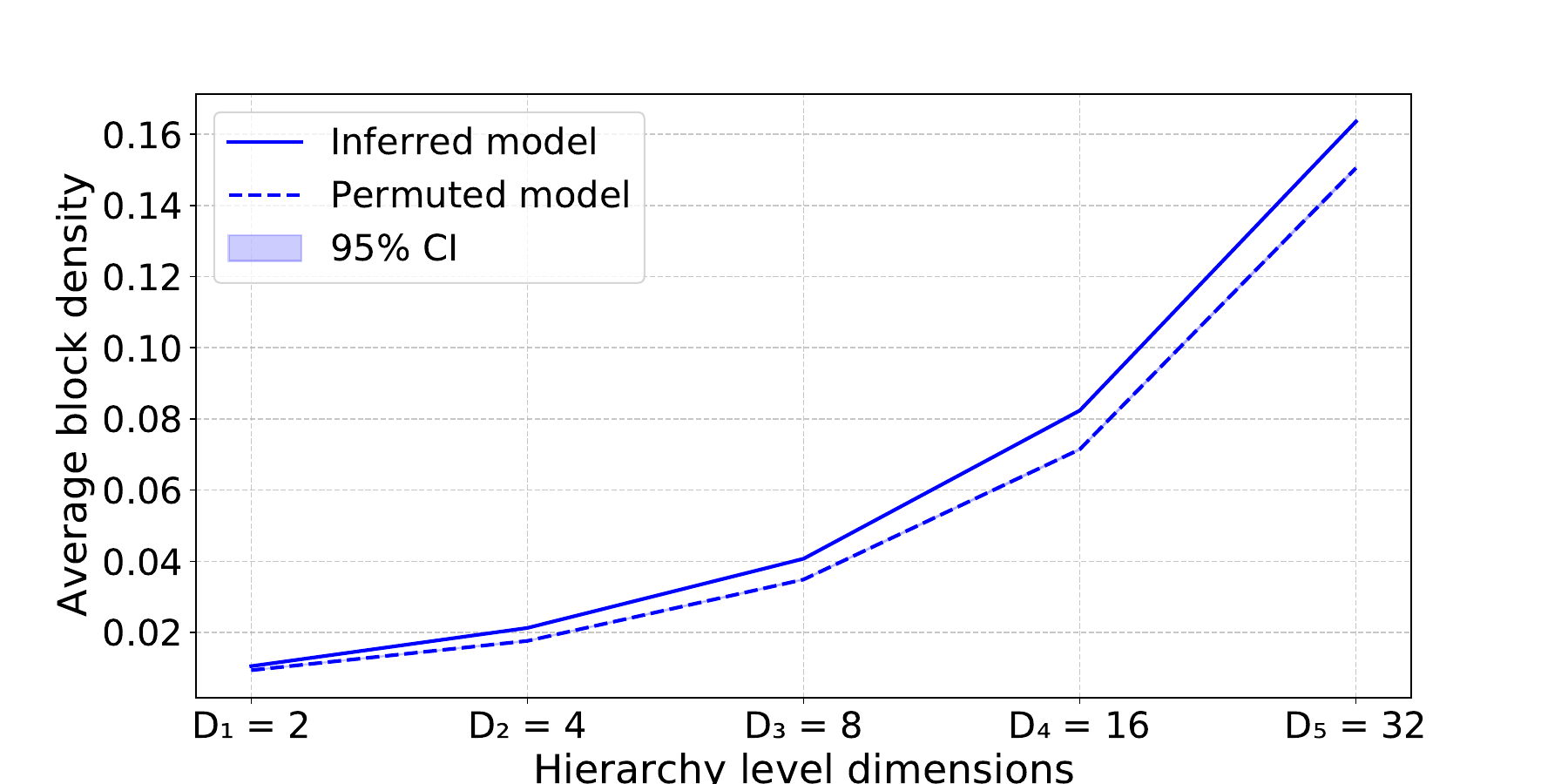}
    \caption{\small Health layer Village \# 3}
\end{subfigure}
\hfill
\begin{subfigure}[t]{0.32\textwidth}
    \includegraphics[width=\textwidth]{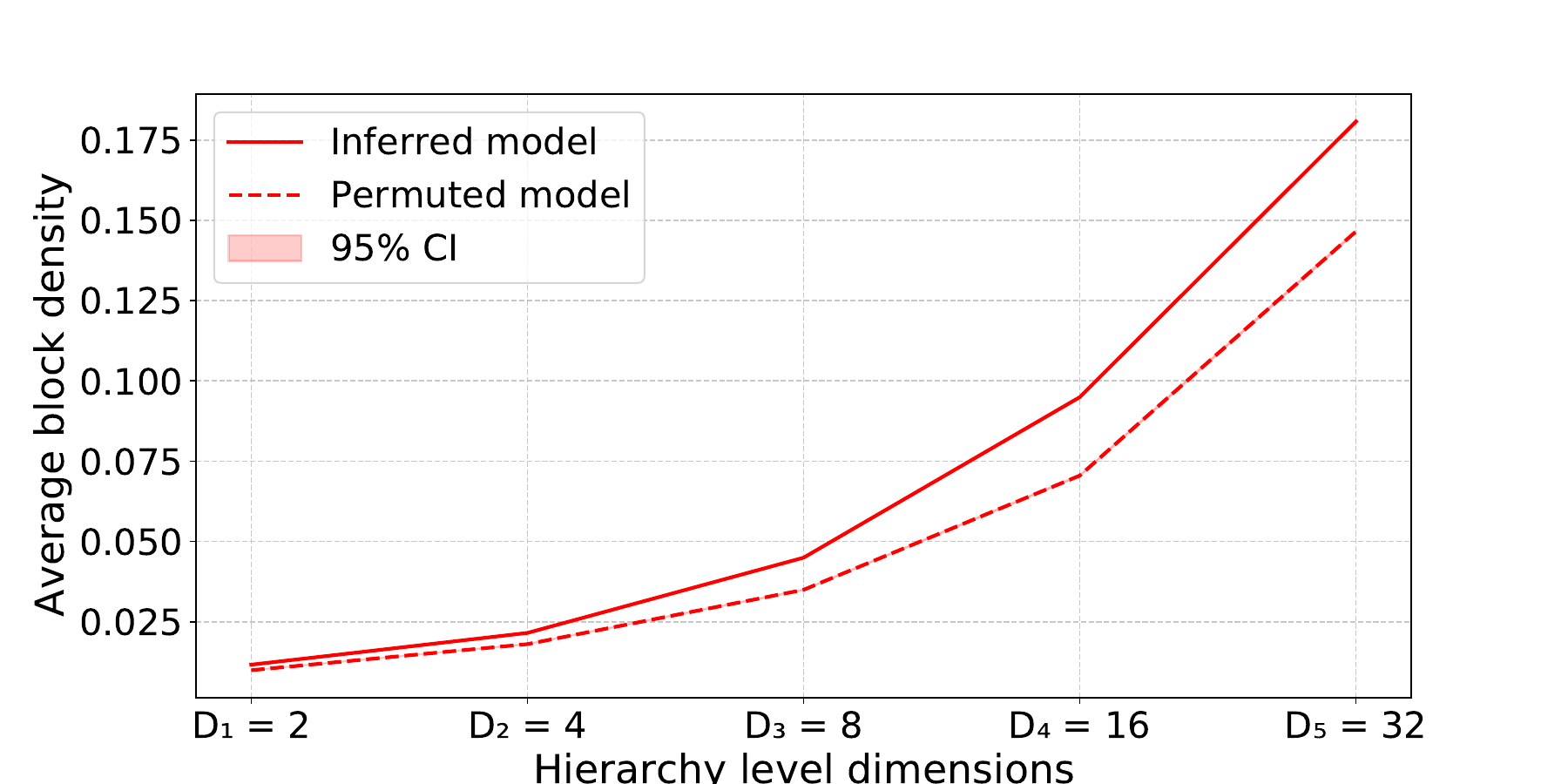}
    \caption{\small Economic layer Village \# 3}
\end{subfigure}
   \centering
\begin{subfigure}[t]{0.32\textwidth}
    \includegraphics[width=\textwidth]{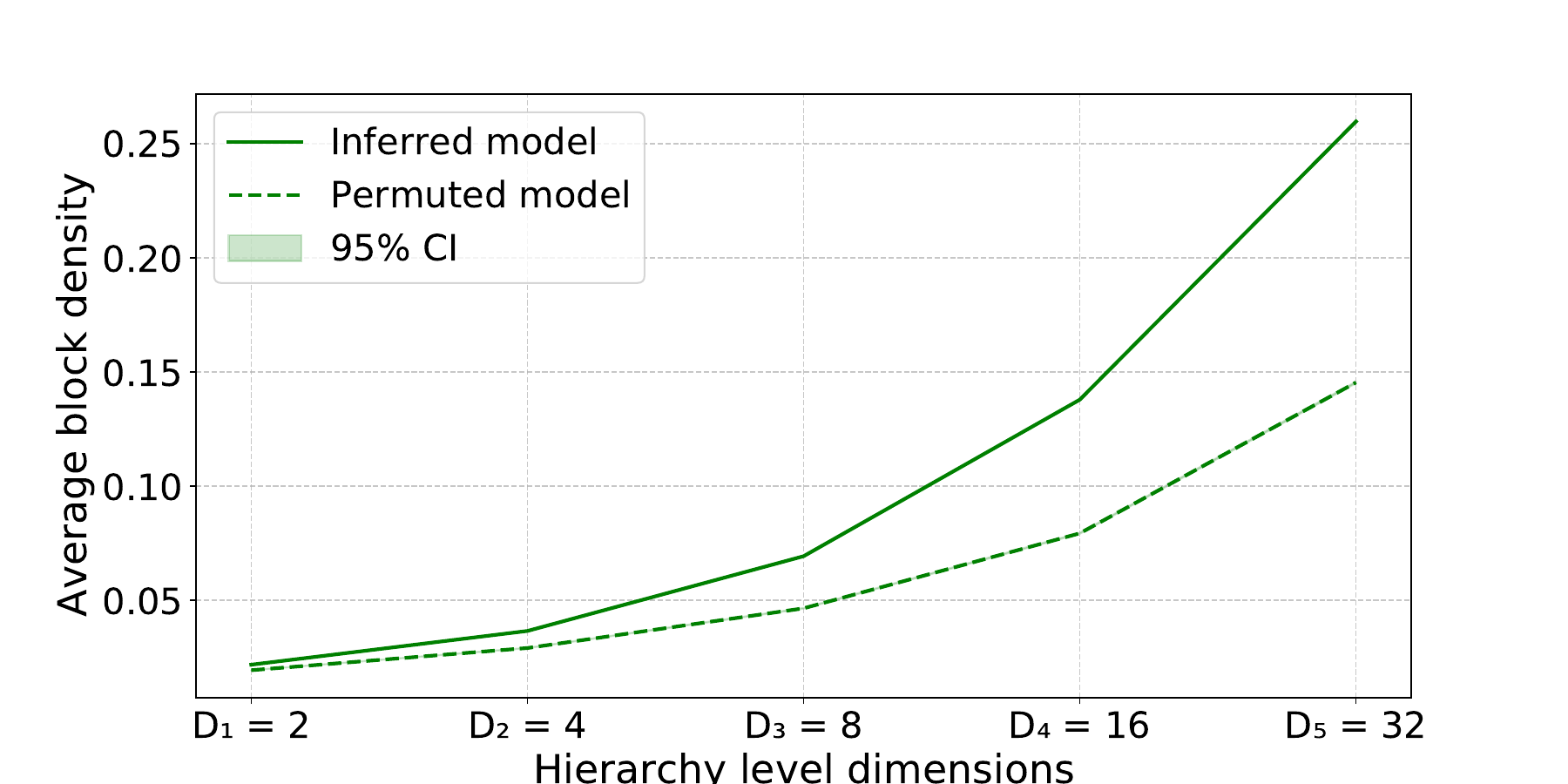}
    \caption{\small Social layer Village \# 51}
\end{subfigure}
\hfill
\begin{subfigure}[t]{0.32\textwidth}
    \includegraphics[width=\textwidth]{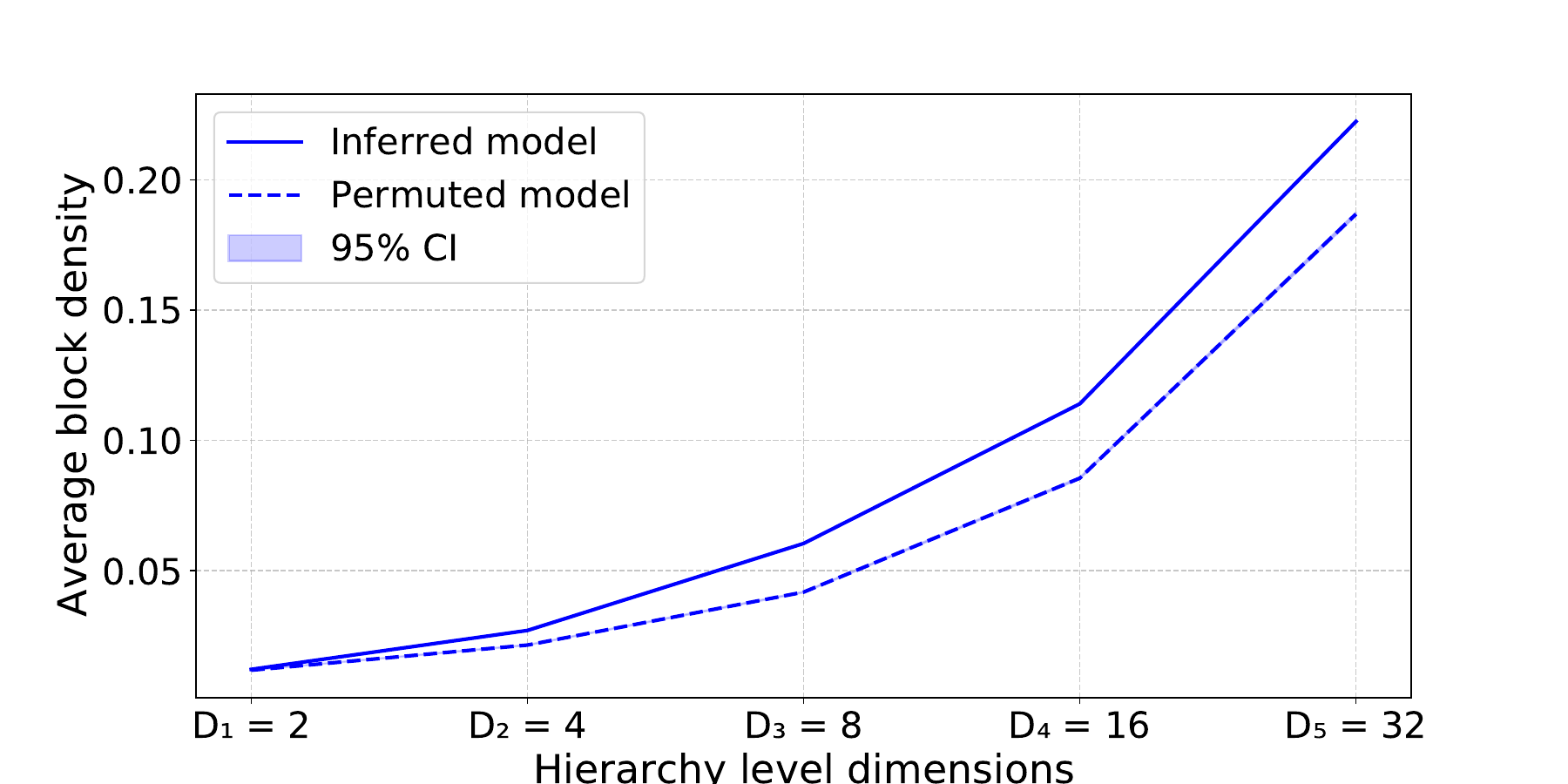}
    \caption{\small Health layer Village \# 51}
\end{subfigure}
\hfill
\begin{subfigure}[t]{0.32\textwidth}
    \includegraphics[width=\textwidth]{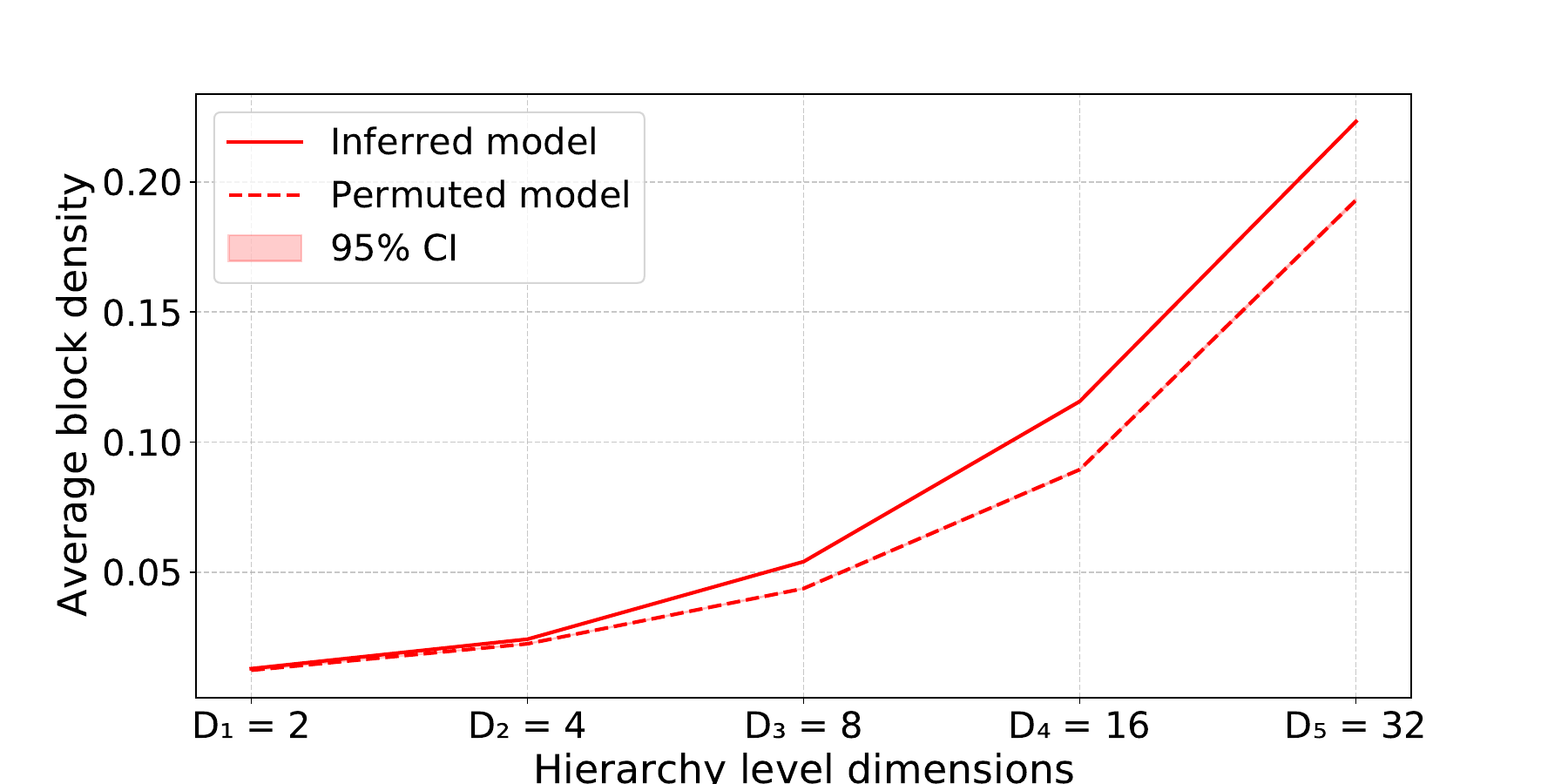}
    \caption{\small Economic layer Village \# 51}
\end{subfigure}
   \centering
\begin{subfigure}[t]{0.32\textwidth}
    \includegraphics[width=\textwidth]{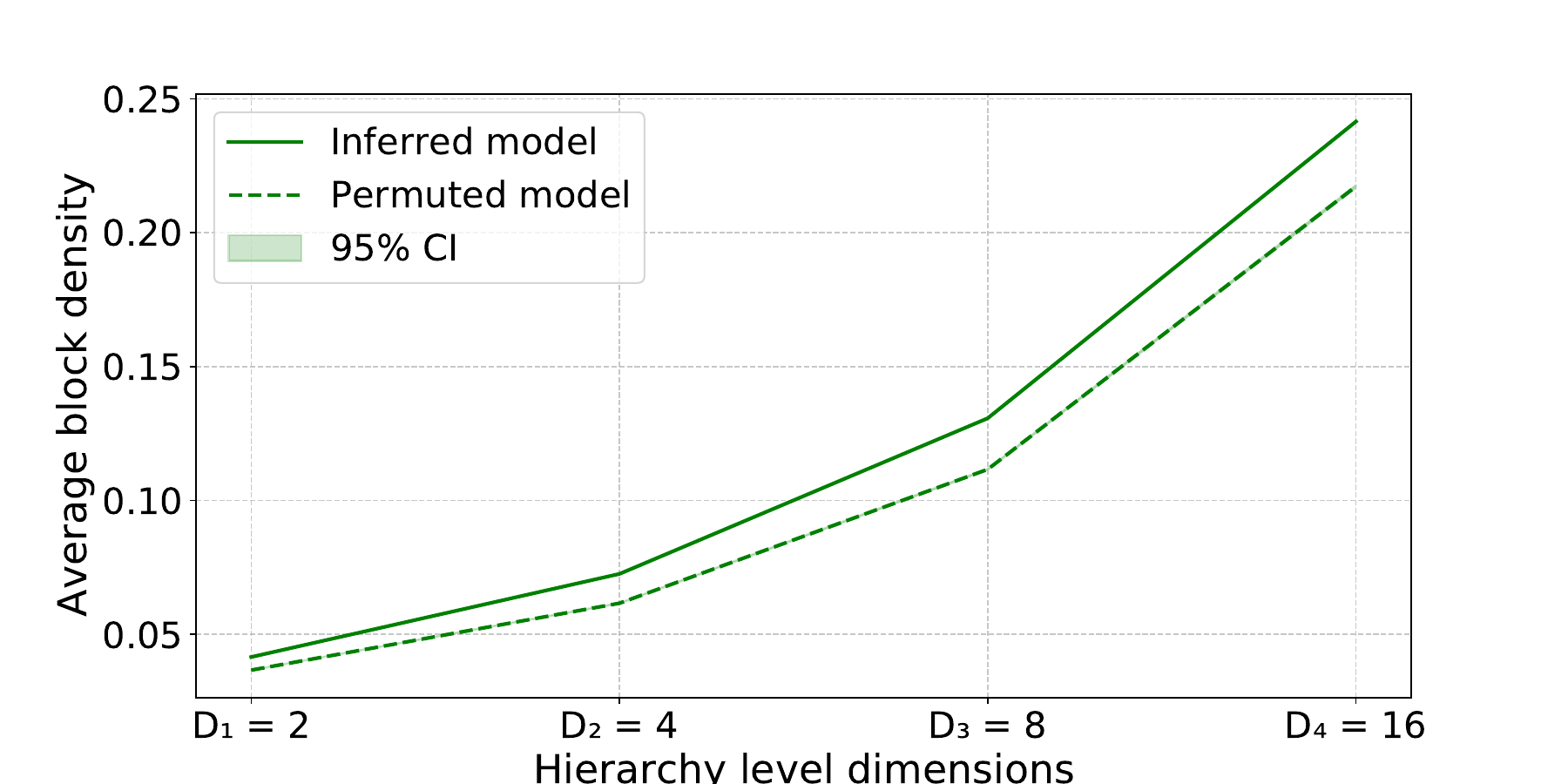}
    \caption{\small Social layer Village \# 175}
\end{subfigure}
\hfill
\begin{subfigure}[t]{0.32\textwidth}
    \includegraphics[width=\textwidth]{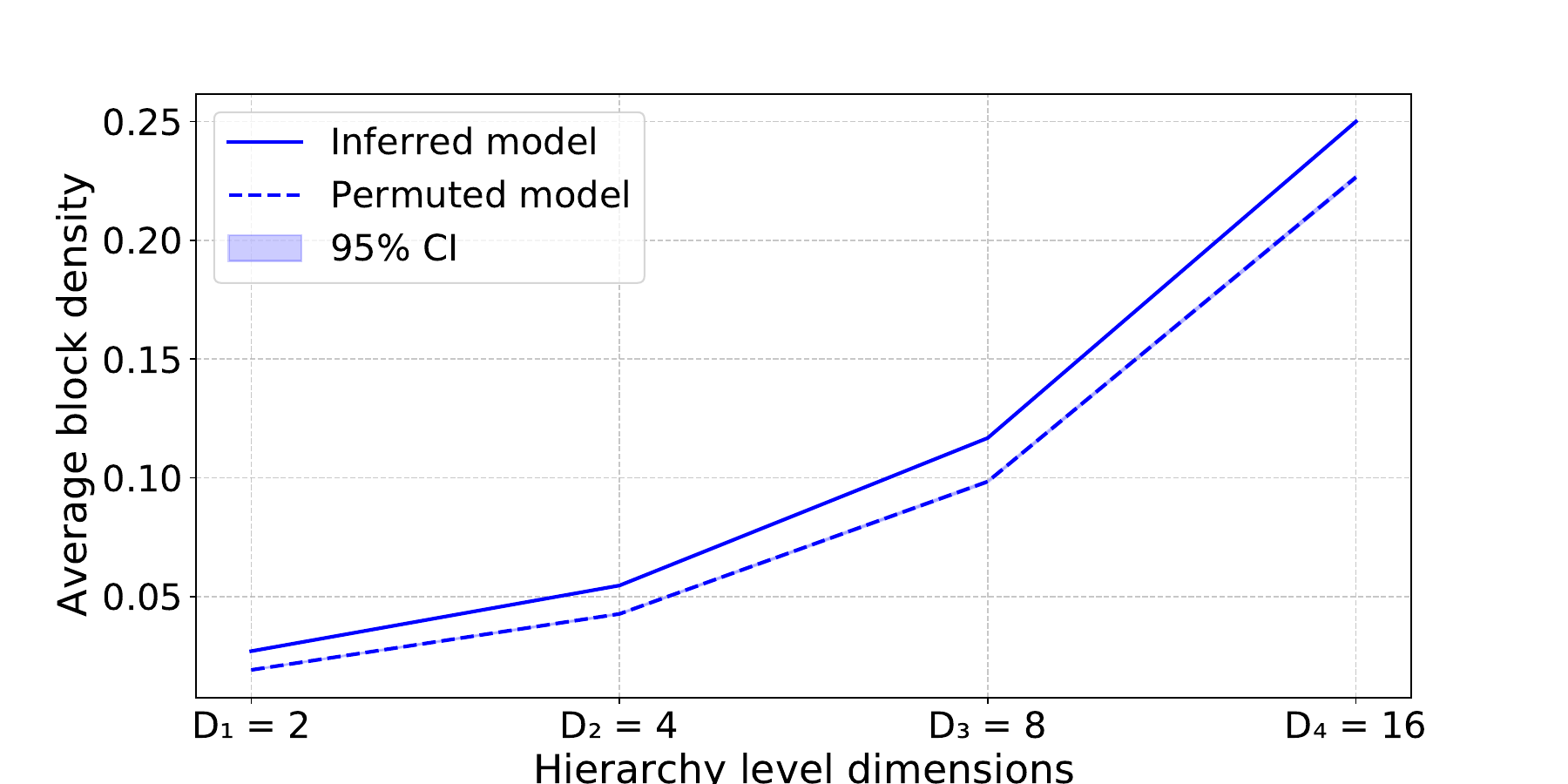}
    \caption{\small Health layer Village \# 175}
\end{subfigure}
\hfill
\begin{subfigure}[t]{0.32\textwidth}
    \includegraphics[width=\textwidth]{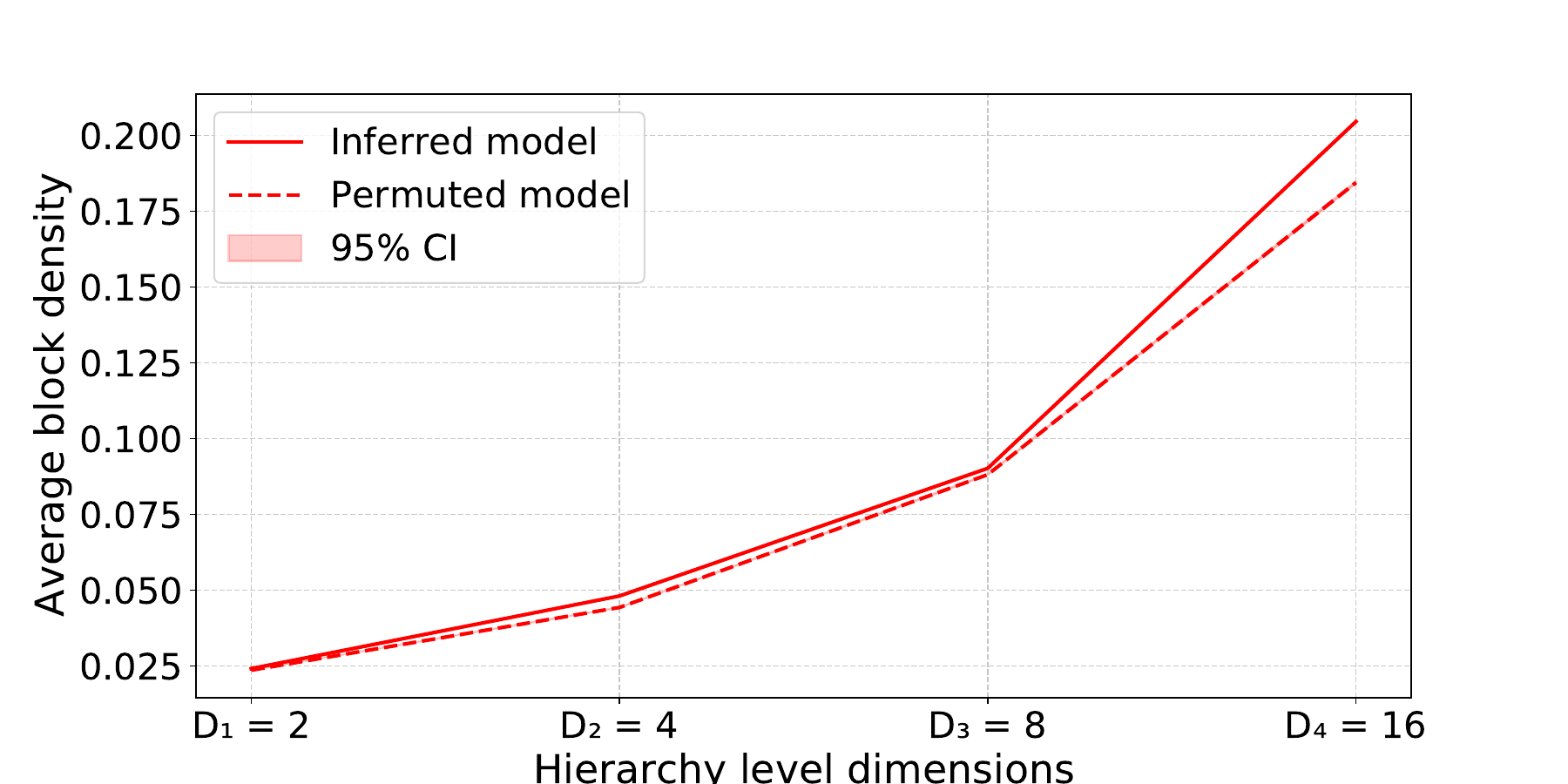}
    \caption{\small Economic layer Village \# 175}
\end{subfigure}
\caption{ \textbf{Villages \# 1 (fist row), \# 3 (second row), \# 51 (third row), \# 175 (fourth row)}: Mean extracted latent group
density, compared across two settings: (i) the inferred hierarchical structure of the model, and (ii) a control in which the latent dimensions (or groups) are randomly permuted,
thereby disrupting the multi-scale correspondence, i.e., parent blocks are assigned randomly to child blocks.}
\label{fig:permute_h}
\end{figure*}

Furthermore, SI Figure \ref{fig:permute_h} reports the mean extracted latent group density, compared across two settings: (i) the inferred hierarchical structure of the model, and (ii) a control in which the latent dimensions (or groups) are randomly permuted, thereby disrupting the multi-scale correspondence, i.e., parent blocks are assigned randomly to child blocks, for various networks. In SI Figure \ref{fig:permute_h} the first, second, and third columns illustrate this analysis for the social, health, and economic layers, respectively. Notably, the observed block densities are consistently higher than those under random permutation, indicating that the model effectively leverages multi-scale structure. This effect is particularly prominent in the higher-resolution regimes, suggesting that finer-grained groups play a more critical role in explaining the observed link structure. We also observe variation across the three layers, with the social layer benefiting most from the inclusion of multi-scale structure.

\begin{figure*}[t!]
\centering

    \centering
    \includegraphics[width=\textwidth]{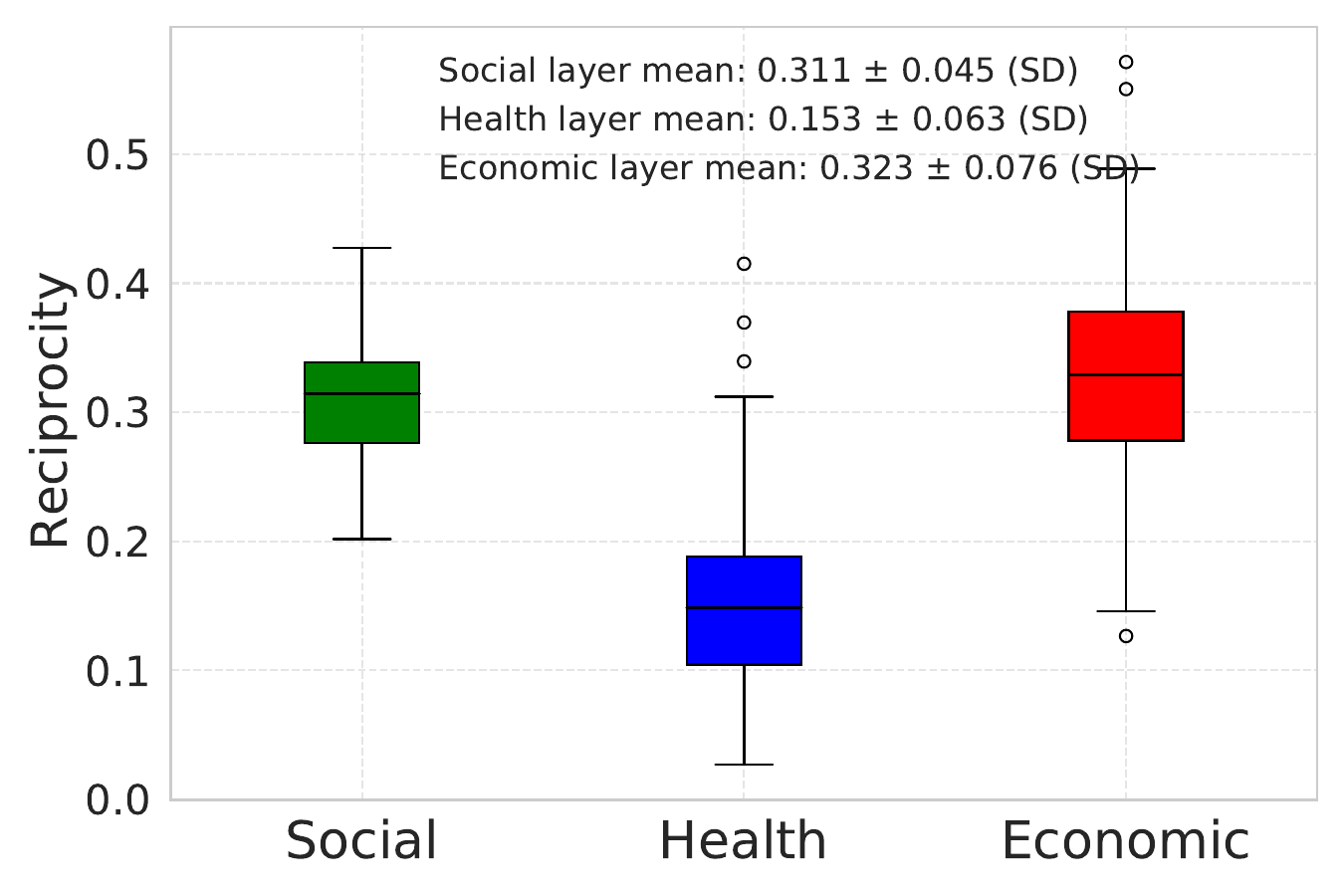}
\caption{\textbf{Layer-specific reciprocity across networks.} 
    Box plots show the distribution of reciprocity values across the 176 networks for the social, health, and economic layers. Reciprocity is highest in the social and economic layers, and substantially lower in the health layer, indicating lower mutuality in health-related interactions.}
\label{fig:side_by_side_figures}
\end{figure*}

\begin{figure*}[t!]

\centering
\begin{subfigure}[t]{1\textwidth}
    \includegraphics[width=\textwidth]{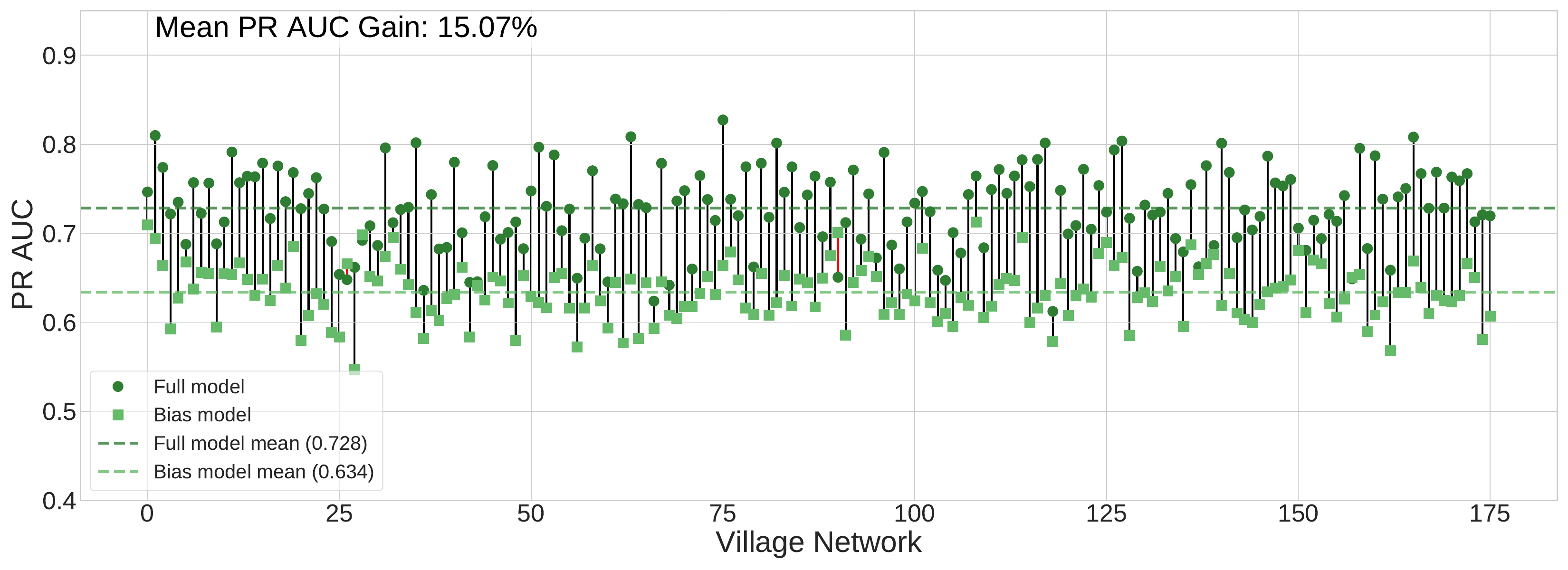}
    \caption{\textbf{PR AUC}: Comparison of link prediction in the \textbf{social layer} using a model capturing dependence and independence (Bias terms only model) versus the \textsc{MLT} model that also incorporates interdependence (Full model).}
\end{subfigure}

\centering
\begin{subfigure}[t]{1\textwidth}
    \includegraphics[width=\textwidth]{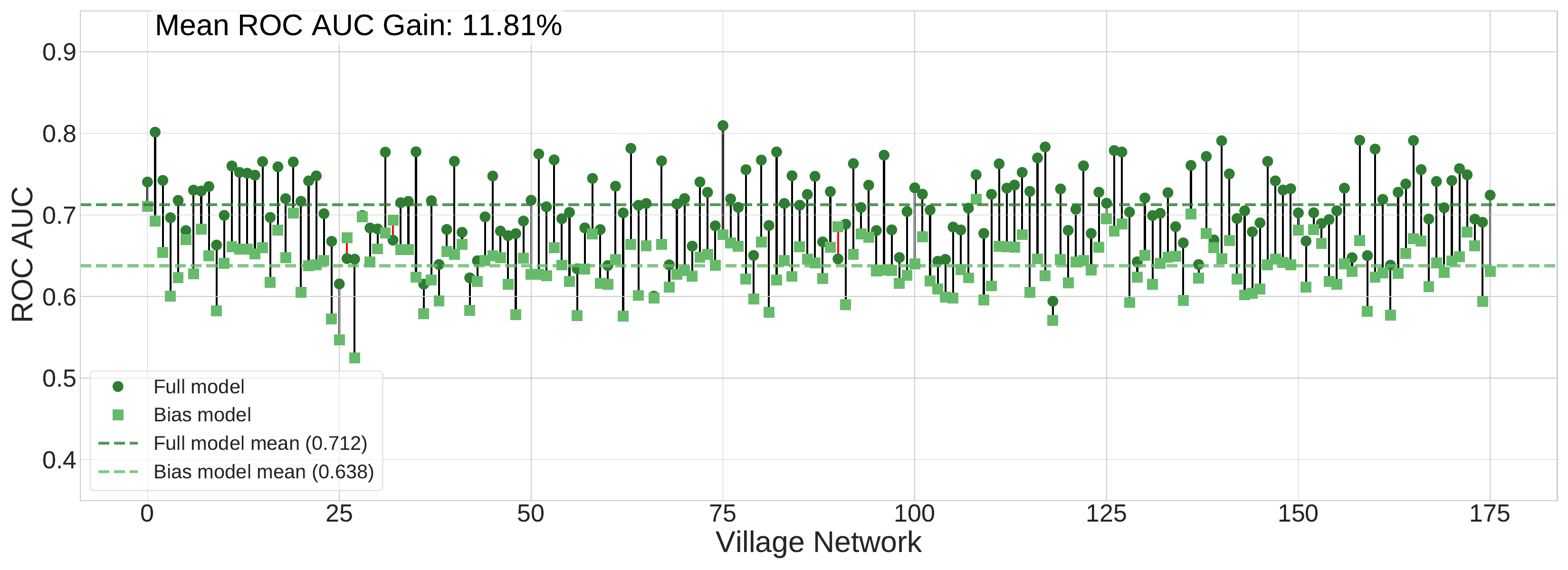}
    \caption{\textbf{ROC AUC}: Comparison of link prediction in the \textbf{social layer} using a model capturing dependence and independence (Bias model) versus the \textsc{MLT} model that also incorporates interdependence (Full model).}
\end{subfigure}
\caption{\textbf{Social layer exchange quantification.} This figure shows the average change in prediction performance across ten-fold cross-validation when comparing a model that captures only dependence and independence with the full \textsc{MLT} model that also includes interdependence, in accordance with Social Exchange Theory. Results are shown for all 176 village networks, using Area Under the PR Curve (top row) and Area Under the ROC Curve (bottom row). Black lines indicate networks with improved predictive performance when modeling interdependence; red lines indicate a decrease.}
\label{fig:SET_social}
\end{figure*}

\begin{figure*}[b!]
\centering
\begin{subfigure}[t]{1\textwidth}
    \includegraphics[width=\textwidth]{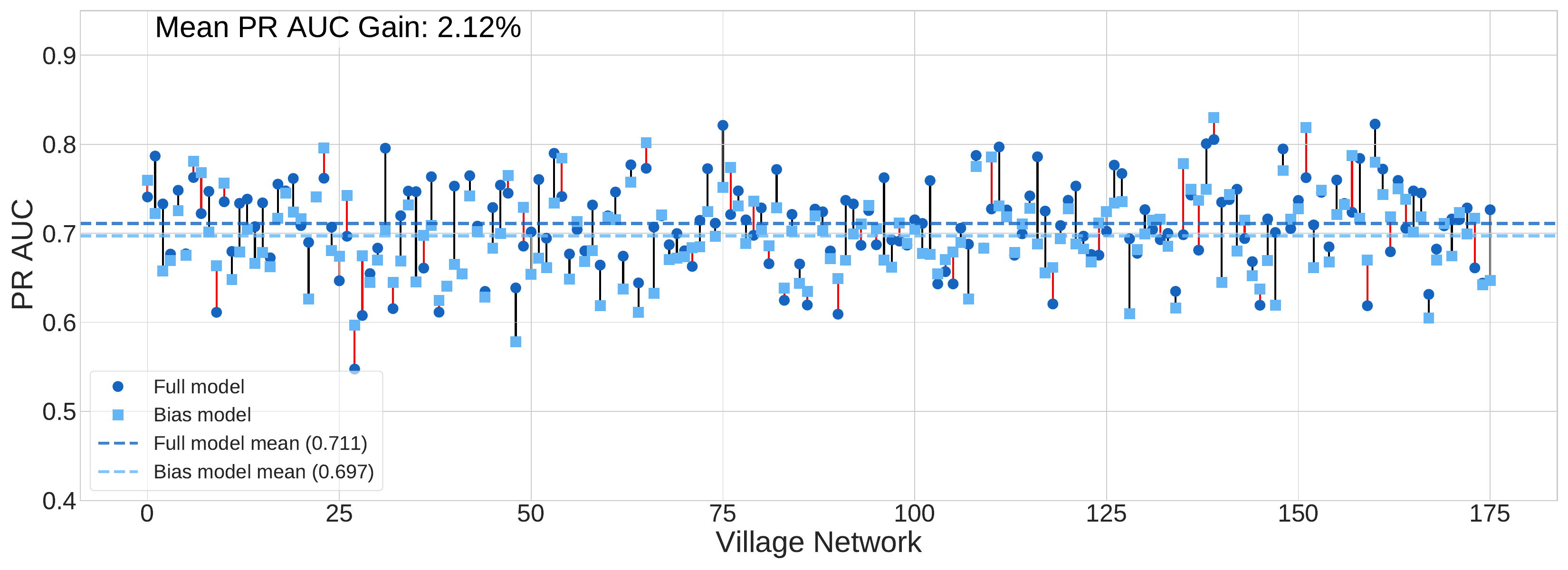}
    \caption{\textbf{PR AUC}: Comparison of link prediction in the \textbf{health layer} using a model capturing dependence and independence (Bias model) versus the \textsc{MLT} model that also incorporates interdependence (Full model).}
\end{subfigure}
\centering
\begin{subfigure}[t]{1\textwidth}
    \includegraphics[width=\textwidth]{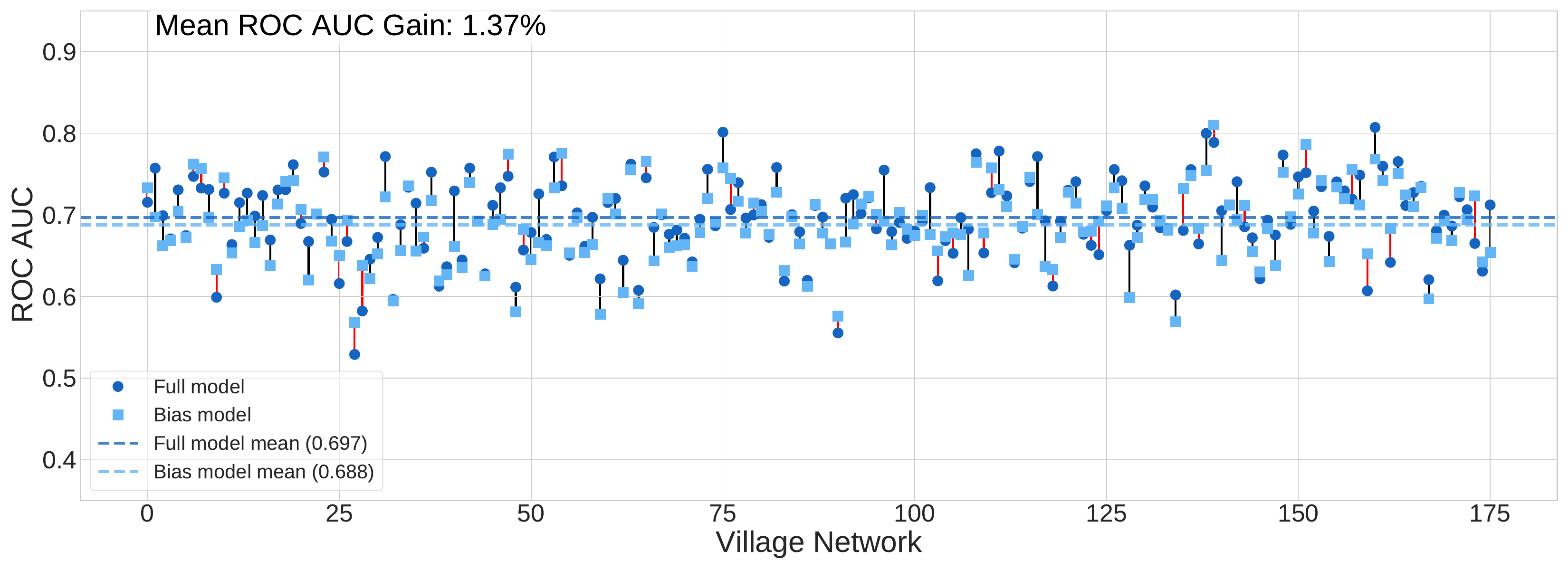}
    \caption{\textbf{ROC AUC}: Comparison of link prediction in the \textbf{health layer} using a model capturing dependence and independence (Bias model) versus the \textsc{MLT} model that also incorporates interdependence (Full model).}
\end{subfigure}

\caption{\textbf{Health layer exchange quantification.} This figure shows the average change in prediction performance across ten-fold cross-validation when comparing a model that captures only dependence and independence with the full \textsc{MLT} model that also includes interdependence, in accordance with Social Exchange Theory. Results are shown for all 176 village networks, using Area Under the PR Curve (top row) and Area Under the ROC Curve (bottom row). Black lines indicate networks with improved predictive performance when modeling interdependence; red lines indicate a decrease.}
\label{fig:SET_heal}
\end{figure*}

\begin{figure*}[t!]
\centering
\begin{subfigure}[t]{1\textwidth}
    \includegraphics[width=\textwidth]{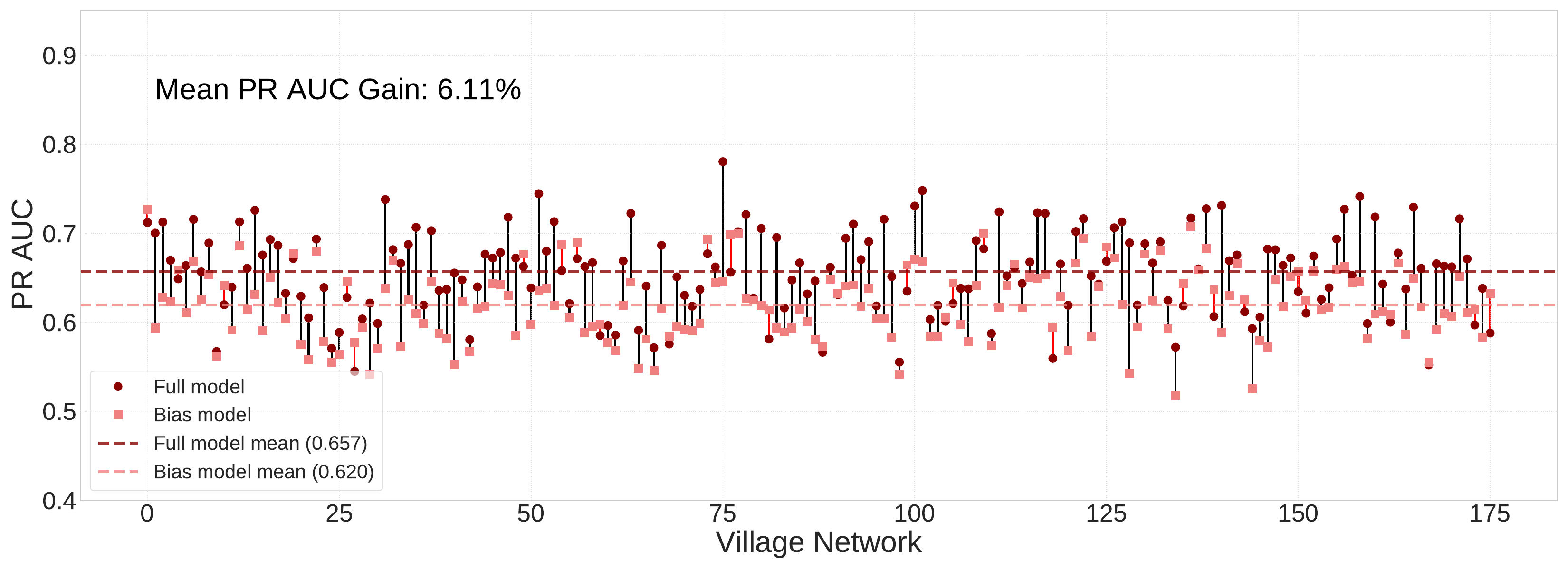}
    \caption{\textbf{PR AUC}: Comparison of link prediction in the \textbf{economic layer} using a model capturing dependence and independence (Bias model) versus the \textsc{MLT} model that also incorporates interdependence (Full model). }
\end{subfigure}
\centering
\begin{subfigure}[t]{1\textwidth}
    \includegraphics[width=\textwidth]{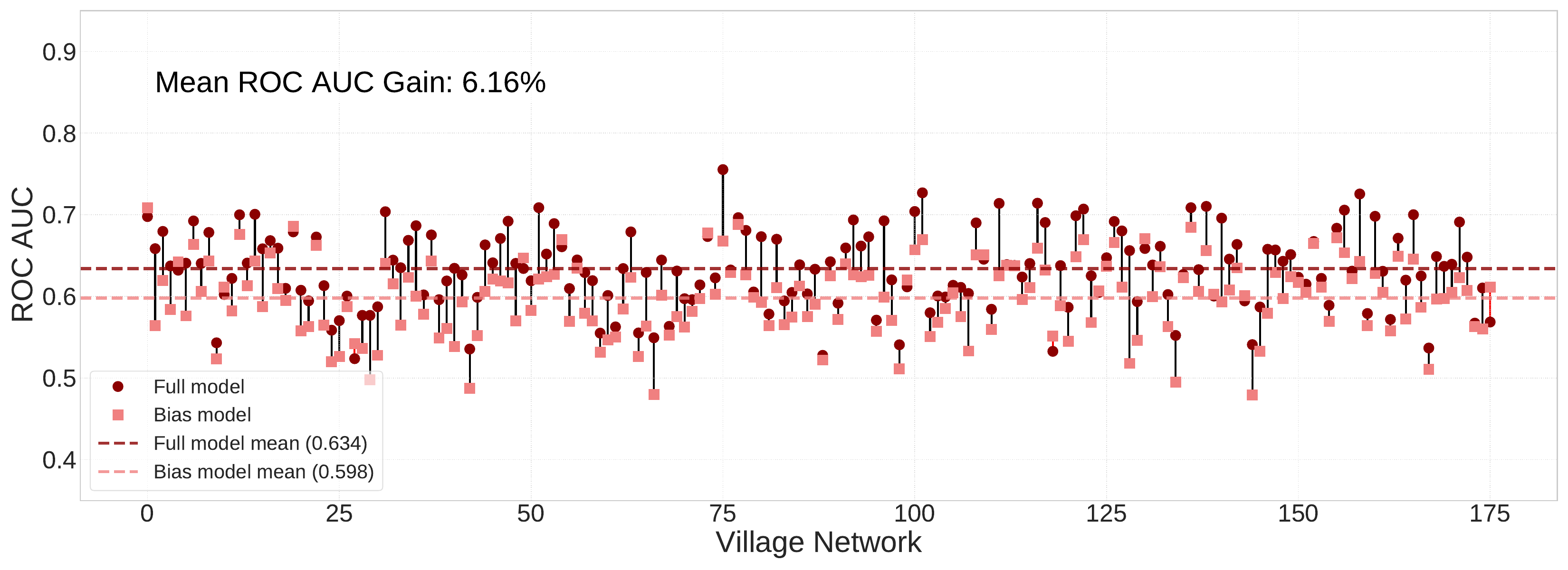}
    \caption{\textbf{ROC AUC}: Comparison of link prediction in the \textbf{economic layer} using a model capturing dependence and independence (Bias model) versus the \textsc{MLT} modelthat also incorporates interdependence (Full model).}
\end{subfigure}
\caption{\textbf{Economic layer exchange quantification.} This figure shows the average change in prediction performance across ten-fold cross-validation when comparing a model that captures only dependence and independence with the full \textsc{MLT} model that also includes interdependence, in accordance with Social Exchange Theory. Results are shown for all 176 village networks, using Area Under the PR Curve (top row) and Area Under the ROC Curve (bottom row). Black lines indicate networks with improved predictive performance when modeling interdependence; red lines indicate a decrease.}
\label{fig:SET_econ}
\end{figure*}

\subsection*{Bootstrapped and observed correlation values between node bias and various centrality measures for the bias-only model versus the full model capturing interdependence} 
\begin{figure*}[!htb]

\centering
\begin{subfigure}[t]{0.32\textwidth}
    \includegraphics[width=\textwidth]{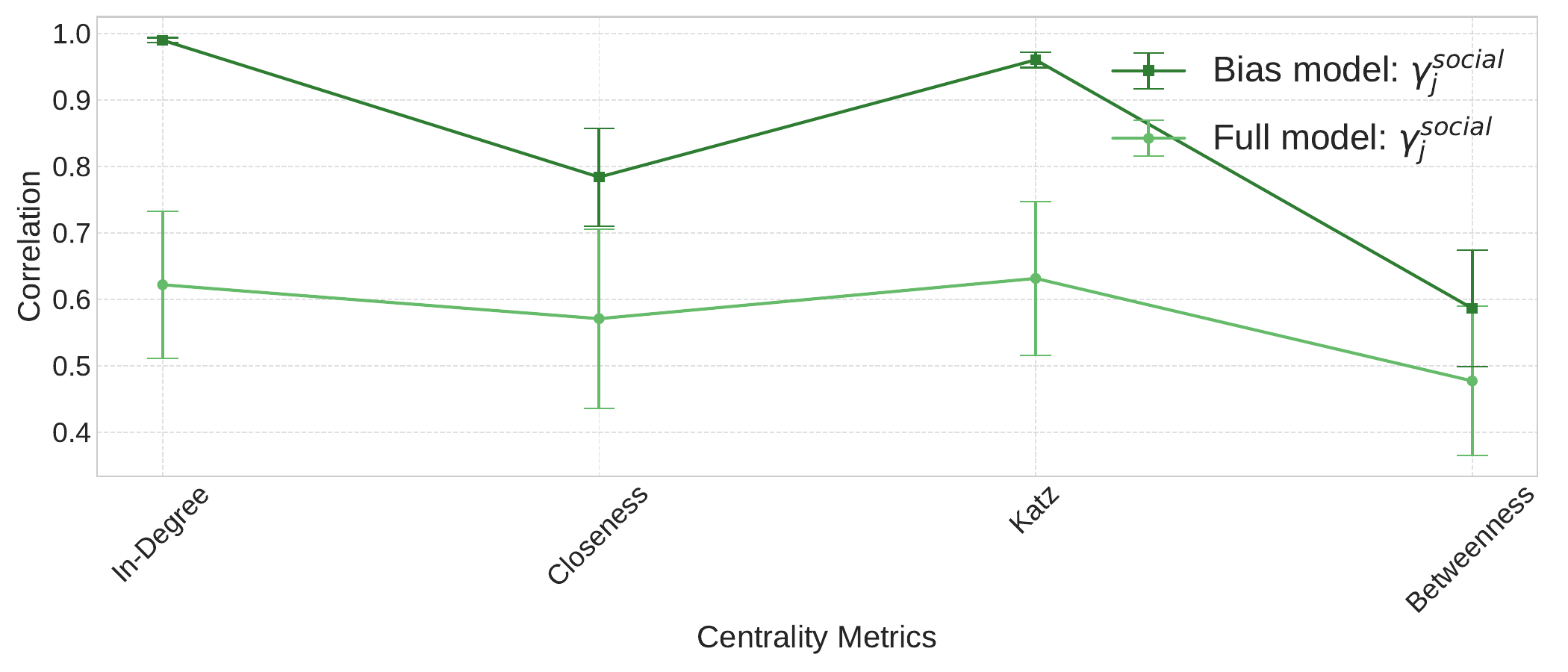}
    \caption{\small Social layer target bias $\gamma_{j}^{1}$ correlation values}
\end{subfigure}
\hfill
\begin{subfigure}[t]{0.32\textwidth}
    \includegraphics[width=\textwidth]{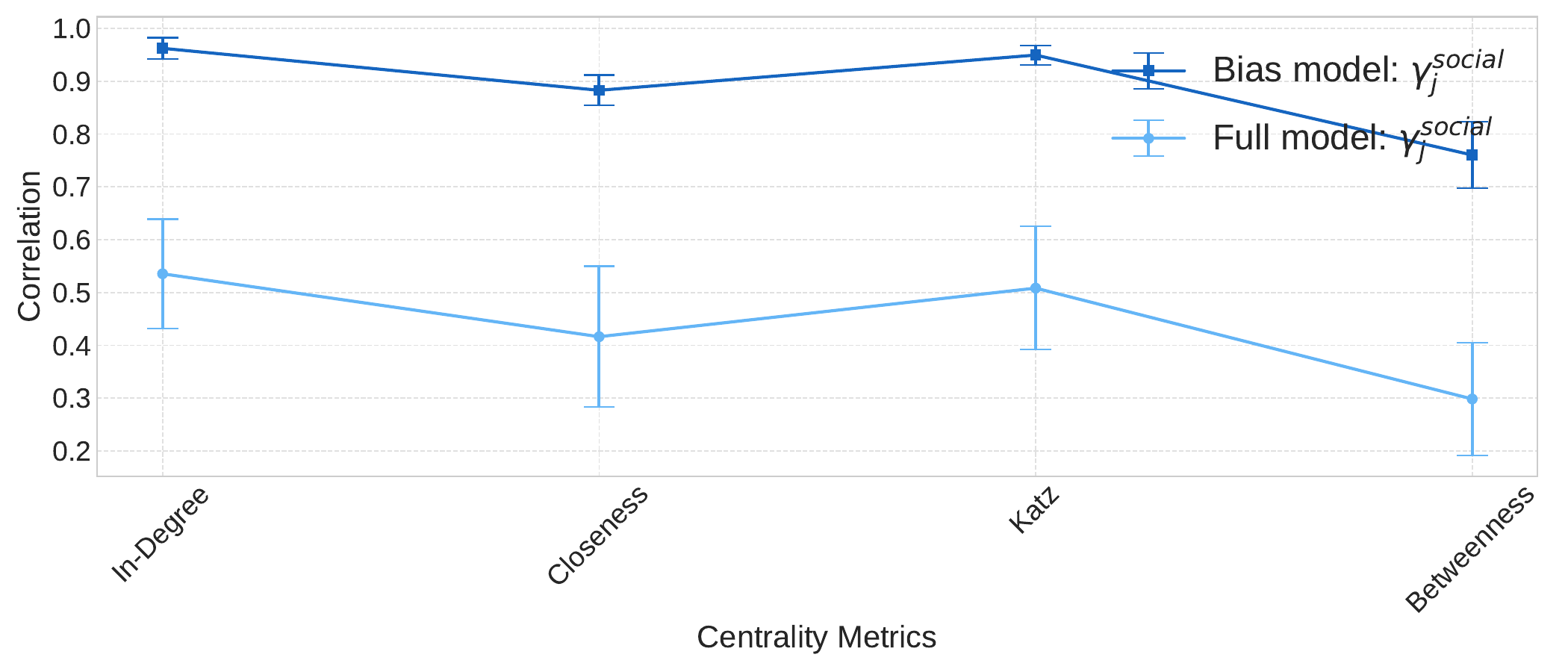}
    \caption{\small Health layer target bias $\gamma_{j}^{2}$ correlation values}
\end{subfigure}
\hfill
\begin{subfigure}[t]{0.32\textwidth}
    \includegraphics[width=\textwidth]{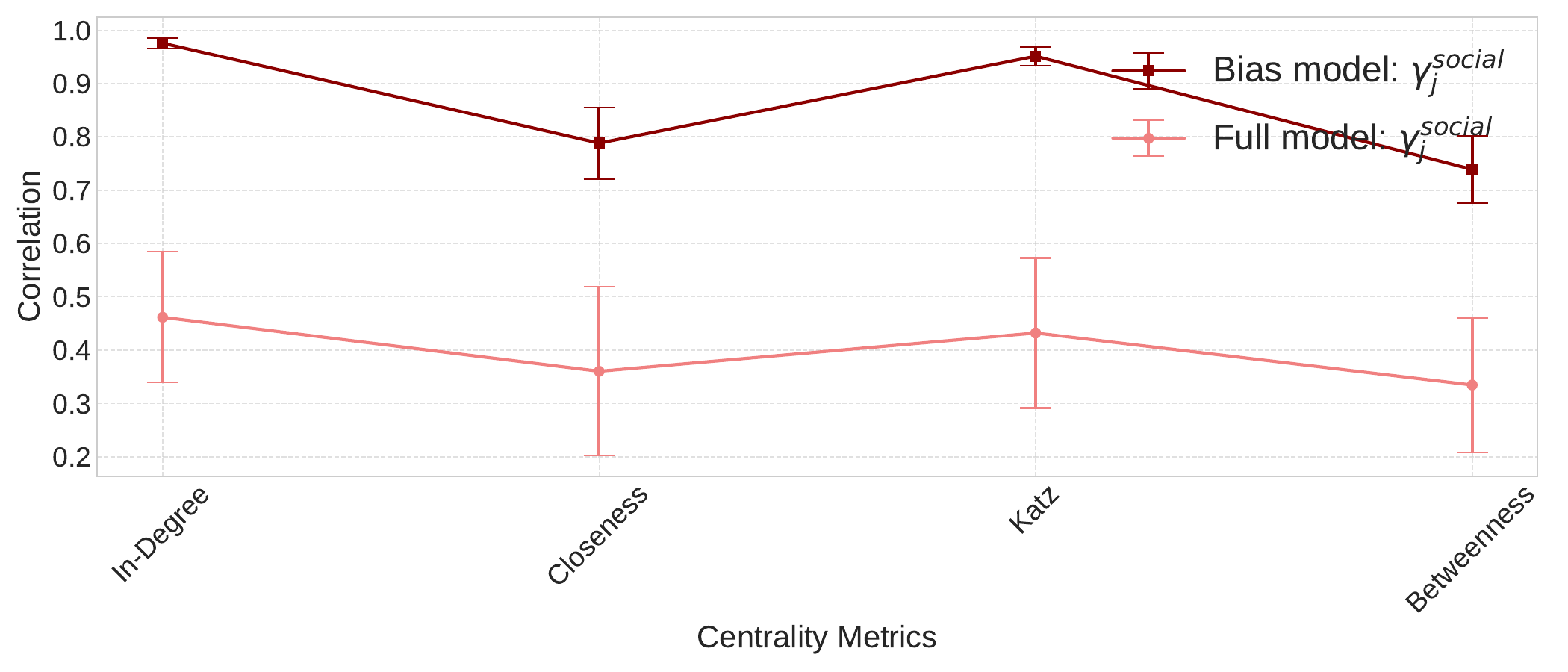}
    \caption{\small Economic layer target bias $\gamma_{j}^{3}$ correlation values}
\end{subfigure}

\centering
\begin{subfigure}[t]{0.32\textwidth}
    \includegraphics[width=\textwidth]{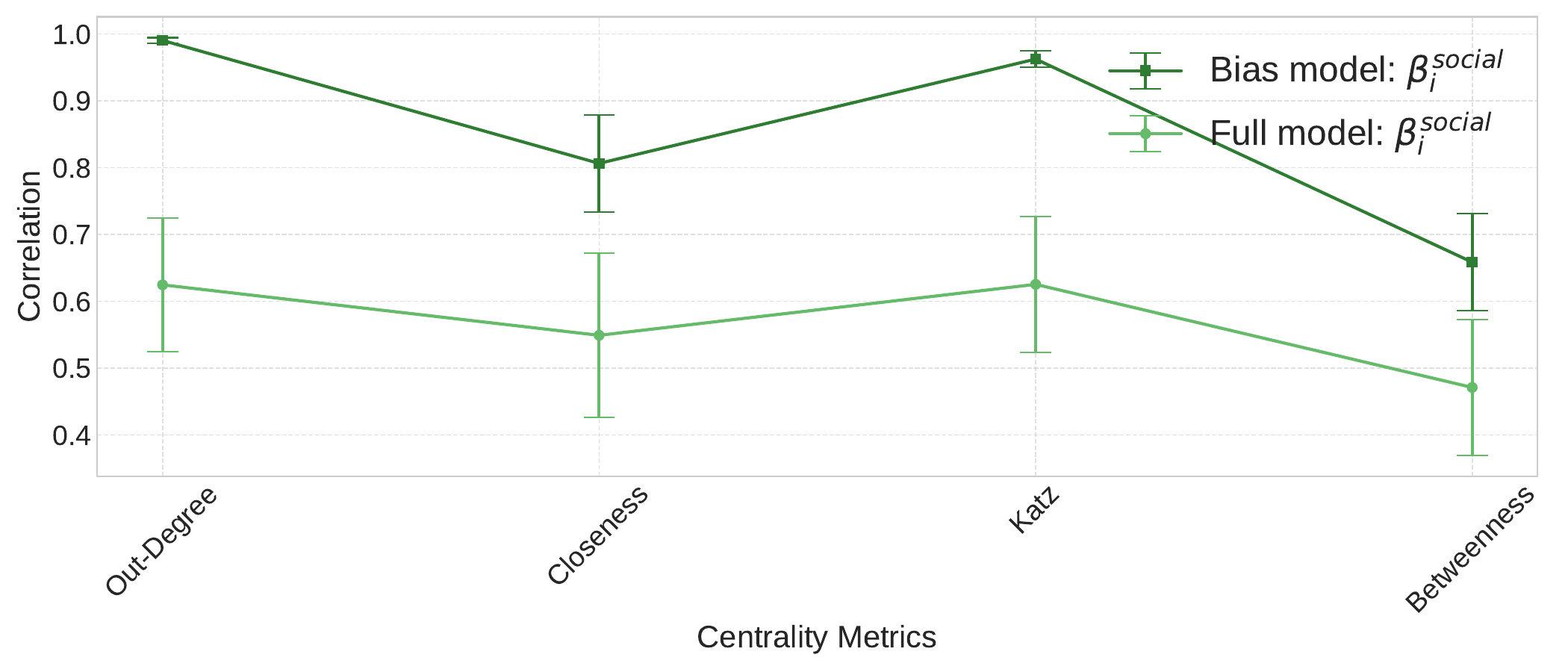}
    \caption{\small Social layer source bias $\beta_{i}^{1}$ correlation values}
\end{subfigure}
\hfill
\begin{subfigure}[t]{0.32\textwidth}
    \includegraphics[width=\textwidth]{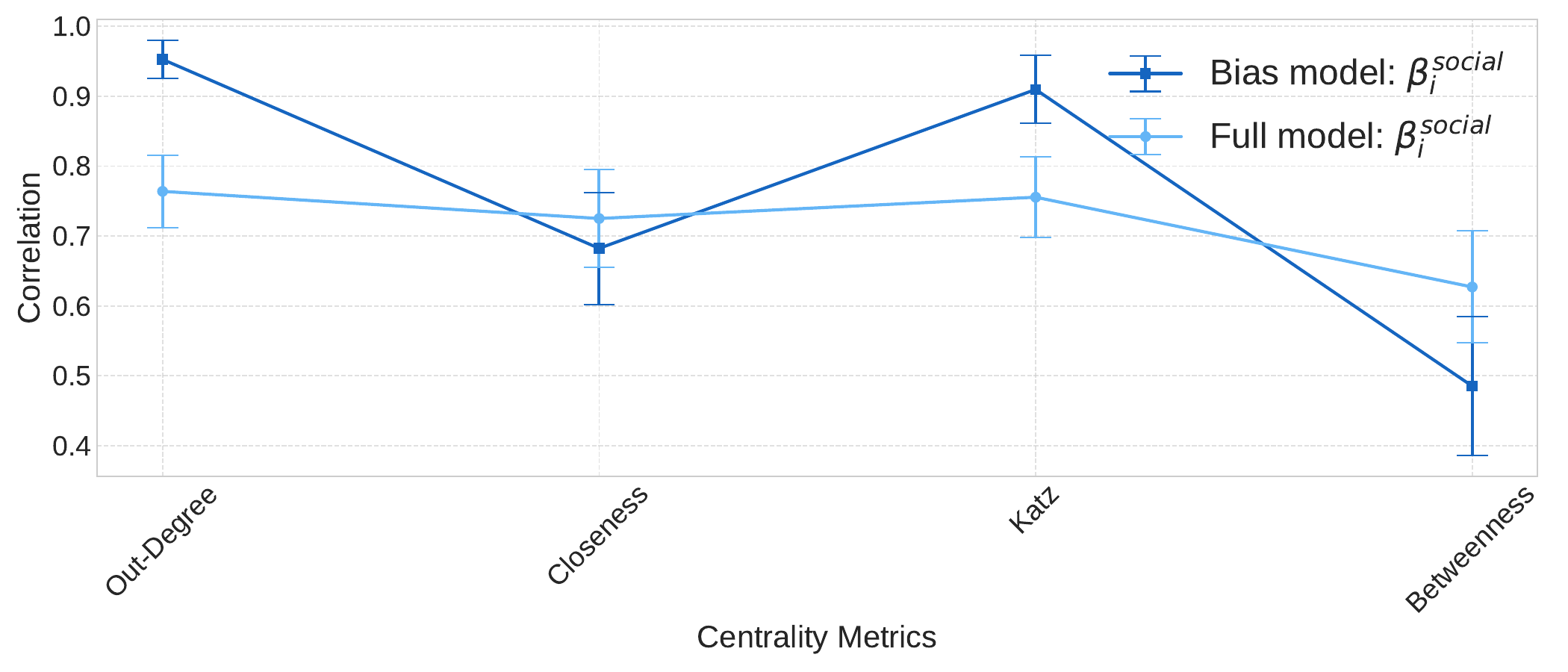}
    \caption{\small Health layer source bias $\beta_{i}^{2}$ correlation values}
\end{subfigure}
\hfill
\begin{subfigure}[t]{0.32\textwidth}
    \includegraphics[width=\textwidth]{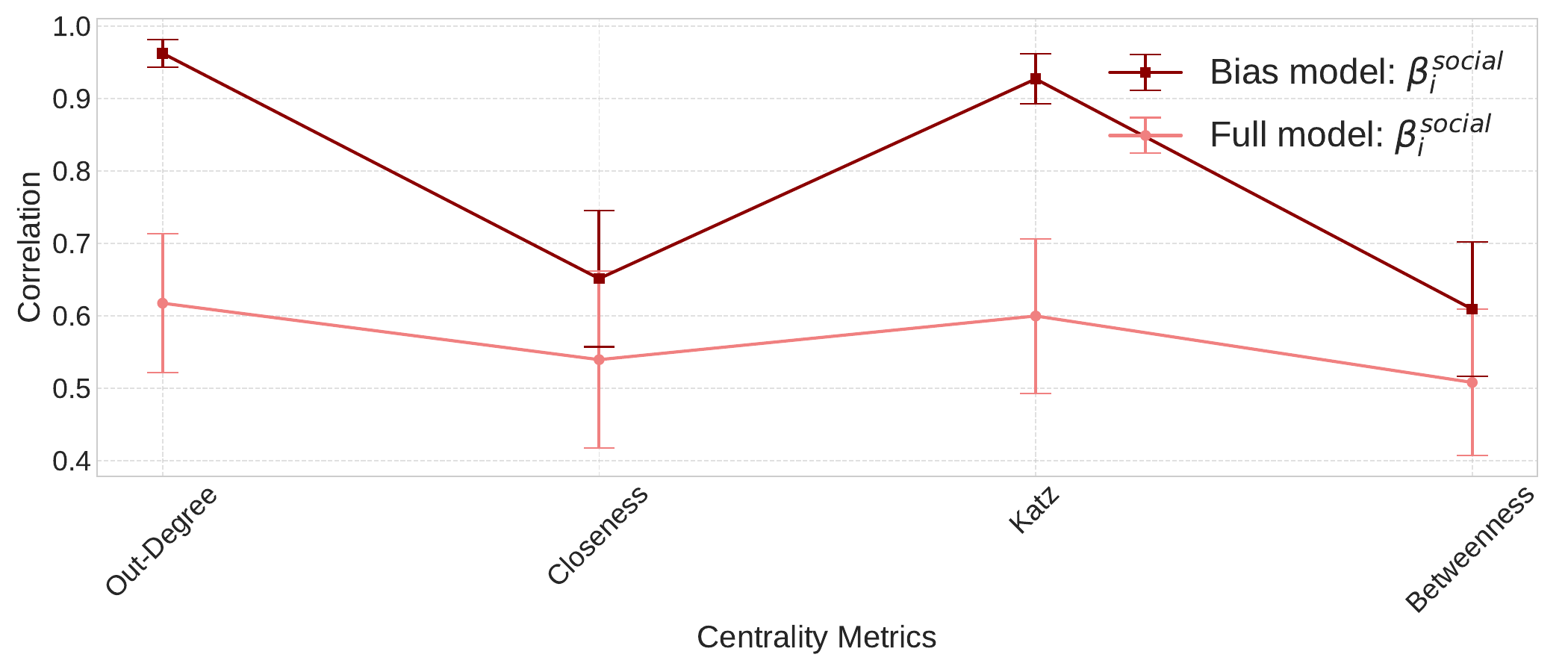}
    \caption{\small Economic layer source bias $\beta_{i}^{3}$ correlation values}
\end{subfigure}
\caption{\textbf{Spearman correlation between layer-specific node biases and centrality measures across model variants.} For each layer of the multiplex network, \textit{social}, \textit{health}, and \textit{economic}, treated as a separate directed network, we compute standard centrality measures (in-degree, out-degree, closeness, Katz, and betweenness) for each node. For the computation of closeness and Katz centralities in the source (sender) case, we reverse the direction of ties to reflect outbound influence. We then calculate the Spearman correlation between these centralities and the model-inferred \textit{source} ($\beta_i^{l}$) and \textit{target} ($\gamma_i^{l}$) biases, averaging results across all $176$ village networks.
The \textit{top row} (panels (a)–(c)) reports correlations for the \textit{target} biases, comparing values from the \textit{bias-only model} (without interdependence) and the \textit{full model} (with interdependence). The \textit{bottom row} (panels (d)–(f)) presents the same comparison for \textit{source} biases. These comparisons reveal how incorporating interdependence alters the relationship between inferred node roles and classical structural centrality measures.
}
\label{fig:RE_core}
\end{figure*}

\begin{figure*}[!htb]

\centering
\begin{subfigure}[t]{0.49\textwidth}
    \includegraphics[width=\textwidth]{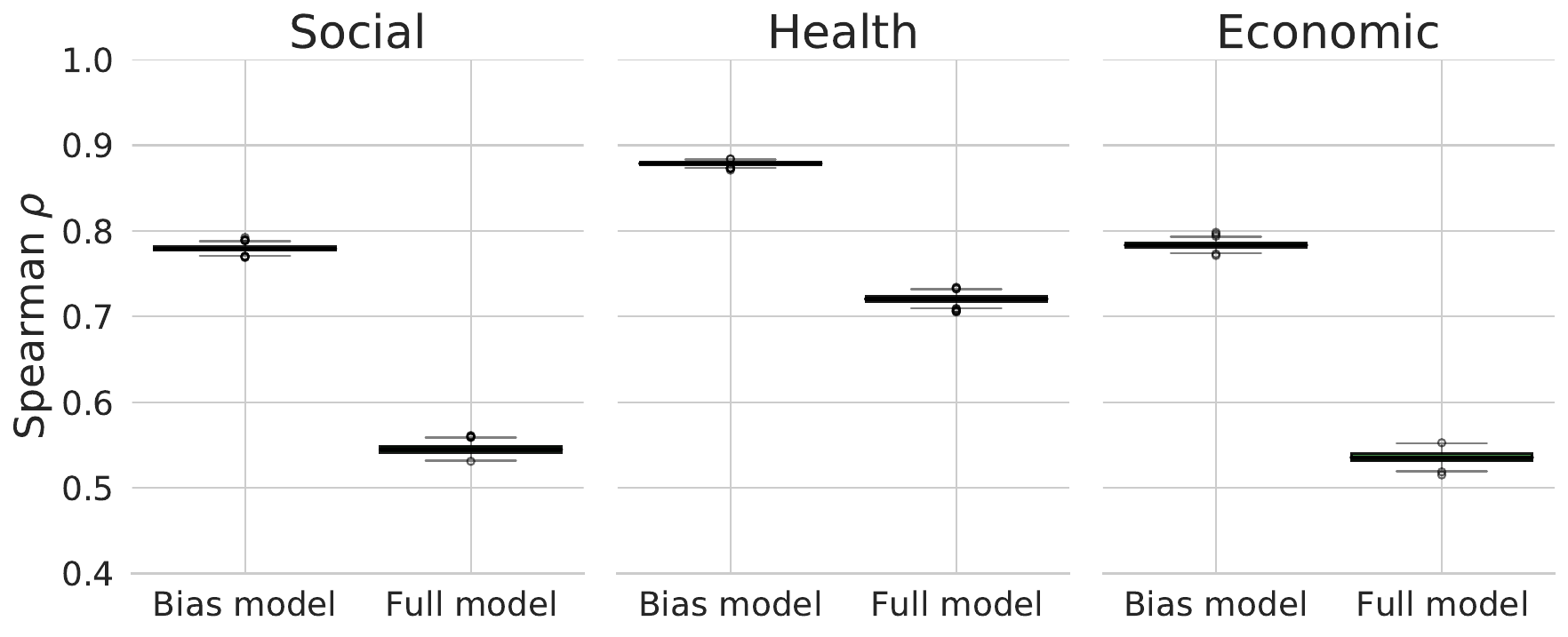}
    \caption{\small Mean bootstrapped correlations between inferred \textbf{target bias} ($\gamma_j^{l}$) and \textbf{closeness centrality} in the social, health, and economic layers, comparing the bias-only model with the full model that accounts for interdependence.}
\end{subfigure}
\hfill
\begin{subfigure}[t]{0.49\textwidth}
    \includegraphics[width=\textwidth]{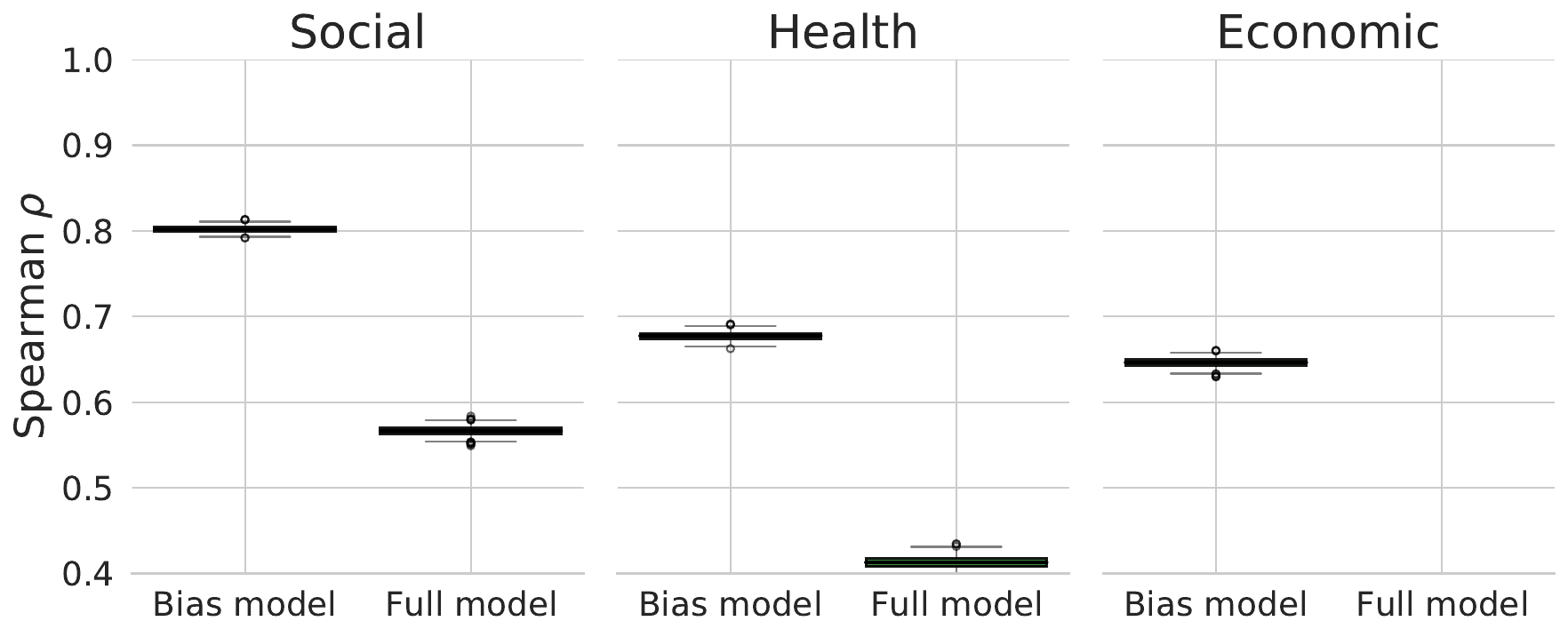}
    \caption{\small Mean bootstrapped correlations between inferred \textbf{source bias} ($\beta_i^{l}$) and \textbf{closeness centrality} in the social, health, and economic layers, comparing the bias-only model with the full model that accounts for interdependence.}
\end{subfigure}
\centering
\begin{subfigure}[t]{0.49\textwidth}
    \includegraphics[width=\textwidth]{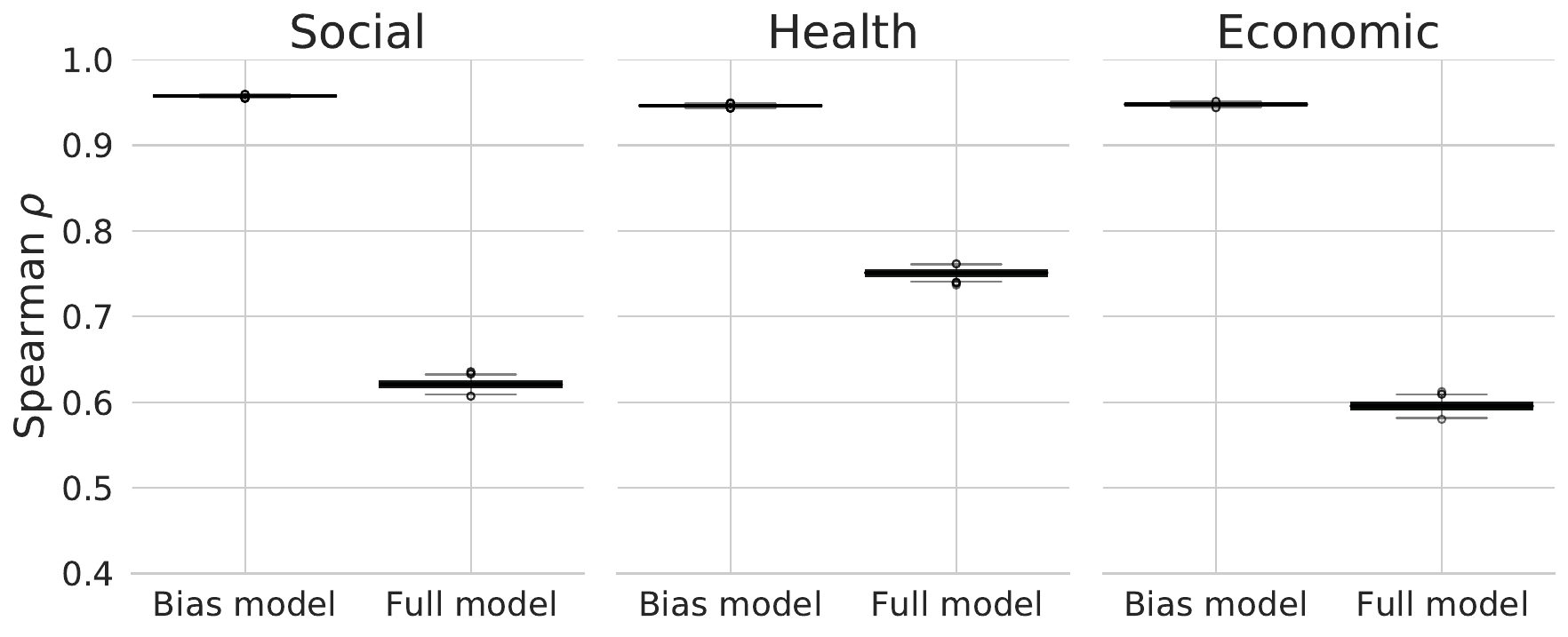}
    \caption{\small Mean bootstrapped correlations between inferred \textbf{target bias} ($\gamma_j^{l}$) and \textbf{Katz centrality} in the social, health, and economic layers, comparing the bias-only model with the full model that accounts for interdependence.}
\end{subfigure}
\hfill
\begin{subfigure}[t]{0.49\textwidth}
    \includegraphics[width=\textwidth]{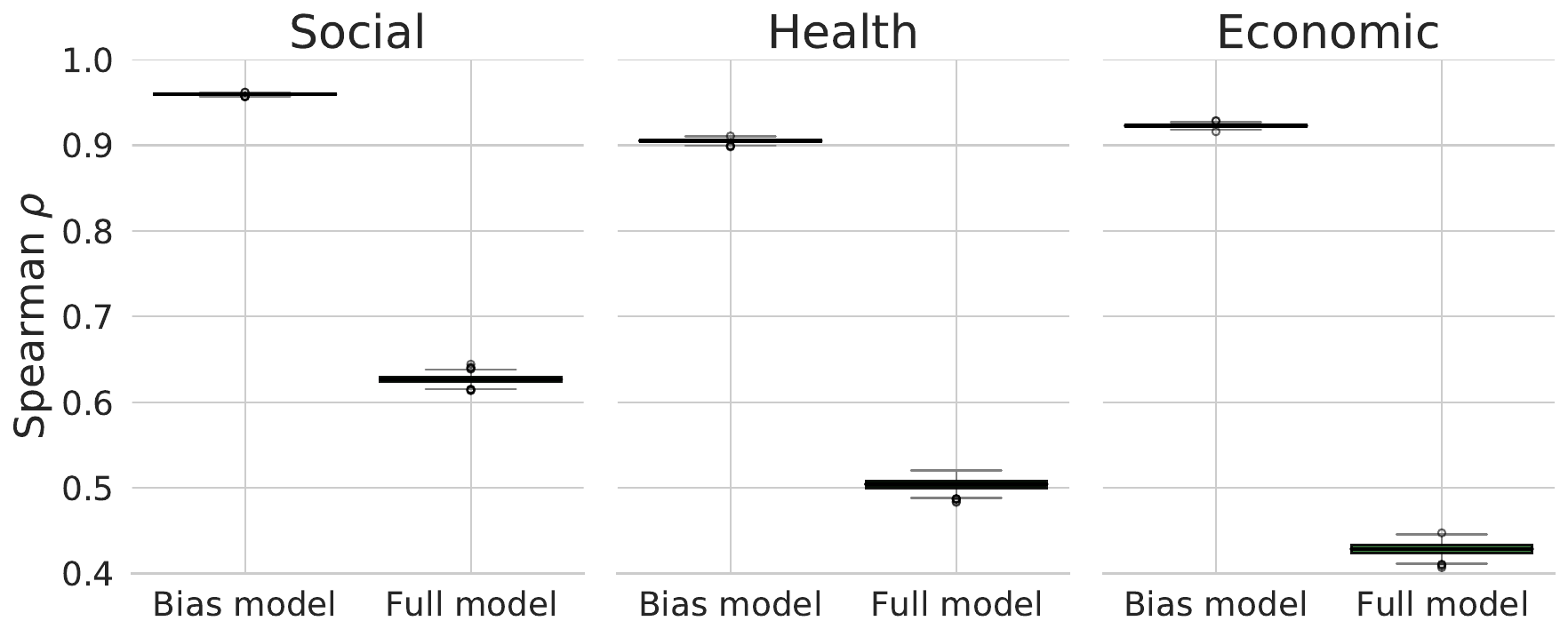}
    \caption{\small Mean bootstrapped correlations between inferred \textbf{source bias} ($\beta_i^{l}$) and \textbf{Katz centrality} in the social, health, and economic layers, comparing the bias-only model with the full model that accounts for interdependence.}
\end{subfigure}
\centering
\begin{subfigure}[t]{0.49\textwidth}
    \includegraphics[width=\textwidth]{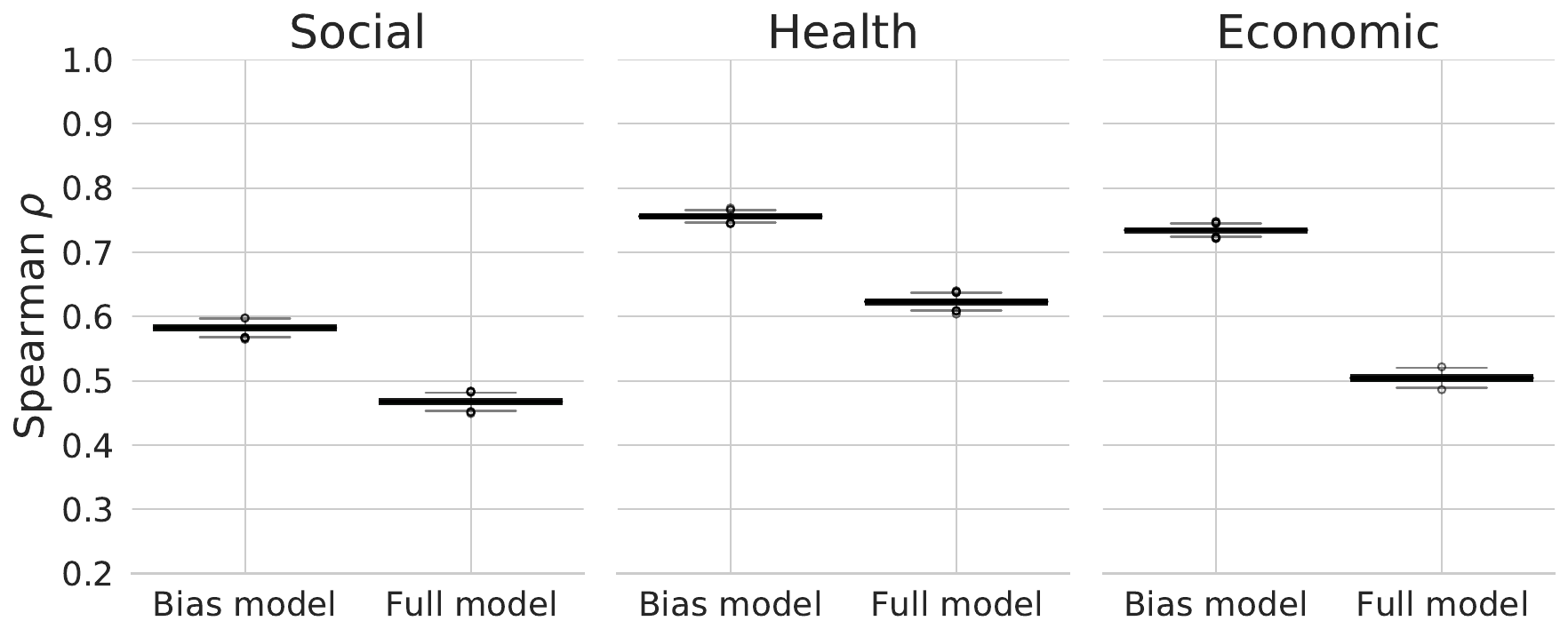}
    \caption{\small Mean bootstrapped correlations between inferred \textbf{target bias} ($\gamma_j^{l}$) and \textbf{betweeness centrality} in the social, health, and economic layers, comparing the bias-only model with the full model that accounts for interdependence.}
\end{subfigure}
\hfill
\begin{subfigure}[t]{0.49\textwidth}
    \includegraphics[width=\textwidth]{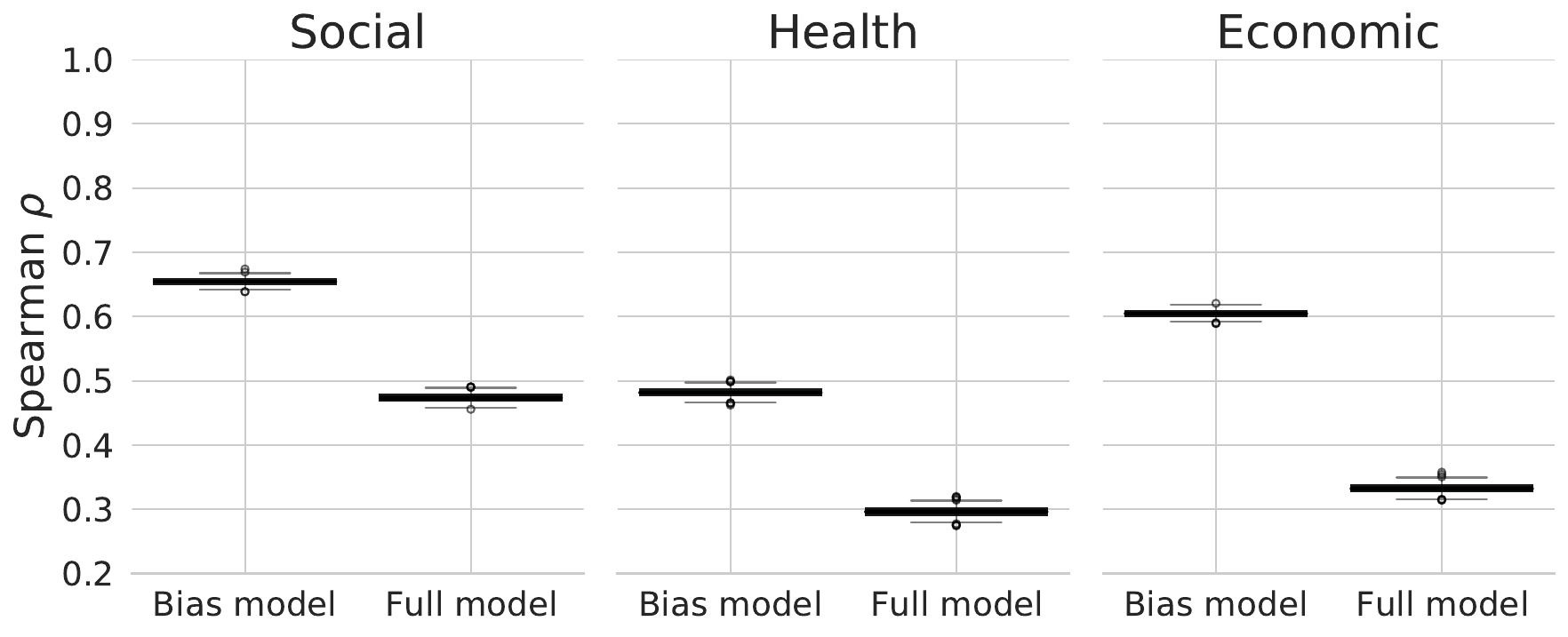}
    \caption{\small Mean bootstrapped correlations between inferred \textbf{source bias} ($\beta_i^{l}$) and \textbf{betweeness centrality} in the social, health, and economic layers, comparing the bias-only model with the full model that accounts for interdependence.}
\end{subfigure}
\caption{\textbf{Bootstrapped Spearman correlations ($\rho$) between the inferred target bias ($\gamma^{l}_{j}$) or source bias ($\beta^{l}_{j}$) and three centrality measures, closeness, Katz, and betweenness, in the social, health, and economic layers.}
For each bootstrap iteration we compute $\rho$ in every one of the 176 village networks, then average these 176 values to obtain a single network-averaged correlation. Repeating this procedure $1\,000$ times yields $1\,000$ mean correlations, whose distribution is shown as a box-and-whisker plot.}
\label{fig:RE_centr}
\end{figure*}

Observed correlation scores are shown in SI Figure~\ref{fig:RE_core}.  
For each layer of the multiplex network, treated as an independent network, we compute the centrality measures such as betweenness, closeness, and Katz centralities and then calculate the Spearman correlation between the layer-specific source biases ($\beta_i^l$), target biases ($\gamma_i^l$), and these measures. (For closeness and Katz centralities in the source role, we reverse tie direction to capture outbound influence.) The \textit{top row} (panels (a)–(c)) reports correlations for the \textit{target} biases, comparing values from the \textit{bias-only model} (without interdependence) and the \textit{full model} (with interdependence).  
The \textit{bottom row} (panels (d)–(f)) presents the same comparison for \textit{source} biases. Bootstrapped correlation scores for additional network centrality measures are provided in SI Figure~\ref{fig:RE_centr}.  
The first row reports results for \textit{closeness} centrality, the second row for \textit{Katz} centrality, and the third row for \textit{betweenness} centrality. These results show that the full model relies more heavily on interdependence to explain the observed link structure, shifting explanatory weight away from individual node activity toward dyadic interactions.

\subsection*{Detailed predictive gain plots across all villages and layers}
We provide detailed plots quantifying the effect of modeling interdependence for each layer across all villages. Results for the social, health, and economic layers are shown in SI Figures~\ref{fig:SET_social}, \ref{fig:SET_heal}, and \ref{fig:SET_econ}, respectively. In each figure, panel~(a) compares the average predictive performance of a model that accounts only for dependence and independence (\textit{Bias model}) with that of the \textit{Full model}, which additionally incorporates interdependence, using Precision-Recall Area Under the Curve (PR AUC) scores. Panel~(b) shows the same comparison using Receiver Operating Characteristic Area Under the Curve (ROC AUC) scores.

\subsection*{Structural Determinants of Predictive Power} 
\begin{figure*}[!htb]
    \centering
\begin{subfigure}[t]{0.24\textwidth}
    \includegraphics[width=\textwidth]{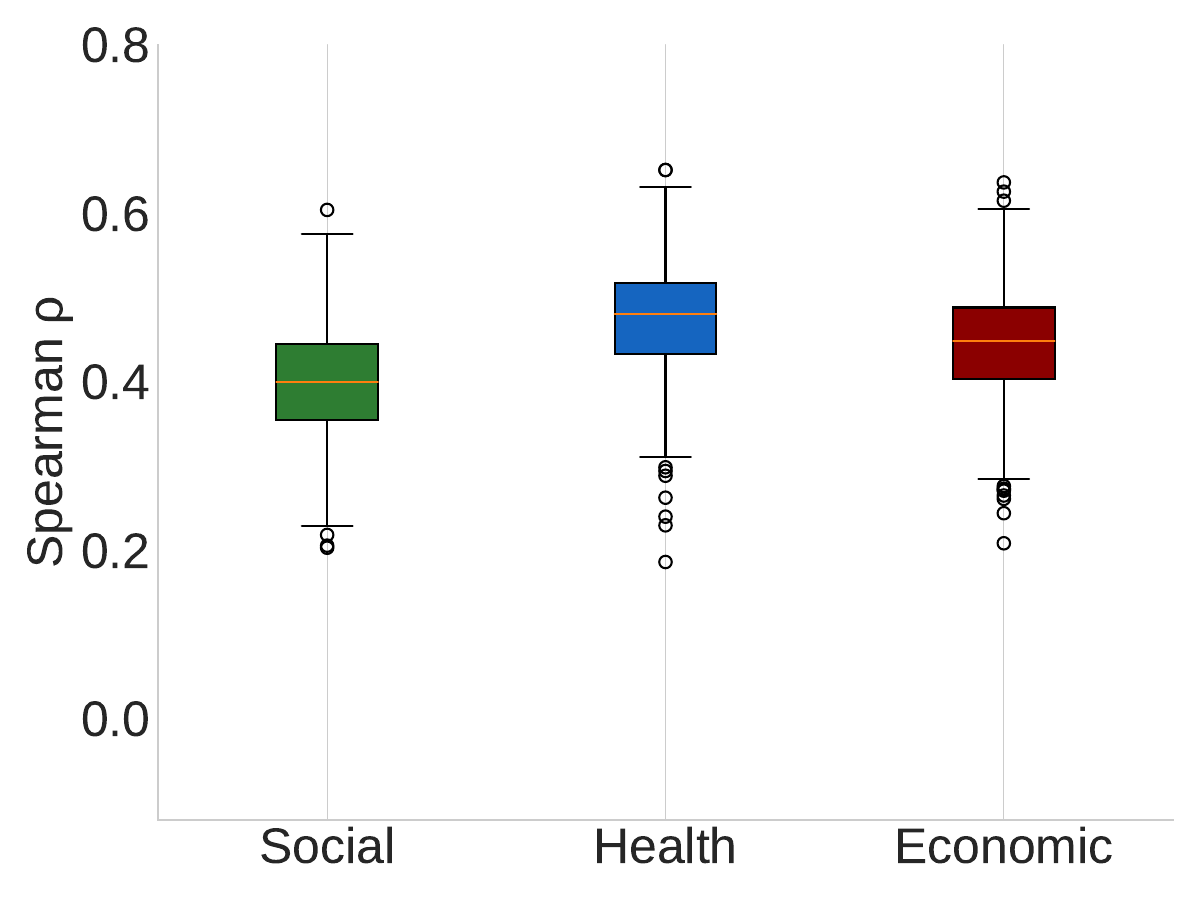}
    \caption{\small  Bootstrapped Spearman correlation between PR-AUC prediction improvement and reciprocity, computed separately for the social, health, and economic layers.}
\end{subfigure}
\hfill
\begin{subfigure}[t]{0.24\textwidth}
    \includegraphics[width=\textwidth]{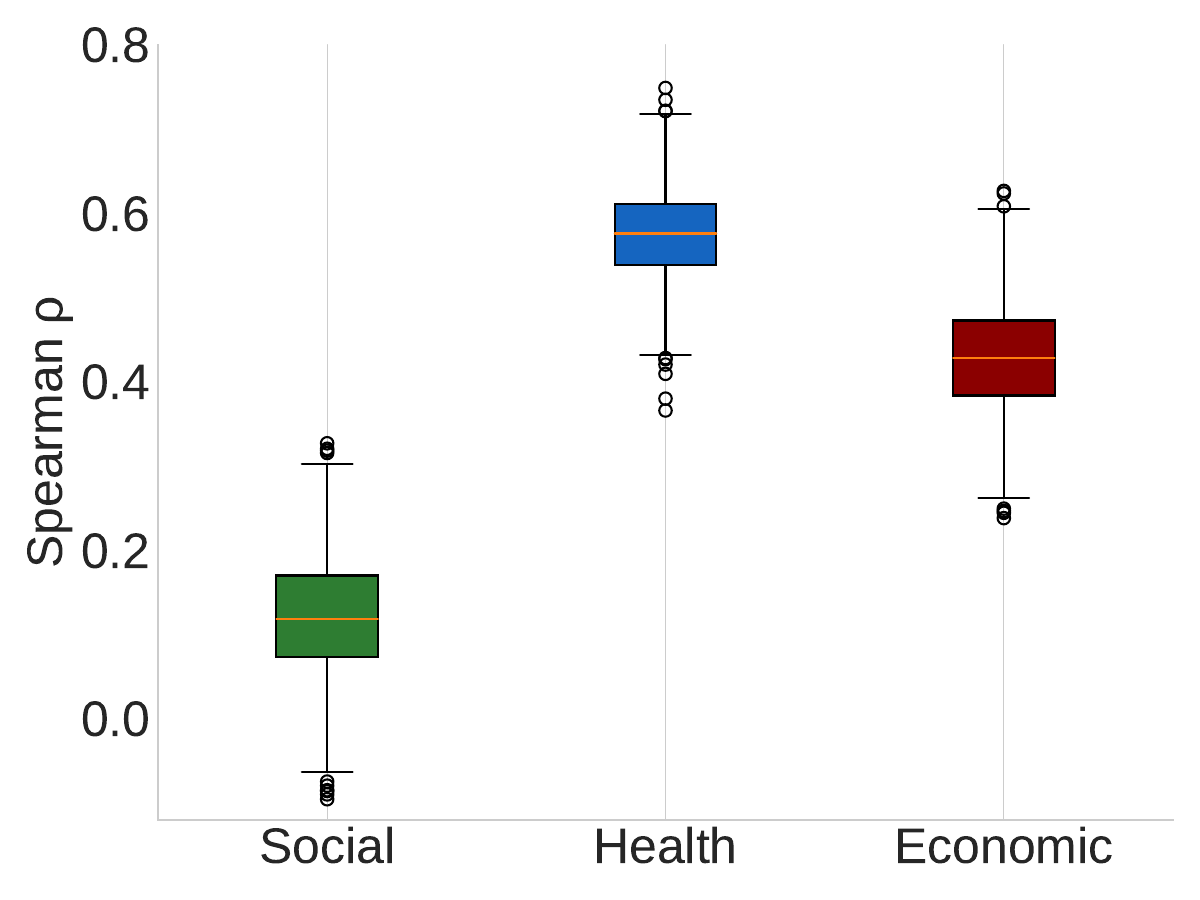}
    \caption{\small Bootstrapped Spearman correlation between PR-AUC prediction improvement and transitivity, computed separately for the social, health, and economic layers.}
\end{subfigure}
\hfill
\begin{subfigure}[t]{0.24\textwidth}
    \includegraphics[width=\textwidth]{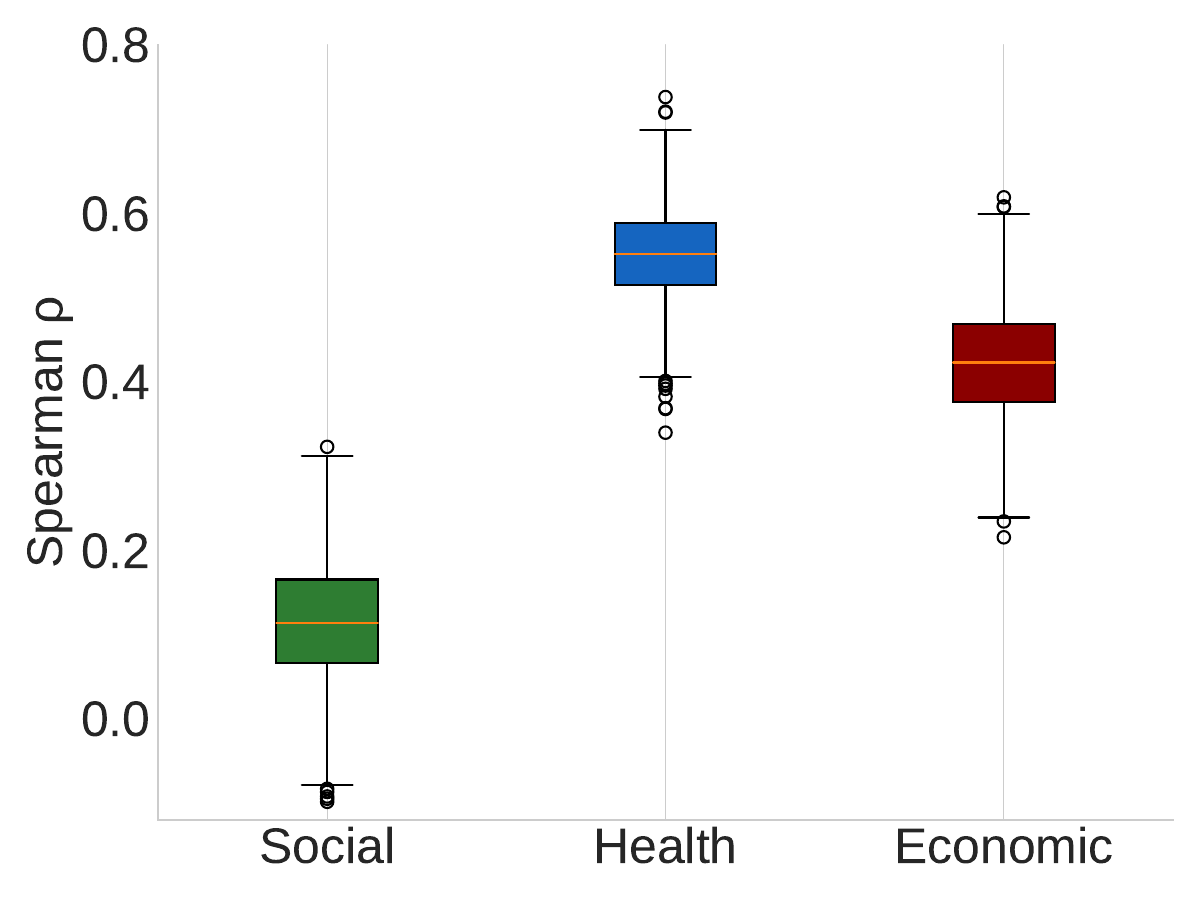}
    \caption{\small   Bootstrapped Spearman correlation between PR-AUC prediction improvement and clustering coefficient, computed separately for the social, health, and economic layers.}
\end{subfigure}
\hfill
\begin{subfigure}[t]{0.24\textwidth}
    \includegraphics[width=\textwidth]{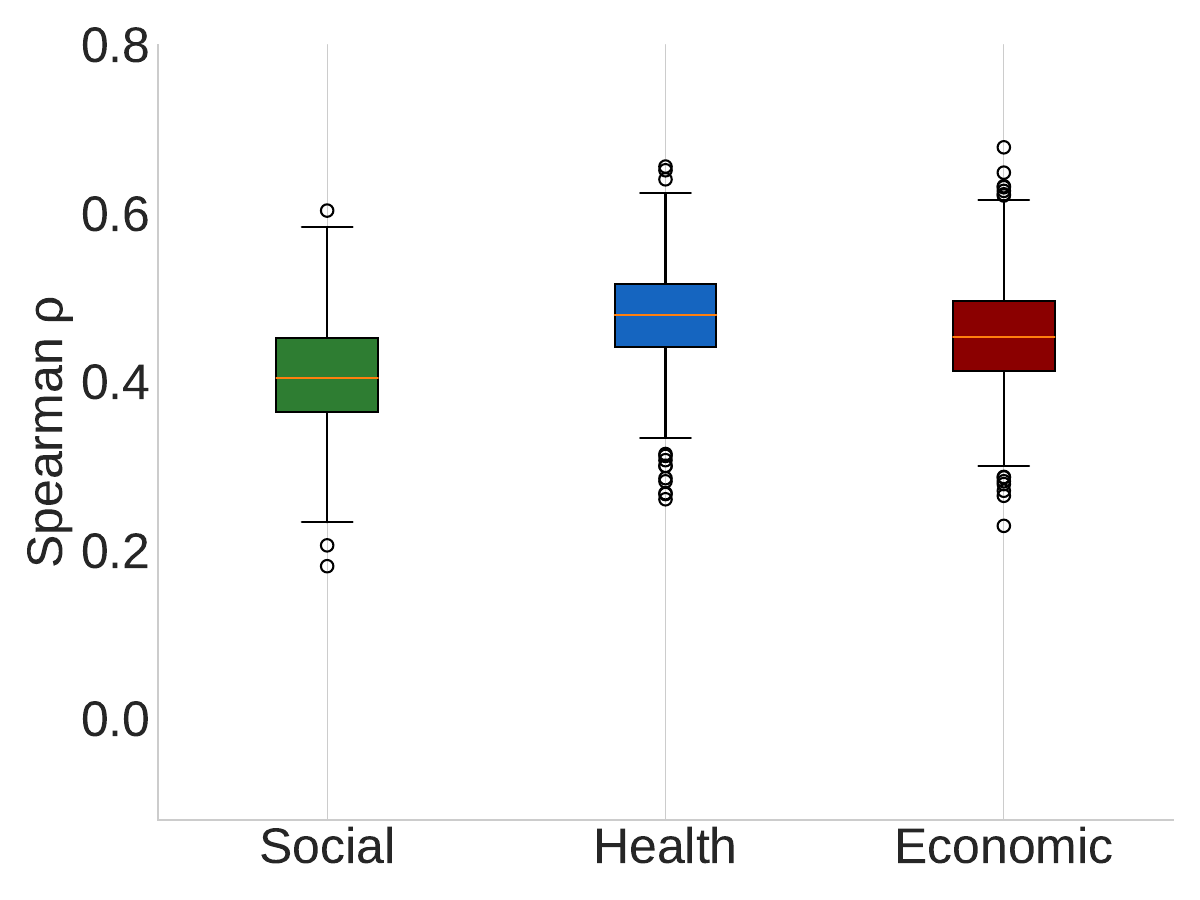}
    \caption{\small Bootstrapped Spearman correlation between PR-AUC prediction improvement and average node degree, computed separately for the social, health, and economic layers.}
\end{subfigure}
    \centering
\begin{subfigure}[t]{0.24\textwidth}
    \includegraphics[width=\textwidth]{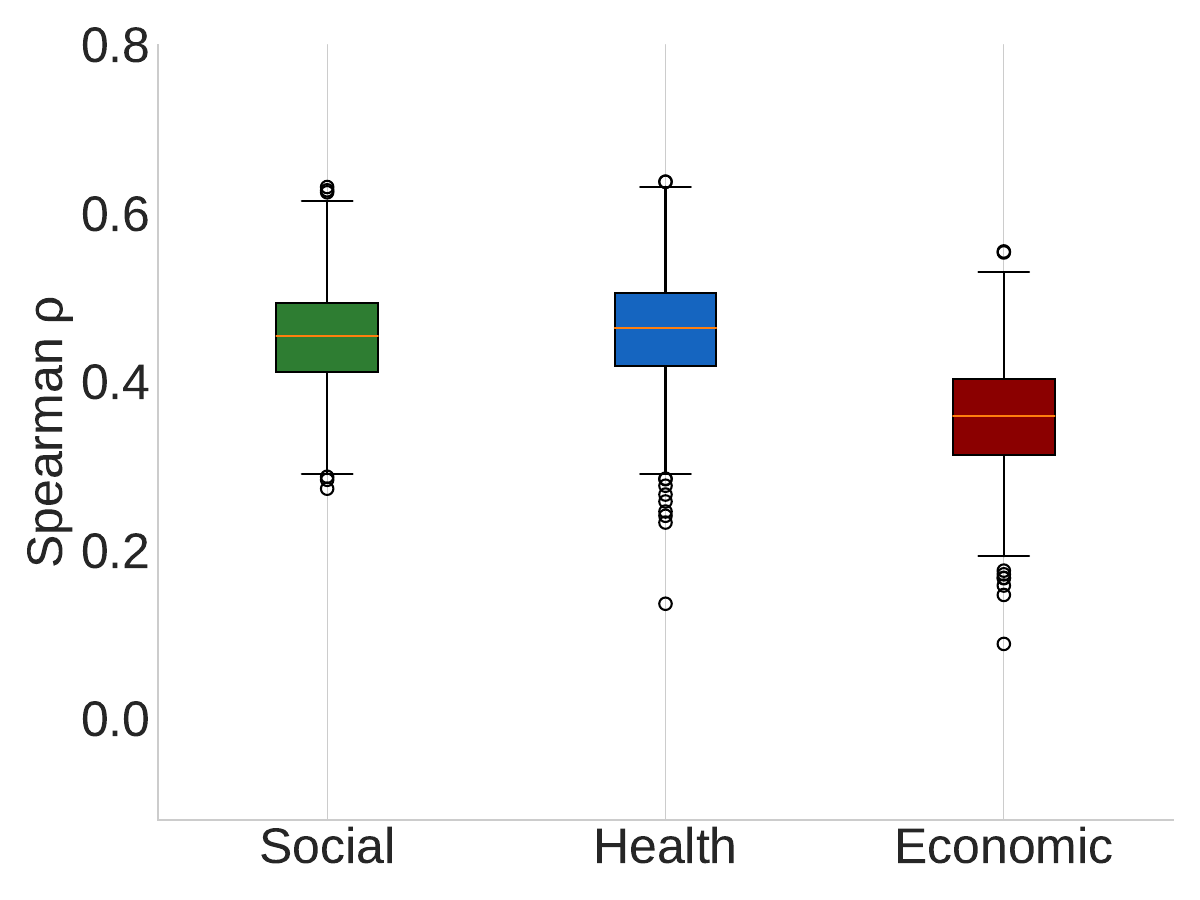}
    \caption{\small  Bootstrapped Spearman correlation between ROC-AUC prediction improvement and reciprocity, computed separately for the social, health, and economic layers.}
\end{subfigure}
\hfill
\begin{subfigure}[t]{0.24\textwidth}
    \includegraphics[width=\textwidth]{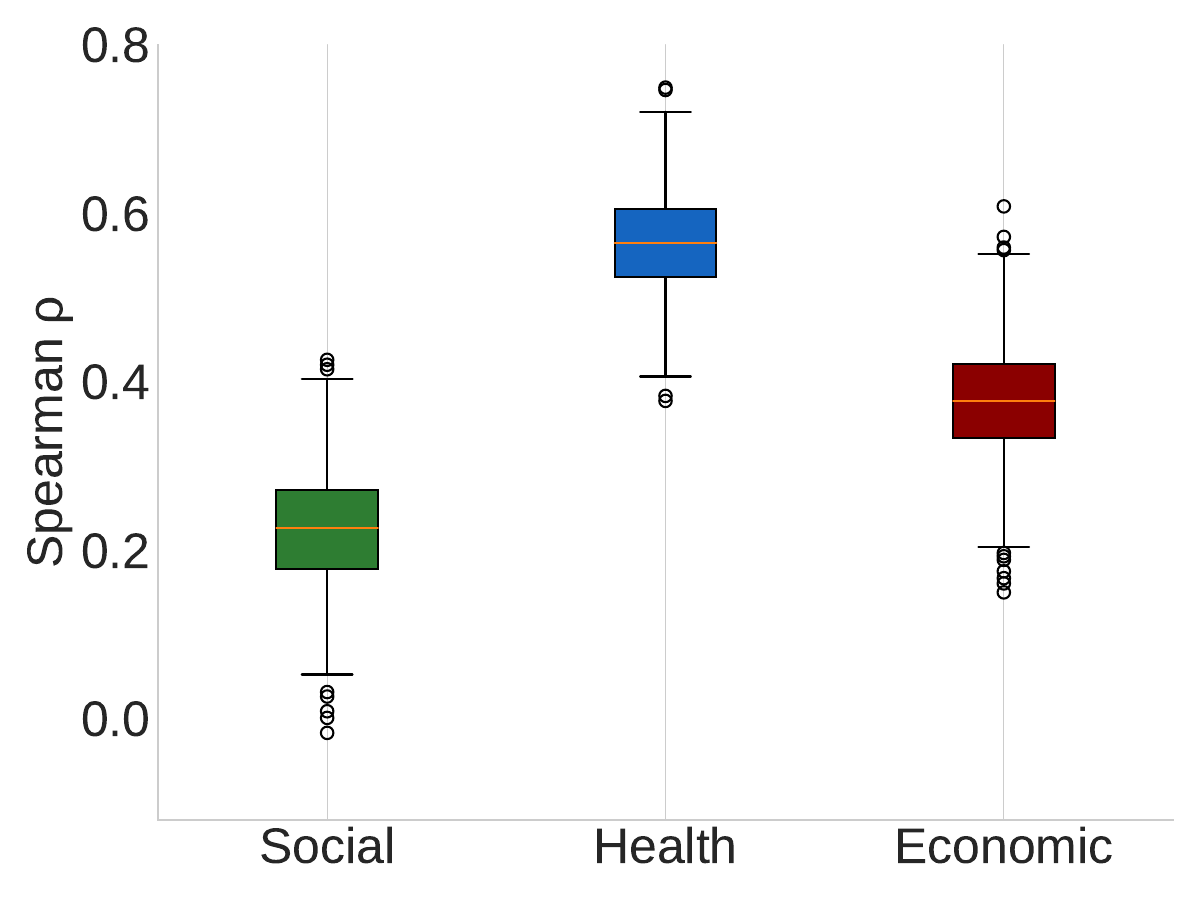}
    \caption{\small  Bootstrapped Spearman correlation between ROC-AUC prediction improvement and transitivity, computed separately for the social, health, and economic layers.}
\end{subfigure}
\hfill
\begin{subfigure}[t]{0.24\textwidth}
    \includegraphics[width=\textwidth]{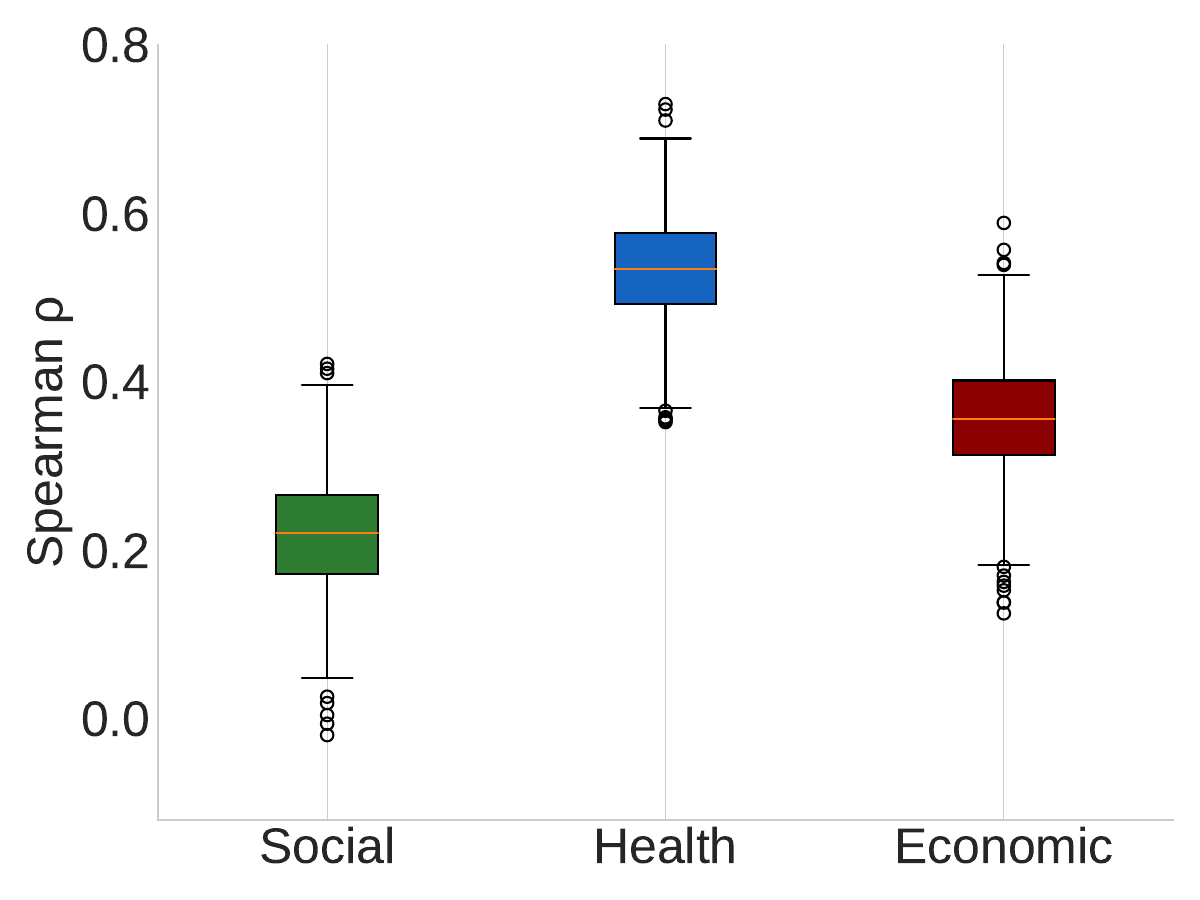}
    \caption{\small  Bootstrapped Spearman correlation between ROC-AUC prediction improvement and clustering coefficient, computed separately for the social, health, and economic layers.}
\end{subfigure}
\hfill
\begin{subfigure}[t]{0.24\textwidth}
    \includegraphics[width=\textwidth]{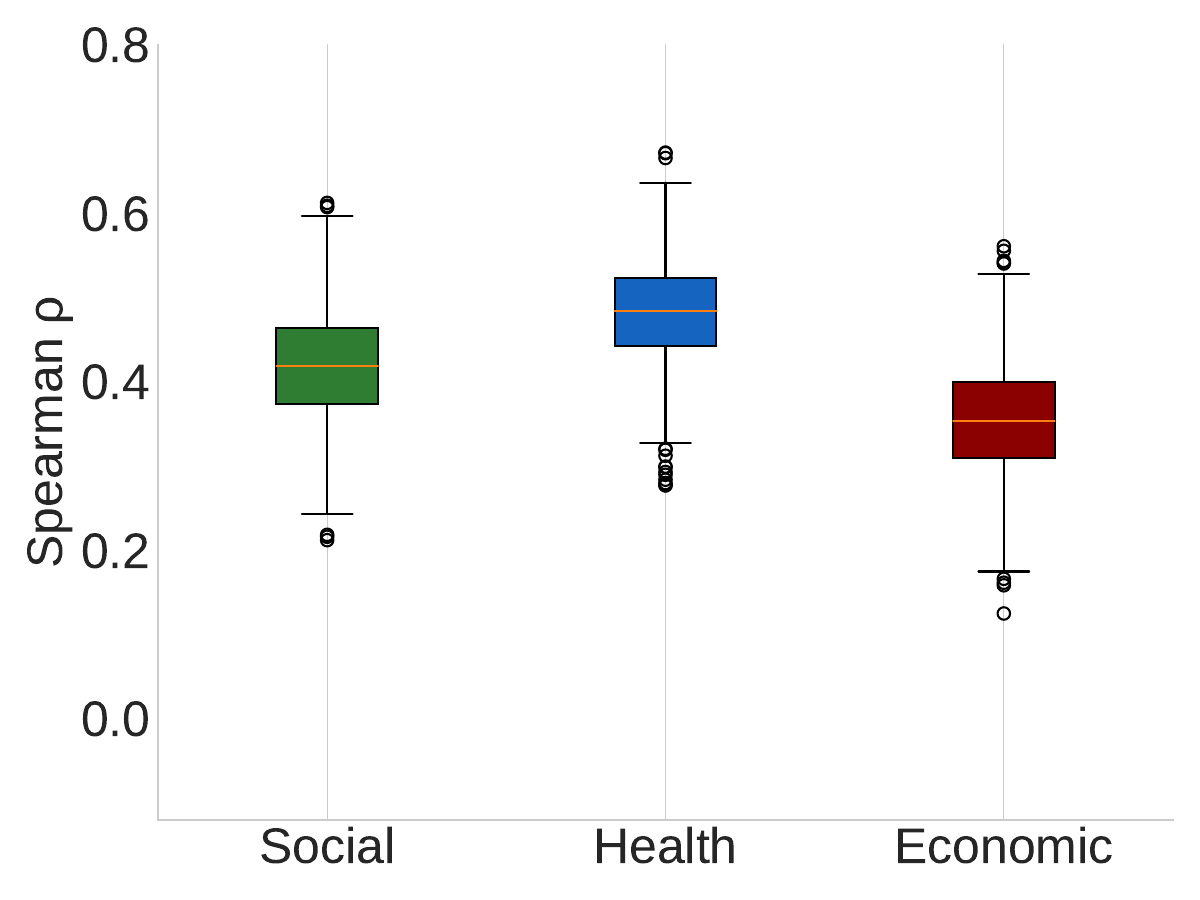}
    \caption{\small Bootstrapped Spearman correlation between ROC-AUC prediction improvement and average node degree, computed separately for the social, health, and economic layers.}
\end{subfigure}
\caption{\textbf{Bootstrapped correlation between prediction improvement and network statistics across 176 networks, over the social, health, and economic layers.} Prediction improvement is defined as the average increase in PR-AUC (top row) and ROC-AUC (bottom row) across ten-fold cross-validation when extending the model to capture interdependence, compared to modeling only dependence and independence. We report Spearman correlation coefficients ($\rho$) between the observed performance increase and key network statistics: reciprocity (panels (a) and (e)), transitivity (panels (b) and (f)), clustering coefficient (panels (c) and (g)), and average node degree (panels (d) and (h)). Correlations are computed separately for the social, health, and economic layers using 1,000 bootstrap samples.}
\label{fig:boot}
\end{figure*}

\begin{figure*}
\centering
\begin{subfigure}[t]{0.24\textwidth}
    \includegraphics[width=\textwidth]{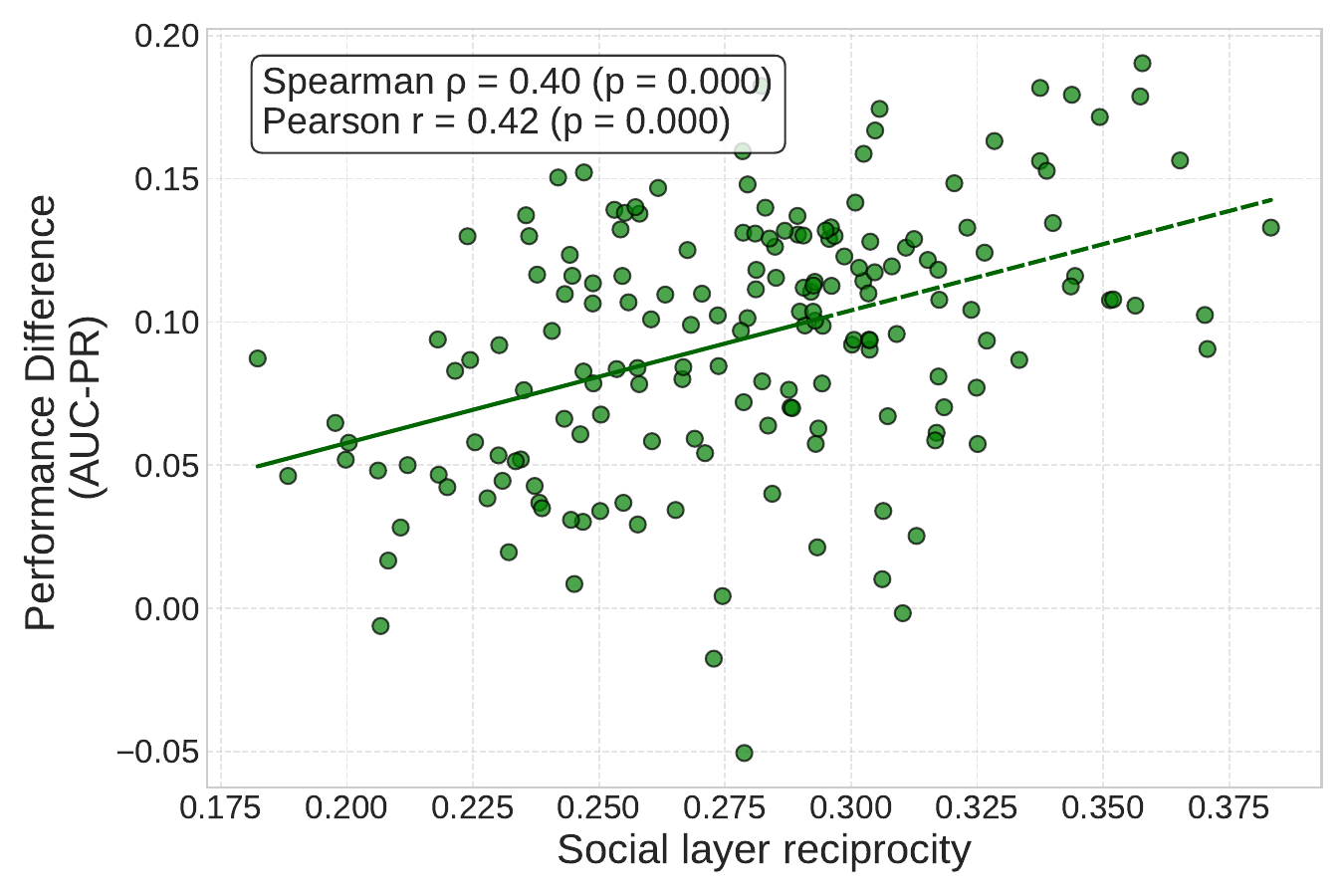}
    \caption{\small Correlation (Pearson and Spearman) between prediction improvement and reciprocity in the social layer.}
\end{subfigure}
\hfill
\begin{subfigure}[t]{0.24\textwidth}
    \includegraphics[width=\textwidth]{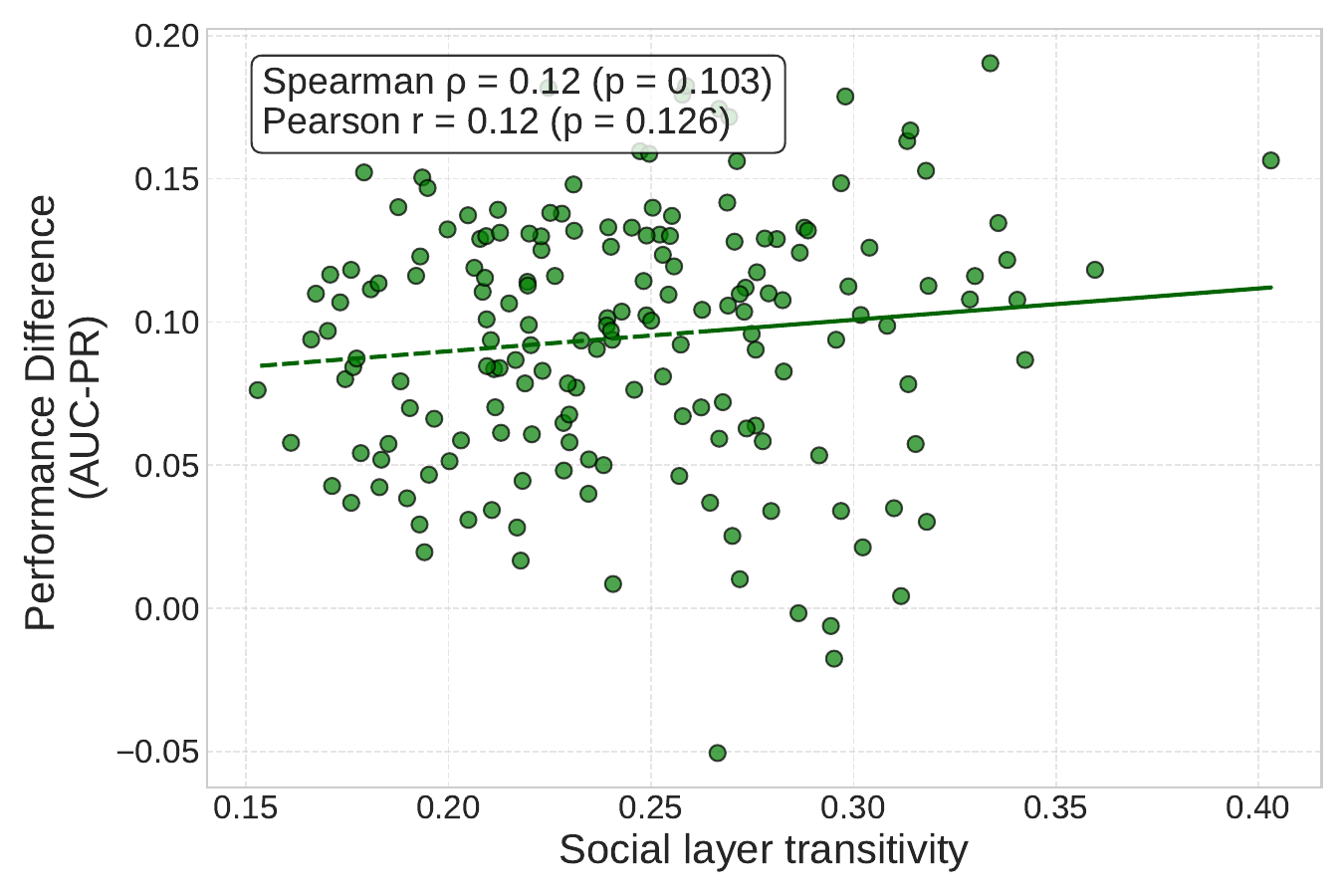}
    \caption{\small  Correlation (Pearson and Spearman) between prediction improvement and transitivity in the social layer.}
\end{subfigure}
\hfill
\begin{subfigure}[t]{0.24\textwidth}
    \includegraphics[width=\textwidth]{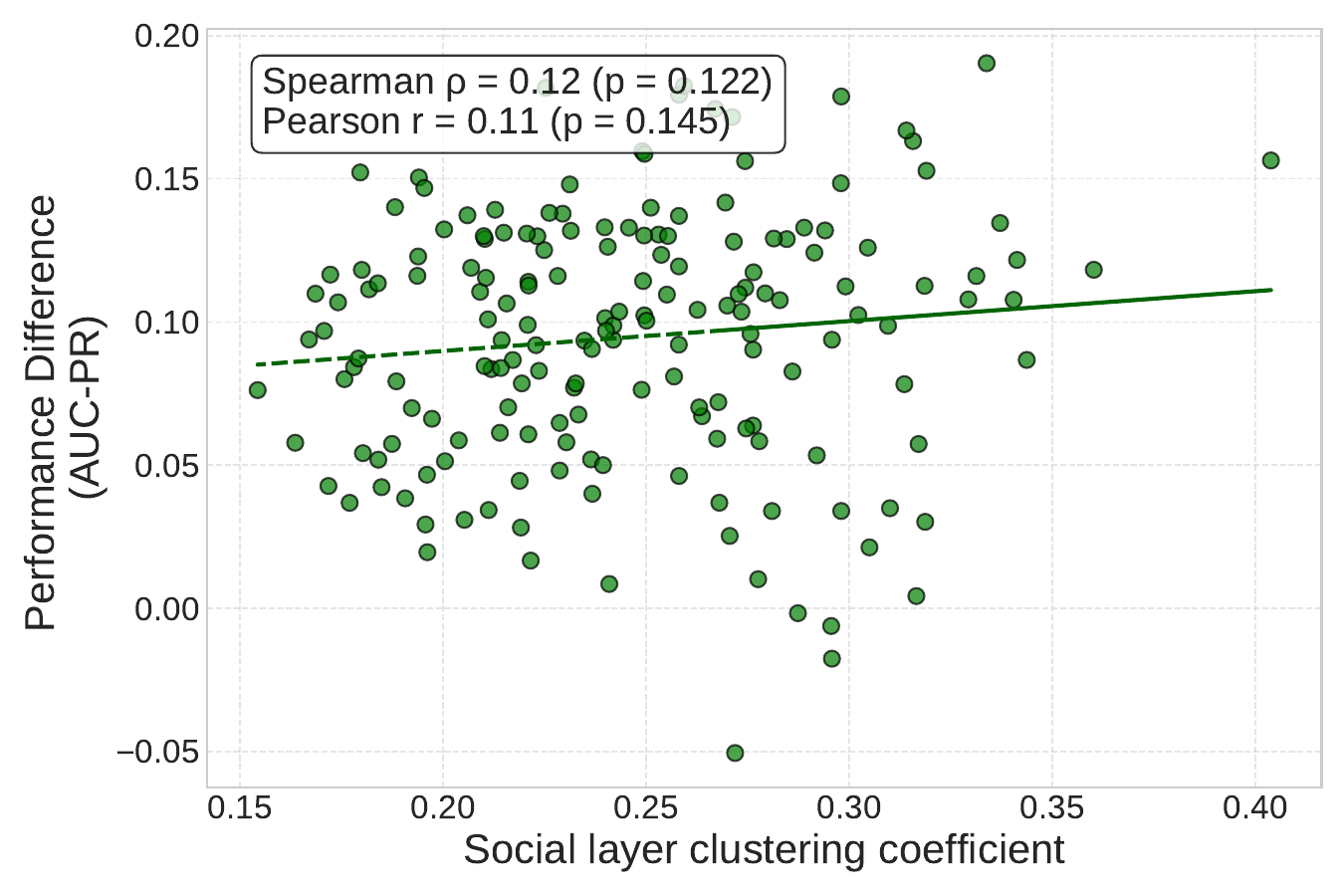}
    \caption{\small  Correlation (Pearson and Spearman) between prediction improvement and clustering in the social layer.}
\end{subfigure}
\hfill
\begin{subfigure}[t]{0.24\textwidth}
    \includegraphics[width=\textwidth]{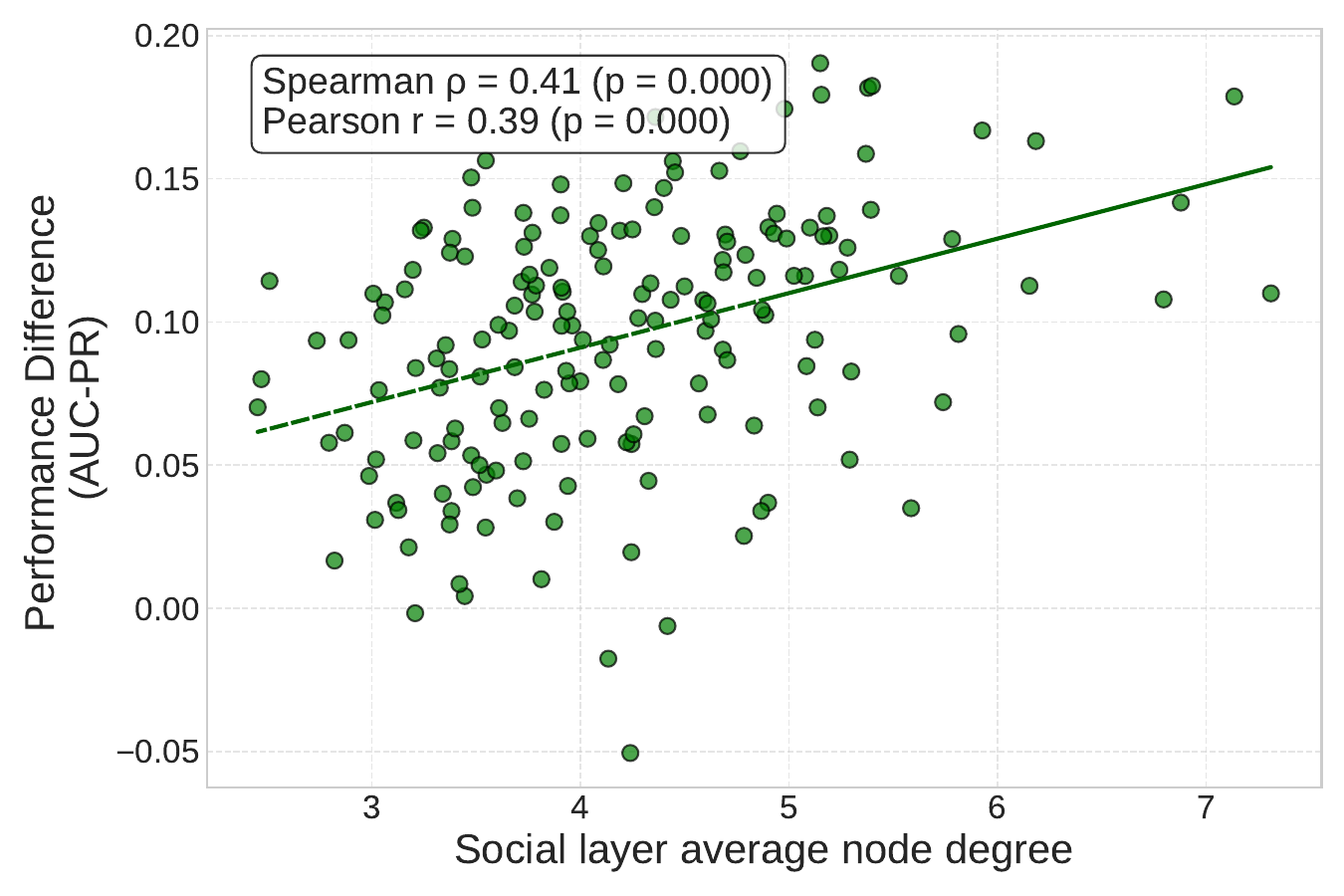}
    \caption{\small  Correlation (Pearson and Spearman) between prediction improvement and average node degree social layer.}
\end{subfigure}
\centering
\begin{subfigure}[t]{0.24\textwidth}
    \includegraphics[width=\textwidth]{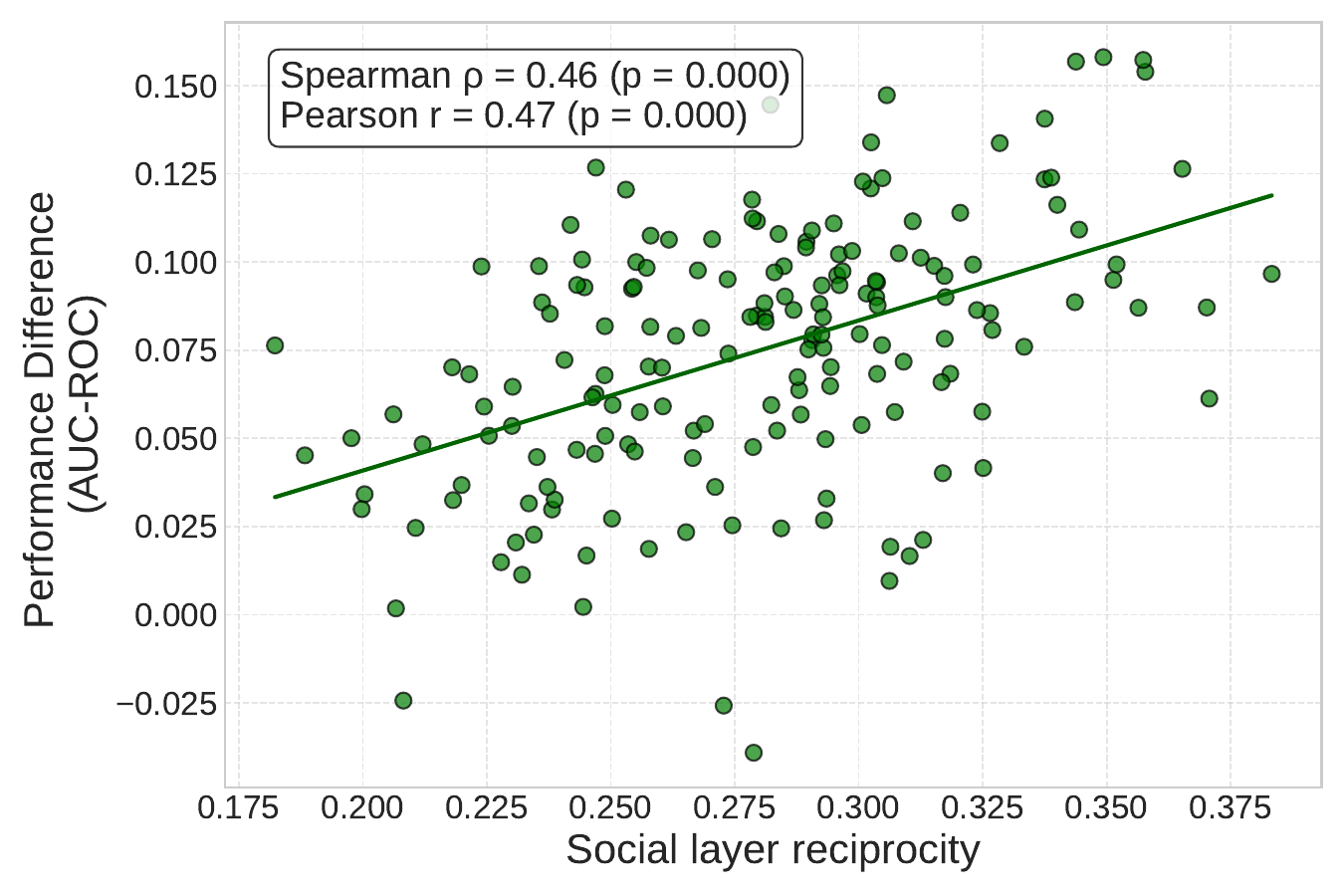}
    \caption{\small Correlation (Pearson and Spearman) between prediction improvement and reciprocity in the social layer.}
\end{subfigure}
\hfill
\begin{subfigure}[t]{0.24\textwidth}
    \includegraphics[width=\textwidth]{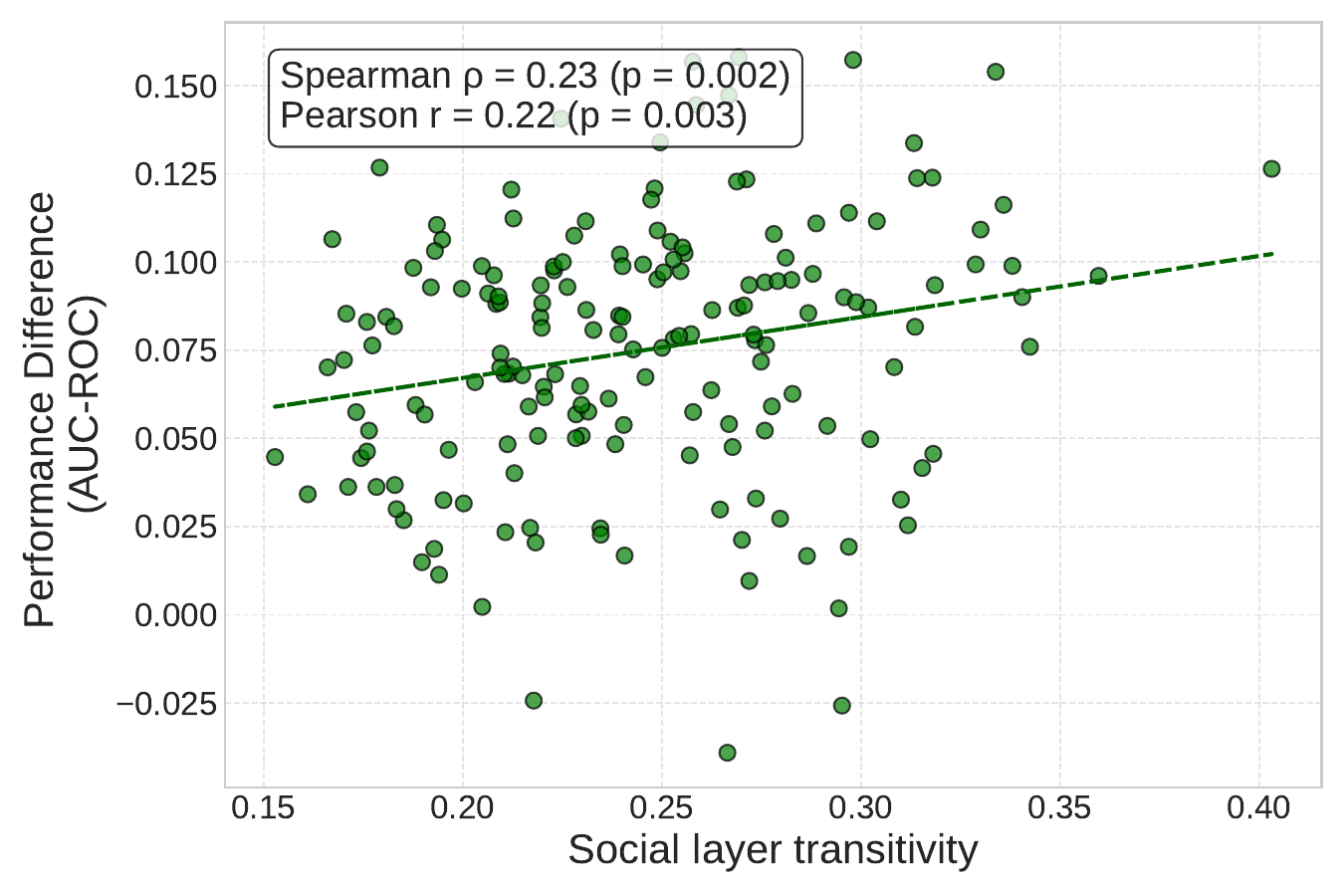}
    \caption{\small  Correlation (Pearson and Spearman) between prediction improvement and transitivity in the social layer.}
\end{subfigure}
\hfill
\begin{subfigure}[t]{0.24\textwidth}
    \includegraphics[width=\textwidth]{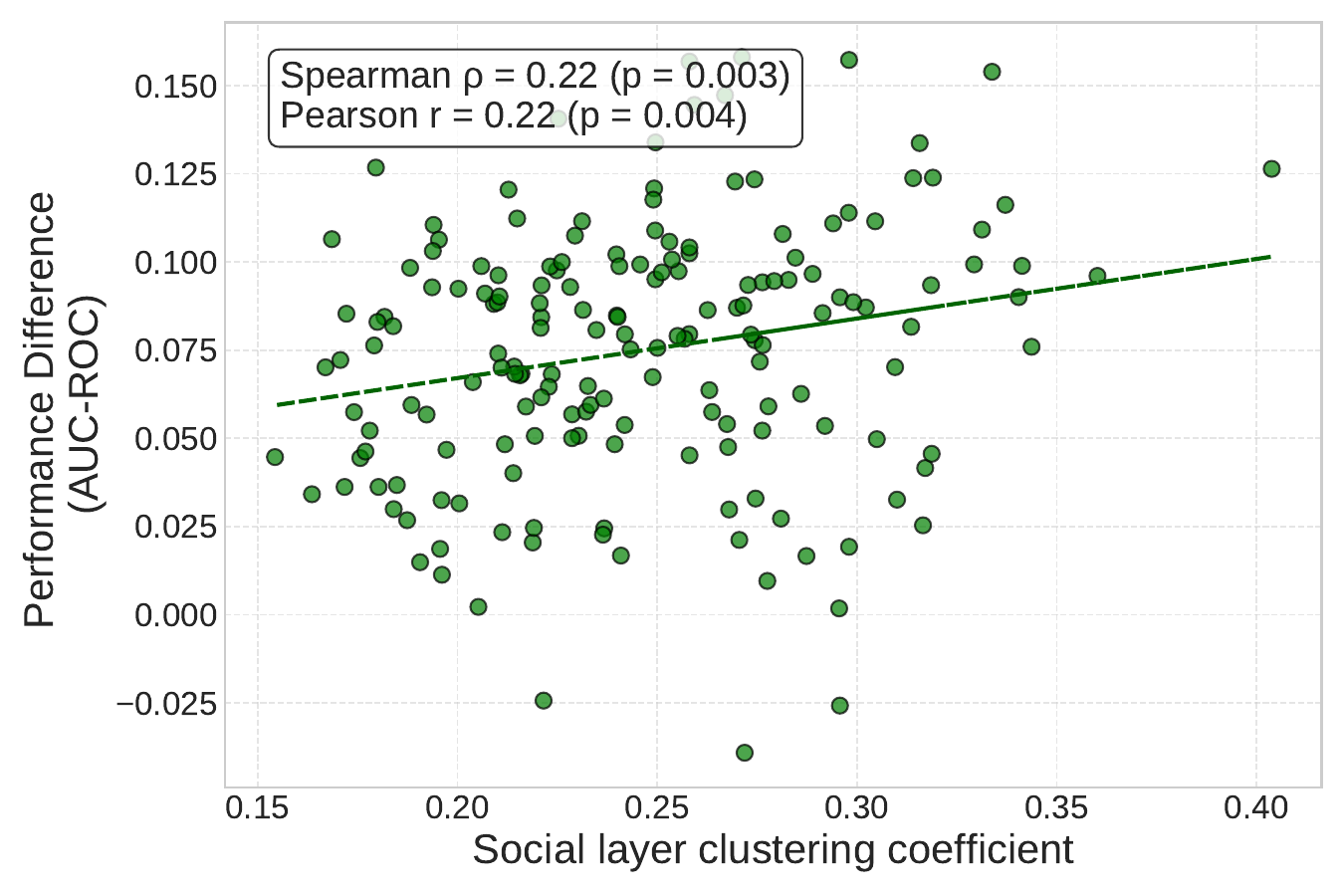}
    \caption{\small  Correlation (Pearson and Spearman) between prediction improvement and clustering in the social layer.}
\end{subfigure}
\hfill
\begin{subfigure}[t]{0.24\textwidth}
    \includegraphics[width=\textwidth]{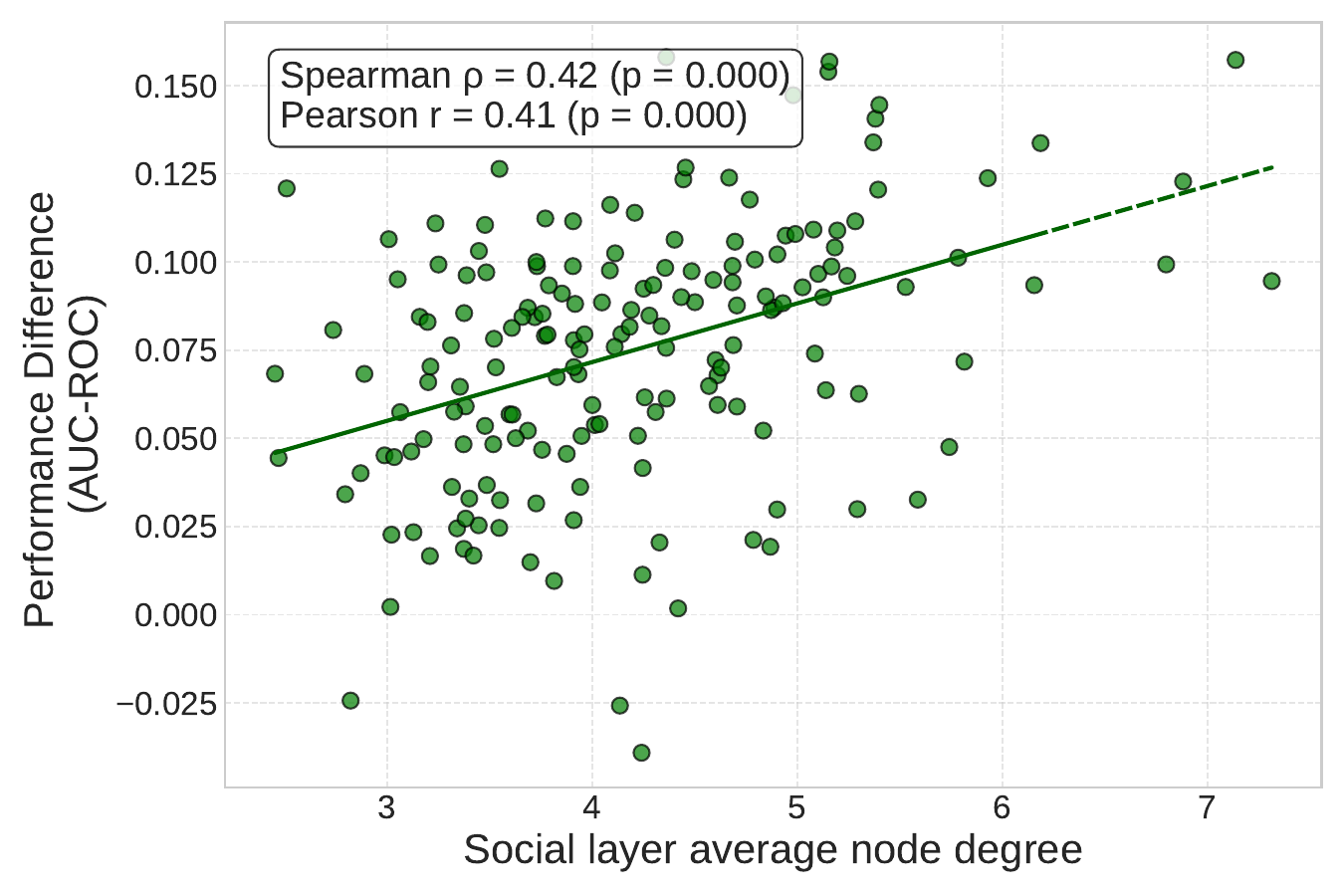}
    \caption{\small  Correlation (Pearson and Spearman) between prediction improvement and average node degree social layer.}
\end{subfigure}

\caption{\textbf{Social layer exchange quantification.} The first row shows the average increase in prediction performance across ten-fold cross-validation when modeling only dependence and independence, versus when extending the model to also capture interdependence, in accordance with Social Exchange Theory, across all 176 village networks. The second row reports both linear (Pearson) and monotonic (Spearman) correlation coefficients between the observed performance increase and key network statistics, reciprocity, transitivity, clustering coefficient, and average node degree, computed over the social layer.}
\label{fig:social_cors}
\end{figure*}

\begin{figure*}
    \centering
\begin{subfigure}[t]{0.24\textwidth}
    \includegraphics[width=\textwidth]{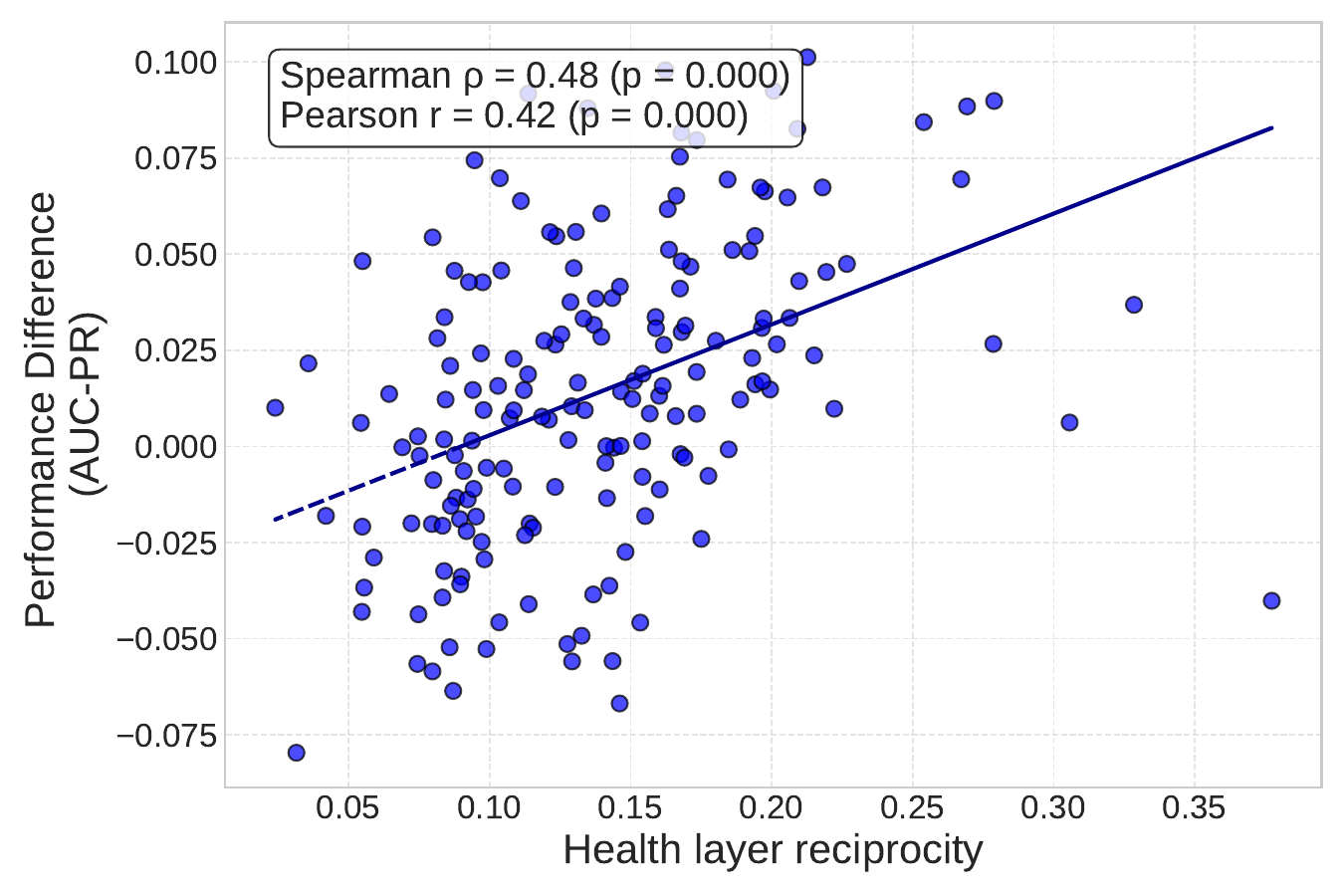}
    \caption{\small Correlation (Pearson and Spearman) between prediction improvement and reciprocity in the health layer.}
\end{subfigure}
\hfill
\begin{subfigure}[t]{0.24\textwidth}
    \includegraphics[width=\textwidth]{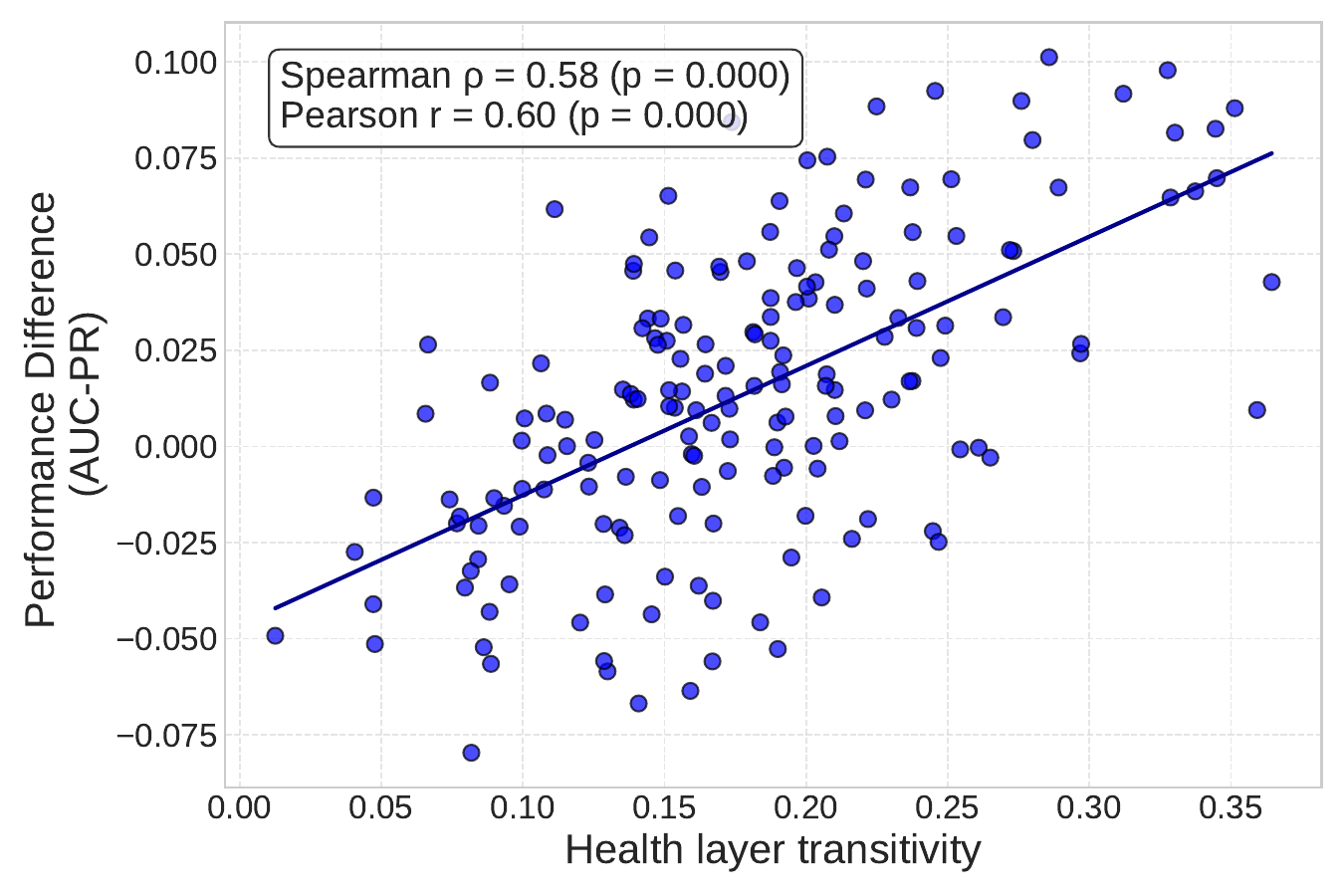}
    \caption{\small  Correlation (Pearson and Spearman) between prediction improvement and transitivity in the health layer.}
\end{subfigure}
\hfill
\begin{subfigure}[t]{0.24\textwidth}
    \includegraphics[width=\textwidth]{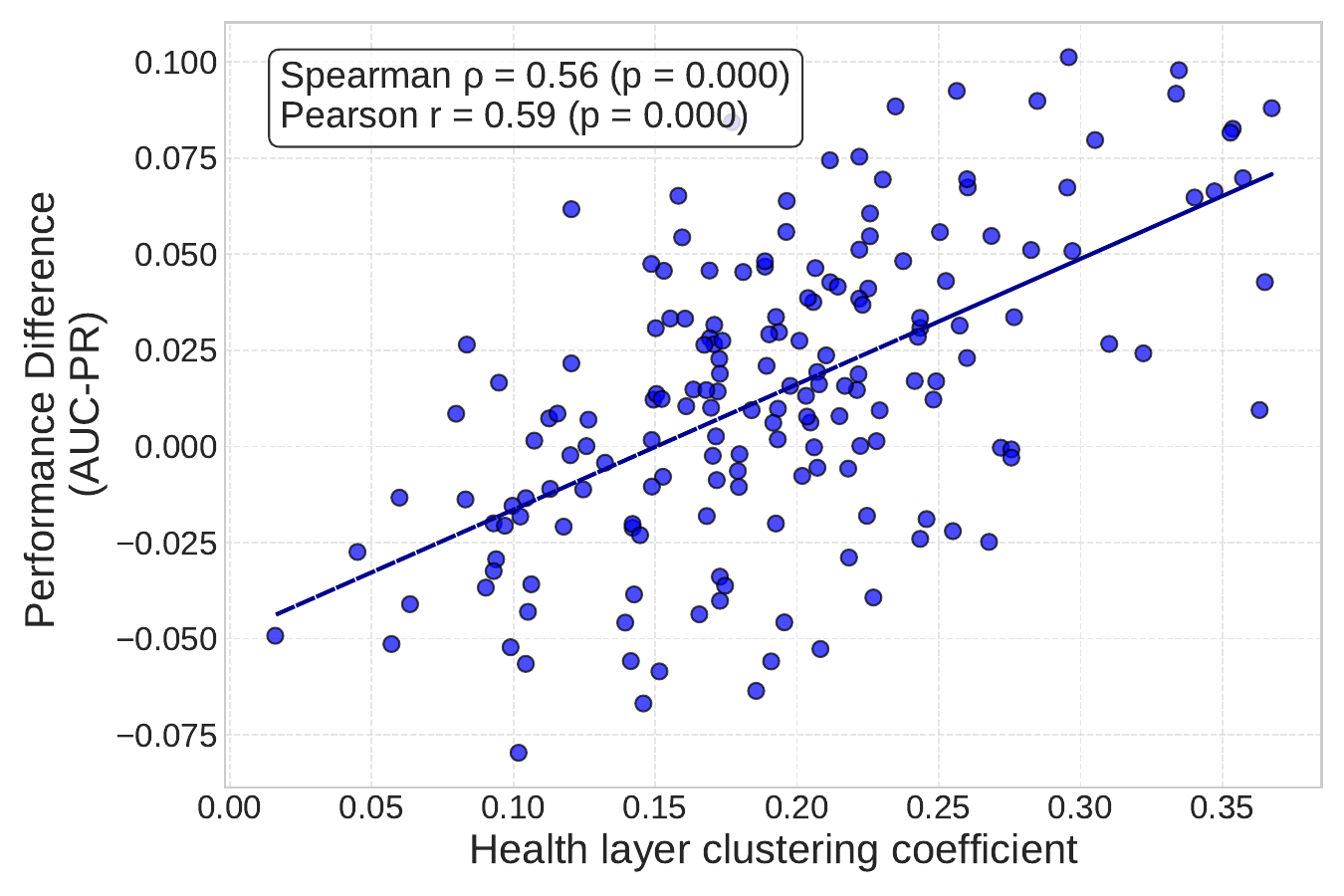}
    \caption{\small  Correlation (Pearson and Spearman) between prediction improvement and clustering in the health layer.}
\end{subfigure}
\hfill
\begin{subfigure}[t]{0.24\textwidth}
    \includegraphics[width=\textwidth]{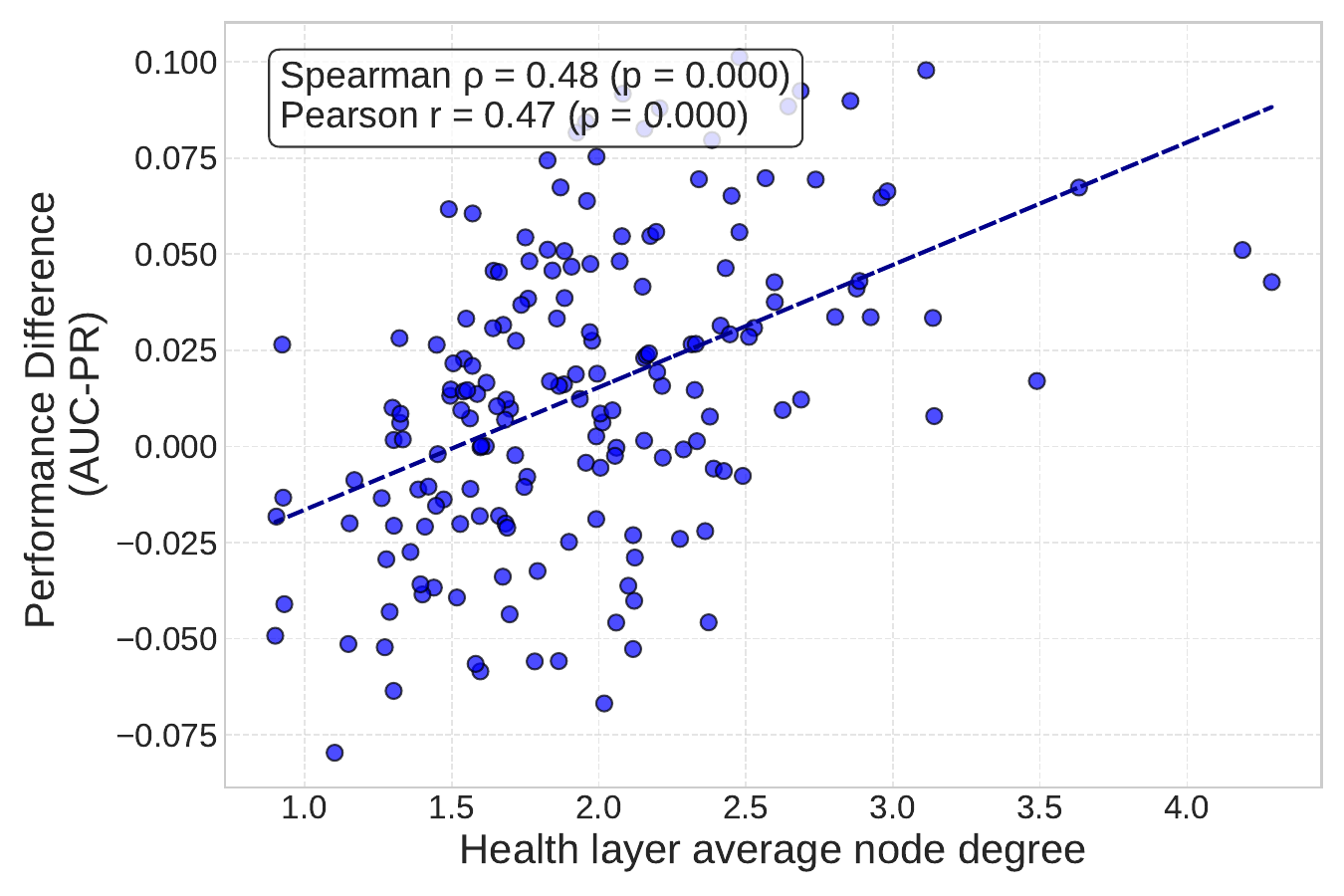}
    \caption{\small  Correlation (Pearson and Spearman) between prediction improvement and average node degree in the health layer.}
\end{subfigure}
    \centering
\begin{subfigure}[t]{0.24\textwidth}
    \includegraphics[width=\textwidth]{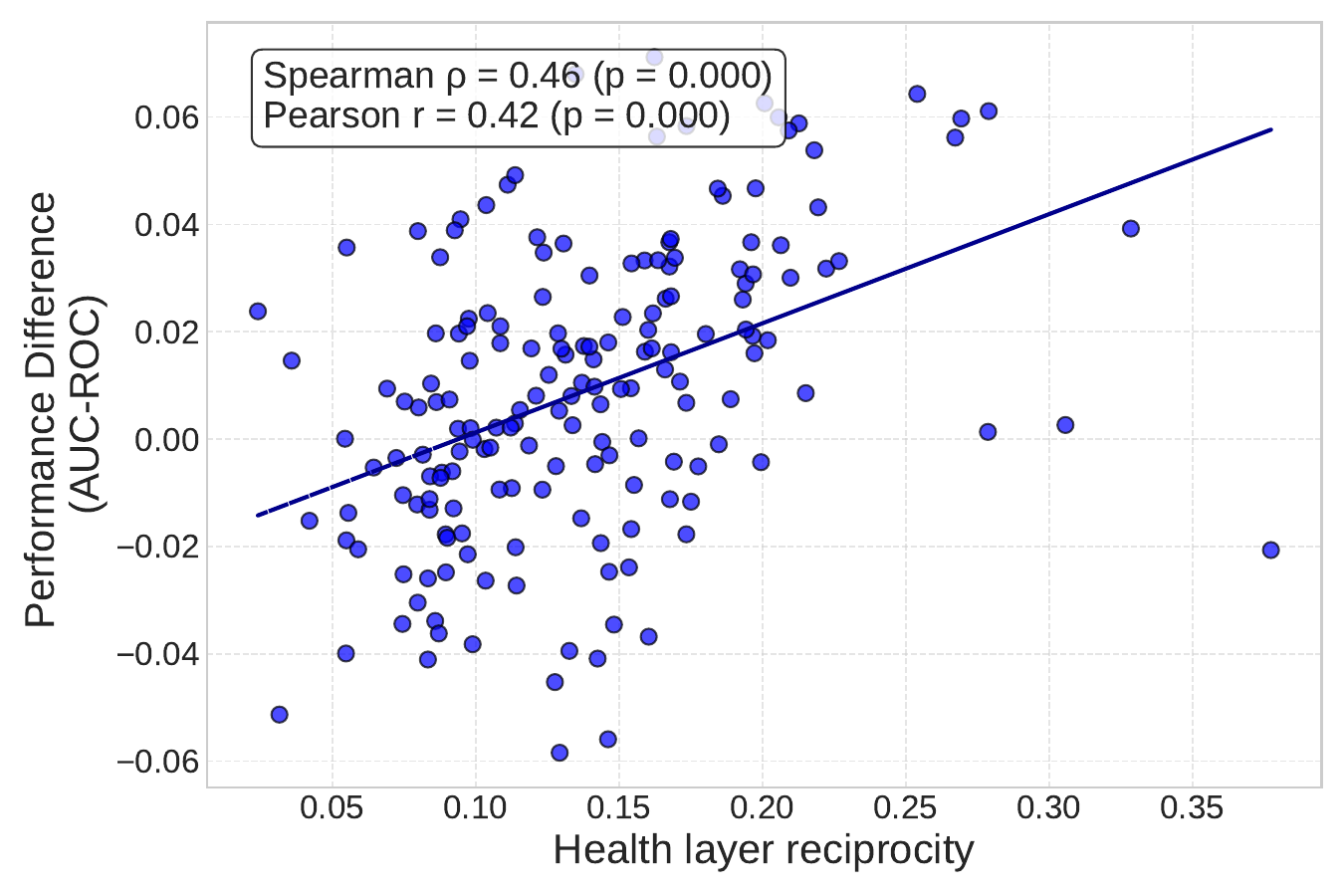}
    \caption{\small Correlation (Pearson and Spearman) between prediction improvement and reciprocity in the health layer.}
\end{subfigure}
\hfill
\begin{subfigure}[t]{0.24\textwidth}
    \includegraphics[width=\textwidth]{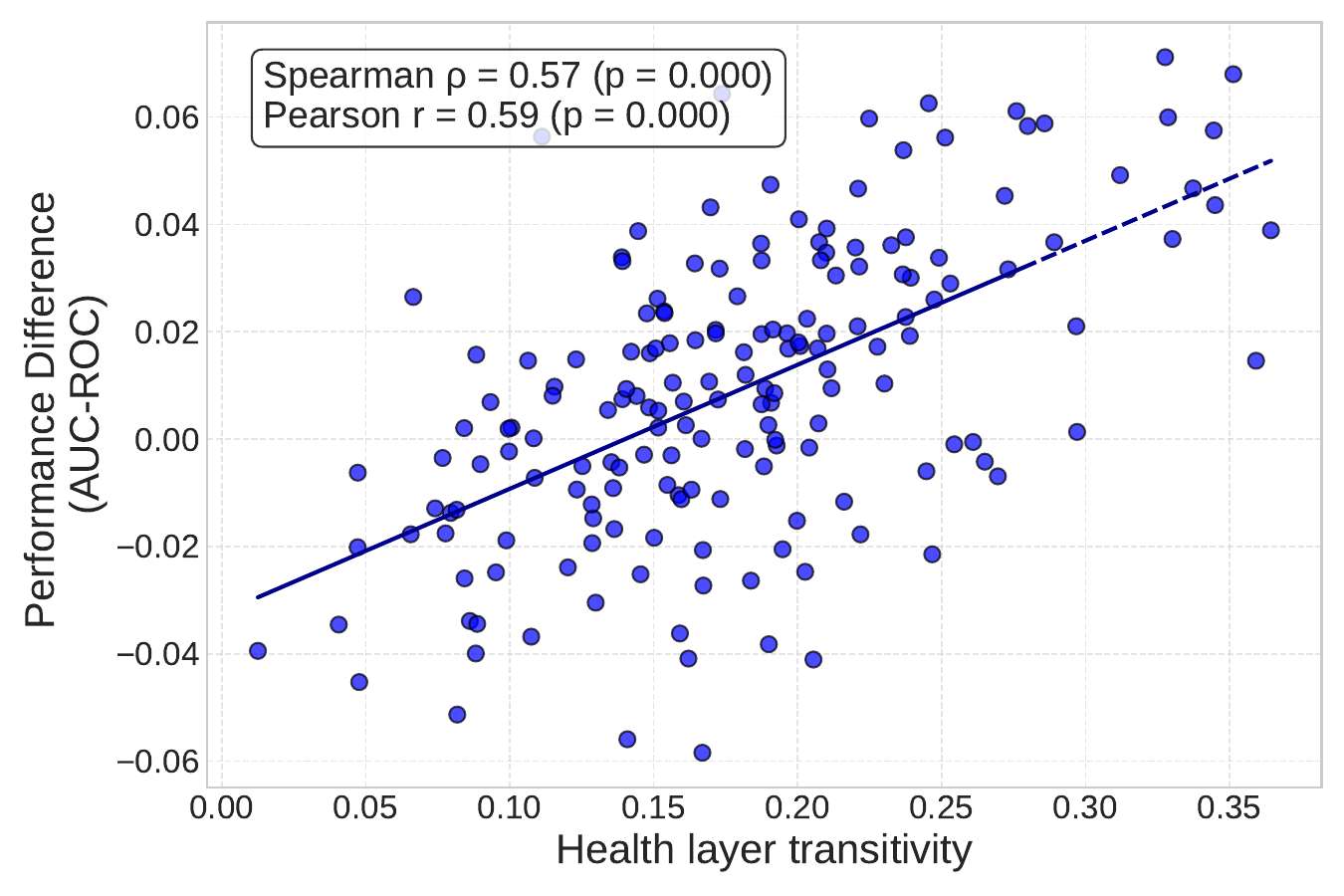}
    \caption{\small  Correlation (Pearson and Spearman) between prediction improvement and transitivity in the health layer.}
\end{subfigure}
\hfill
\begin{subfigure}[t]{0.24\textwidth}
    \includegraphics[width=\textwidth]{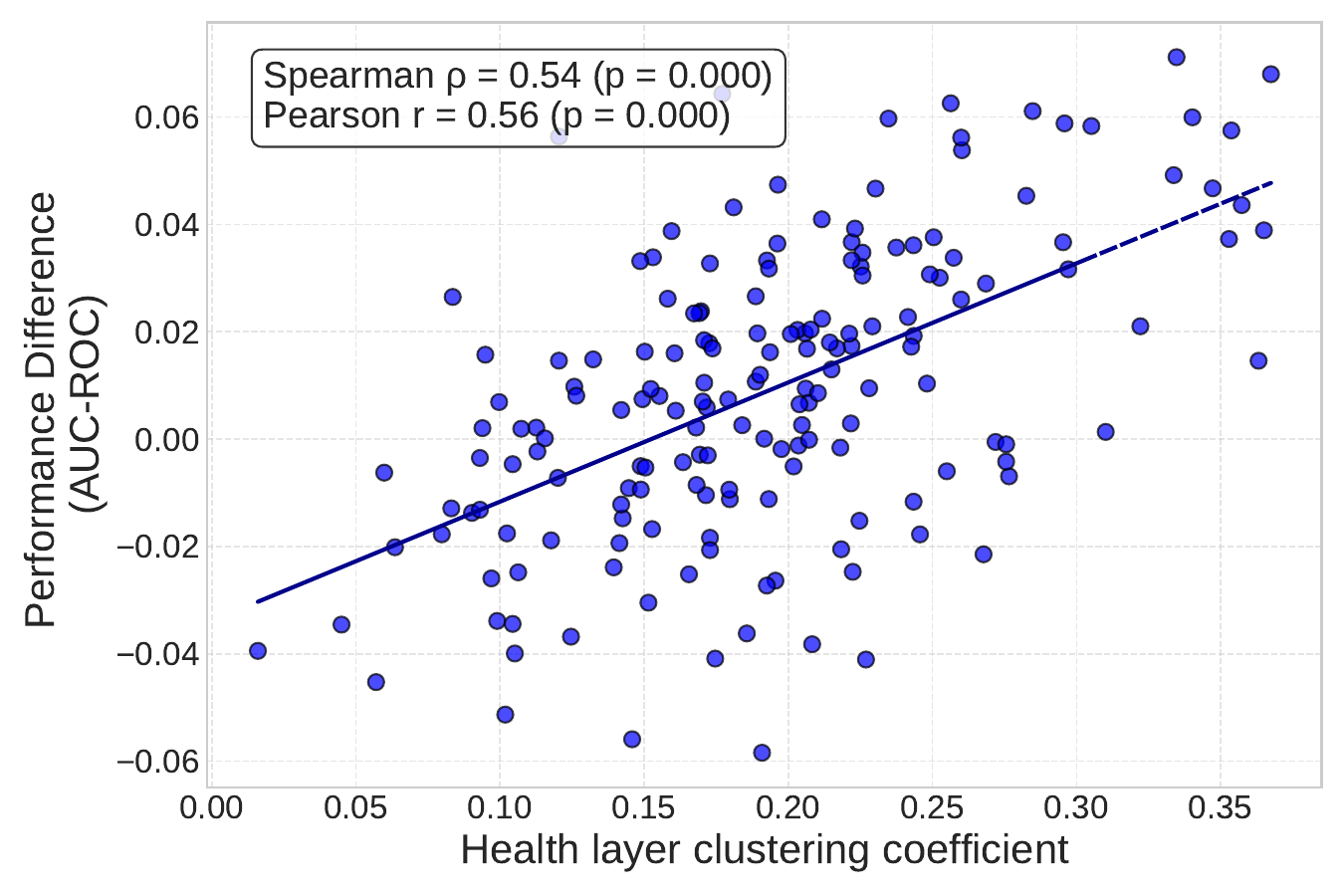}
    \caption{\small  Correlation (Pearson and Spearman) between prediction improvement and clustering in the health layer.}
\end{subfigure}
\hfill
\begin{subfigure}[t]{0.24\textwidth}
    \includegraphics[width=\textwidth]{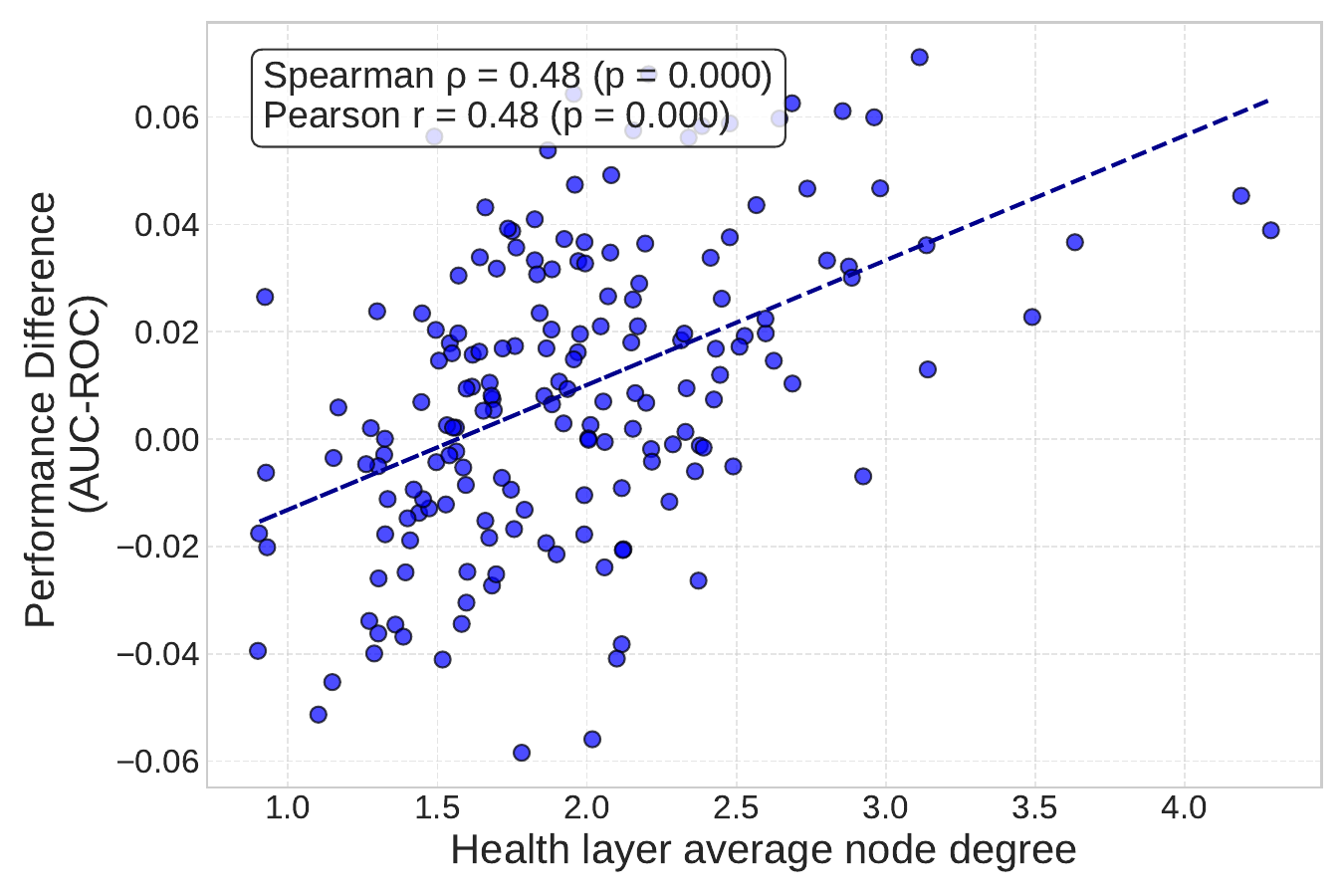}
    \caption{\small  Correlation (Pearson and Spearman) between prediction improvement and average node degree in the health layer.}
\end{subfigure}
\caption{\textbf{Health layer exchange quantification.} The first row shows the average increase in prediction performance across ten-fold cross-validation when modeling only dependence and independence, versus when extending the model to also capture interdependence, in accordance with Social Exchange Theory, across all 176 village networks. The second row reports both linear (Pearson) and monotonic (Spearman) correlation coefficients between the observed performance increase and key network statistics, reciprocity, transitivity, clustering coefficient, and average node degree, computed over the health layer.}
\label{fig:heal_cors}
\end{figure*}

\begin{figure*}
\centering
\begin{subfigure}[t]{0.24\textwidth}
    \includegraphics[width=\textwidth]{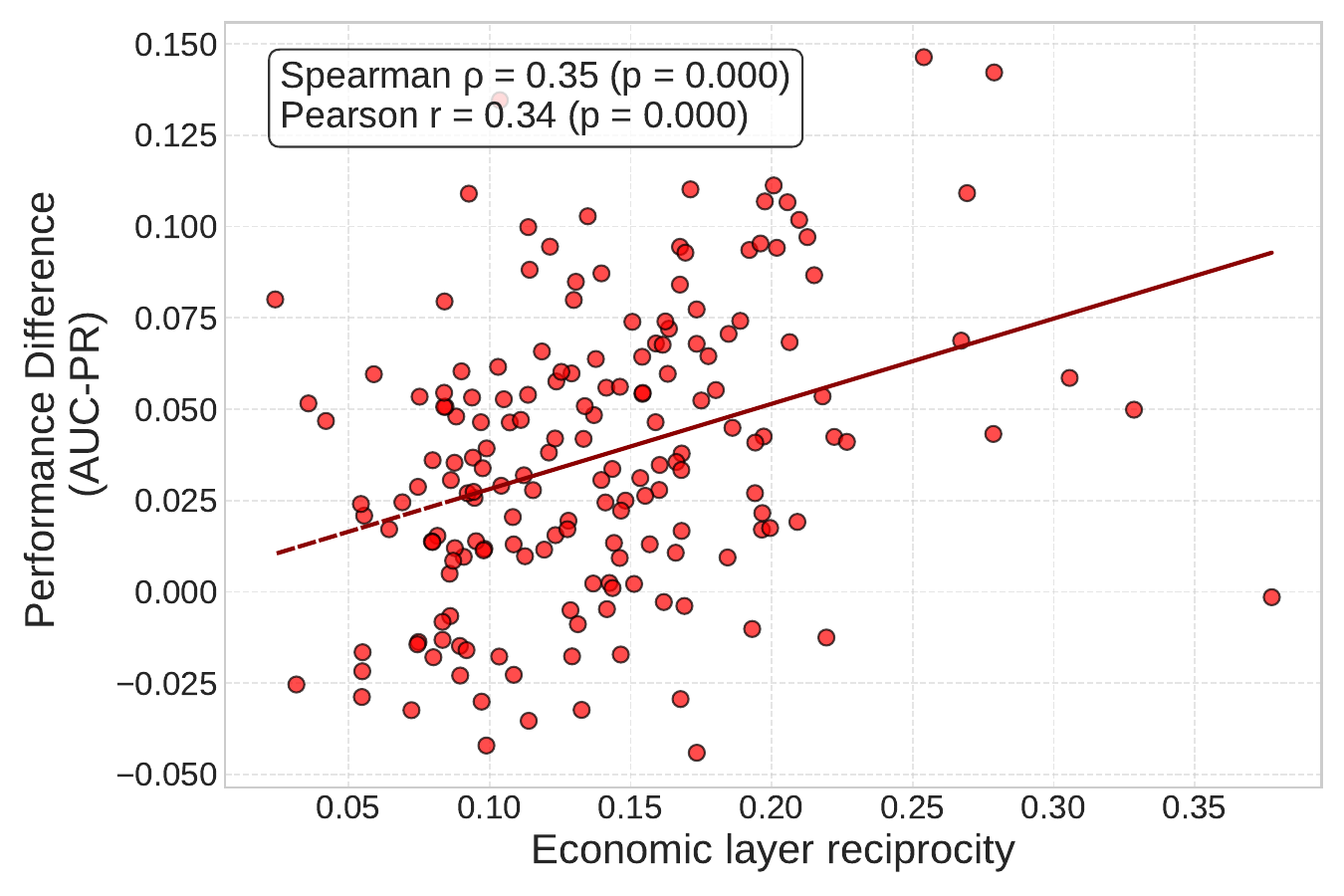}
    \caption{\small  Correlation (Pearson and Spearman) between prediction improvement and reciprocity in the economic layer.}
\end{subfigure}
\hfill
\begin{subfigure}[t]{0.24\textwidth}
    \includegraphics[width=\textwidth]{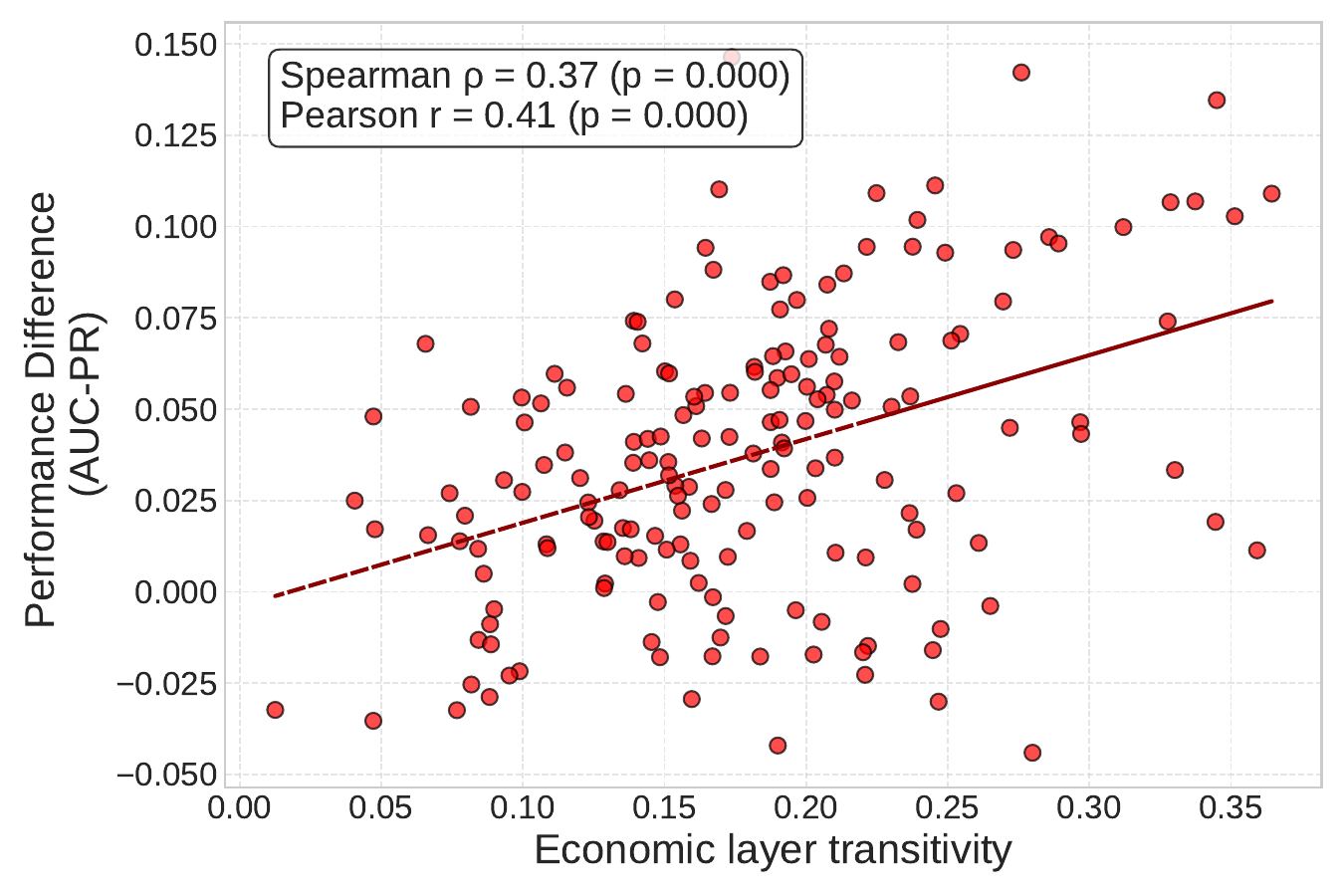}
    \caption{\small  Correlation (Pearson and Spearman) between prediction improvement and transitivity in the economic layer.}
\end{subfigure}
\hfill
\begin{subfigure}[t]{0.24\textwidth}
    \includegraphics[width=\textwidth]{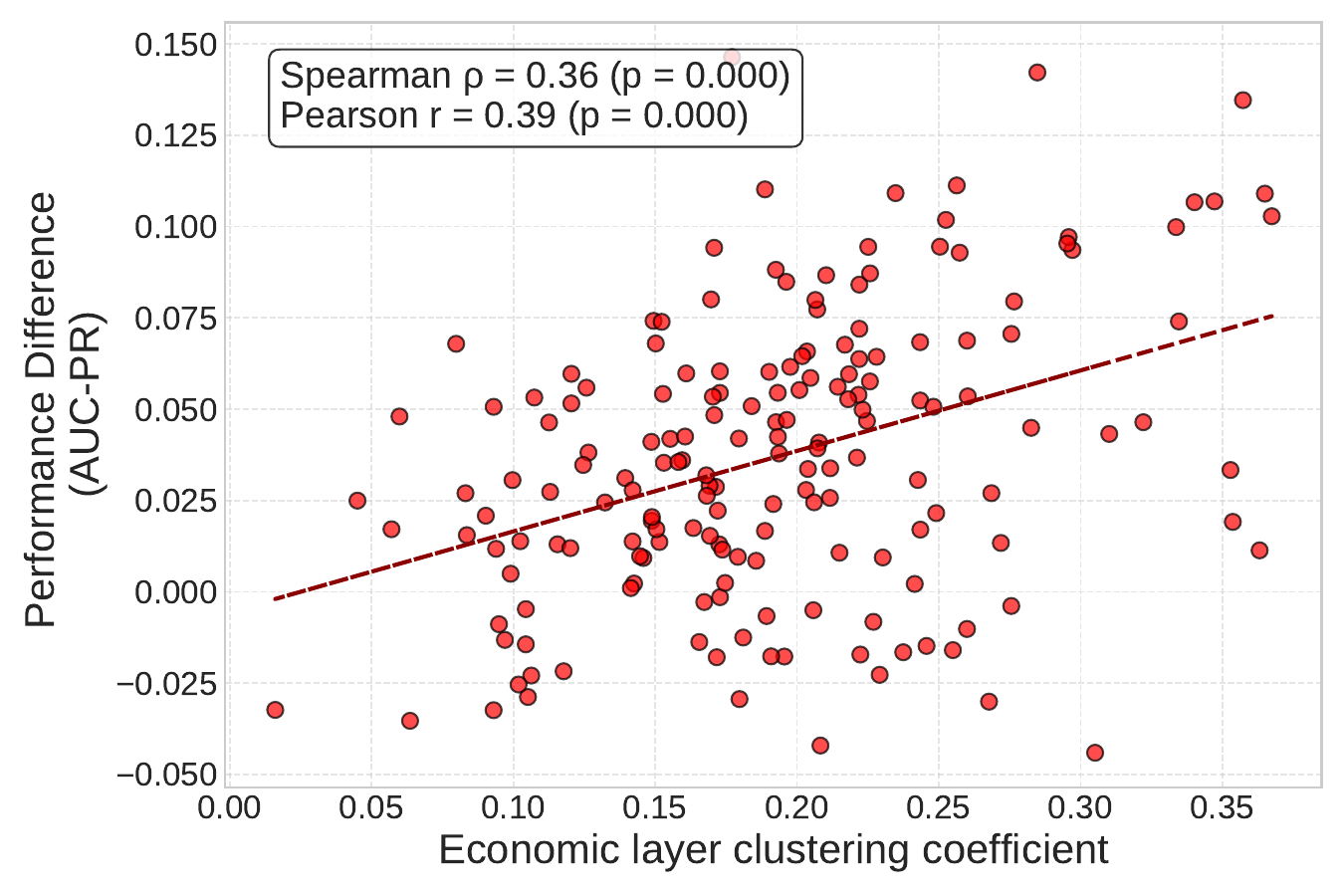}
    \caption{\small  Correlation (Pearson and Spearman) between prediction improvement and clustering in the economic layer.}
\end{subfigure}
\hfill
\begin{subfigure}[t]{0.24\textwidth}
    \includegraphics[width=\textwidth]{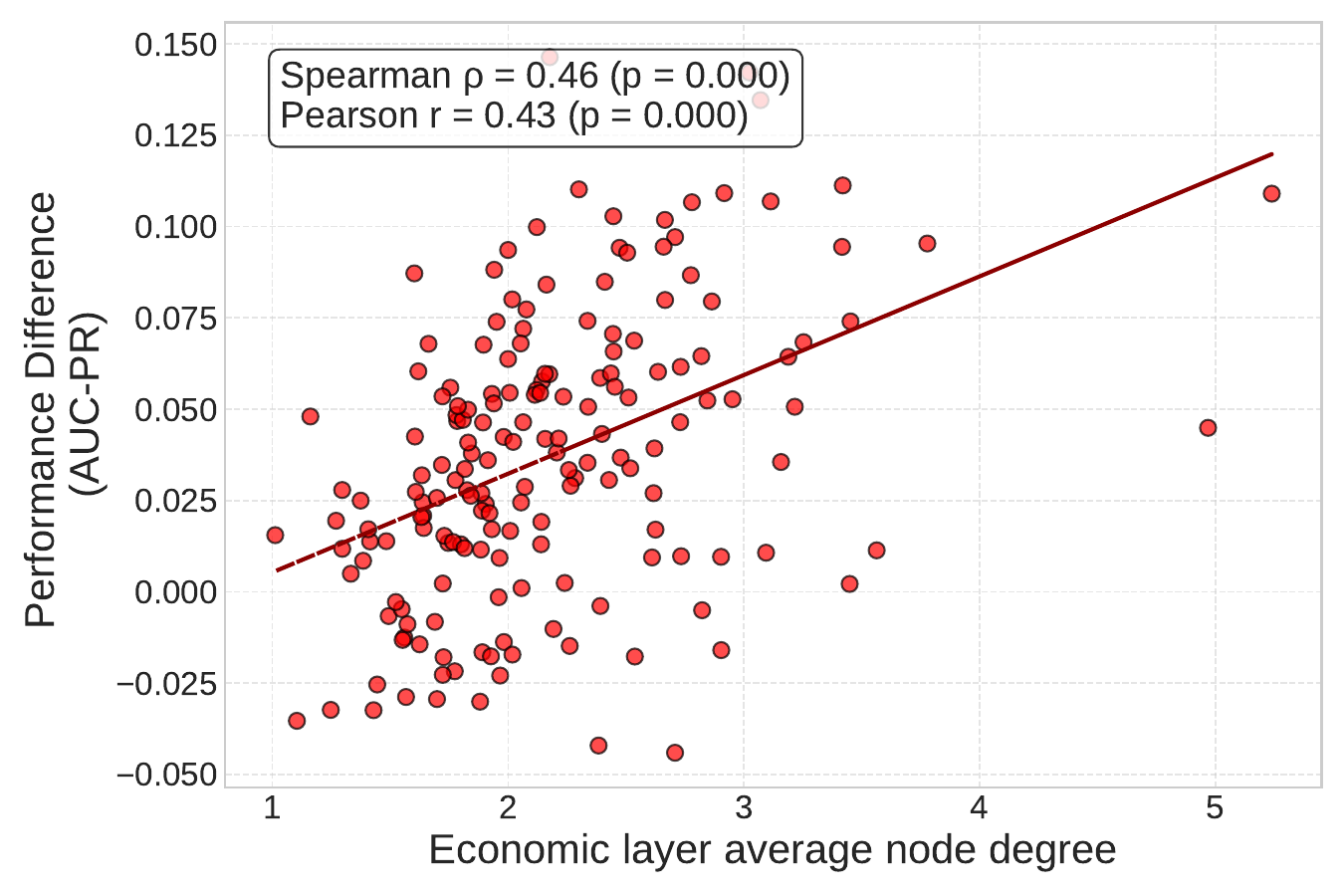}
    \caption{\small  Correlation (Pearson and Spearman) between prediction improvement and average node degree in the economic layer.}
\end{subfigure}
\centering
\begin{subfigure}[t]{0.24\textwidth}
    \includegraphics[width=\textwidth]{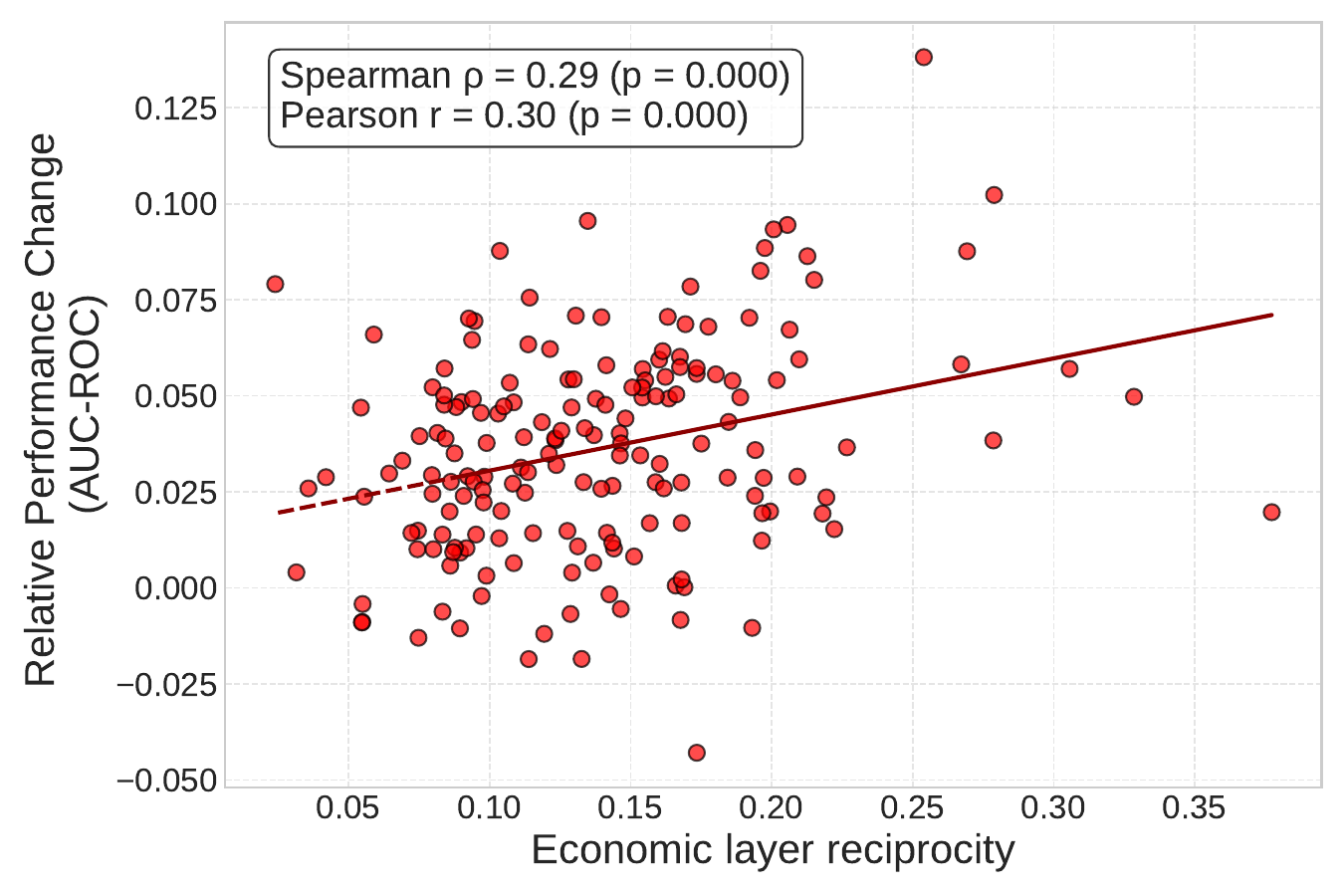}
    \caption{\small  Correlation (Pearson and Spearman) between prediction improvement and reciprocity in the economic layer.}
\end{subfigure}
\hfill
\begin{subfigure}[t]{0.24\textwidth}
    \includegraphics[width=\textwidth]{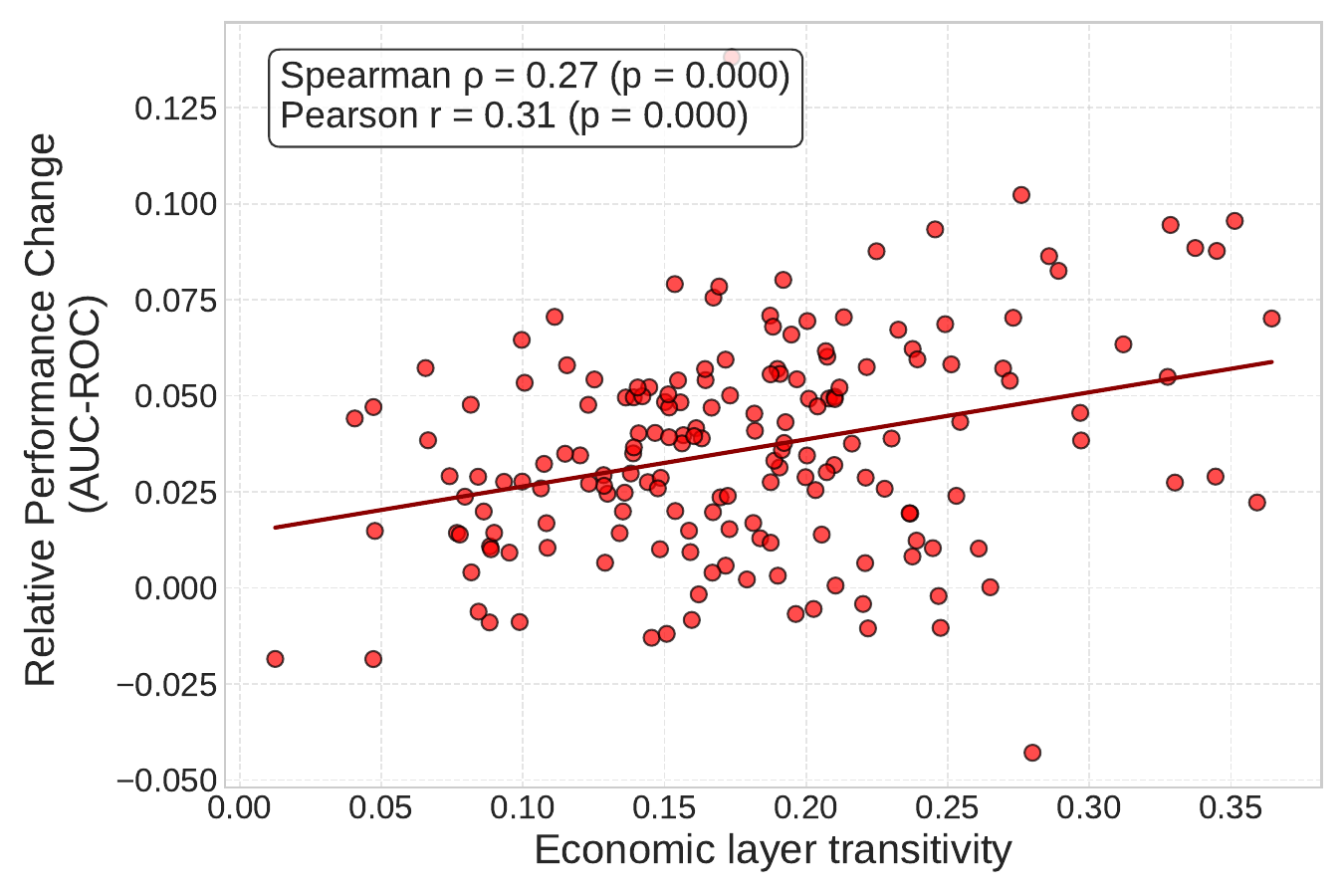}
    \caption{\small  Correlation (Pearson and Spearman) between prediction improvement and transitivity in the economic layer.}
\end{subfigure}
\hfill
\begin{subfigure}[t]{0.24\textwidth}
    \includegraphics[width=\textwidth]{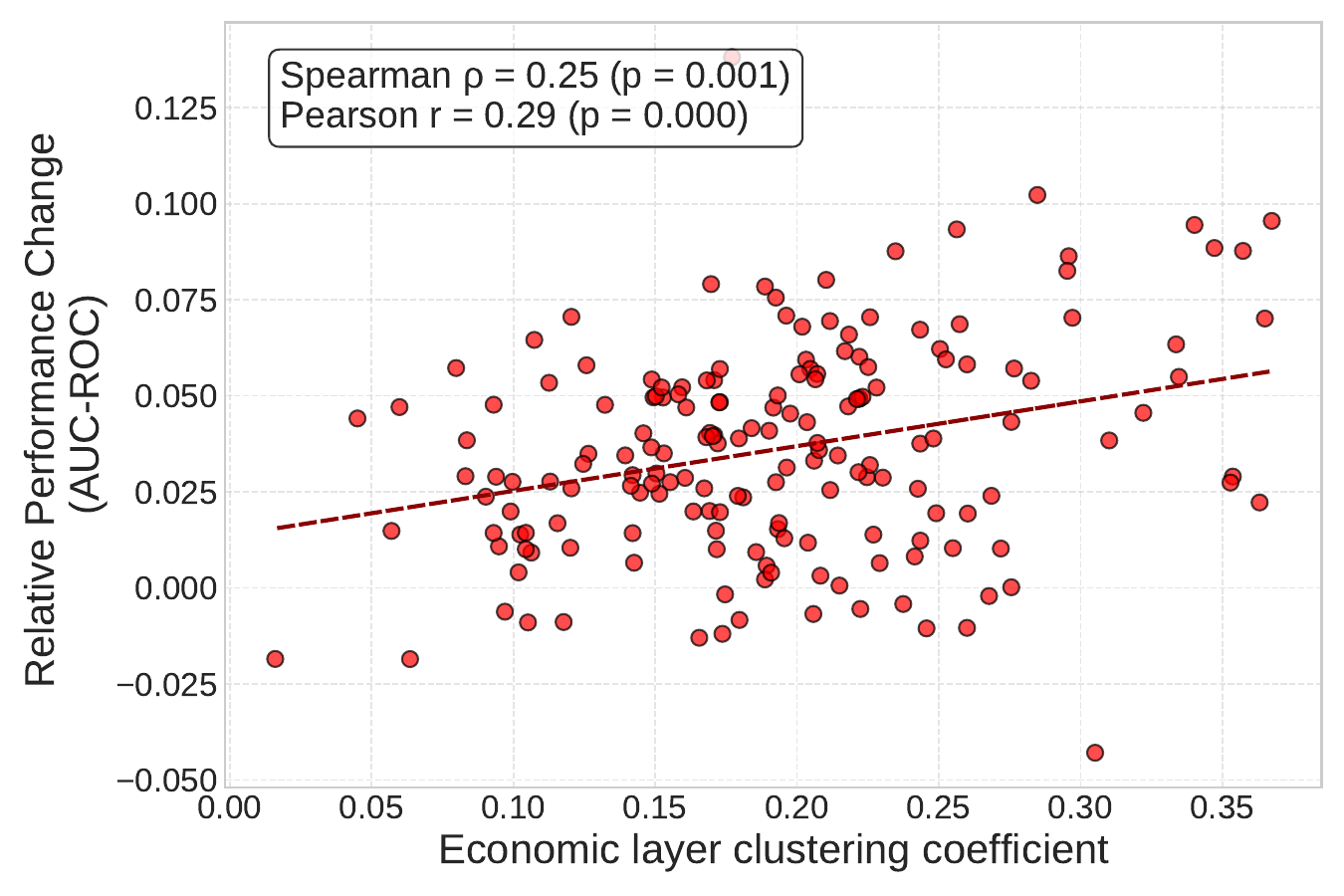}
    \caption{\small  Correlation (Pearson and Spearman) between prediction improvement and clustering in the economic layer.}
\end{subfigure}
\hfill
\begin{subfigure}[t]{0.24\textwidth}
    \includegraphics[width=\textwidth]{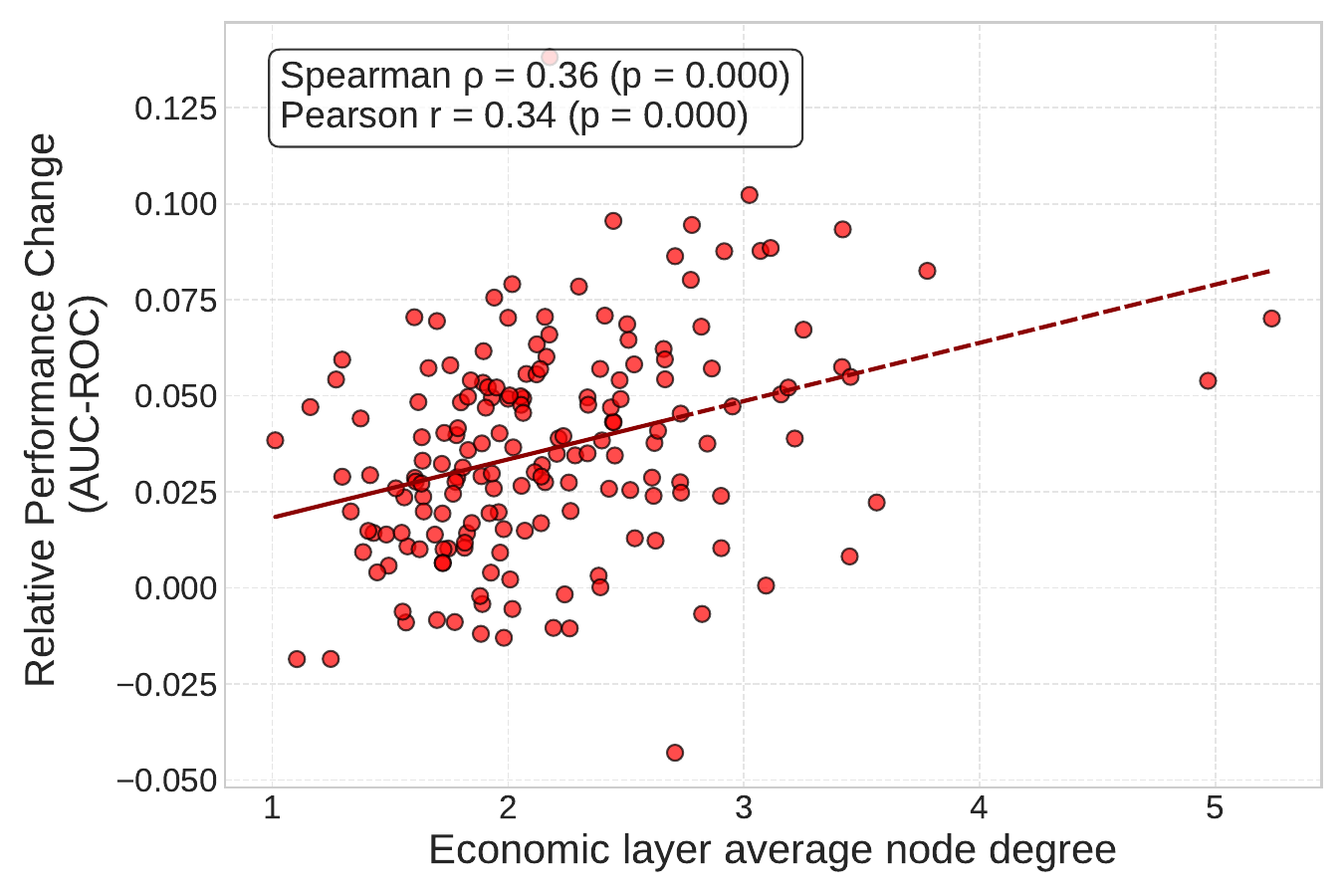}
    \caption{\small  Correlation (Pearson and Spearman) between prediction improvement and average node degree in the economic layer.}
\end{subfigure}

\caption{\textbf{Economic layer exchange quantification.} The first row shows the average increase in prediction performance across ten-fold cross-validation when modeling only dependence and independence, versus when extending the model to also capture interdependence, in accordance with Social Exchange Theory, across all 176 village networks. The second row reports both linear (Pearson) and monotonic (Spearman) correlation coefficients between the observed performance increase and key network statistics, reciprocity, transitivity, clustering coefficient, and average node degree, computed over the economic layer.}
\label{fig:econ_cors}
\end{figure*}

Structural features of networks significantly influence how well interdependence can predict behavior with health and economic layers benefiting more from local structures than the social layer. To investigate further the effect of structural features, we compute Spearman correlation coefficients ($\rho$) between the prediction scores  difference and key network statistics, using 1,000 bootstrap samples across the social, health, and economic layers (SI Figure~\ref{fig:boot}). We report results for both PR-AUC raw score difference (top row) and ROC-AUC raw score difference (bottom row). \textbf{PR-AUC Metric:} Reciprocity (Panel a) shows a significant positive correlation with performance increase across all layers, while transitivity and clustering coefficient (Panels b and c) are significantly associated with improvement only in the health and economic layers. Average node degree (Panel d) is positively correlated with prediction improvement across all layers. \textbf{ROC-AUC Metric:} Similar patterns are observed: reciprocity (Panel e) is significantly correlated across all layers, while transitivity and clustering (Panels f and g) show significant correlations mostly in the health and economic layers. Average node degree (Panel h) remains a strong positive predictor across all domains. Full bootstrapped correlation values and confidence intervals follow. Furthermore non-bootstrapped (observed) Pearson and Spearman correlation values are provided separately in SI Figures~\ref{fig:social_cors}–\ref{fig:econ_cors}.

We provide bootstrapped detailed correlation values based on the raw performance differences between the full and bias-only models, measuring how predictive improvements from modeling interdependence relate to various network statistics, using PR AUC and ROC AUC scores. \textbf{PR AUC Metric: } We observe significant positive correlations between prediction improvement and layer reciprocity (Panel a) across all three layers: 
$\tilde{\rho}_{\text{social}} = 0.400$, 95\% CI $[0.271,\ 0.524]$; 
$\tilde{\rho}_{\text{health}} = 0.481$, 95\% CI $[0.351,\ 0.584]$; and 
$\tilde{\rho}_{\text{economic}} = 0.449$, 95\% CI $[0.310,\ 0.561]$. Transitivity (Panel b) is significantly correlated with performance only in the health and economic layers: 
$\tilde{\rho}_{\text{social}} = 0.119$, 95\% CI $[-0.024,\ 0.269]$; 
$\tilde{\rho}_{\text{health}} = 0.576$, 95\% CI $[0.469,\ 0.677]$; 
$\tilde{\rho}_{\text{economic}} = 0.429$, 95\% CI $[0.300,\ 0.553]$. Similarly, clustering coefficient (Panel c) shows significant associations in the health and economic layers only: 
$\tilde{\rho}_{\text{social}} = 0.114$, 95\% CI $[-0.031,\ 0.262]$; 
$\tilde{\rho}_{\text{health}} = 0.551$, 95\% CI $[0.438,\ 0.661]$; 
$\tilde{\rho}_{\text{economic}} = 0.422$, 95\% CI $[0.292,\ 0.551]$. Average node degree (Panel d) is positively and significantly associated with performance across all layers: 
$\tilde{\rho}_{\text{social}} = 0.404$, 95\% CI $[0.288,\ 0.526]$; 
$\tilde{\rho}_{\text{health}} = 0.479$, 95\% CI $[0.362,\ 0.581]$; 
$\tilde{\rho}_{\text{economic}} = 0.453$, 95\% CI $[0.334,\ 0.578]$. \textbf{ROC AUC Metric: } We find similar patterns when considering ROC AUC improvement. Reciprocity (Panel e) remains significantly correlated across all layers: 
$\tilde{\rho}_{\text{social}} = 0.454$, 95\% CI $[0.329,\ 0.573]$; 
$\tilde{\rho}_{\text{health}} = 0.464$, 95\% CI $[0.330,\ 0.575]$; 
$\tilde{\rho}_{\text{economic}} = 0.359$, 95\% CI $[0.223,\ 0.482]$. Transitivity (Panel f) is significant in the health and economic layers, and less in the social layer: 
$\tilde{\rho}_{\text{social}} = 0.225$, 95\% CI $[0.080,\ 0.371]$; 
$\tilde{\rho}_{\text{health}} = 0.564$, 95\% CI $[0.446,\ 0.674]$; 
$\tilde{\rho}_{\text{economic}} = 0.377$, 95\% CI $[0.236,\ 0.506]$. Clustering coefficient (Panel g) shows the same layer-specific pattern: 
$\tilde{\rho}_{\text{social}} = 0.220$, 95\% CI $[-0.007,\ 0.366]$; 
$\tilde{\rho}_{\text{health}} = 0.533$, 95\% CI $[0.414,\ 0.650]$; 
$\tilde{\rho}_{\text{economic}} = 0.356$, 95\% CI $[0.217,\ 0.495]$. Finally, average node degree (Panel h) is significantly associated with predictive improvement across all layers: 
$\tilde{\rho}_{\text{social}} = 0.418$, 95\% CI $[0.303,\ 0.548]$; 
$\tilde{\rho}_{\text{health}} = 0.484$, 95\% CI $[0.366,\ 0.592]$; 
$\tilde{\rho}_{\text{economic}} = 0.354$, 95\% CI $[0.216,\ 0.489]$. To assess whether the observed correlation patterns differ significantly across layers, we compare the bootstrap distributions using a one-sided Mann–Whitney U test for each metric. All pairwise differences were found to be statistically significant ($p < 0.0001$), confirming that the ordering of median correlations across layers (e.g., $\tilde{\rho}_{\text{health}} > \tilde{\rho}_{\text{economic}} > \tilde{\rho}_{\text{social}}$) reflects robust differences rather than sampling variability. Non-bootstrapped correlation values are provided in SI Figures \ref{fig:social_cors}, \ref{fig:heal_cors}, \ref{fig:econ_cors} including both linear (Pearson) and monotonic (Spearman) correlation coefficients between the observed performance increase (both in terms PR AUC (top-row) and ROC AUC (bottom-row)) and the various network statistics, for the social, health and economic layers, respectively.

\subsection*{Non-bootstrapped correlation values between layer mean activity and performance increase when modeling interdependence}

\begin{figure*}[h!]
\centering
\begin{subfigure}[t]{0.24\textwidth}
    \includegraphics[width=\textwidth]{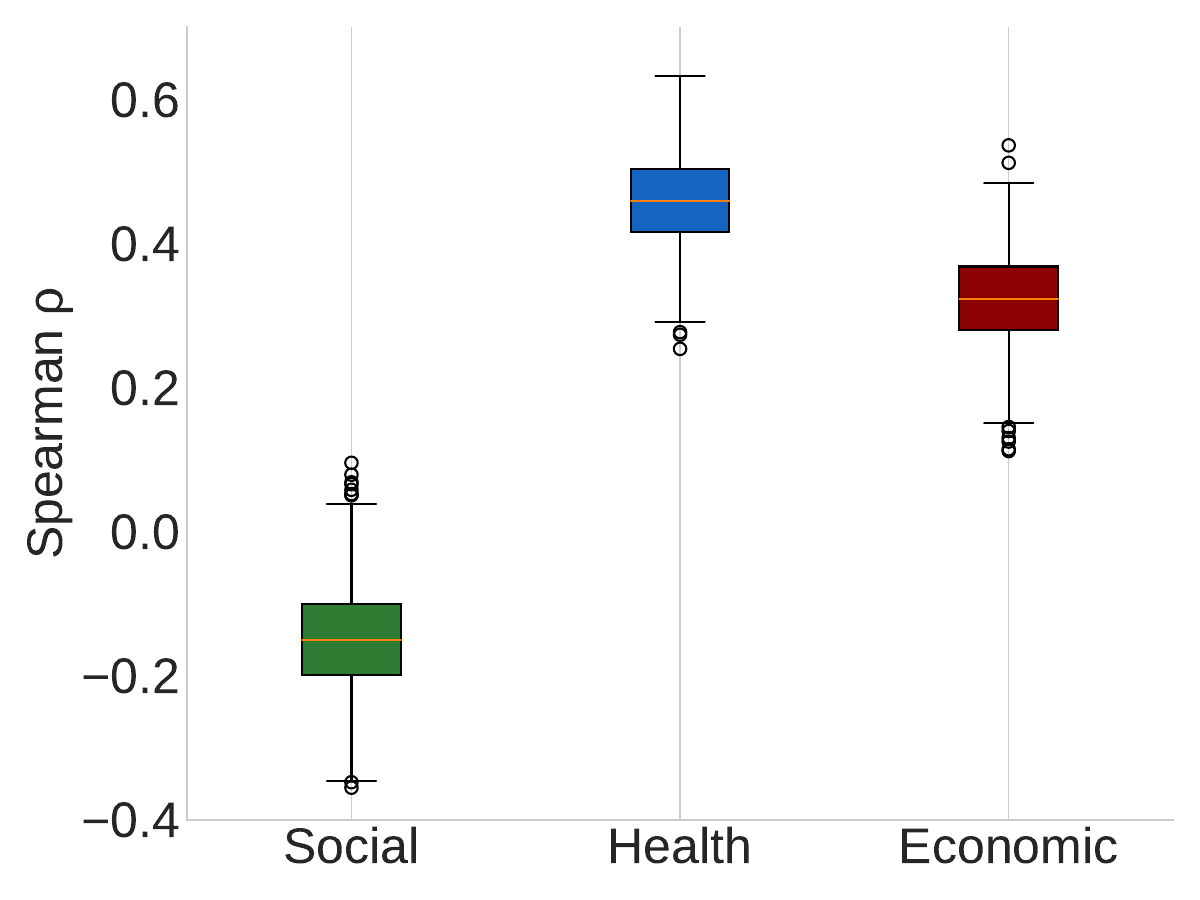}
    \caption{\small Bootstrapped Spearman correlation between PR-AUC prediction improvement and mean source activity $\Bar{\mathbf{Z}}[l]$ in each network, computed separately for the social, health, and economic layers.}
\end{subfigure}
\hfill
\begin{subfigure}[t]{0.24\textwidth}
    \includegraphics[width=\textwidth]{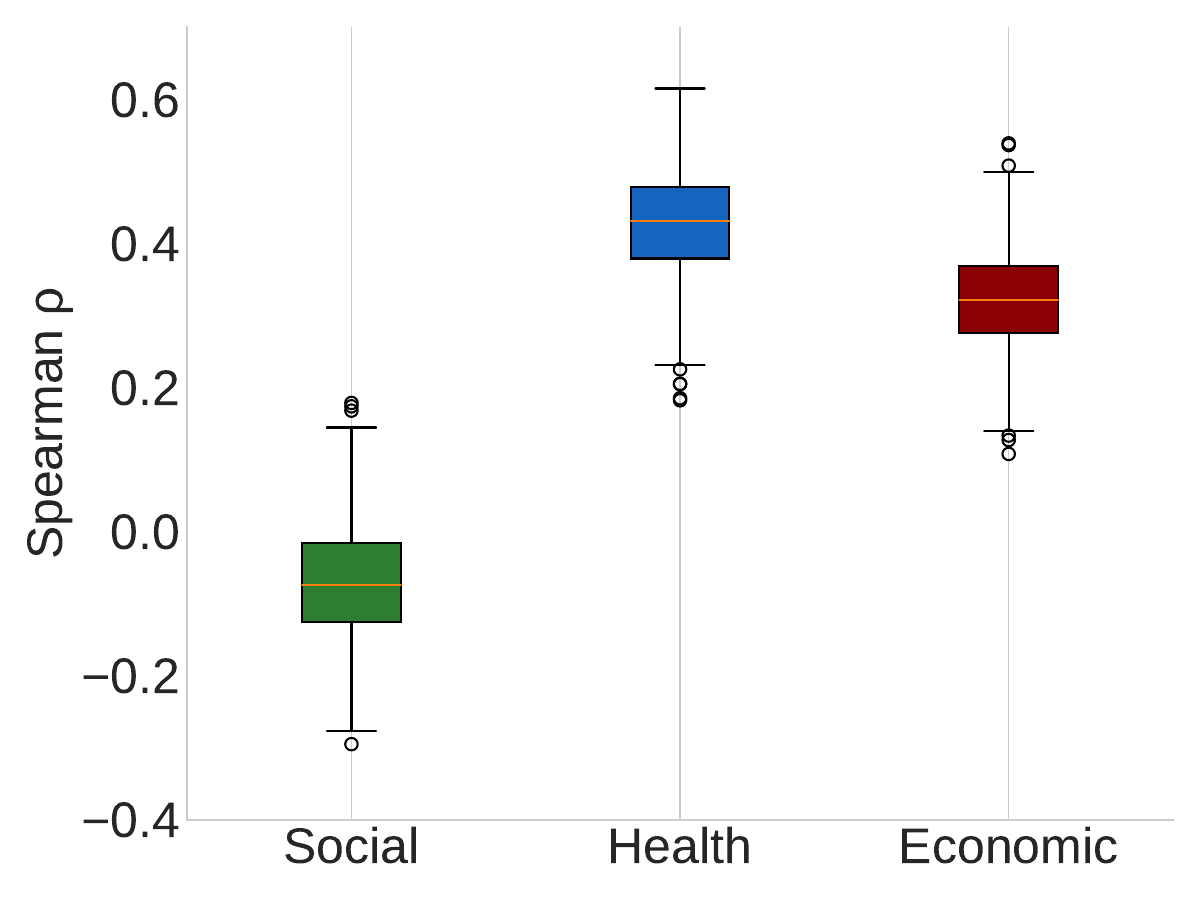}
    \caption{\small Bootstrapped Spearman correlation between PR-AUC prediction improvement and mean target activity $\Bar{\mathbf{W}}[l]$ in each network, computed separately for the social, health, and economic layers.}
\end{subfigure}
\hfill
\begin{subfigure}[t]{0.24\textwidth}
    \includegraphics[width=\textwidth]{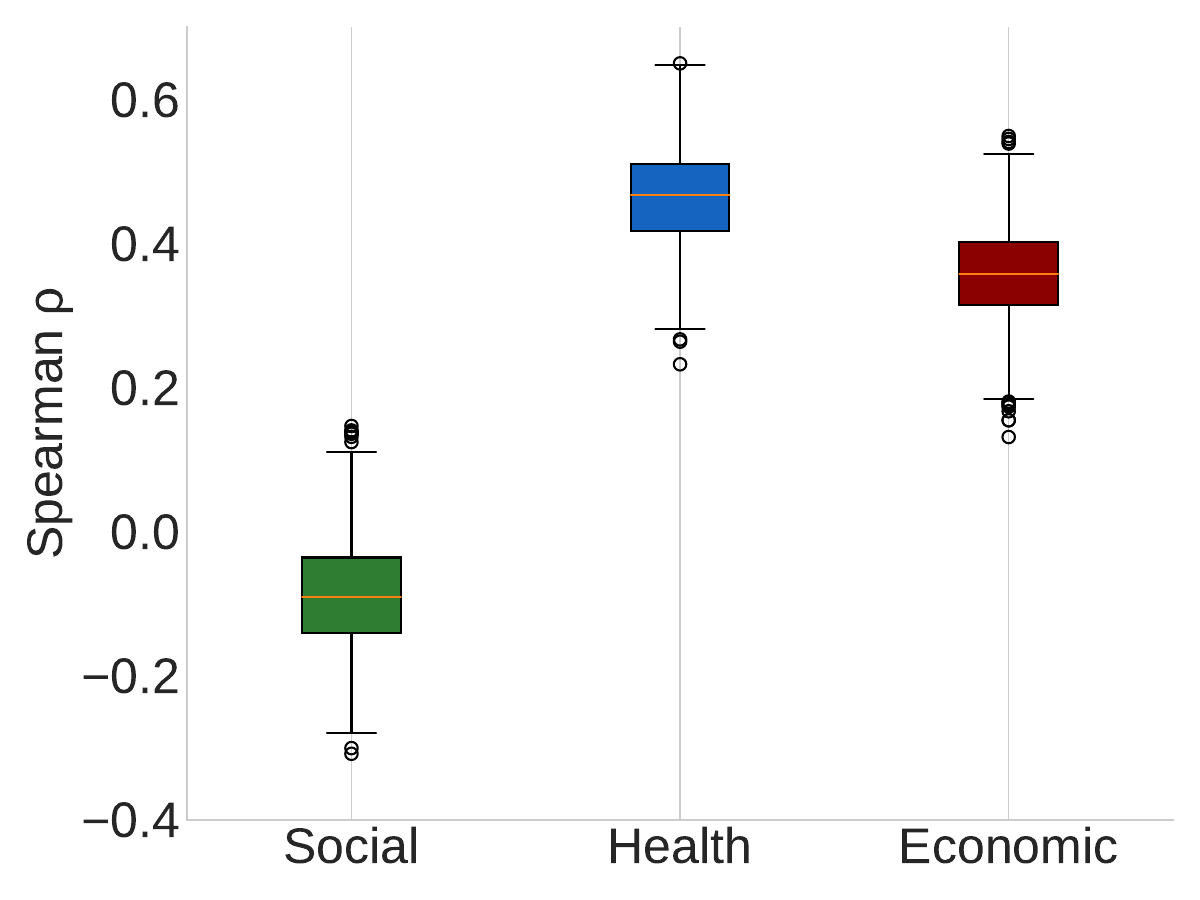}
    \caption{\small Bootstrapped Spearman correlation between ROC-AUC prediction improvement and mean source activity $\Bar{\mathbf{Z}}[l]$ in each network, computed separately for the social, health, and economic layers.}
\end{subfigure}
\hfill
\begin{subfigure}[t]{0.24\textwidth}
    \includegraphics[width=\textwidth]{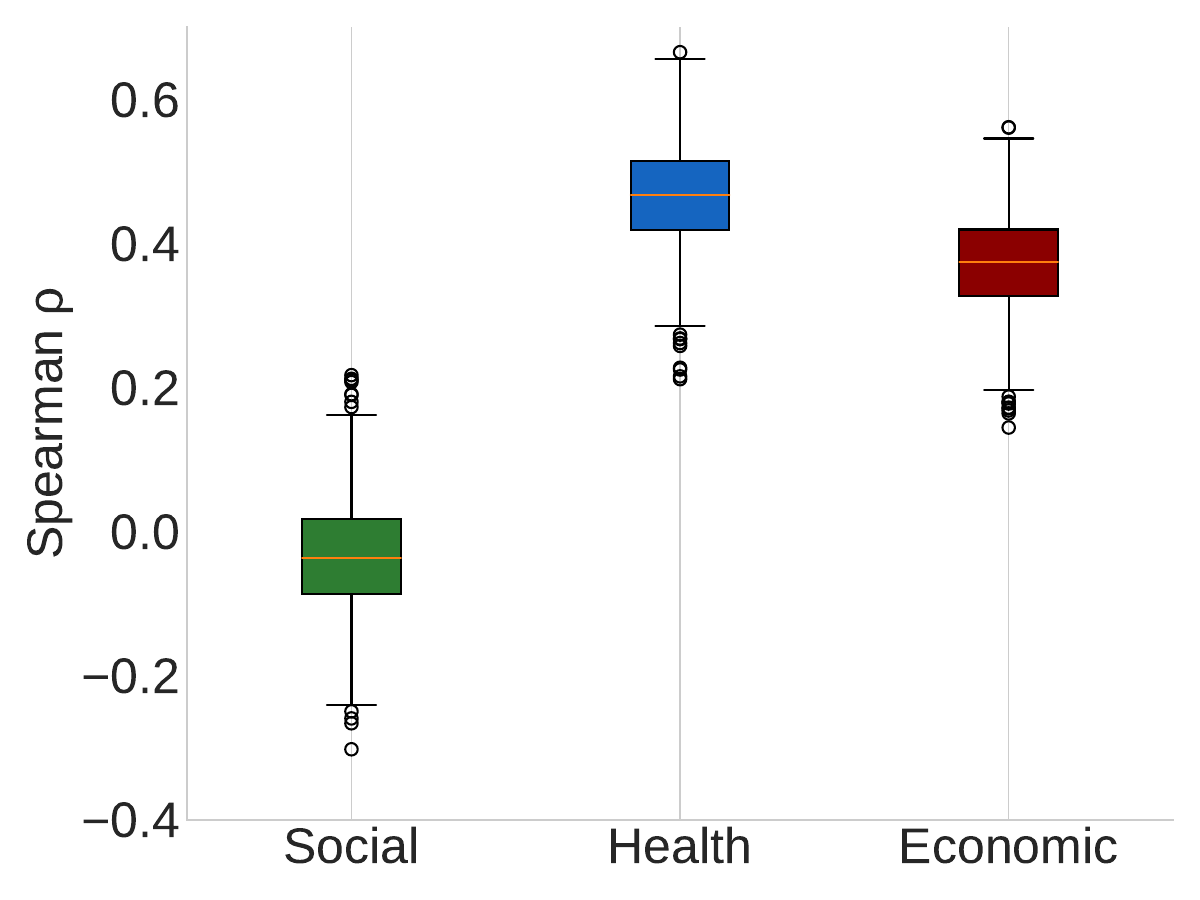}
    \caption{\small Bootstrapped Spearman correlation between ROC-AUC prediction improvement and mean target activity $\Bar{\mathbf{W}}[l]$ in each network, computed separately for the social, health, and economic layers.}
\end{subfigure}
\caption{\textbf{Relationship between average layer activity and performance improvement across all 176 networks.} Prediction improvement is defined as the average increase in PR-AUC (panels (a) and (b)) and ROC-AUC (panels (c) and (d)) across ten-fold cross-validation when extending the model to capture interdependence, compared to modeling only dependence and independence. We report Spearman correlation coefficients ($\rho$) between the observed performance change and average network activity for each layer, measured as average source activity $\Bar{\mathbf{Z}}[l] = \frac{1}{N} \sum_{i=1}^N z_i[l]$ and average target activity $\Bar{\mathbf{W}}[l] = \frac{1}{N} \sum_{j=1}^N w_j[l]$, using 1,000 bootstrap samples. Panel (i) shows the correlation between average source activity $\Bar{\mathbf{Z}}[l]$ and PR-AUC improvement; panel (j) shows the same for average target activity $\Bar{\mathbf{W}}[l]$. Panels (k) and (l) report the corresponding correlations for ROC-AUC improvement, using average source and target activity, respectively, across the social, health, and economic layers.}
\label{fig:mean_cor}
\end{figure*}

In health and economic ties, interdependent exchange depends on behavioral roles and engagement, whereas in social ties, it is structurally embedded. Combining the model's trade-off formulation with the performance increase from modeling interdependence, we compute the Spearman $\rho$ correlation (using 1,000 bootstrap samples) between network-averaged node-level engagement in each layer (social, health, and economic) and the corresponding predictive raw score improvement across all 176 village networks. Specifically, we define engagement as $\Bar{\mathbf{Z}}[l] = \frac{1}{N} \sum_{i=1}^N z_i[l]$ and $\Bar{\mathbf{W}}[l] = \frac{1}{N} \sum_{j=1}^N w_j[l]$, capturing average source and target activations respectively, and distinguishing directional roles in exchange. This allows us to investigate how layer-specific engagement relates to the expression of interdependence. Results are presented in SI Figure~\ref{fig:mean_cor}, where panels (a) and (b) show correlations for mean source activity under PR- and ROC-AUC metrics, respectively. We find a significant positive correlation between average node-level activity and the predictive improvement from modeling interdependence, focusing here on PR-AUC (with similar patterns obtained for ROC-AUC in panels (c) and (d)). We observe a strong and significant correlation in the \textit{health} layer ($\tilde{\rho}_{\text{health}} = 0.458$, 95\% CI: [0.321, 0.574]), and a moderate yet significant effect in the \textit{economic} layer ($\tilde{\rho}_{\text{economic}} = 0.323$, 95\% CI: [0.185, 0.447]). In contrast, no significant correlation is found in the \textit{social} layer ($\tilde{\rho}_{\text{social}} = -0.150$, 95\% CI: [–0.288, 0.009]). 

A similar pattern holds for target-side activity. In the \textit{health} layer, the correlation remains strong ($\tilde{\rho}_{\text{health}} = 0.431$, 95\% CI: [0.286, 0.564]); the \textit{economic} layer shows a moderate effect ($\tilde{\rho}_{\text{economic}} = 0.321$, 95\% CI: [0.169, 0.455]); and again, no significant correlation appears in the \textit{social} layer ($\tilde{\rho}_{\text{social}} = -0.074$, 95\% CI: [–0.222, 0.086]). These findings suggest that interdependence in the health and economic layers yields predictive benefits primarily when the network is actively engaged in those types of ties. In contrast, the social layer, despite exhibiting the highest and most variable engagement, shows no such relationship between activity and performance, indicating that its value may not be contingent on behavioral prioritization. 

\begin{figure*}[h!]
\centering
\begin{subfigure}[t]{0.16\textwidth}
    \includegraphics[width=\textwidth]{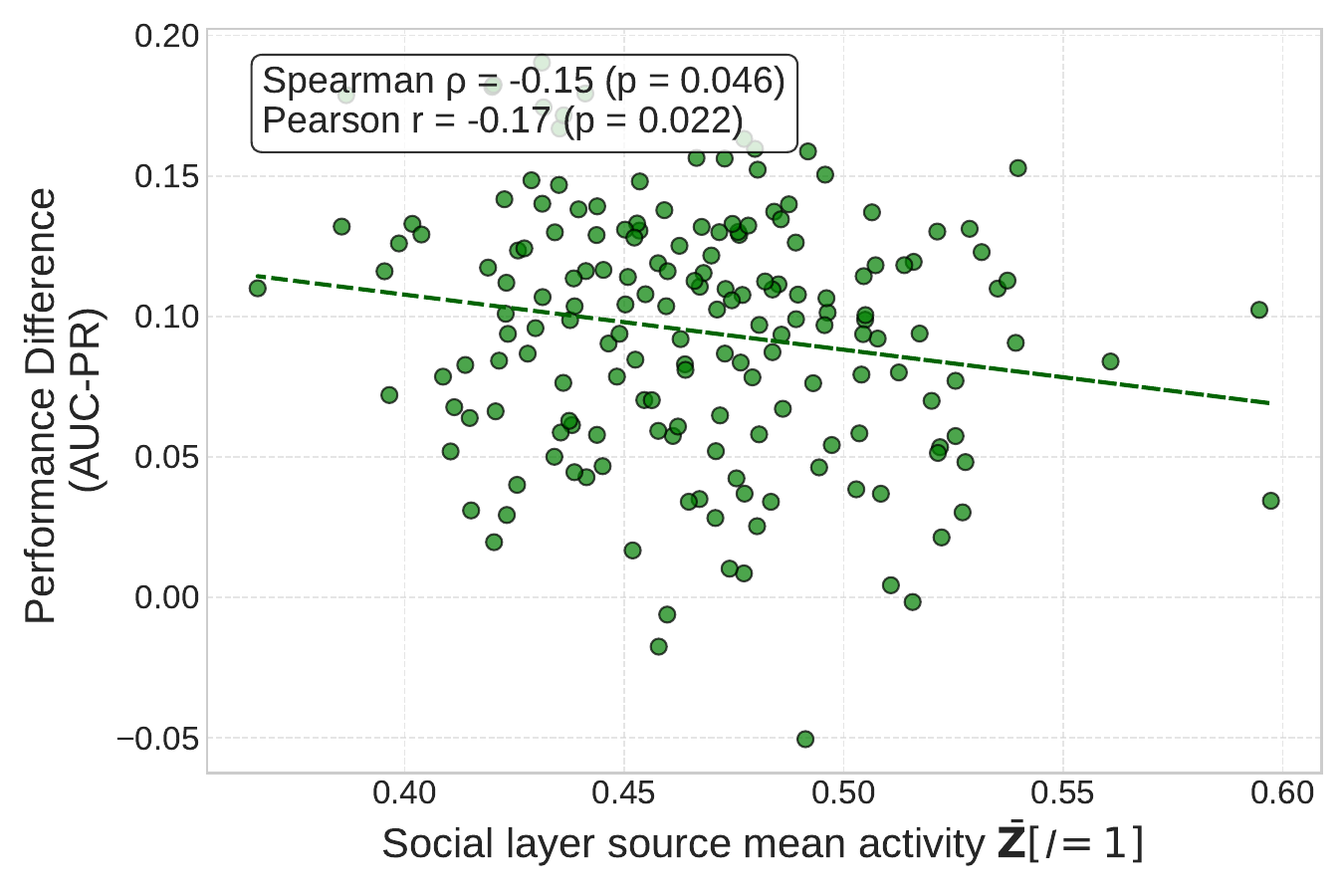}
    \caption{\small Social layer source mean activity and performance}
\end{subfigure}
\hfill
\begin{subfigure}[t]{0.16\textwidth}
    \includegraphics[width=\textwidth]{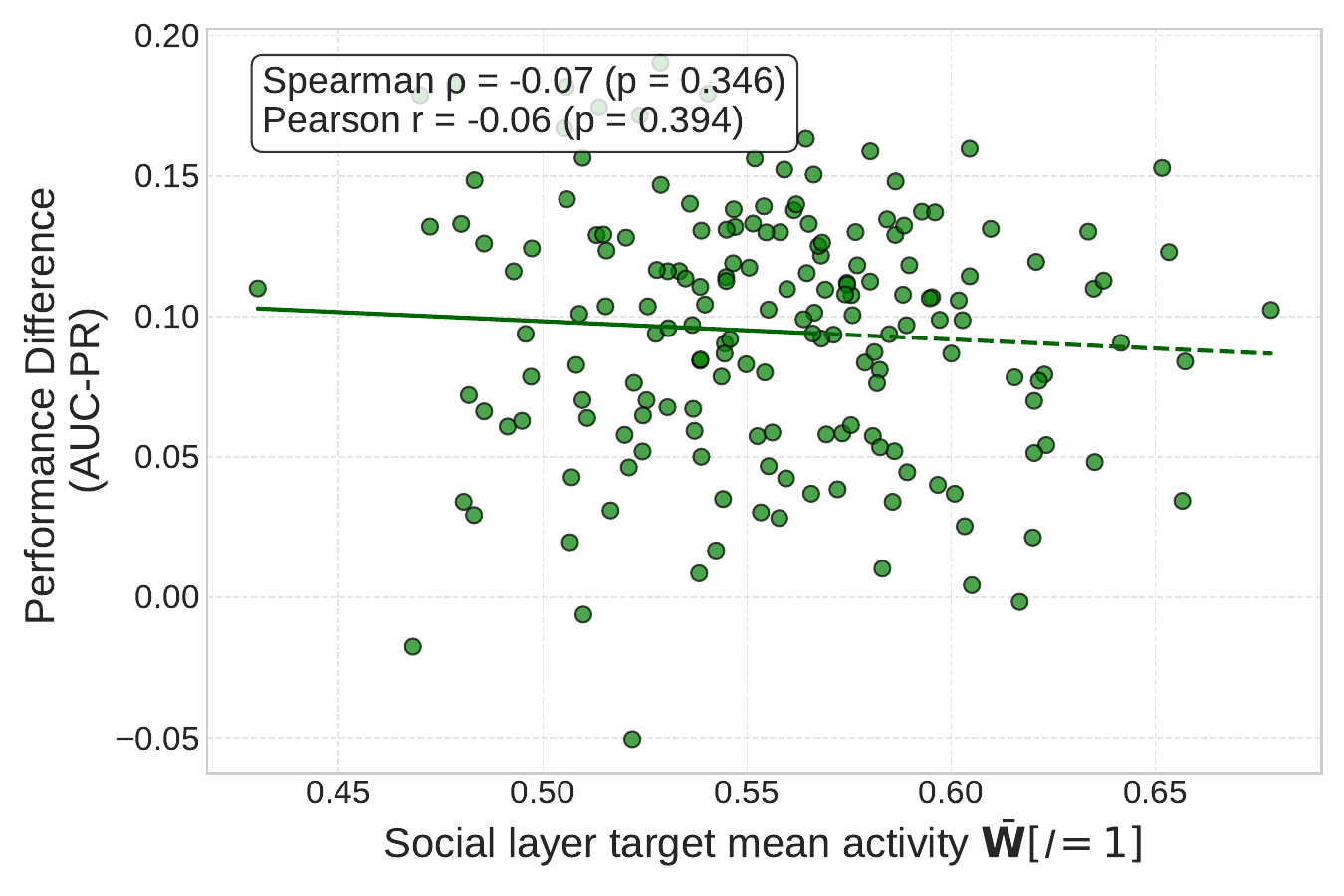}
    \caption{\small Social layer target mean activity and performance}
\end{subfigure}
\hfill
\begin{subfigure}[t]{0.16\textwidth}
    \includegraphics[width=\textwidth]{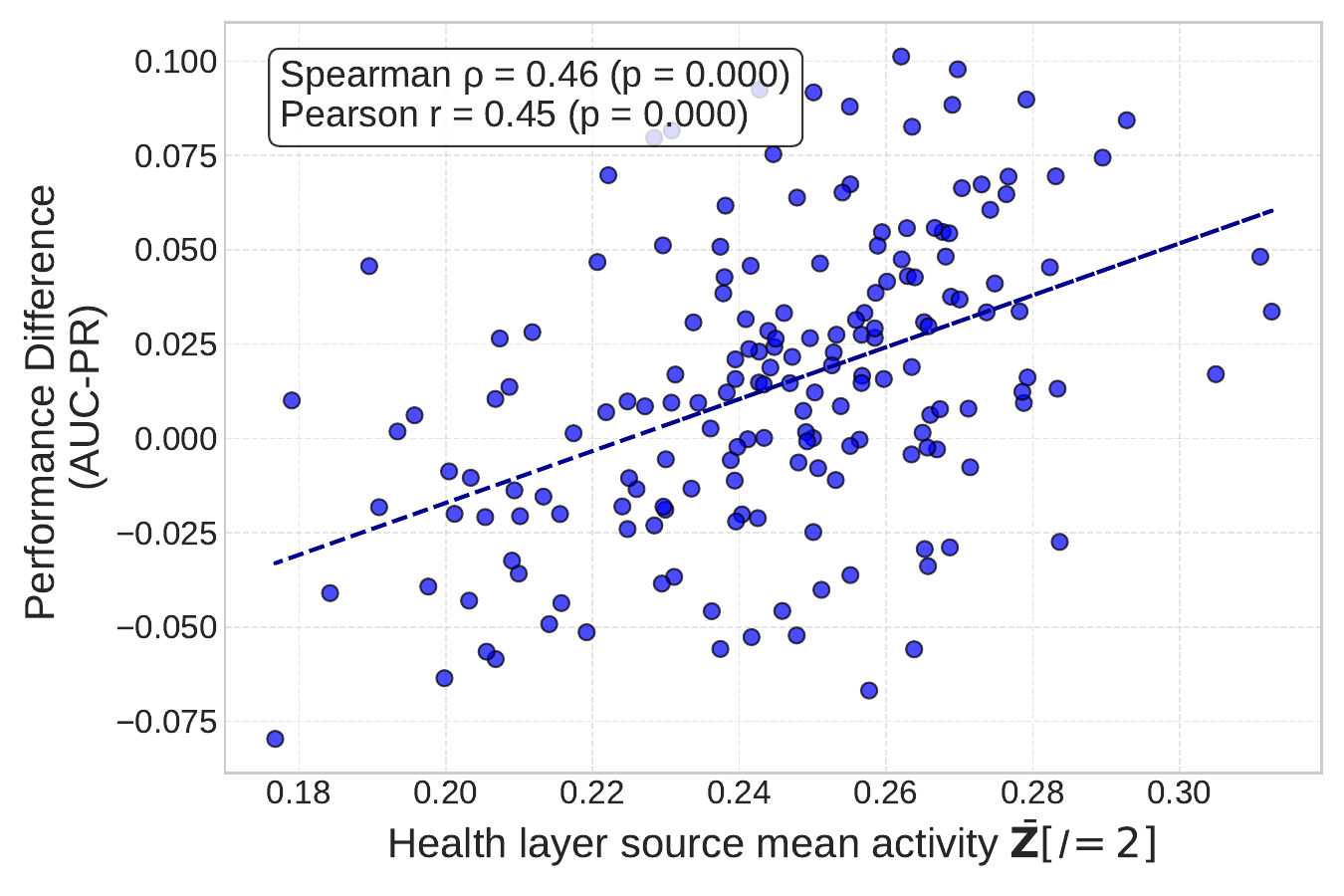}
    \caption{\small Health layer source mean activity and performance}
\end{subfigure}
\hfill
\begin{subfigure}[t]{0.16\textwidth}
    \includegraphics[width=\textwidth]{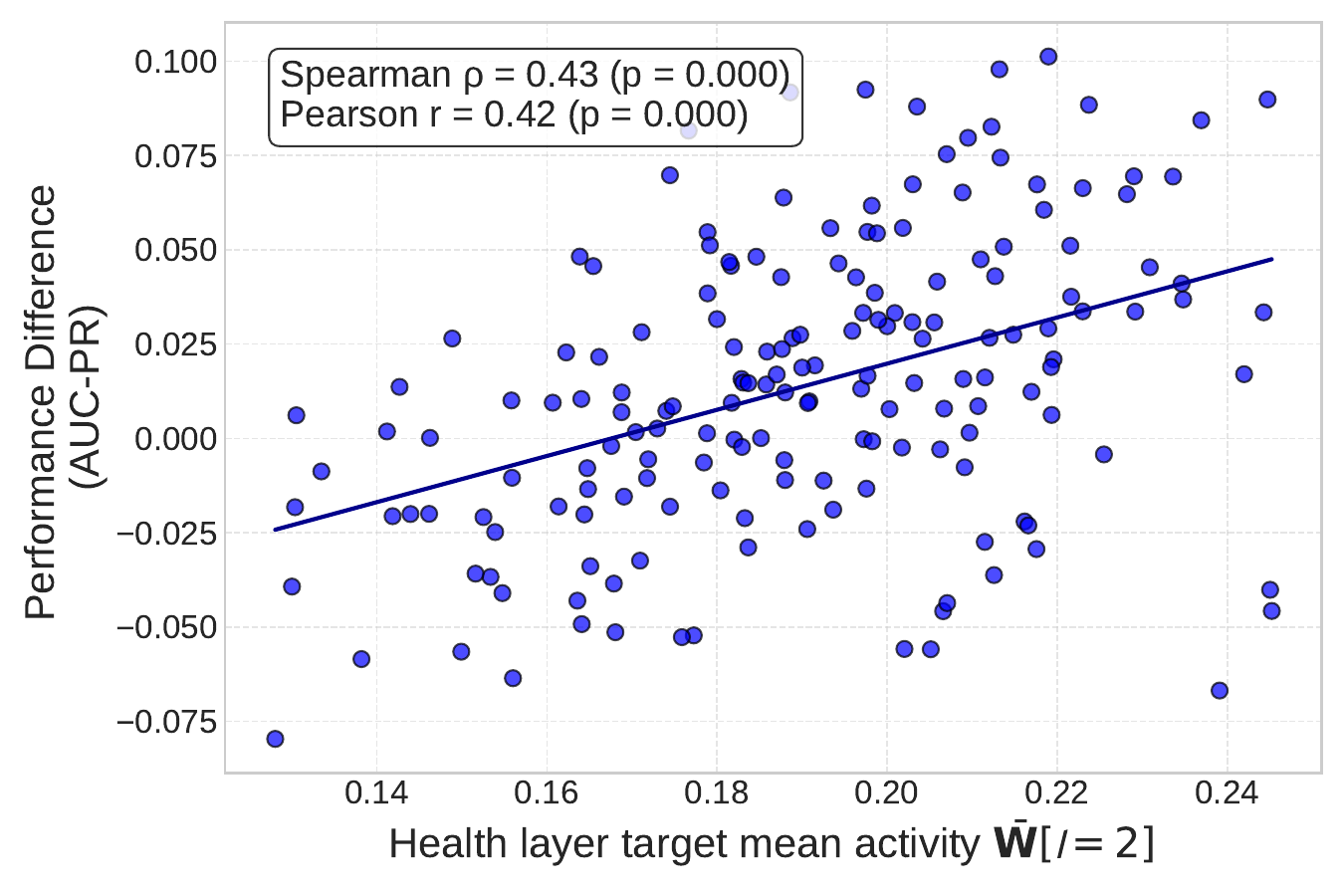}
    \caption{\small Health layer target mean activity and performance}
\end{subfigure}
\hfill
\begin{subfigure}[t]{0.16\textwidth}
    \includegraphics[width=\textwidth]{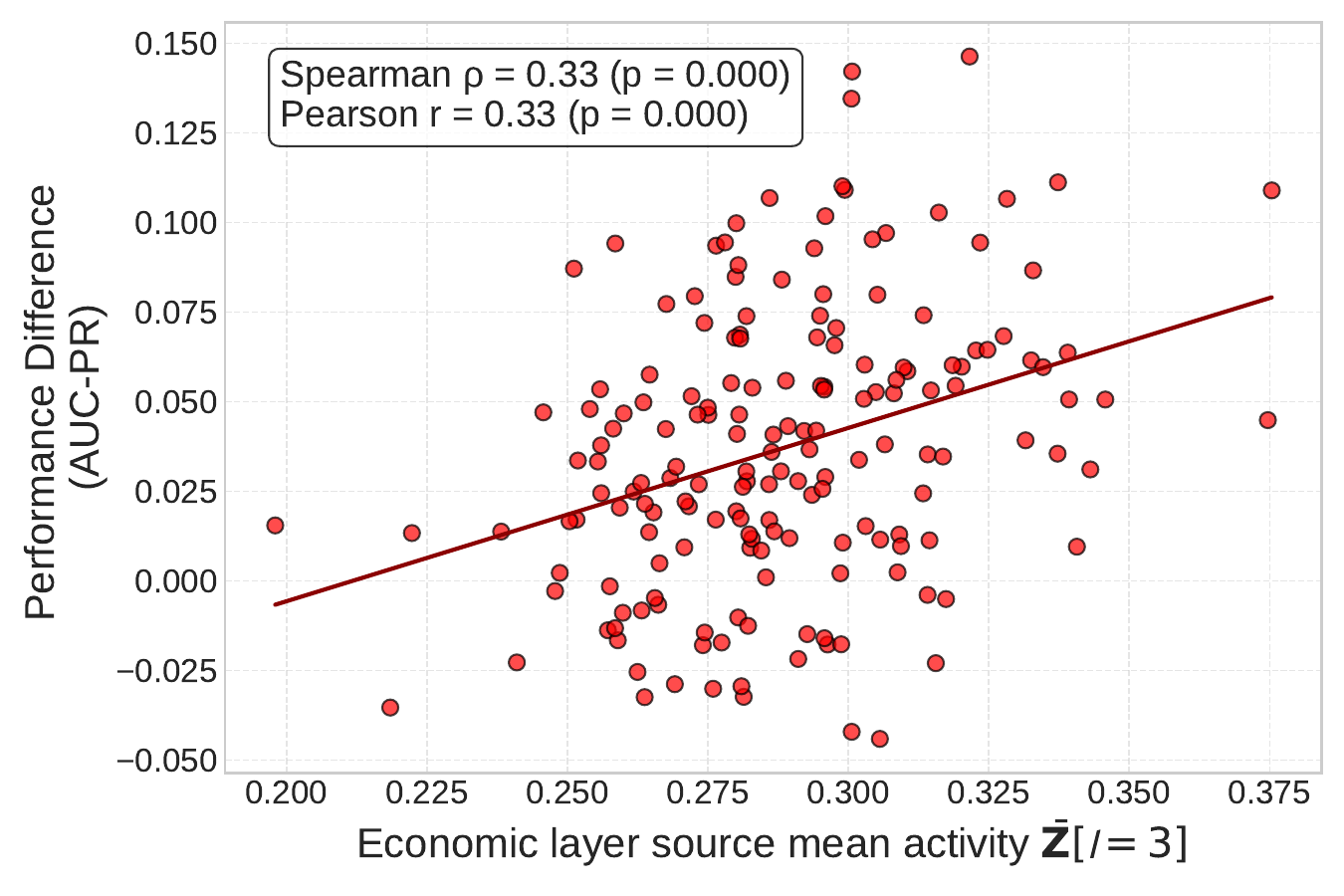}
    \caption{\small Economic layer source mean activity and performance}
\end{subfigure}
\hfill
\begin{subfigure}[t]{0.16\textwidth}
    \includegraphics[width=\textwidth]{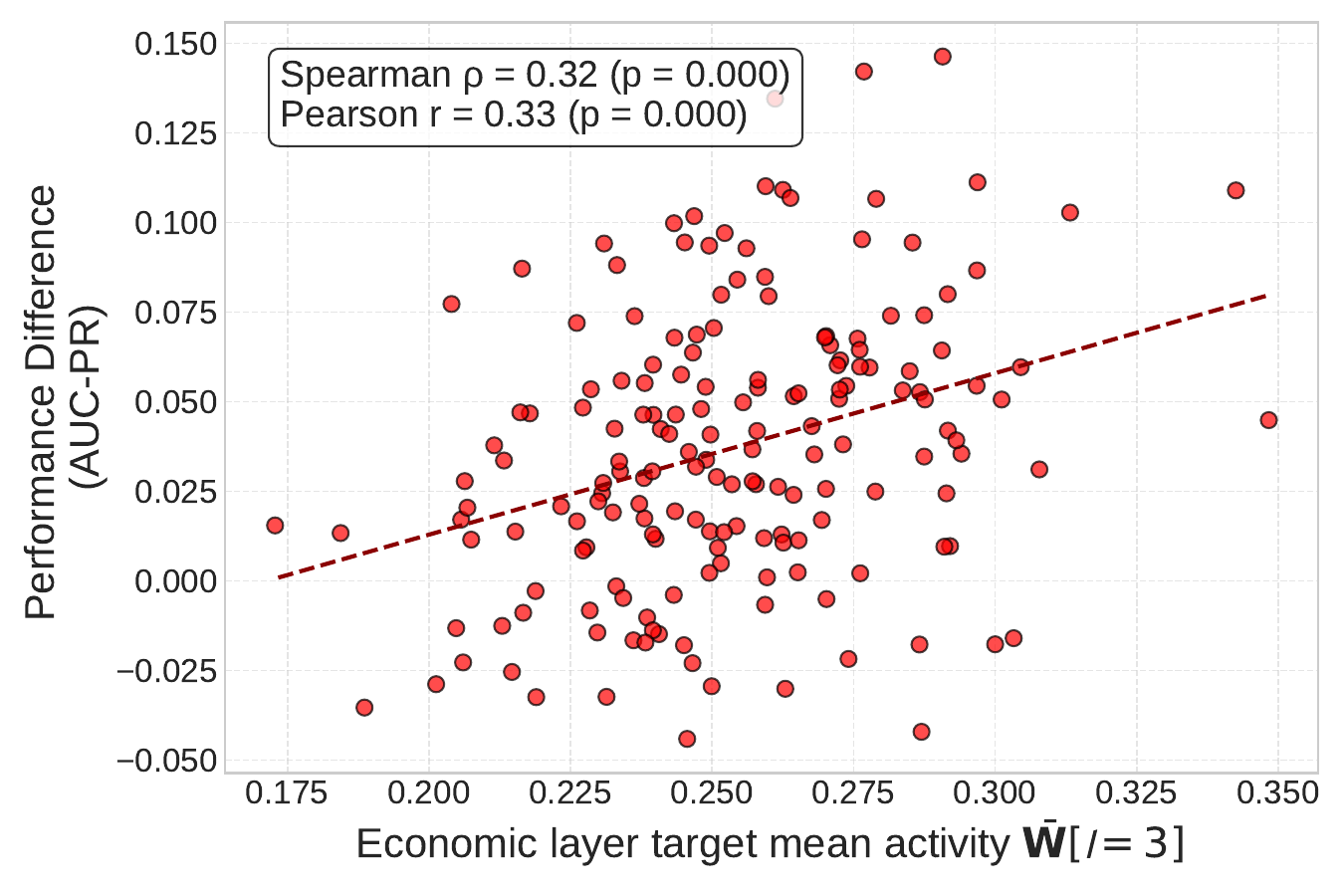}
    \caption{\small Economic layer target mean activity and performance}
\end{subfigure}
\centering
\begin{subfigure}[t]{0.16\textwidth}
    \includegraphics[width=\textwidth]{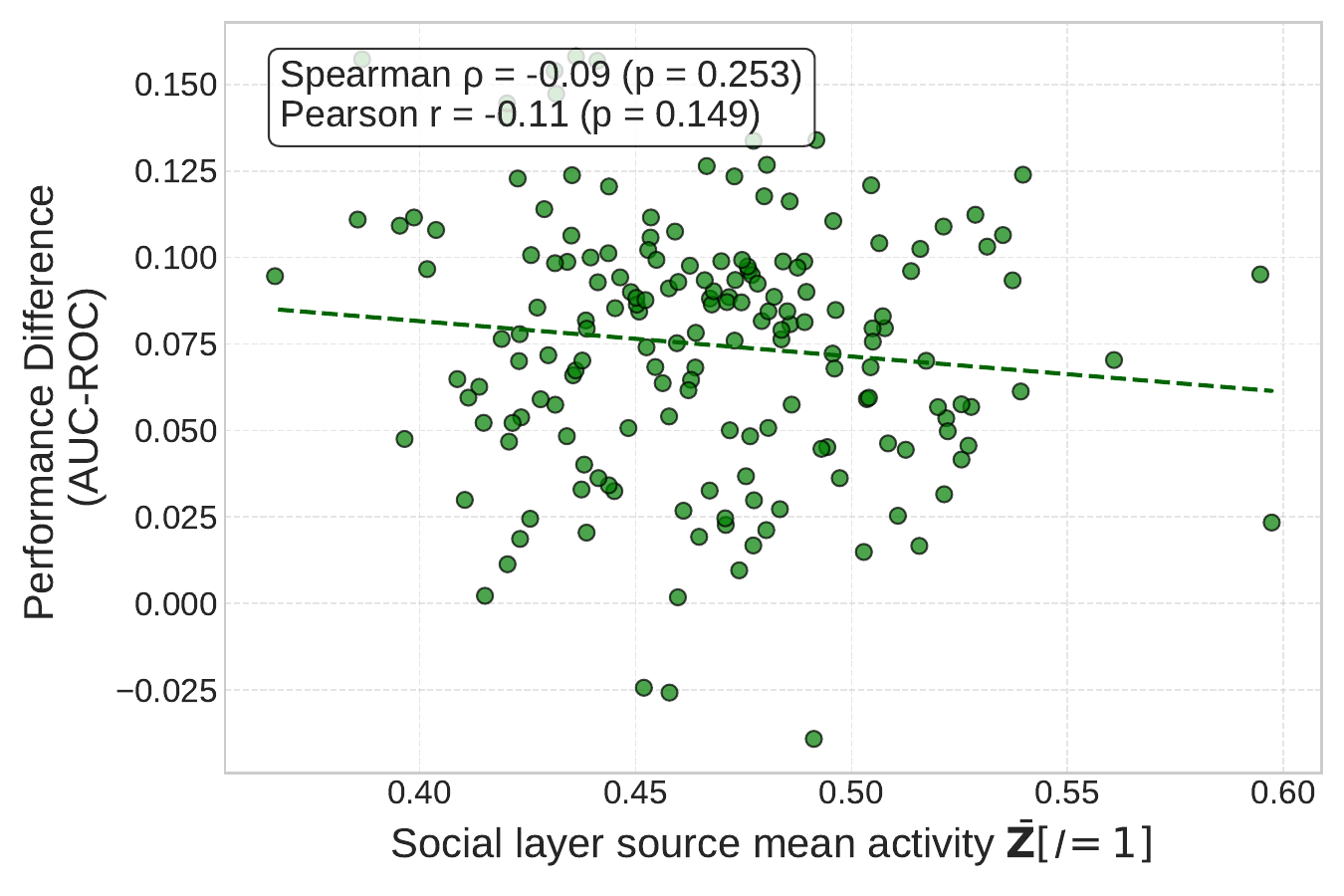}
    \caption{\small Social layer source mean activity and performance}
\end{subfigure}
\hfill
\begin{subfigure}[t]{0.16\textwidth}
    \includegraphics[width=\textwidth]{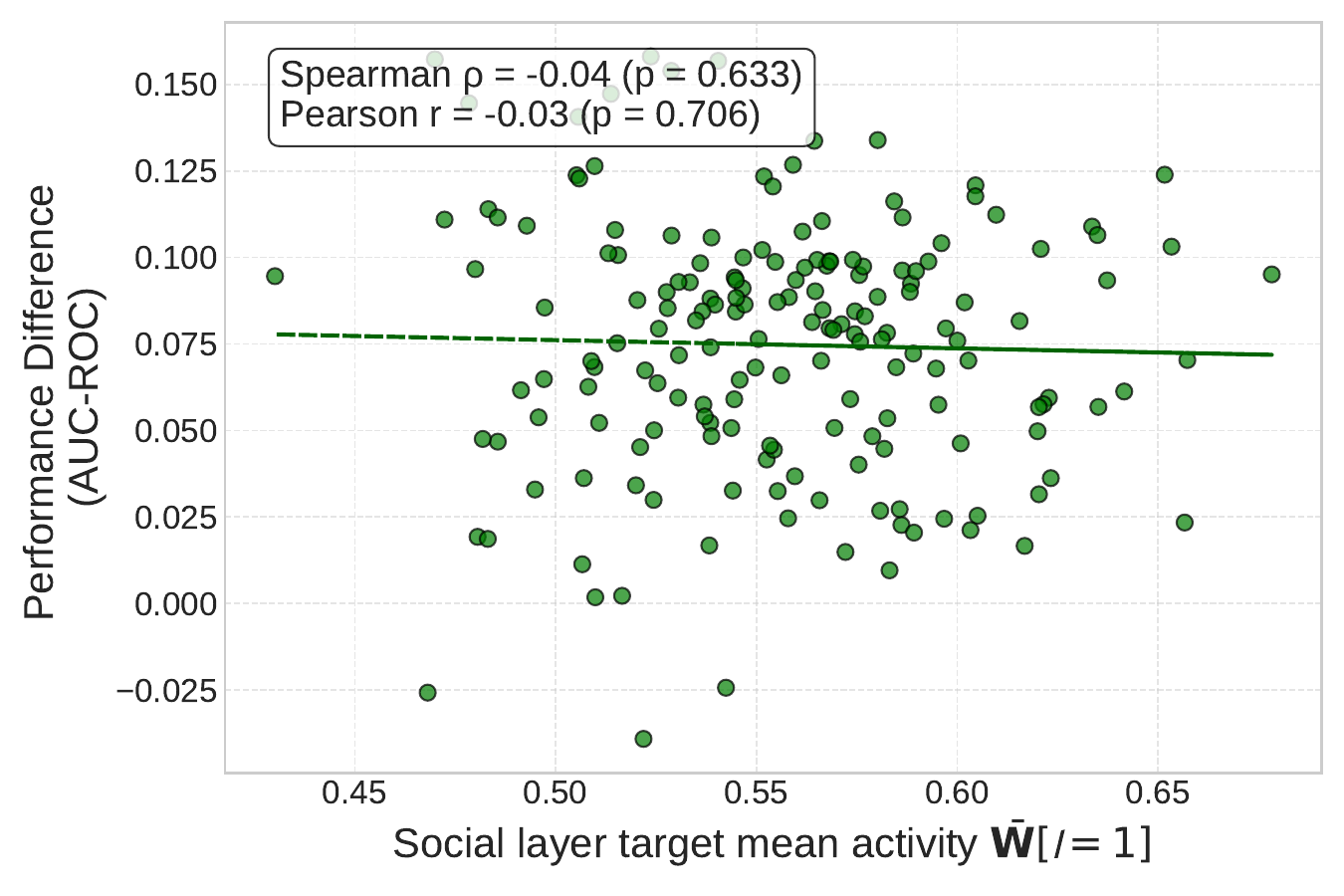}
    \caption{\small Social layer target mean activity and performance}
\end{subfigure}
\hfill
\begin{subfigure}[t]{0.16\textwidth}
    \includegraphics[width=\textwidth]{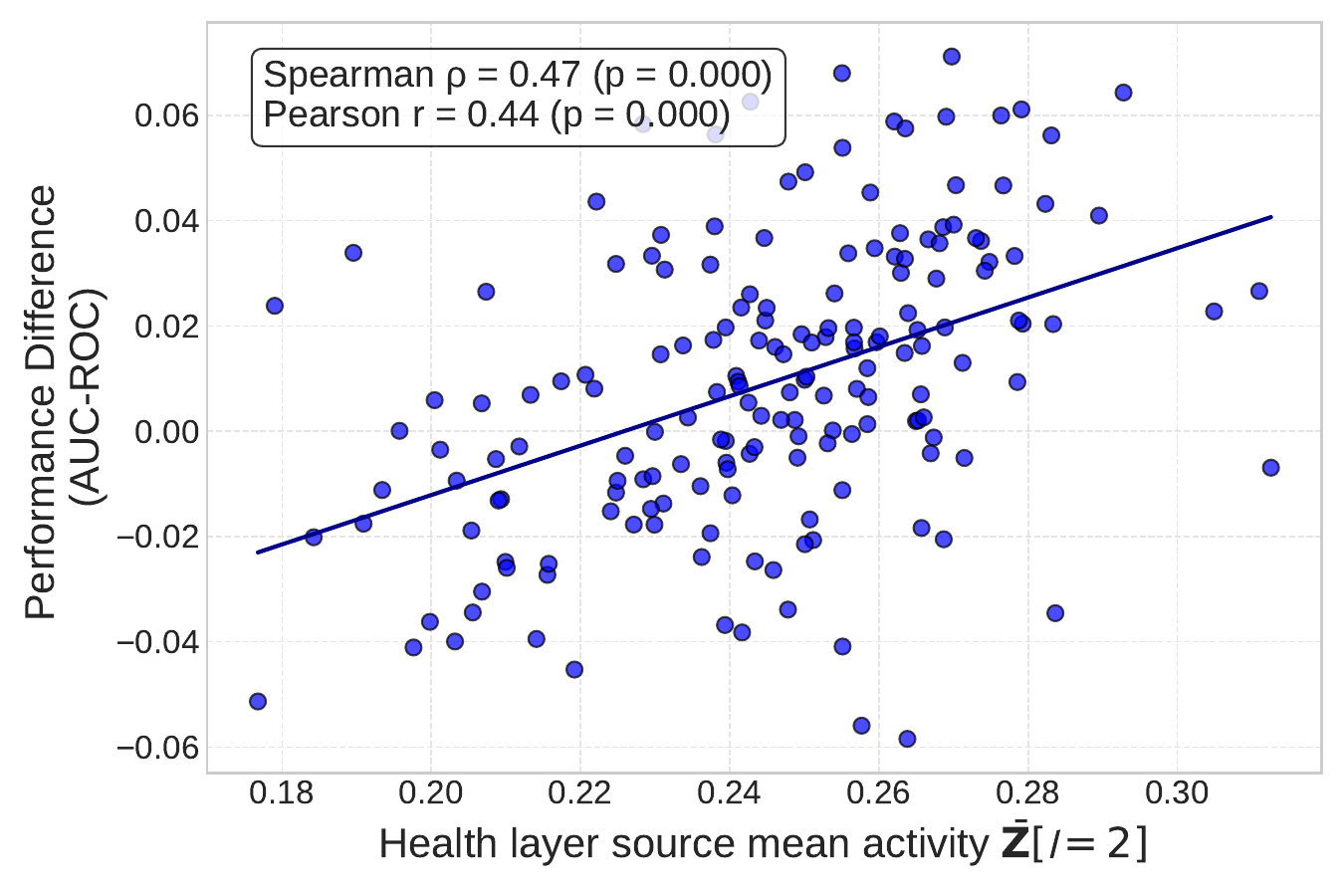}
    \caption{\small Health layer source mean activity and performance}
\end{subfigure}
\hfill
\begin{subfigure}[t]{0.16\textwidth}
    \includegraphics[width=\textwidth]{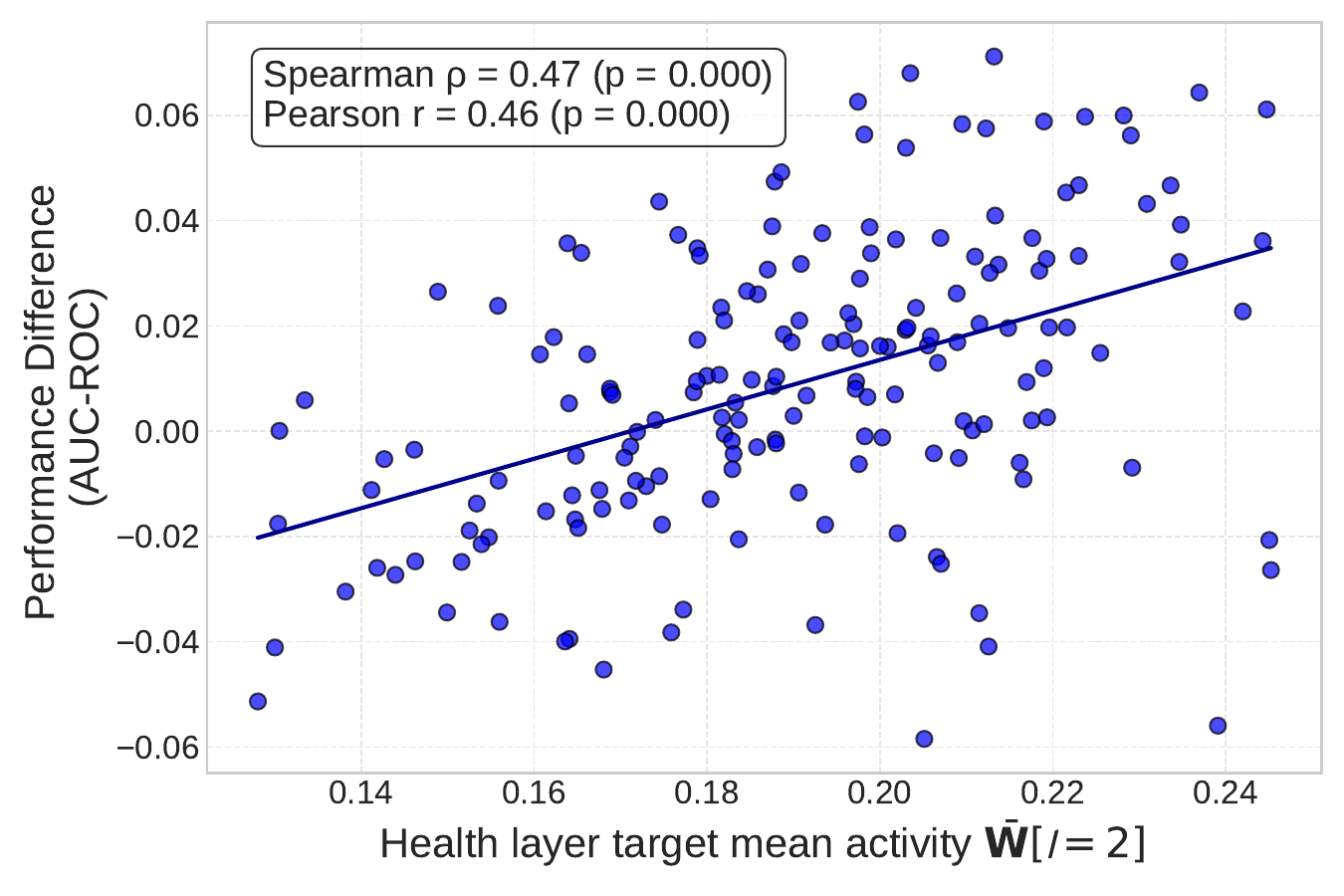}
    \caption{\small Health layer target mean activity and performance}
\end{subfigure}
\hfill
\begin{subfigure}[t]{0.16\textwidth}
    \includegraphics[width=\textwidth]{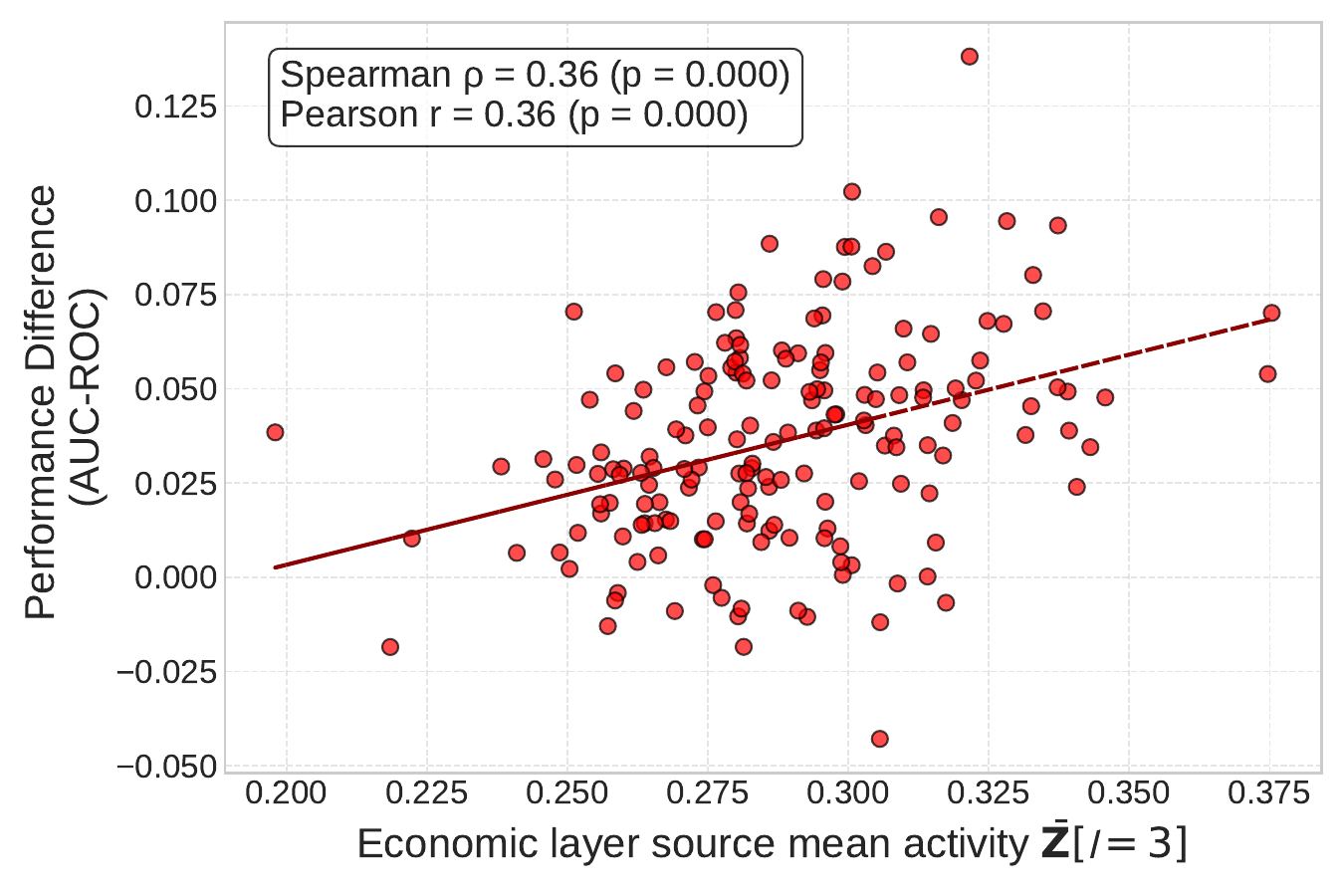}
    \caption{\small Economic layer source mean activity and performance}
\end{subfigure}
\hfill
\begin{subfigure}[t]{0.16\textwidth}
    \includegraphics[width=\textwidth]{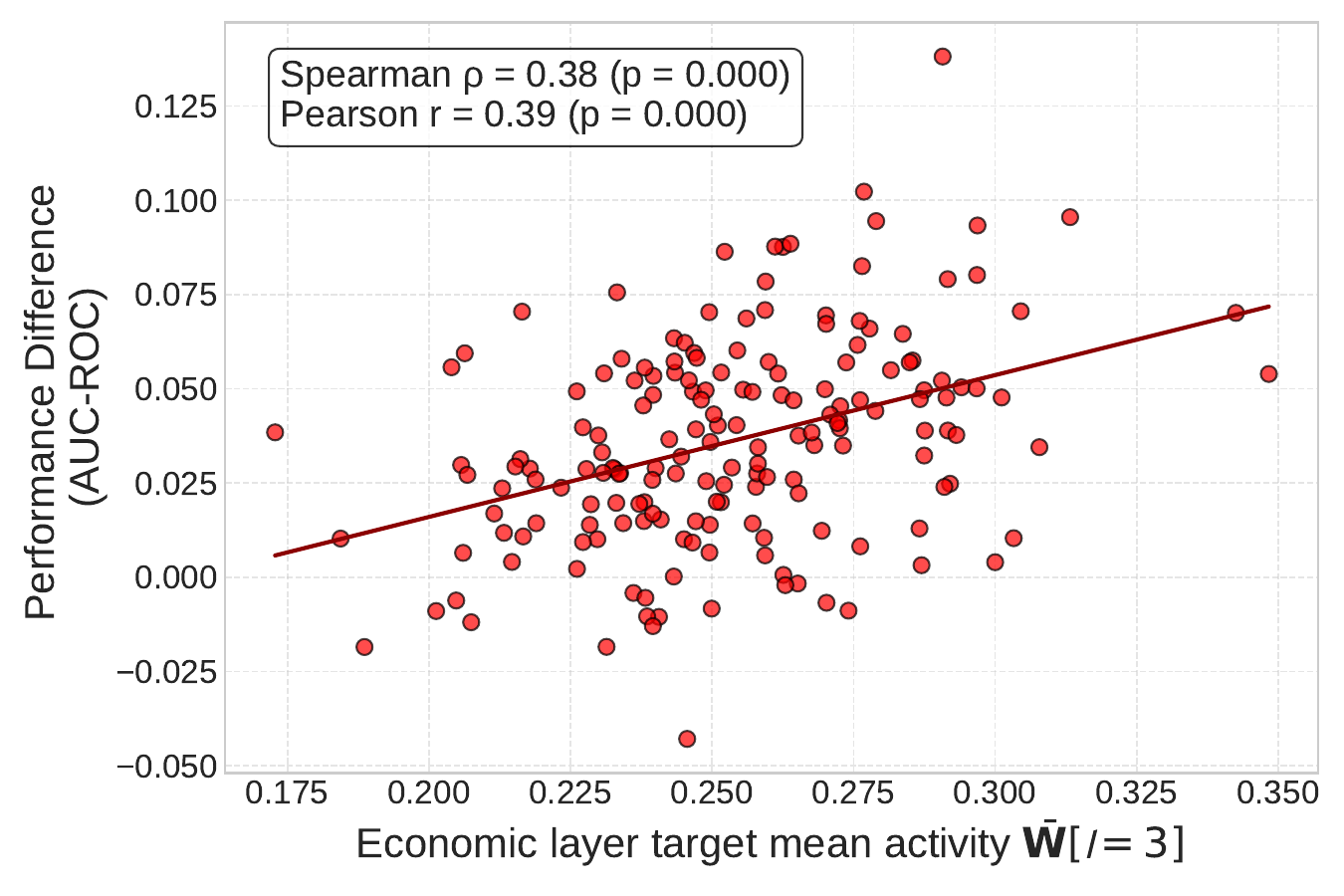}
    \caption{\small Economic layer target mean activity and performance}
\end{subfigure}
\caption{\textbf{Relationship between average layer activity and performance change across all $176$ networks.} We report both linear (Pearson) and monotonic (Spearman) correlation coefficients between the observed performance change and the average network activity for each layer, measured as average source activity $\Bar{\mathbf{Z}}[l] = \frac{1}{N} \sum_{i=1}^N z_i[l]$ and average target activity $\Bar{\mathbf{W}}[l] = \frac{1}{N} \sum_{j=1}^N w_j[l]$. Panels (a) and (b) show results for the social layer, panels (c) and (d) for the health layer, and panels (e) and (f) for the economic layer.}
\label{fig:mean_cor_sup}
\end{figure*}

To test whether the correlation in the health layer is significantly stronger than in the economic layer, we conduct a one-sided Mann--Whitney U test and find a statistically significant difference ($\tilde{\rho}_{\text{health}} > \tilde{\rho}_{\text{economic}}$, $p < 0.001$) for both sender and receiver roles. Results suggest that \textit{social interdependence is structurally embedded}, manifesting regardless of behavioral engagement, whereas \textit{instrumental interdependence} (health and economic) depends more directly on the role-specific activity levels. These insights align with core tenets of Social Exchange Theory where instrumental exchange is conditional on active role performance, while social exchange is sustained through generalized obligation and normative expectation. Observed (non-bootstrapped) correlation values are reported in SI Figure~\ref{fig:mean_cor_sup}.

Finally, non-bootstrapped (observed) correlation values are presented in SI Figure~\ref{fig:mean_cor_sup}, showing Pearson and Spearman coefficients between performance improvements (PR AUC and ROC AUC) and average node activity across the social, health, and economic layers.

\end{document}